\documentclass[a4paper,11pt]{article}
\pdfoutput=1

\usepackage{amsmath}
\usepackage{jheppub}

\usepackage[utf8]{inputenc}
\usepackage[T1]{fontenc}
\usepackage[kerning,spacing]{microtype}

\usepackage{graphicx}

\usepackage{amsbsy}
\usepackage{amssymb}
\usepackage{mathtools}
\usepackage{xspace}

\usepackage{dcolumn}
\usepackage{bm}
\usepackage{slashed}
\usepackage{color}
\usepackage{subfig}
\usepackage[font=small]{caption}

\usepackage{booktabs}

\newcommand{\Higgs}{\ensuremath{\mathrm{H}}\xspace}
\newcommand{\Wboson}{\ensuremath{\mathrm{W}}\xspace}
\newcommand{\Zboson}{\ensuremath{\mathrm{Z}}\xspace}
\newcommand{\Wjjj}{\Wboson\!+\! 3\! jets\xspace}
\newcommand{\Hj}{\Higgs\!+\! 1\! jet\xspace}
\newcommand{\Hjp}{\Higgs\!+\! 1\! jet\phantom{s}\xspace}
\newcommand{\Hjj}{\Higgs\!+\! 2\! jets\xspace}
\newcommand{\Hjjj}{\Higgs\!+\! 3\! jets\xspace}
\newcommand{\Hjjjj}{\Higgs\!+\! 4\! jets\xspace}
\newcommand{\Hnj}{\Higgs\!+\! $n$\! jets\xspace}
\newcommand{\pTHjj}{p_{{T},\,\Higgs j_1j_2}}
\newcommand{\dphiHjj}{\Delta\phi_{\,\Higgs,\,j_1j_2}}
\newcommand{\ystar}[2]{y_{#1,#2}^\ast}
\newcommand{\hthatprime}{\hat{H}^\prime_T}
\newcommand{\alphaS}{\alpha_\mathrm{s}}
\newcommand{\tot}{\mathrm{tot}}
\newcommand{\GeV}{\ensuremath{\mathrm{GeV}}\xspace}
\newcommand{\TeV}{\ensuremath{\mathrm{TeV}}\xspace}

\newcommand\sss{\mathchoice%
{\displaystyle}%
{\scriptstyle}%
{\scriptscriptstyle}%
{\scriptscriptstyle}%
}

\def\d{\hbox{d}}
\def\CA{C_{\sss\rm A}}
\def\NF{N_{\sss\rm F}}

\title{Phenomenological analysis of Higgs boson
  production through gluon fusion\\in association with jets}

\author[a]{Nicolas Greiner,}
\author[b]{Stefan H\"oche,}
\author[c]{Gionata Luisoni,}
\author[d]{Marek Sch\"onherr,}
\author[c]{Jan-Christopher Winter,}
\author[c]{Valery Yundin\,}
\affiliation[a]{DESY Theory Group, Notkestr.~85, D-22607 Hamburg, Germany}
\affiliation[b]{SLAC National Accelerator Laboratory, Menlo Park, CA 94025, USA}
\affiliation[c]{Max-Planck-Institut f\"ur Physik, F\"ohringer Ring 6, D-80805 M\"unchen, Germany}
\affiliation[d]{Physik-Institut, Universit\"at Z\"urich, Wintherturerstrasse 190, CH-8057 Z\"urich, Switzerland}

\preprint{{\small%
  DESY 15-081\\
  \hphantom{.}\hfill MPP-2015-108\\
  \hphantom{.}\hfill SLAC-PUB-16298\\
  \hphantom{.}\hfill ZH-TH 15/15\\
  \hphantom{.}\hfill MCNET-15-11}
}


\abstract{We present a detailed phenomenological analysis of the production of a Standard Model Higgs boson
in association with up to three jets. We consider the gluon fusion channel using an effective theory in the large top-quark mass limit.
Higgs boson production in gluon fusion constitutes an irreducible background to the vector boson fusion (VBF) process;
hence the precise knowledge of its characteristics is a prerequisite for any measurement in the VBF channel.
The calculation is carried out at next-to-leading order (NLO) in QCD in a fully automated way by combining the two programs \textsc{GoSam} and \textsc{Sherpa}.
We present numerical results for a large variety of observables for
both standard cuts and VBF selection cuts.
We find that for all jet multiplicities the NLO corrections are sizeable. This is particularly true in the presence 
of kinematic selections enhancing the VBF topology, which are based on vetoing additional jet activity.
In this case, precise predictions for the background can be made using our calculation by taking the difference 
between the inclusive \Hjj and the inclusive \Hjjj result.
}

\begin{document}

\maketitle

\section{Introduction}

The discovery of a Higgs boson~\cite{Aad:2012tfa,Chatrchyan:2012ufa} during Run~I of the Large Hadron Collider (LHC) 
has ushered in an era of precision measurements to determine the nature of the new particle. A major step was the 
analysis of its spin structure~\cite{Aad:2013xqa,Chatrchyan:2013iaa,Chatrchyan:2013mxa}, resulting in a very good 
agreement of the measurement with the Standard Model prediction. A second major step was the measurement of different 
production times decay rates by ATLAS~\cite{Aad:2013wqa,Aad:2014eha,Aad:2014lwa,Aad:2014aba,Aad:2014fia,Aad:2015vsa,ATLAS:2014aga}
and CMS~\cite{Chatrchyan:2014nva,Chatrchyan:2014vua,Khachatryan:2014ira,Khachatryan:2014jba}.
All experimental results point towards the mechanism of electroweak symmetry breaking~\cite{Englert:1964et,Higgs:1964pj,Guralnik:1964eu,Kibble:1967sv}
being indeed as intimately linked to the generation of fermion masses as predicted by the Standard Model. 
This hypothesis will be further scrutinized during Run~II of the LHC, where the Vector Boson Fusion (VBF) 
channel will play a leading role. In this production mode, a Higgs boson is created by annihilation 
of virtual \Wboson or \Zboson bosons, radiated off the initial-state (anti-)quarks in a $t$-channel scattering process with no 
color exchange at leading order~\cite{Cahn:1983ip,Kane:1984bb}. The experimental signature thus consists 
of two forward jets, separated by a large rapidity gap with no hadronic activity~\cite{Rainwater:1997dg,
  Rainwater:1998kj,Rainwater:1999sd,Plehn:1999nw,Eboli:2000ze}.
This topology is the key to identifying the signal among an overwhelming number of backgrounds,
which include Higgs boson production through gluon fusion accompanied by two or more jets. Despite the latter 
production mechanism of the Higgs boson being dominant, it can be largely suppressed compared to the VBF channel by requiring 
a large rapidity separation between the leading two jets (called the tagging jets), a large invariant
mass of the corresponding dijet system, and an additional veto on jet activity in the central rapidity region. Higher-order QCD and EW corrections are known
to alter the signal rate only insignificantly under these cuts~\cite{Han:1992hr,Figy:2003nv,Bolzoni:2010xr,Ciccolini:2007jr,Ciccolini:2007ec}.
In the present article, we investigate to which extent QCD corrections alter the background.

We also present a comprehensive analysis of \Hjjj production in its own right. 
The high phenomenological relevance of the gluon fusion channel has
spurred an unprecedented effort in the theoretical community.
For Higgs boson production in conjunction with up to two jets, the NLO
corrections have been available for some time~\cite{Dawson:1990zj,Djouadi:1991tka,deFlorian:1999zd,Campbell:2010cz,Campbell:2006xx,Ravindran:2002dc,
Dixon:2009uk,DelDuca:2004wt,Dixon:2004za,Badger:2004ty,Ellis:2005qe,Ellis:2005zh,Berger:2006sh,Badger:2006us,Badger:2007si,Glover:2008ffa,Badger:2009hw,Badger:2009vh,vanDeurzen:2013rv}.
In addition it has been shown how to include parton shower resummation
on top of the fixed-order result. Especially, in the case of Higgs
boson plus two jets, this has been demonstrated
recently~\cite{Campbell:2012am,Hoeche:2014lxa}.
The first computation of Higgs boson production in association with
three jets was accomplished just two years ago~\cite{Cullen:2013saa}.
The development of the improved reduction algorithm
\textsc{Ninja}~\cite{Mastrolia:2012bu,vanDeurzen:2013saa,Peraro:2014cba}
to compute the virtual corrections then allowed a first
phenomenological analysis that was published in
Ref.~\cite{Butterworth:2014efa}.
For the Higgs boson plus one jet final state, the full NNLO QCD
results were computed
lately~\cite{Boughezal:2013uia,Chen:2014gva,Boughezal:2015dra,Boughezal:2015aha}
whereas the NNLO results for inclusive Higgs boson production have
been around for a
decade~\cite{Harlander:2002wh,Anastasiou:2005qj,Grazzini:2008tf}.
The frontier concerning the latter however has been pushed further
with a seminal calculation of the NNNLO corrections that has just been finalized~\cite{Anastasiou:2013mca,Kilgore:2013gba,Ball:2013bra,
Li:2013lsa,Duhr:2013msa,Anastasiou:2014vaa,Anastasiou:2014lda,Li:2014bfa,Ahmed:2014uya,Anastasiou:2015ema}.

In this article we focus on the behavior of the NLO results under different scale choices, and on the scaling
with increasing number of jets. We also test for potential high-energy effects, which may require 
resummation~\cite{Andersen:2009he,Andersen:2011hs}. We note that the scaling of jet cross sections 
with increasing number of jets is comparable to \Wboson plus jets production~\cite{Bern:2013gka,Bern:2014fea},
once the number of jets is large enough. This can be understood using jet calculus~\cite{Gerwick:2012hq,Gerwick:2012fw}.
We note that the effect could be tested experimentally to a high accuracy by measuring the ratio of jet rate ratios 
in different processes. Systematic uncertainties should cancel to a large extent in the analyses.

In our calculations we use an effective gluon-to-Higgs coupling, where the top quark is treated 
as an infinitely heavy particle. Although this requires the top mass to be much larger than the Higgs mass, 
the approximation has been shown to work very well~\cite{DelDuca:2001eu,DelDuca:2001fn,Campanario:2013mga}.

This article is organized as follows: Section~\ref{sec:setup} presents the technical prerequisites for our
calculation. Section~\ref{sec:gf} discusses the properties of our results under generic cuts. Section~\ref{sec:vbf} finally focuses
on the experimentally most interesting case of VBF background predictions, and on the behavior under different
selection cuts for the tagging jets. Section~\ref{sec:conclusions} contains our conclusions and an outlook.

\section{Calculational setup}
\label{sec:setup}
The calculation of the NLO corrections is performed by combining the two automated programs
\textsc{GoSam} \cite{Cullen:2011ac,Cullen:2014yla} for the generation and evaluation of the virtual one-loop
amplitudes, and the Monte Carlo event generator \textsc{Sherpa} \cite{Gleisberg:2008ta}.
The two are linked using the Binoth Les Houches Accord \cite{Binoth:2010xt,Alioli:2013nda}, a standard for event
and parameter passing between one-loop programs and Monte-Carlo generators.

\subsection{Virtual corrections}
\label{sec:setup:virt}
The \textsc{GoSam} framework is based on an algebraic generation of $d$-dimensional integrand using a Feynman
diagrammatic approach, employing
\textsc{QGraf}~\cite{Nogueira:1991ex} and
\textsc{Form}~\cite{Vermaseren:2000nd,Kuipers:2012rf} for the diagram generation, and
\textsc{Spinney}~\cite{Cullen:2010jv}, {\textsc{Haggies}}~\cite{Reiter:2009ts} and
\textsc{Form} to write an optimized Fortran output. For the reduction of the tensor integrals
we used \textsc{Ninja}~\cite{Mastrolia:2012bu,vanDeurzen:2013saa,Peraro:2014cba}, an automated package for integrand reduction via Laurent expansion.
 Alternatively one can use other reduction
techniques such as integrand reduction in the OPP method~\cite{Ossola:2006us,Mastrolia:2008jb,Ossola:2008xq} as implemented in
\textsc{Samurai}~\cite{Mastrolia:2010nb} or methods of tensor integral reduction as implemented in
\textsc{Golem95}~\cite{Heinrich:2010ax,Binoth:2008uq,Cullen:2011kv}.
The resulting scalar integrals are evaluated using \textsc{OneLoop}~\cite{vanHameren:2010cp}. As already anticipated, with respect to the very first computation~\cite{Cullen:2013saa}, the integrand reduction performed with \textsc{Ninja} allows an improved timing and also a better stability in the evaluation of the virtual. This was a crucial aspect for both the analysis presented here and also the previous one presented in~\cite{Butterworth:2014efa}.

\subsection{Real emission and phase space integration}
\label{sec:setup:rest}
The calculation of tree-level matrix elements, real emission contributions and subtraction terms
in the Catani-Seymour framework \cite{Catani:1996vz}, as well as phase space integration
have been performed using \textsc{Sherpa} \cite{Gleisberg:2008ta}. We have used the matrix element
generator \textsc{Comix} \cite{Gleisberg:2008fv,Hoeche:2014xx}.
We emphasize that this is different to the results obtained previously \cite{Cullen:2013saa,Butterworth:2014efa},
where we used a combination of \textsc{MadGraph} 4 \cite{Stelzer:1994ta,Alwall:2007st},
\textsc{MadDipole} \cite{Frederix:2008hu,Frederix:2010cj} and \textsc{MadEvent} \cite{Maltoni:2002qb}
for the calculation of real emission matrix elements, subtraction terms
and phase space integration of the real emission contribution. We recalculated the results obtained in \cite{Cullen:2013saa,Butterworth:2014efa}
with the \textsc{Sherpa} setup and found complete agreement. This is a very strong
consistency check on both results.

\subsection{Definitions relevant to the calculation}
\label{sec:setup:defs}
In the approximation of an infinitely large top mass, the Higgs coupling to gluons, which at LO
is mediated by a top-quark loop, becomes independent of $m_t$, and can be described by an effective operator~\cite{Wilczek:1977zn}, as
\begin{equation}
\mathcal{L}_{\rm eff} =- \frac{c_i}{4} \, \Higgs \, \mbox{tr} \left(G_{\mu \nu} G^{\mu \nu} \right ) .
\label{Eq:EffL}
\end{equation}
In the $\overline{\mbox{MS}}$ scheme, the coefficient $c_i$
is given by~\cite{Djouadi:1991tka,Dawson:1990zj}
\begin{equation}
c_i = -\frac{\alphaS}{3 \pi v} \left ( 1 + \frac{11}{4 \pi} \alphaS\right )+ \mathcal{O}(\alphaS^3) ,
\end{equation}
in terms of the Higgs vacuum expectation value $v$, set to $v=246$ \GeV.  The
operator~\eqref{Eq:EffL} leads to new Feynman rules, with vertices
involving the Higgs field and up to four gluons.

In the absence of accompanying jets, i.e.~in the Higgs boson
production process, it is natural to evaluate the strong coupling
associated with the effective vertex at a scale equal to the Higgs boson
mass $m_{\Higgs}$. When further jets are present, the natural scale choice is
more ambiguous. One possibility is to keep the two powers of the
strong coupling in the effective vertex fixed to $m_{\Higgs}$, while the
other powers of $\alphaS$ are computed at a different scale
$\mu_{R}$. When choosing this method at NLO, an additional finite correction
has to be added to the virtual contribution, taking into account
the fact that the strong couplings in the Born contribution are
evaluated at different scales \cite{Hamilton:2012np}.

In the case of Higgs boson
production associated with $N$ jets the general formula for computing
the cross section is
\begin{equation}
\frac{\d\sigma}{\d\Phi}=\alphaS^{N}\left(\mu_{R}\right)\alphaS^{2}\left(m_{\Higgs}\right)B+
\alphaS^{N+1}\left(\mu_{R}\right)\alphaS^{2}\left(m_{\Higgs}\right)\left[V(\mu_R)+2\beta_{0}\log\left(\frac{m_{\Higgs}^{2}}{\mu_{R}^{2}}\right)B+R\right] ,
\end{equation}
where $B$, $V$ and $R$ are the Born, the virtual and the real
contribution respectively, and $\beta_{0}$ is the one loop beta
function coefficient:
\begin{equation}
\beta_{0}=\frac{11\CA-2\NF}{12\pi}.
\end{equation}
In Section~\ref{sec:numericalresults} we will discuss and compare
different settings for the renormalization and factorization scale and
their impact on the theoretical predictions.

\subsection{Ntuples generation and usage}
\label{sec:setup:ntuples}
In order to simplify our analysis, the numerical results have been
produced and stored in the form of \textsc{Root} Ntuples. This format is
particularly useful for changing cuts and observables, as they can be
directly extracted from the given set of Ntuples without having to
generate new results. It also simplifies changing PDFs and scale
choices, as this can be achieved without having to reevaluate the
matrix elements, which is the most time consuming contribution.  The
writing of the Ntuples is implemented in \textsc{Sherpa}, and we refer to
\cite{Bern:2013zja} for more details. Table~\ref{tab:ntuples_tab}
summarizes the total number of Ntuple files which were generated for
the analyses presented in the next sections. The files are available
upon request.







%

\begin{table}[t!]
  \centering\small
  \begin{tabular}{lp{17mm}rrr}
    \toprule\\[-8pt]
    Sample && Nr. of files & Events per file & Space per file\\[4pt]
    \midrule\\[-10pt]
    & \Hjp & \multicolumn{3}{l}{For 8 and 13 \TeV individually}\\[5pt]
    \midrule\\[-7pt]
    Born                 && 51  & 5 million & 0.5 GB \\[5pt]
    Integrated dipoles   && 51  & 5 million & 1.2 GB \\[5pt]
    Virtual              && 101 & 5 million & 1.0 GB \\[5pt]
    Real and Subtraction && 101 & 5 million & 1.0 GB \\[5pt]
    \midrule\\[-17pt]
    \midrule\\[-12pt]
    Total:       & \multicolumn{3}{l}{Events: $\sim$ 1.5 billion } & Size: $\sim$ 290 GB\\[5pt]
    \midrule\\[-10pt]
    & \Hjj & \multicolumn{3}{l}{For 8 and 13 \TeV individually}\\[5pt]
    \midrule\\[-7pt]
    Born                 && 51  & 5 million & 0.6 GB \\[5pt]
    Integrated dipoles   && 51  & 5 million & 1.3 GB \\[5pt]
    Virtual              && 201 & 100'000  &  14 MB \\[5pt]
    Real and Subtraction && 101 & 5 million & 1.5 GB \\[5pt]
    \midrule\\[-17pt]
    \midrule\\[-12pt]
    Total:       & \multicolumn{3}{l}{Events: $\sim$ 1.0 billion } & Size: $\sim$ 250 GB\\[5pt]
    \midrule\\[-10pt]
    & \Hjjj & \multicolumn{3}{l}{For 8 and 13 \TeV individually}\\[5pt]
    \midrule\\[-7pt]
    Born                 && 51  & 5 million & 0.7 GB \\[5pt]
    Integrated dipoles   && 51  & 5 million & 1.3 GB \\[5pt]
    Virtual              && 301 & 25'000   & 3.9 MB \\[5pt]
    Real and Subtraction && 601 & 5 million & 1.9 GB \\[5pt]
    \midrule\\[-17pt]
    \midrule\\[-12pt]
    Total:       & \multicolumn{3}{l}{Events: $\sim$ 3.5 billion } & Size: $\sim$ 1.25 TB\\[5pt]
    \bottomrule
  \end{tabular}
  \caption{\label{tab:ntuples_tab}
    Events and size per file for the different Ntuple samples.}
\end{table}


\subsection{Kinematic requirements and parameter settings}
\label{sec:numericalresults}
The following numerical results have been calculated  for center of mass
energies of 8 and 13 \TeV. We discuss first a set of basic cuts and second a set
of VBF cuts. In both cases jets are clustered using the anti-kt algorithm \cite{Cacciari:2005hq,Cacciari:2008gp}
as implemented in the FastJet package \cite{Cacciari:2011ma}. If not specified differently, the jet
radius has been set to $R=0.4$ and the PDF set was CT10nlo \cite{Lai:2010vv}.
In the basic selection, the following set of jet cuts is applied
\begin{equation}
 p_T > 30 \;\GeV, \quad |\eta| < 4.4 \;.
 \label{cuts:basic}
\end{equation}
For the VBF analysis additional cuts on the jets are imposed, namely
\begin{equation}
 m_{j_1 j_2} > 400 \;\GeV,\quad \left|\Delta y_{j_1,j_2}\right| > 2.8\;.
 \label{cuts:vbf}
\end{equation}
If not stated otherwise the tagging jets $j_1$ and $j_2$ are defined as ordered in $p_T$,
i.e.~the jets with the highest and second highest transverse momentum.
In sections \ref{sec:gf:hej} and \ref{sec:vbf} we will also discuss different tagging schemes.

Concerning the choice of renormalization and factorization scale we define our central scale to be given by
\begin{equation}
\mu_{F}=\mu_{R}=\frac{\hthatprime}{2}=\frac{1}{2}\left(\sqrt{m_{\Higgs}^{2}+p_{T,\Higgs}^{2}}+\sum_{i}|p_{T,i}|\right),
\end{equation}
where the sum runs over all partons accompanying the Higgs boson in
the event. This scale was shown to be sensible in other multi-leg calculations~\cite{Berger:2009zg}.
However, in the Higgs-production process it is not obvious that this scale should be adopted for all strong couplings.
Because the top quark has been integrated out, one could argue that the strong coupling associated to the effective Higgs-gluon vertex 
should be fixed to the natural scale of this vertex, and that this would be of the order of $m_{\Higgs}$. 
Furthermore it is not obvious whether one should vary this scale together with the scales for the other powers of
$\alphaS$ when performing the usual scale variation to assess the theoretical uncertainties.
We note that there is no 'correct' choice and that the differences between the choices are formally of higher order.
It is therefore interesting to investigate their effect on phenomenological results.

We consider three different scale choices, defined as
\begin{subequations}
\label{scales}
\begin{flalign}
\label{scales:A}
\textrm{A}:& \quad \alphaS\biggl(x\cdot \frac{\hthatprime}{2}\biggr)^3\alphaS\left(x \cdot m_{\Higgs}\right)^2 \\
\label{scales:B}
\textrm{B}:& \quad \alphaS\biggl(x\cdot \frac{\hthatprime}{2}\biggr)^5 \\
\label{scales:C}
\textrm{C}:& \quad \alphaS\left(x \cdot m_{\Higgs}\right)^5.
\end{flalign}
\end{subequations}
The presence of the factor $x$ indicates that this scale is varied in
the range $x \in [0.5\,\ldots\,2]$. A variant of scale A, where the two
powers of $\alphaS$ evaluated at $m_{\Higgs}$ do not change with varying $x$
was used in previous computations of \Higgs+ 3
jets~\cite{Cullen:2013saa,Butterworth:2014efa}. This however leads to
a somewhat artificial reduction of the scale uncertainty, and
therefore we do not consider the corresponding scale in the following.

\section{Higgs boson plus jets phenomenology}
\label{sec:gf}
In this section we discuss the results that have been obtained with
the set of basic gluon fusion cuts as described in Eq.~\eqref{cuts:basic}.

\subsection{Cross sections, scale dependence and technicalities}
\label{sec:gf:xsecs}

\begin{figure}[t!]
  \centering
  \includegraphics[width=0.49\textwidth]{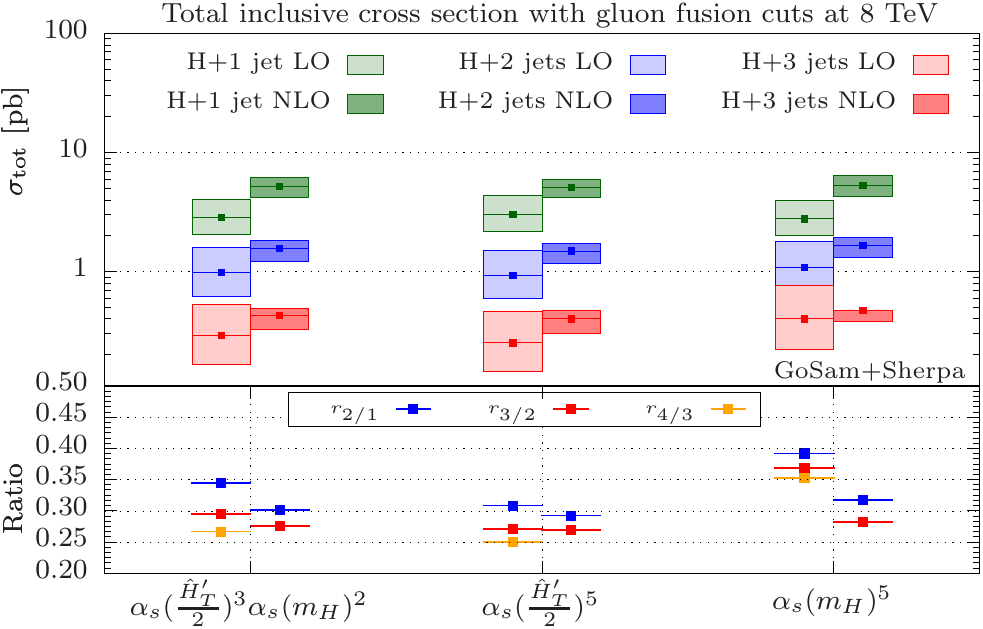}
  \hfill
  \includegraphics[width=0.49\textwidth]{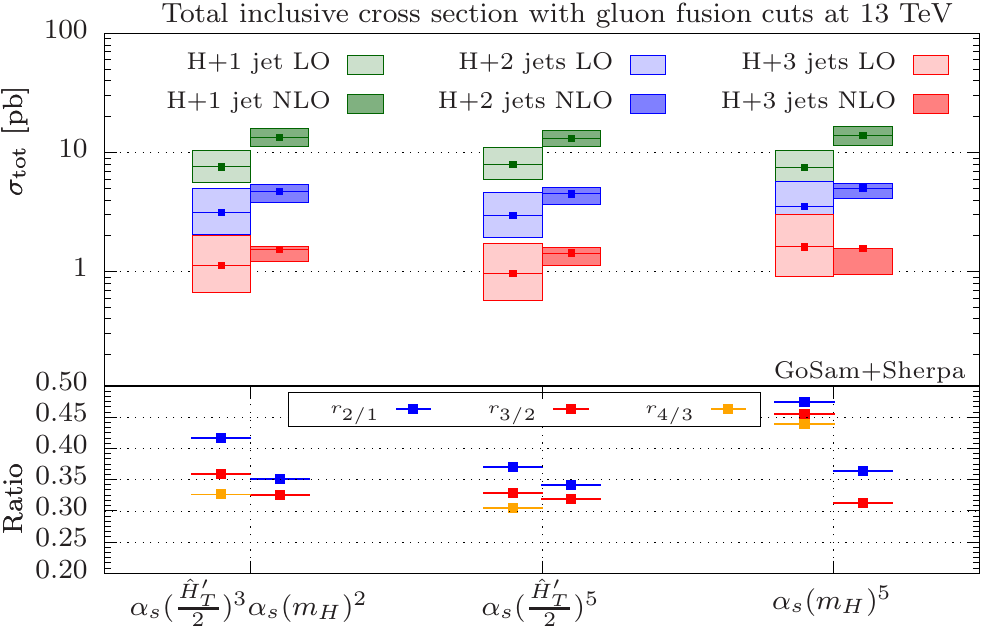}
  \caption{\label{fig:XsecAndScaleOptions}%
    Total cross sections for \Hj (green), \Hjj (blue) and \Hjjj (red) for
    LO and NLO. In the lower part of the plot the ratios $r_{2/1}$ 
    (blue), $r_{3/2}$ (red) and $r_{4/3}$ (orange) are shown. On the left plot the results
    have been obtained for 8 \TeV, the right plot is for 13 \TeV.}
\end{figure}

We start our analysis by presenting the results for the inclusive
cross sections using basic gluon fusion cuts and the different scale
choices shown above.
Fig.~\ref{fig:XsecAndScaleOptions} shows the total cross sections for \Hj, \Hjj and \Hjjj for both LO
and NLO at 8 \TeV (left plot) and 13 \TeV (right plot). The cross sections are calculated for the three
different scale choices A, B, and C defined in Eq.~\eqref{scales}. The upper part of the plot displays the
LO and NLO results for the \Hnj process for the central scale and the variations around
the central scale as defined in the corresponding scale choices. The lower plot shows the ratios
\begin{equation}
r_{n/n-1}=\sigma_{\tot}(\Higgs+nj)/\sigma_{\tot}(\Higgs+(n-1)j)\,,
\end{equation}
for $n+2,3,4$ at LO and $n=2,3$ at NLO accuracy.

Independent of the scale choice and order in perturbation theory we see that the ratio $r_{2/1}$
is larger than the ratio $r_{3/2}$. One might expect this behavior for two reasons: 
On the one hand the \Hjj process 
has new kinematical topologies compared to \Hj. In particular, it is possible that the Higgs boson 
is (almost) at rest in a \Hjj final state, while it must always carry some transverse momentum 
in a \Hj final state, at least when computed at LO. This means that the phase space is larger
in the \Hjj case. On the other hand, new partonic channels open up for the \Hjj process which do not exist 
in the \Hj case. Adding a third jet neither opens additional phase nor further partonic channels.

Regarding the scale choices we observe that except for scale C the ratios are largely independent
of the choice and that the ratios between LO and NLO are very similar, in particular for scale B
there is a very good agreement.
 Scale C works well
for the NLO results but yields an increased ratio for the LO results. This is an indication that
this is not a sensible choice.
Going from 8 \TeV to 13 \TeV, the jet rate ratios increase, as may be expected due to the increase 
in phase space for jet production.

One interesting point in this context is the fact that for scale B the ratio $r_{3/2}$ is very similar
between LO and NLO. Given the additional evidence from \Wboson/\Zboson+jets calculations, which also points towards 
dynamical scales being more suitable for multi-leg processes, we choose scale B~\eqref{scales:B} as our default.
It is also striking that the NLO ratio is almost identical in the $r_{2/1}$ and in the $r_{3/2}$
case. This hints at the possibility to extrapolate to the $r_{4/3}$ case.
The idea has been worked out in Ref.~\cite{Bern:2014fea} for the case of \Wboson production in association with jets,
and it is supported by jet calculus~\cite{Gerwick:2012hq,Gerwick:2012fw}.
However, in our case the fit to $r_{n/n-1}$ that was performed in Ref.~\cite{Bern:2014fea} would be
trivial, with the first nontrivial check of staircase scaling being the ratio $r_{4/3}$ itself.
Fig.~\ref{fig:XsecAndScaleOptions} also suggests that the quality of the fit will depend on
the scale choice. Furthermore, it remains to be seen whether the good agreement for the ratios 
will hold for differential distributions.

Nevertheless a very interesting experimental opportunity opens up at this point: Let us assume a set of hard
processes that contain all possible kinematical topologies, and also all possible partonic channels, 
like \Wboson/\Zboson + 2 jets production and \Hjj production. Then it should be possible to test the universality
of staircase scaling to a very good precision by measuring the ratio of jet rate ratios $r_{n/n-1}$ for
different hard processes, like \Wboson/\Zboson and \Higgs production. The differences between the dominant partonic 
production mechanisms at leading, next-to-leading and next-to-next-to-leading order will be reduced at higher
jet multiplicity, and the ratio should be largely independent of $n$. A major deviation from this behavior 
would very likely signal the presence of new physics.

A more detailed summary of the total cross sections and their ratios is presented in Table \ref{hj3nlo_tab_xsecs}.
It lists the total cross sections for the one-, two- and three-jet process as well as it theoretical
uncertainties from scale variations, the global
K-factors, cross section ratios, labelled with $r$, and jet fractions denoted by $f$.







%

\begin{table}[t!]
  \centering\scriptsize
  \begin{tabular}{clrp{0.3mm}lrp{0.3mm}lrp{0.3mm}lr}
    \toprule\\[-8pt]
    Sample     & \multicolumn{6}{c}{Cross sections in pb for Higgs boson plus}\\[4pt]
    \phantom{$K$-factor} & $\ge1$ jets & $f_2$ && $\ge2$ jets & $f_3$ && $\ge3$ jets & $f_4$ && $\ge4$ jets & $r_{n+1/n}$\\[4pt]
    \midrule\\[-7pt]
    \small{8 \TeV} \\[0pt]
    \midrule\\[-10pt]
    \multicolumn{4}{l}{LO \scriptsize(NLO PDFs)}\\[0pt]
    \midrule\\[-10pt]
    \Hjp             & $3.020~^{+44\%}_{-28\%}$ &         &&                          &        &&                          &         &&                        & $0.308$ \\[8pt]
    \Hjj             &                          & $1.00$  && $0.931~^{+63\%}_{-36\%}$ &        &&                          &         &&                        & $0.271$ \\[8pt]
    \Hjjj            &                          &         &&                          & $1.00$ && $0.252~^{+82\%}_{-42\%}$ &         &&                        &         \\[5pt]
    \midrule\\[-10pt]
    \multicolumn{4}{l}{NLO}\\[0pt]
    \midrule\\[-10pt]
    \Hjp             & $5.096~^{+17\%}_{-17\%}$ & $0.183$ && $0.930~^{+63\%}_{-36\%}$ &         &&                          &         &&                           & $0.292$ \\[8pt]
    \Hjj             &                          &         && $1.490~^{+17\%}_{-21\%}$ & $0.169$ && $0.252~^{+82\%}_{-42\%}$ &         &&                           & $0.269$ \\[8pt]
    \Hjjj            &                          &         &&                          &         && $0.401~^{+17\%}_{-25\%}$ & $0.157$ && $0.063~^{+101\%}_{-47\%}$ & $(0.157)$ \\[20pt]
    $K_1$, $K_2$, $K_3$   & $1.69$                   &         && $1.60$                   &         && $1.59$                   & \\[5pt]
    \midrule\\[-14pt]
    \midrule\\[-7pt]
    \small{13 \TeV} \\[0pt]
    \midrule\\[-10pt]
    \multicolumn{4}{l}{LO \scriptsize(NLO PDFs)}\\[0pt]
    \midrule\\[-10pt]
    \Hjp             & $7.968~^{+38\%}_{-26\%}$ &         &&                          &        &&                          &         &&                        & $0.371$ \\[8pt]
    \Hjj             &                          & $1.00$  && $2.954~^{+58\%}_{-34\%}$ &        &&                          &         &&                        & $0.329$ \\[8pt]
    \Hjjj            &                          &         &&                          & $1.00$ && $0.972~^{+76\%}_{-40\%}$ &         &&                        &         \\[5pt]
    \midrule\\[-10pt]
    \multicolumn{4}{l}{NLO}\\[0pt]
    \midrule\\[-10pt]
    \Hjp             & $13.19~^{+15\%}_{-15\%}$ & $0.288$ && $2.953~^{+58\%}_{-34\%}$ &         &&                          &         &&                          & $0.341$   \\[8pt]
    \Hjj             &                          &         && $4.500~^{+13\%}_{-18\%}$ & $0.216$ && $0.971~^{+76\%}_{-40\%}$ &         &&                          & $0.319$   \\[8pt]
    \Hjjj            &                          &         &&                          &         && $1.437~^{+11\%}_{-22\%}$ & $0.206$ && $0.296~^{+94\%}_{-45\%}$ & $(0.206)$ \\[20pt]
    $K_1$, $K_2$, $K_3$   & $1.66$                   &         && $1.52$                   &         && $1.48$                   & \\[5pt]
    \bottomrule
  \end{tabular}
  \caption{\label{hj3nlo_tab_xsecs} Cross sections in pb for the
    various parton-level Higgs boson plus jet samples used in this
    study with scale choice B (i.e. all scales are evaluated at
    $\hthatprime/2$). The upper and lower parts of the table show
    the LO and NLO results, respectively, together with their
    uncertainties (in percent) from varying scales by factors of two,
    up (subscript position) and down (superscript position).
    NLO-to-LO $K$-factors, $K_n$, for the inclusive 1-jet ($n=1$),
    2-jets ($n=2$) and 3-jets ($n=3$) bin, the cross section ratios
    $r_{2/1}$, $r_{3/2}$, $r_{4/3}$ and $m$-jet fractions, $f_m$, are
    given in addition. Since the predictions for \Hjjjj are only
    LO accurate, $f_4$ and $r_{4/3}$ coincide.}
\end{table}

\begin{figure}[t!]
  \centering
  \includegraphics[width=0.49\textwidth,trim=15pt 0 20pt 0]{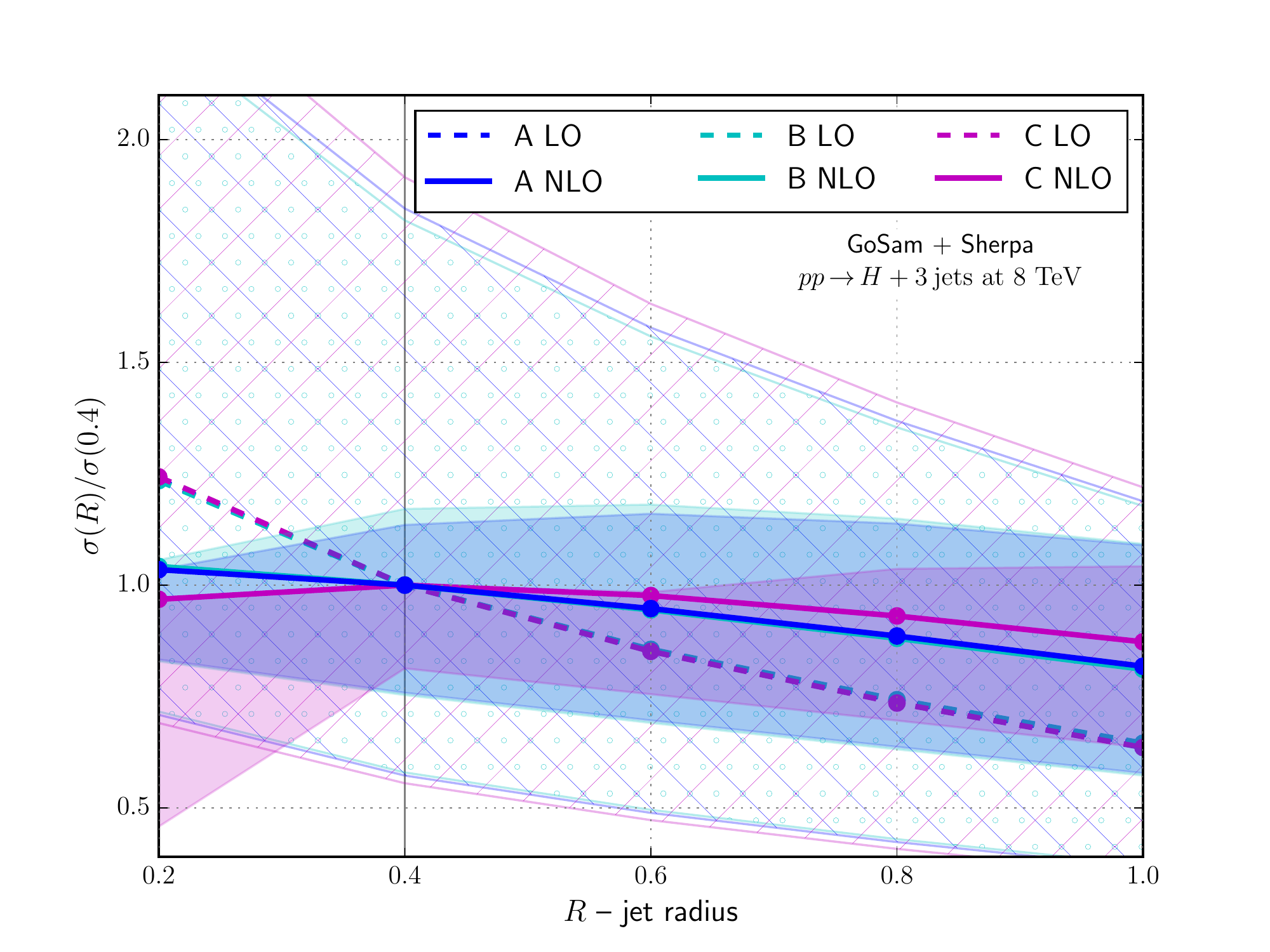}
  \hfill
  \includegraphics[width=0.49\textwidth,trim=15pt 0 20pt 0]{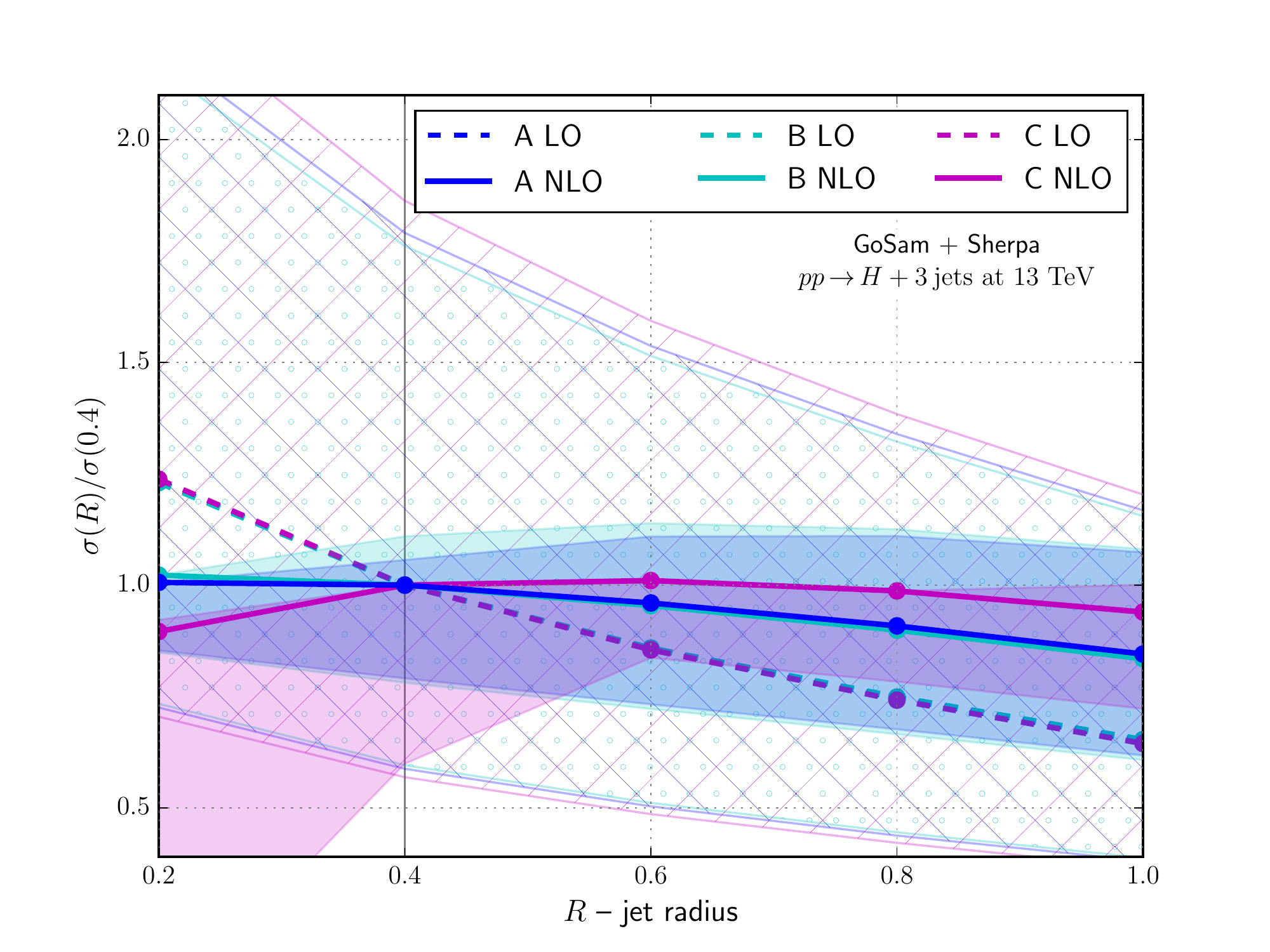}
  \caption{Total cross section as a function of the jet radius $R$ normalized to the $R=0.4$ result.
  The left plots shows the results for 8 \TeV, the right plot shows the 13 \TeV result.}
  \label{fig:jetradius}
\end{figure}

It is interesting to investigate the NLO result and its stability as a function of the jet radius, $R$.
The corresponding results are shown in Fig.~\ref{fig:jetradius} for 8 \TeV (left) and 13 \TeV (right).
The cross sections are normalized to $\sigma_{\tot}(R=0.4)$. At leading order, the cross section decreases 
with increasing $R$, because the probability for two partons to be clustered into a single jet increases,
leading to a rejection of the event. From this basic consideration it is clear 
that the dependence on the jet radius is stronger for \Hjjj than for \Hjj, simply because with
an increasing number of jets, the average distance in $R$-space between two jets becomes smaller.

At NLO the situation is more involved.
The additional parton present in the real corrections leads, on average, to softer partons than at LO.
For small values of $R$ the partons will not be clustered which means that there is an increased probability
that the event will not pass the cuts because there are not enough jets above the $p_T$ threshold.
If the jet radius is increased, partons that are relatively close in $R$-space but where each of them
is not above the threshold, get clustered into a jet. If the radius is increased further we end up with
the same situation as for the leading order result and the cross section decreases again. Therefore we expect
a change of shape and the interplay between the behavior of the leading order results with the one
of the real emission contribution leads to a stabilization of the $R$-dependence. In general we see that
the effects between different scale choices are very small, also largely owing to the fact that the plots
in Fig.~\ref{fig:jetradius} show ratios, so differences due to different runnings and factors of $\alphaS$
cancel out.

\begin{figure}[t!]
  \centering
  \includegraphics[width=0.49\textwidth]{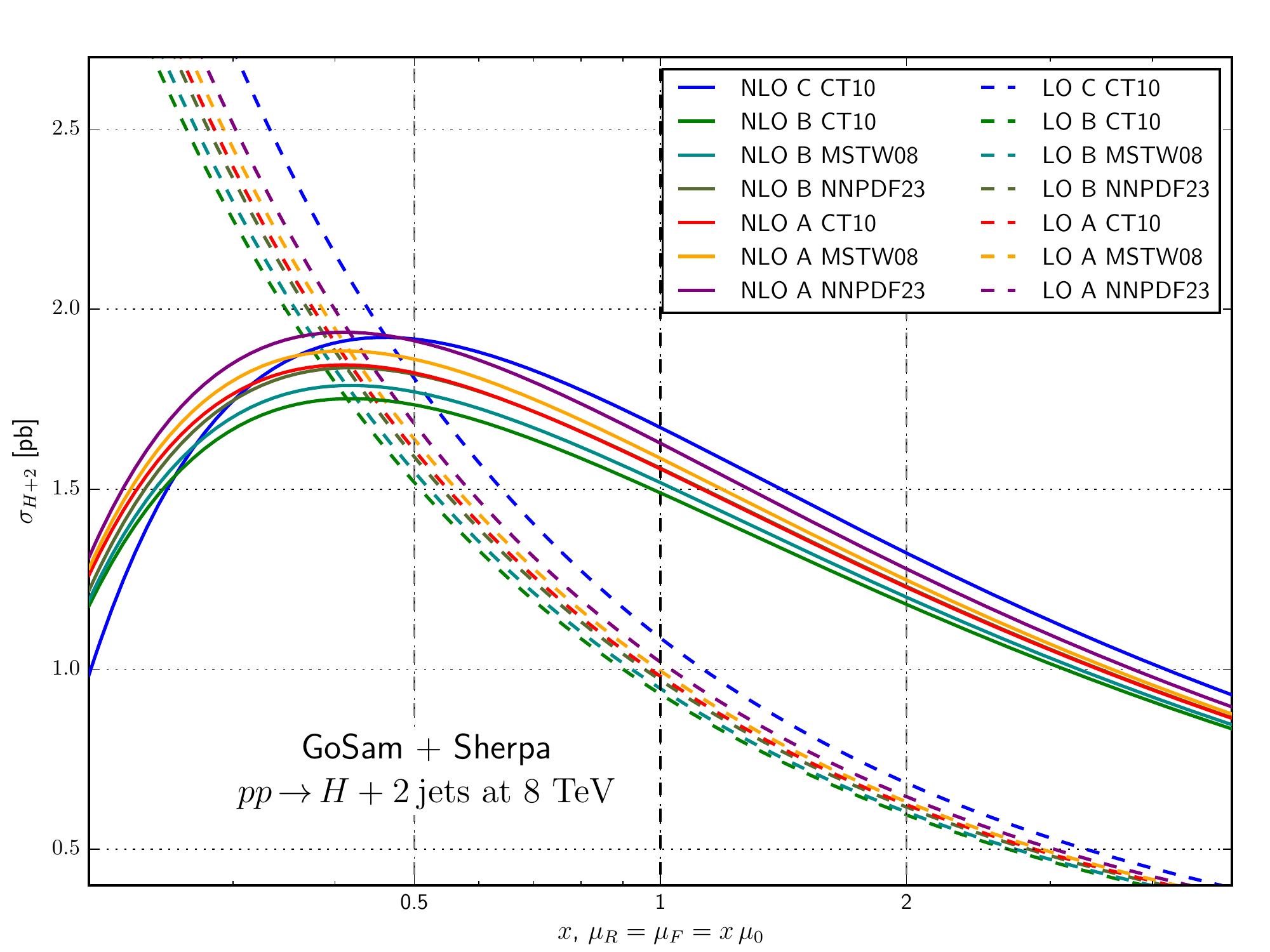}
  \hfill
  \includegraphics[width=0.49\textwidth]{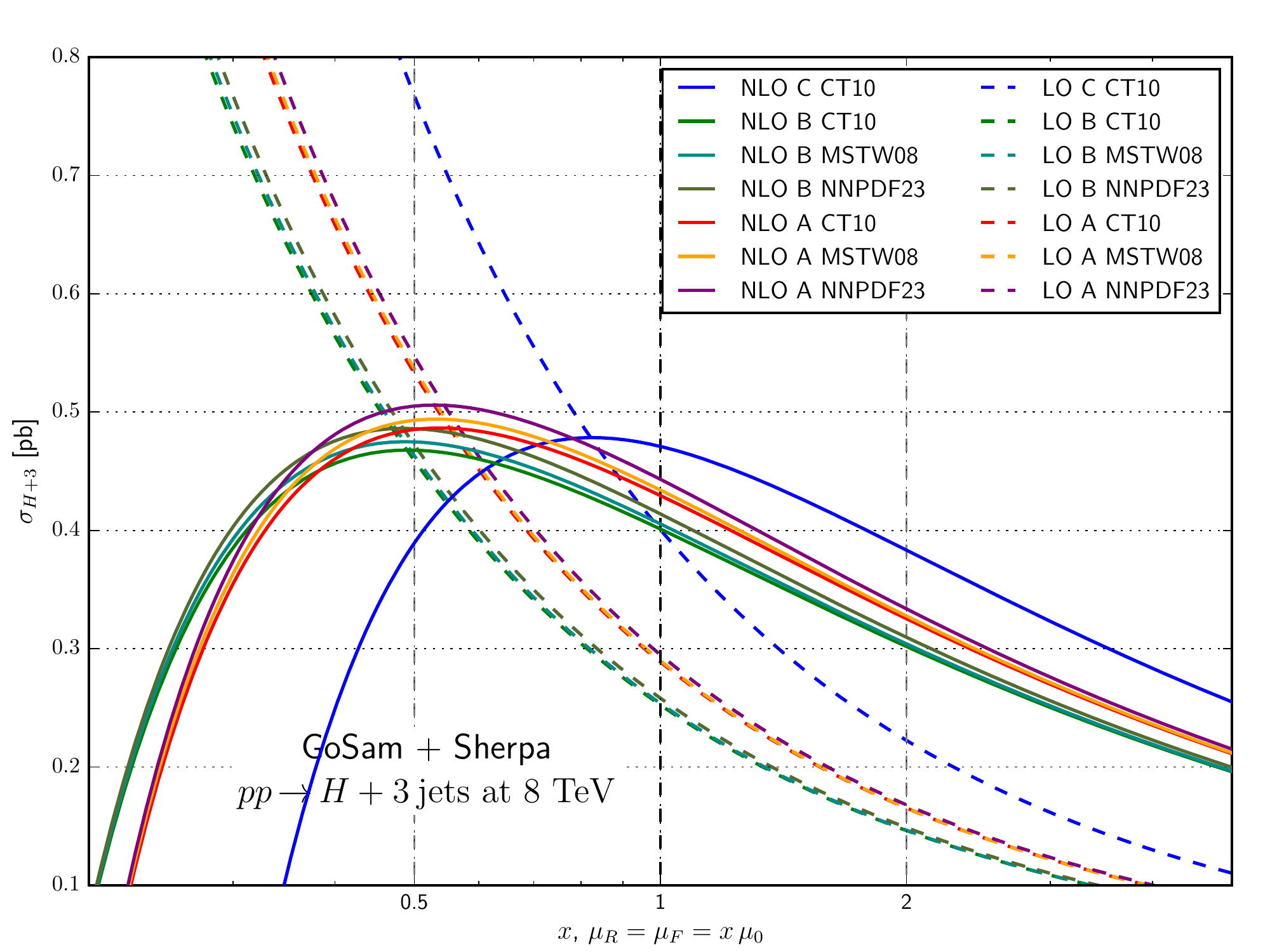}
  \\
  \centering
  \includegraphics[width=0.49\textwidth]{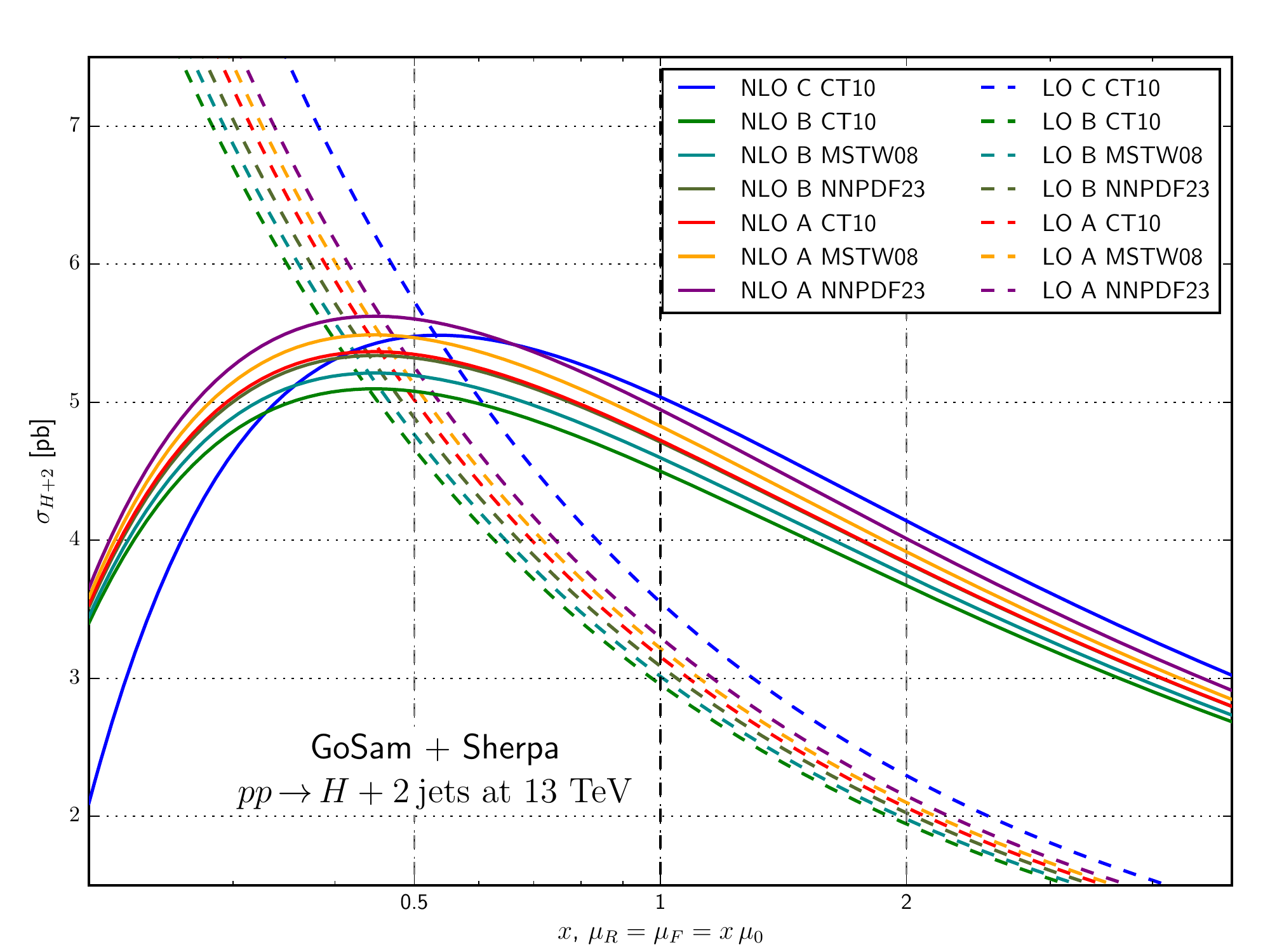}
  \hfill
  \includegraphics[width=0.49\textwidth]{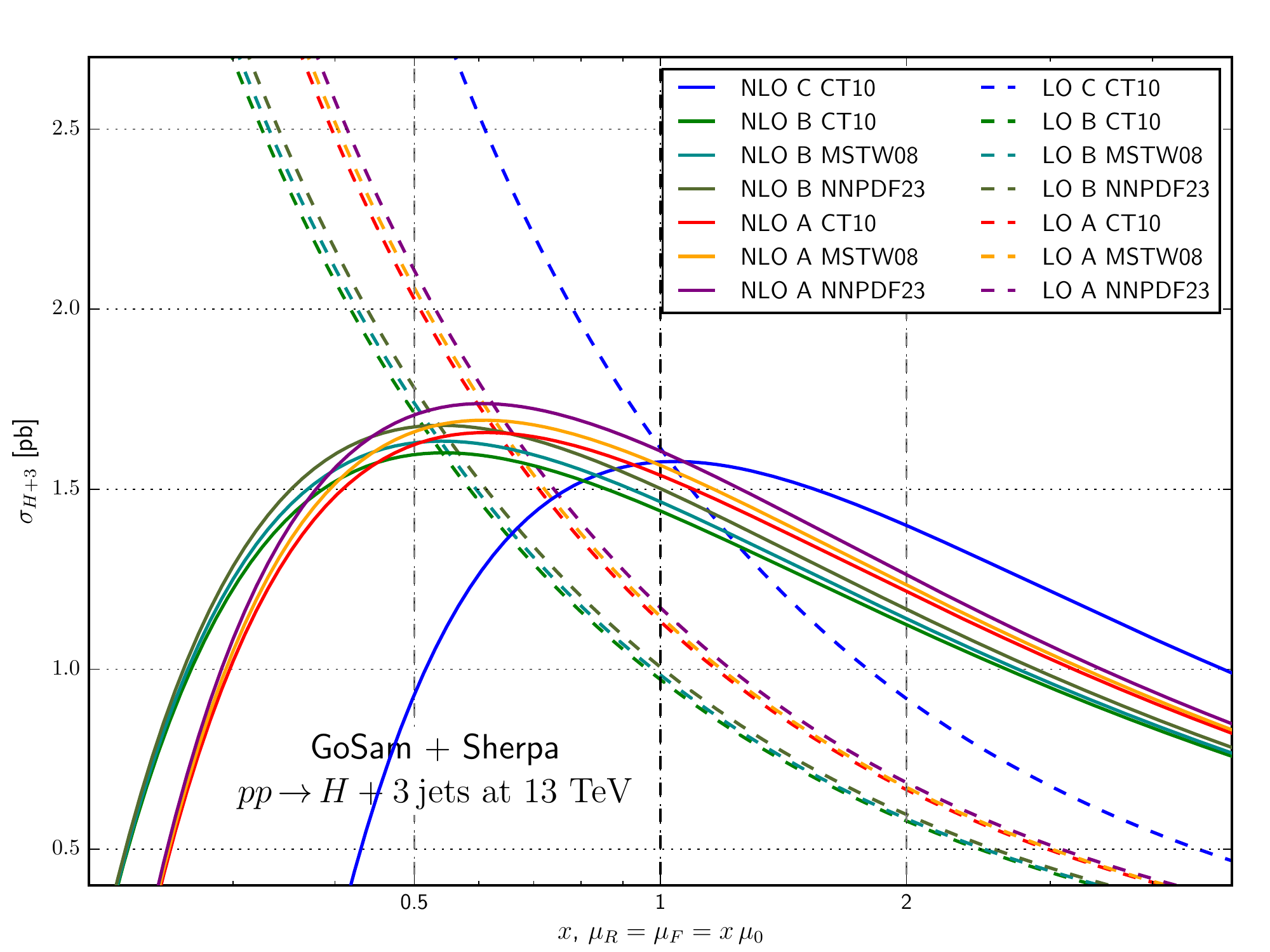}
  \caption{Scale dependence of the total cross section for scale choices A, B and C
  for different PDF sets with a center of mass energy of 8 \TeV (upper row) and 13 \TeV (lower row),
  and for \Hjj (left column) and \Hjjj (right column).}
  \label{fig:pdf}
\end{figure}

Fig.~\ref{fig:pdf} shows the scale dependence of the total cross
section for different PDF sets and different scale choices. The upper
plots show the 8 \TeV result, the lower plots show the 13 \TeV
results. We also show this result for both \Hjj and \Hjjj.

Independent of the jet multiplicity we observe the typical change of
shape when going from LO to NLO. Both processes have their maximum
approximately at half of the central scale, except for scale C in the
3 jet case, which is shifted towards higher values.  This means that
taking the central scale and varying up and down by a factor of two
yields a reliable estimation of the theoretical uncertainties. The
plots show the results for three different PDF sets, CT10nlo
\cite{Lai:2010vv}, MSTW08 \cite{Martin:2009iq} and NNPDF23
\cite{Ball:2012cx}.  This is compared to the scale choices A, B and
C. For scale C we show the result only for one PDF set. Although the
different PDF sets lead to slightly different results, their effect is
considerably smaller than a different choice of the scale.  If we
compare the 2-jet result and the 3-jet result we see that for the
former the effects of scale and PDF choice leads to a broader range of
results, i.e.~the spread between the single curves is bigger compared
to the 3-jet case, where the curves (except for scale C) seem to be
more bundled. For the 2-jet process the scale C is in quite good
agreement with the other scales, which is clearly not the case for the
3-jet process.  This indicates that this scale (i.e.~the Higgs mass)
is smaller than the other scales hence shifting curves to higher
values of x.  Another interesting point is that for both jet
multiplicities the shapes are almost independent of the center of mass
energy, the plots for 8 \TeV and 13 \TeV are very similar. Only for
scale choice C we find a visible deviation, which is not surprising as
it is a fixed scale and therefore does not account for a change in the
center of mass energy.

\begin{figure}[t!]
  \centering
  \includegraphics[width=0.49\textwidth]{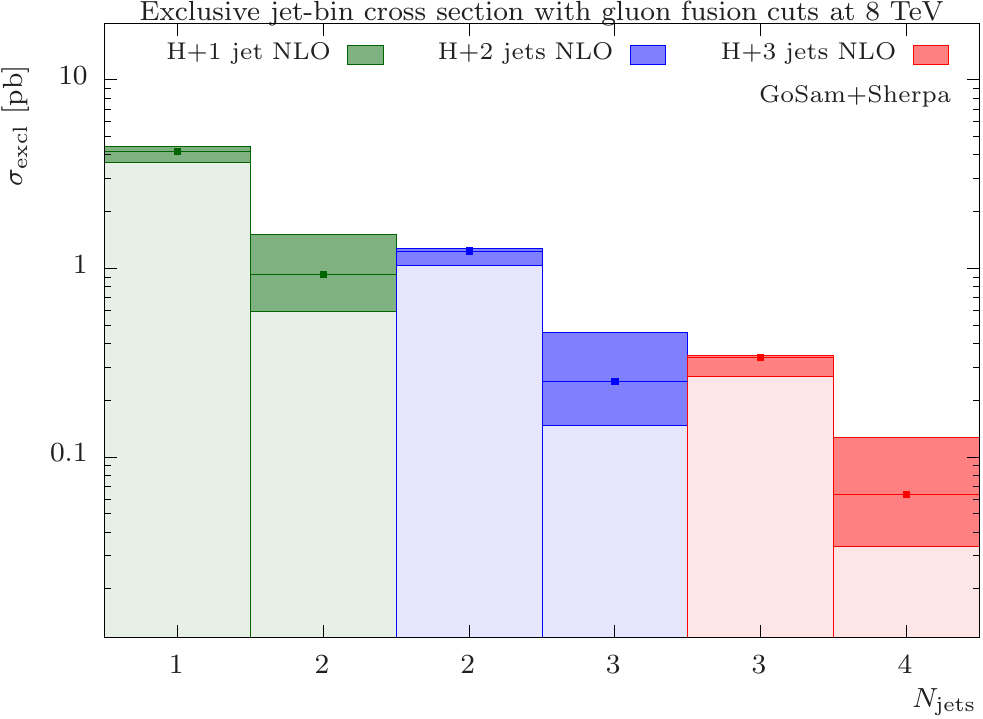}
  \hfill
  \includegraphics[width=0.49\textwidth]{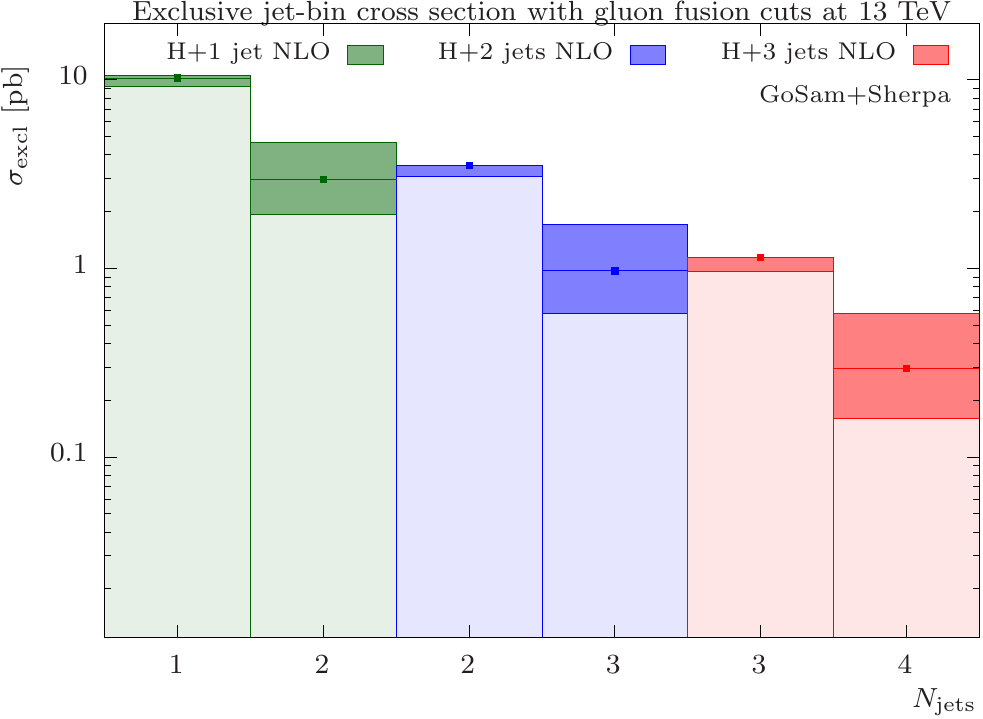}
  \caption{Exclusive jet cross sections for \Hnj $(n=1,2,3$) for 8 \TeV (left) and 13 \TeV (right).
  For each of the three processes two bins are plotted. The first contains its exclusive
  NLO contribution, the second the contribution to the n+1 process, i.e. the real emission
  with LO accuracy. The darker shaded areas denote the uncertainty from scale variation.}
  \label{fig:nj}
\end{figure}

Fig.~\ref{fig:nj} shows the exclusive jet cross sections for Higgs plus one, two and three
jets for both 8 and 13 \TeV. At NLO a \Hnj process contributes of course to two jet
multiplicities, to the n-jet bin and to the $(n+1)$-jet bin. The contribution to the n-jet
bin is given at NLO accuracy, whereas the $(n+1)$-jet contribution is only present at
LO accuracy, as it is given by the real emission contribution.
For each subprocess, the left bin contains the exclusive contribution to the n-jet bin,
the right plot its contribution to the $(n+1)$-jet bin. As we have used the same set
of NLO PDFs for both LO and NLO contributions the real emission contribution of the
n-jet process is exactly equal to the LO contribution of the $(n+1)$-jet process.
The dark shaded areas on the different jet-bins denote the theoretical uncertainty stemming
from the scale variation. As the contribution to the $(n+1)$-jet bin is only given at
leading order accuracy the error bars are substantially larger compared to the n-jet bin,
which is given at NLO accuracy.
It is important to mention that for all three subprocesses the $(n+1)$-jet contribution constitutes
a substantial fraction of the total cross section. 
This is particularly important when investigating observables that
separate the two contributions, as in the case of vetoed cross
sections.  We will return to this point in Sec.~\ref{sec:vbf}, when
dealing with VBF topologies.

\begin{figure}[t!]
  \centering
  \subfloat[\label{sfig:higgs12pT}]{%
    \includegraphics[width=0.49\textwidth]{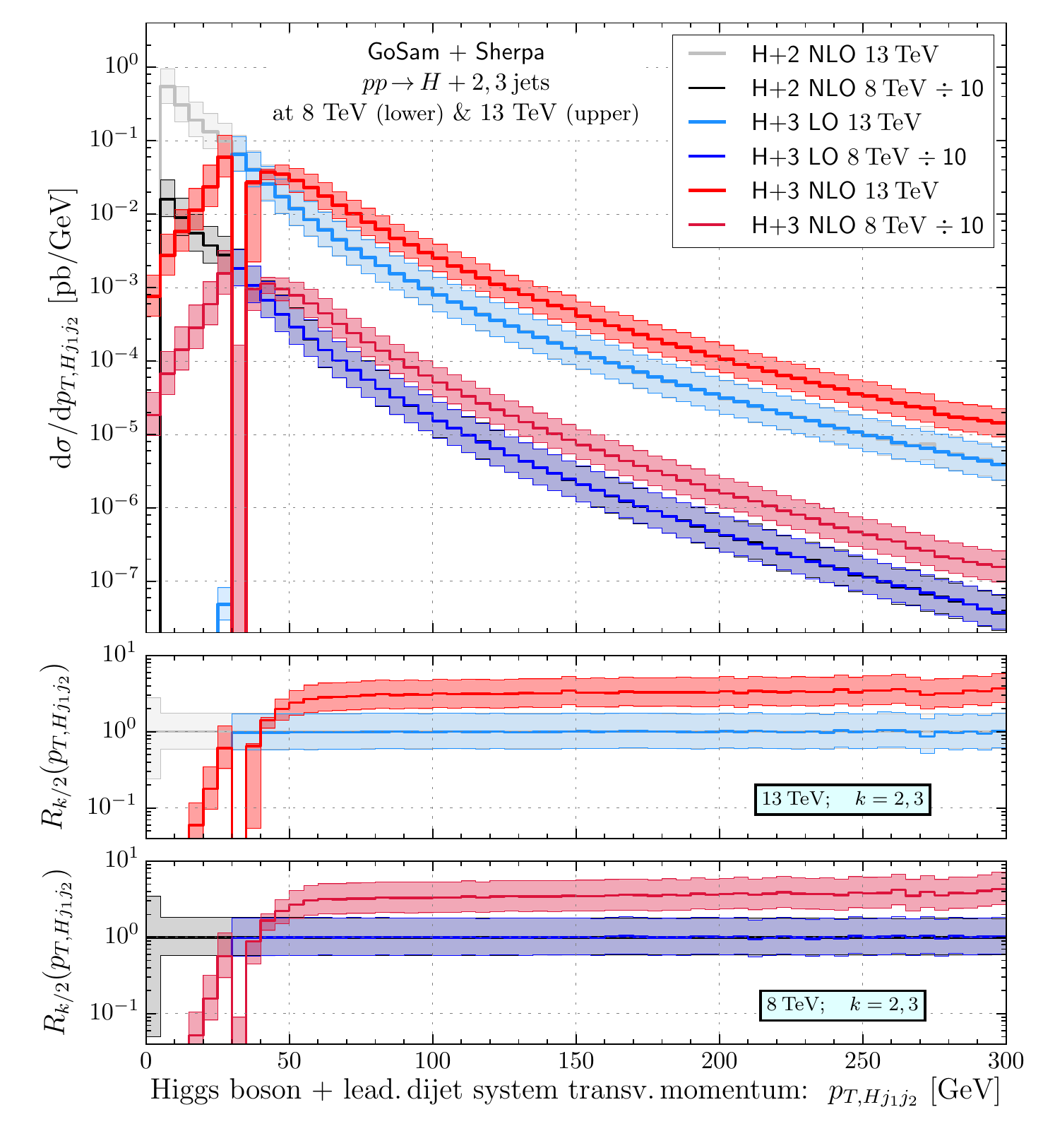}
  }\hfill
  \subfloat[\label{sfig:higgs12dphi}]{%
    \includegraphics[width=0.49\textwidth]{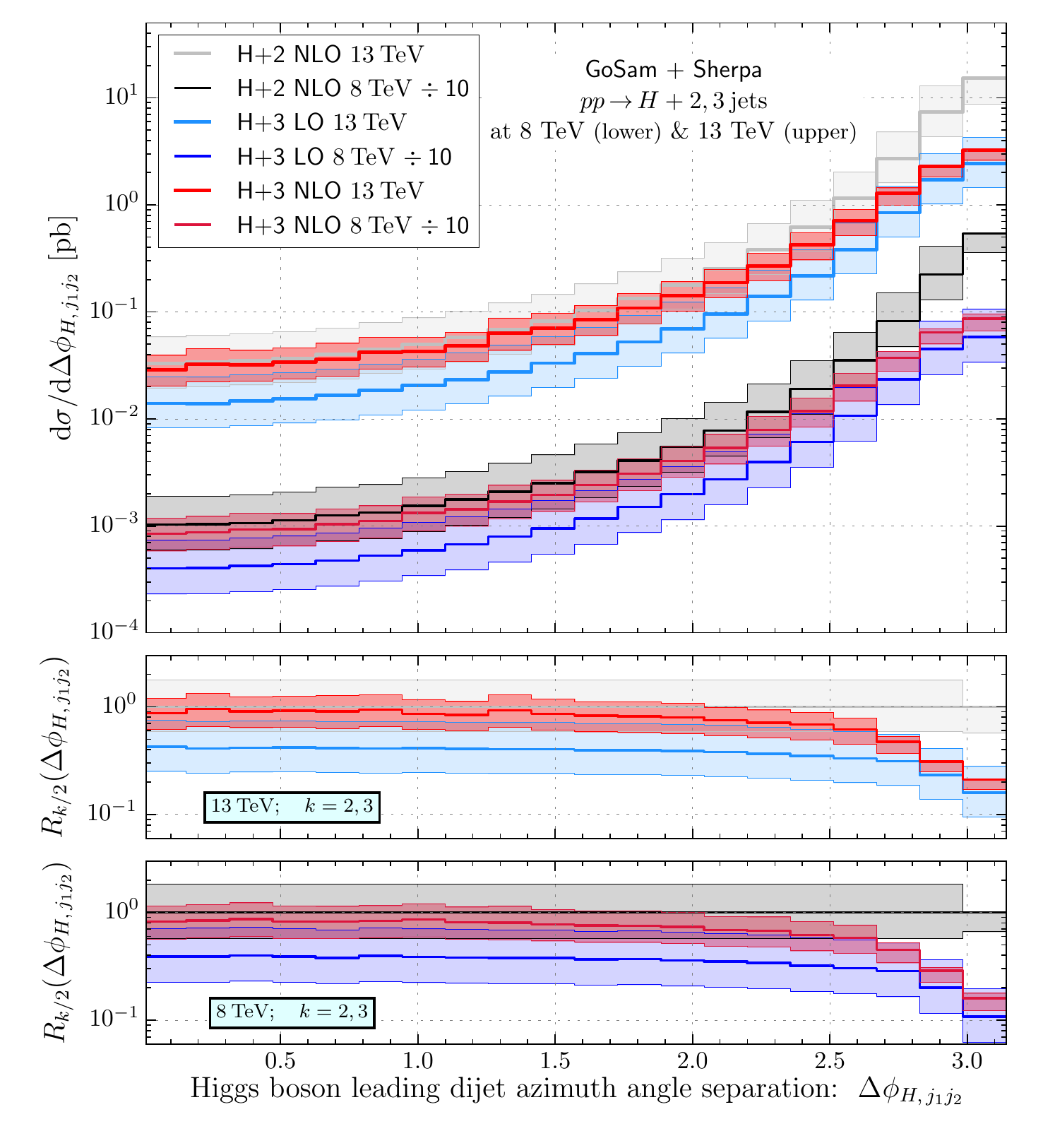}
  }
  \caption{\label{fig:higgs12}%
    The transverse momentum distribution of the combined Higgs boson
    plus associated leading dijet system (\ref{sfig:higgs12pT}), and
    the distribution of their relative azimuthal angle, $\dphiHjj$,
    (\ref{sfig:higgs12dphi}) at both LHC center-of-mass energies of
    13~\TeV as well as 8~\TeV.}
\end{figure}

Another interesting class of observables which is somewhat related to
VBF topologies since it requires at least two jets in the final state,
is the one combining the Higgs boson and the two leading-$p_T$ jets
momenta. These observables are particularly interesting because they
are directly sensitive to any additional radiation beyond the two
leading jets. We start with the transverse momentum distribution of
the system consisting of the Higgs and the two leading jets $\pTHjj$.
Fig.~\ref{fig:higgs12} shows this observable, computed using \Hjj NLO
and \Hjjj LO and NLO, both at 8 and 13 \TeV. To better disentangle the
two energies, the distributions for 8 \TeV were divided by a factor of
$10$. The lower part of the plot shows the LO and NLO \Hjjj curves
normalized to the NLO \Hjj predictions separately for the two
center of mass energies.

We are using the same PDF set for both LO and NLO calculations, hence
the NLO \Hjj and the LO \Hjjj curves are identical. They predict the
transverse momentum spectrum of a third parton at LO. In the \Hjjj
case, the jet-$p_T$ threshold at 30 \GeV is made explicit. In the \Hjj 
sample instead the third parton can become arbitrarily soft, leading to a
distribution which diverges for $\pTHjj\to 0$, with the divergence
being canceled by the virtual corrections that contribute to the first bin
of the spectrum only.

Our \Hjjj NLO calculation adds additional, well-known features to this spectrum:
Sudakov suppression effects around the jet-$p_T$ threshold, and NLO K factors
at larger values of $\pTHjj$. Kinematical configurations with 
four jets or with three jets and an unresolved parton lead to $p_T$ balancing
as well as $p_T$ enhancing effects such that we find large, $\mathcal{O}(3)$, 
fairly constant corrections for higher $\pTHjj$ as well as a contribution
to the spectrum below the $p_T$ threshold. Universal parts of the virtual corrections
in combination with kinematical effects in the radiative corrections lead to a depletion
right above the threshold. The reduction of scale uncertainty when going from \Hjjj at LO
to \Hjjj at NLO is small at large $\pTHjj$, anticipating a large influence of 
higher-multiplicity processes. This will be discussed in Sec.~\ref{sec:inclusivevsexclusive}.
In contrast to other \Hjjj NLO distributions, we find a much more
symmetric uncertainty, which is again a consequence of the dominance of the
four-jet contributions that feature a LO scale variation%
\footnote{On a more quantitative level, we note that the
  `BVI' contributions of the $\pTHjj$ and $p_{T,j_3}$
  distributions are exactly the same while the `RS' ones of the former
  are considerably harder than those of the latter. This leads to a
  different cancellation pattern when combining the scale varied `BVI'
  and `RS' predictions (they work in opposite directions). For the
  $\pTHjj$, we then find the `RS' uncertainties to be the
  dominating ones for $\pTHjj\gtrsim 60\;\GeV$.}.

Figure~\ref{fig:higgs12} (right) shows the azimuthal separation between 
the Higgs boson and the two leading transverse momentum jets, $\dphiHjj$.
The peak at $\pi$ is common to all curves and indicates that the Higgs boson is
preferably recoiling against the jet-jet system. In fact, this is the only possible
kinematical configuration in \Hjj at LO, therefore the spread to smaller values of
$\dphiHjj$ indicates the size of the higher-order corrections.
Correspondingly, the curves for \Hjj at NLO show a very large scale dependence
over the full spectrum, the only exception being the bin at $\pi$, which contains
the singular real emission configurations and the virtual corrections.
In contrast to $\pTHjj$, the LO predictions for \Hjjj do not overlap 
with the NLO results for \Hjj, as real radiation can become relatively soft in the \Hjj case,
while still contributing to the spectrum at $\dphiHjj<3$,
i.e.~outside the bin at $\pi$. In other words the gap between the \Hjjj and \Hjj curves
is filled by events that contribute to the $\pTHjj$ spectrum below 30 \GeV.
The apparent difference in scale uncertainty between \Hjjj at LO and \Hjj at NLO shown in the 
ratio plots is an artifact of the different central value of these predictions.
Our \Hjjj NLO calculation allows to describe $\dphiHjj$ at true NLO accuracy.
For both collider energies the relative size of the corrections is more than 
a factor of $2$ for $\dphiHjj\to0$ and decreases to about $40\%$ for
$\dphiHjj\to\pi$.

\begin{figure}[t!]
  \centering
  \subfloat[\label{sfig:hjdphi}]{%
    \includegraphics[width=0.49\textwidth]{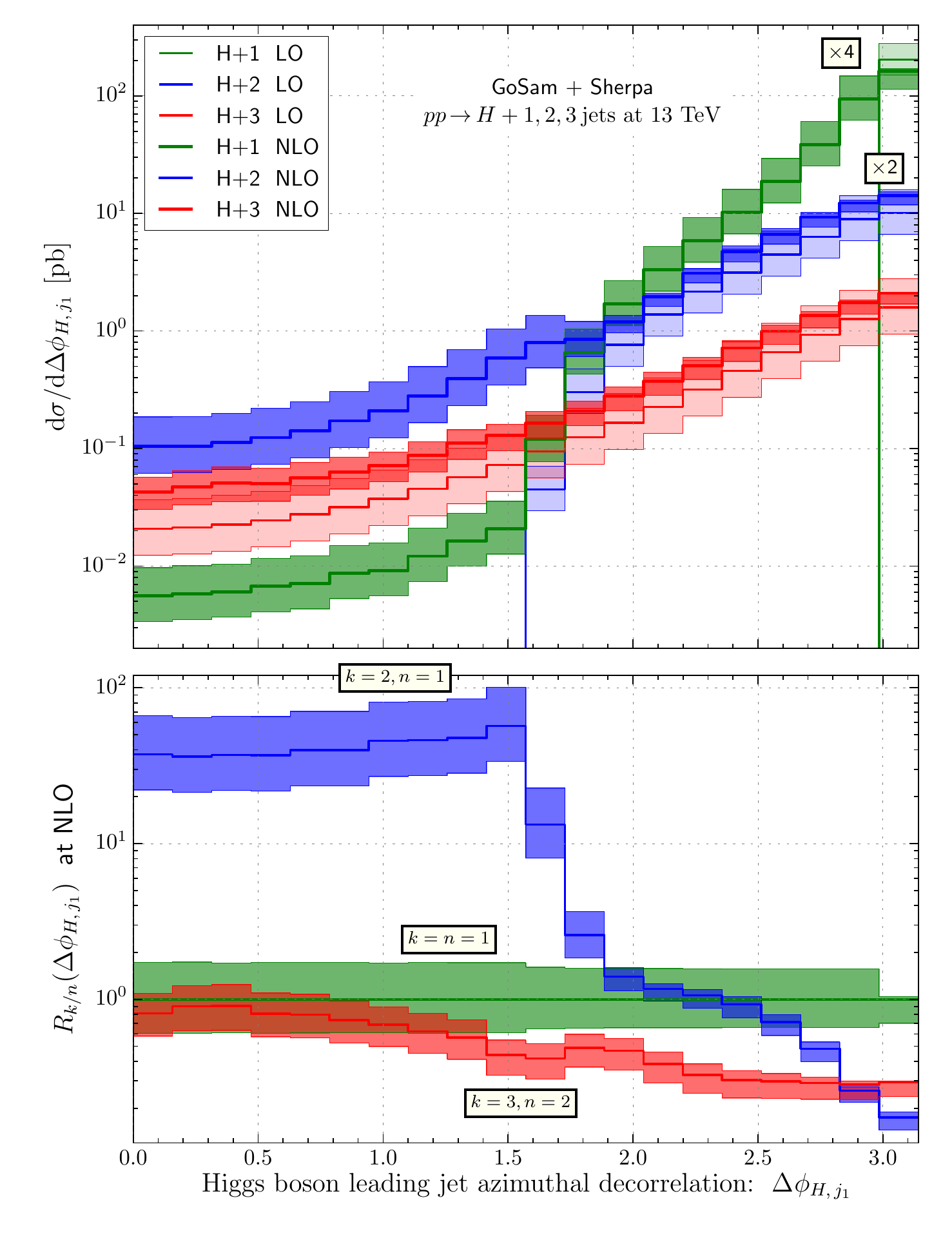}
  }\hfill
  \subfloat[\label{sfig:mTs}]{%
    \includegraphics[width=0.49\textwidth]{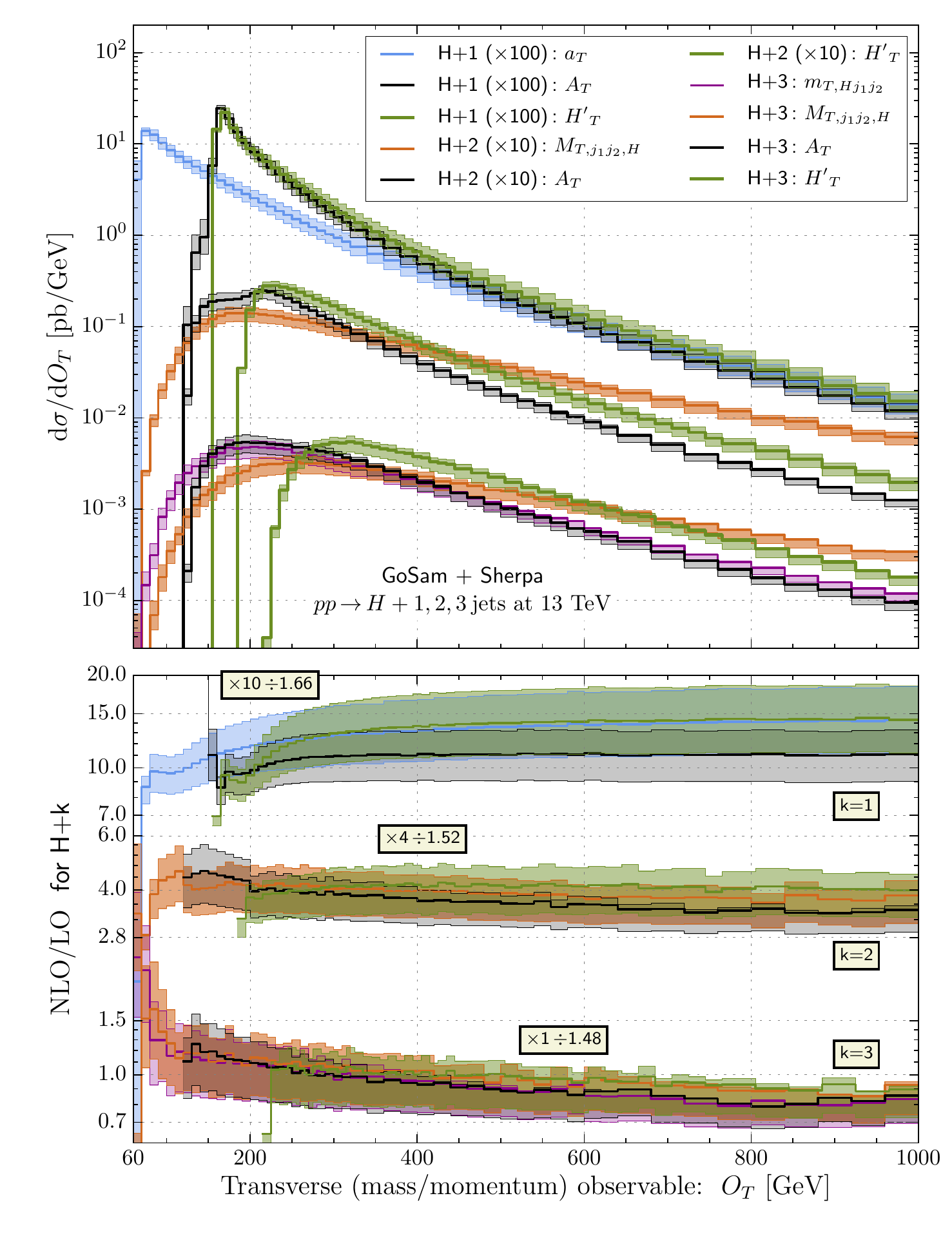}}
  \caption{\label{fig:higgsjetdphi}%
    Azimuthal separation between the Higgs boson and the hardest jet
    (\ref{sfig:hjdphi}) for \Hj, \Hjj and \Hjjj production at 13~\TeV
    LHC center-of-mass energy. Note that the results for the \Hj and
    \Hjj channels have been multiplied by additional factors of 4 and
    2, respectively. While both the LO and NLO predictions are shown
    in the main plot, the lower subplot depicts differential ratios,
    $R_{k/n}$, between various \Higgs\!+~\!jet cross sections at NLO
    only.
    Alternative transverse mass/momentum observables (\ref{sfig:mTs})
    in comparison to the $H^\prime_T$ observable as derived from our
    default scale choice. NLO results are shown for all three \Hnj
    channels (separated by additional factors of 10) and an LHC
    collision energy of 13 \TeV. For each observable, the differential
    K-factor is given in the associated ratio plot where again
    separation factors have been applied to enhance the visibility of
    the results.}
\end{figure}

We conclude this section discussing two more technicalities that are
of relevance to all Higgs boson plus jets final states. The left panel
of Fig.~\ref{fig:higgsjetdphi} shows the azimuthal separation between
the Higgs boson and the leading $p_T$ jet for the three different
final state multiplicities considered. Increasing the number of jets has
a very dramatic effect on this observable. In fact for \Hj, deviations
from a pure back-to-back configuration are only possible through real
radiation. The distribution is therefore only LO accurate and displays
a large uncertainty band. For \Hjj the azimuthal separation in a
Born-like event cannot exceed 90 degrees, because the Higgs boson and
the leading jet cannot recoil together against a second softer
jet. This however becomes possible in the presence of a third jet
originating from the real radiation contribution. For this reason the
\Hjj curve features a reduced uncertainty band for
$\pi/2<\Delta\phi_{H,j_1}<\pi$ and a larger uncertainty band for
smaller angles. This also explains the large drop of the \Hj curve for
angles $\Delta\phi_{H,j_1}<\pi/2$. The difference between the LO \Hjj
distribution, which vanishes at $\Delta\phi_{H,j_1}=\pi/2$ and the NLO
\Hj is due to real-radiation events which have the hardest jet with
very large pseudorapidity ($|\eta|>4.4$), therefore falling outside
the applied jet cuts. Thus the events populating the distribution of
\Hj below $\pi/2$ have large missing transverse energy. A full NLO
accuracy is finally reached over the whole kinematical range for
\Hjjj. Despite the several jets present in the final state, there is a
clear preference for the Higgs boson to recoil against the hardest jet.

As the $\hthatprime$ scale plays a central role in our
calculations, we are interested in how it compares to other reasonable
choices characterizing the transverse activity of Higgs boson
occurrences in association with jets. Using the generalized
transverse mass definitions,
$m_{T,1\ldots_n}=\sqrt{H^2_{T,1\ldots n}-(\sum_i \vec p_{T,i})^2}$
and
$M_{T,A,B}=\sqrt{m^2_A+2\,\big(p_{T,B}\,(m^2_A+p^2_{T,A})^{1/2}-\vec p_{T,B}\cdot\vec p_{T,A}\big)}$,
given in Ref.~\cite{Lykken:2011uv}, we can construct
observables that may serve as a proxy for alternative scale choices.
At the level of observables, i.e.~where jets rather than partons are
used to build the variable, we can investigate to what extent the
alternative choices lead to deviations from the $\hthatprime$
default. We then obtain at least a qualitative understanding of the
possible size of theoretical uncertainties that arise from a
variation of the scale's functional form. This theoretical scale
dependence should be kept in mind as it may turn out to be important
in the interpretation of \Hnj cross section measurements.

The results of this comparison are presented in the right panel of
Fig.~\ref{fig:higgsjetdphi} where for the three different jet
multiplicity final states, the main plot shows the transverse
observables including $H'_T$ (as green curves) while the ratio plot
shows the NLO vs.~LO K-factors in dependence of the respective
observable. These K-factors are noticed to show a great amount of
overlap supporting the fact that the higher-order corrections to the
different observables lead to rather similar effects. As before the
bands indicate the size of standard scale B variations by factors of
two. Note that for better visibility, the \Higgs\!+~\!$1,2,3$\!~jets
results are separated by additional factors of 10 (by factors of 10
and 4) in the main (ratio) plot. The differential K-factors of the
ratio plot are moreover divided by their respective inclusive \Hnj
K-factors. The black curves, $A_T\equiv M_{T,\Higgs,j_1\ldots j_n}$,
represent a choice whose threshold is always given by $m_\Higgs$ and
does not shift upwards with increasing jet multiplicity as it occurs
for $H'_T$. In the \Hj case, the differences are small while for
\Higgs\!+~\!$2,3$\!~jets, they are more pronounced as $A_T$ dives into
the softer region and thus gives rise to softer tails.
In the \Higgs\!+~\!$2,3$\!~jets cases, we also show
the effects of choosing a more VBF-inspired $M_T$ scale (orange
curves). Although the soft region is covered even more widely,
$M_{T,j_1j_2,\Higgs}$ provides much harder tails compared to $A_T$ with no
overlap among them. Again, softer tails but similar coverage of the
soft region can be achieved by switching to $m_{T,\Higgs j_1j_2}$
(purple curve) which we only demonstrate for \Hjjj. For \Hj events, an
$m_T$ based choice such as $a_T\equiv m_{T,\Higgs j_1\ldots j_n}$
(depicted by the lightblue shaded curve) would then offer a scale
setting very similar to $H'_T$ but neglecting the Higgs boson
mass. This is depicted by the curve shaded in lightblue. Based on the
overall behavior of the alternative scales apart from
$M_{T,j_1j_2,H}$, they can be anticipated to yield larger
\Higgs\!+\!~jets cross sections, for some cases they may even be
outside the uncertainty range as the standard scale variation bands do
not overlap in all cases. This is where further investigation will
have to step/set in.

\subsection{Single-particle observables}
\label{sec:gf:singleobs}
\begin{figure}[t!]
  \centering
  \subfloat[Ratio\label{fig:H3_higgs_scales:scaleRatio}]{\includegraphics[width=0.49\textwidth,height=74mm]{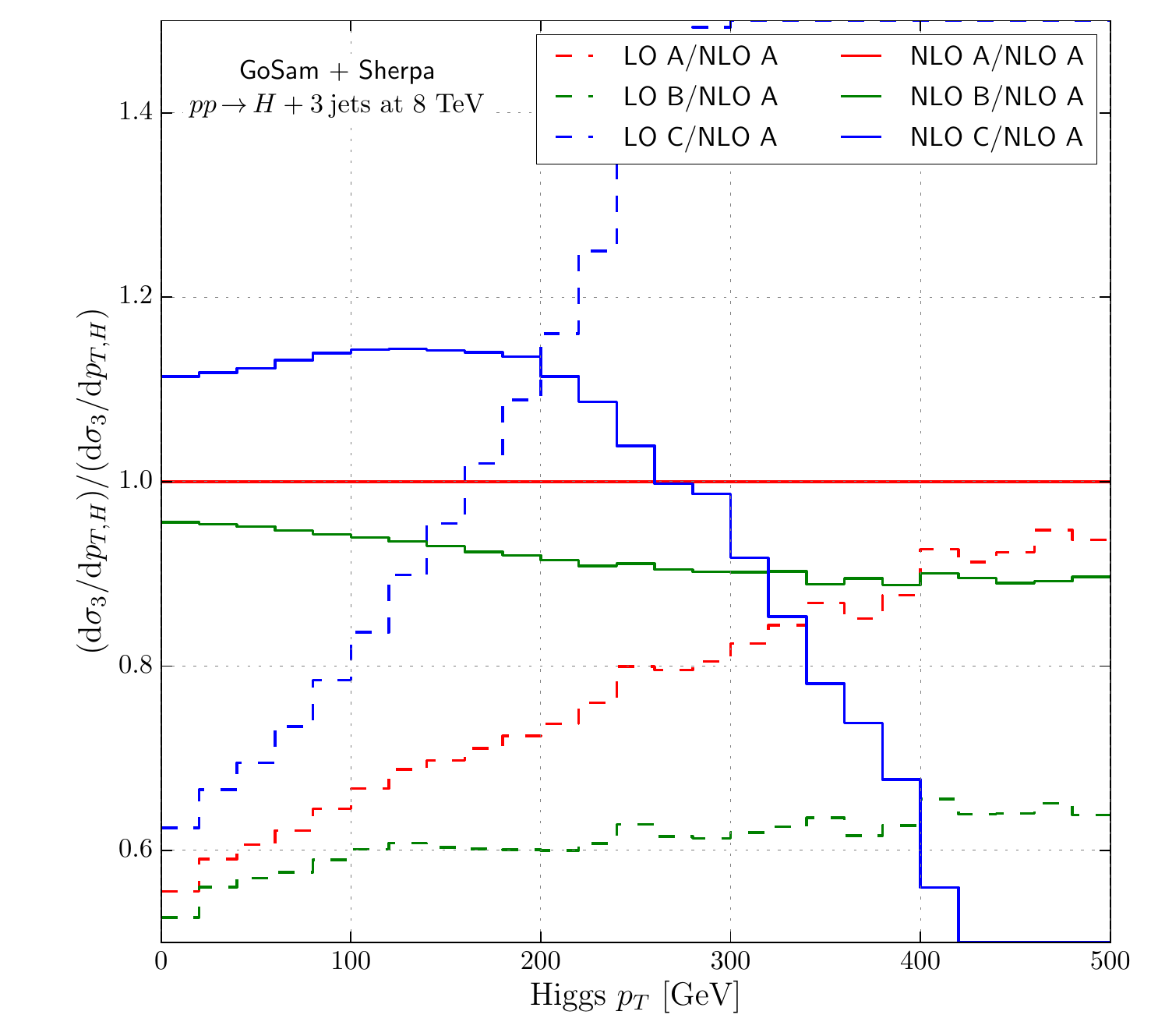}}
  \hfill
  \subfloat[Scale choice A \eqref{scales:A}\label{fig:H3_higgs_scales:scaleA}]{\includegraphics[width=0.49\textwidth]{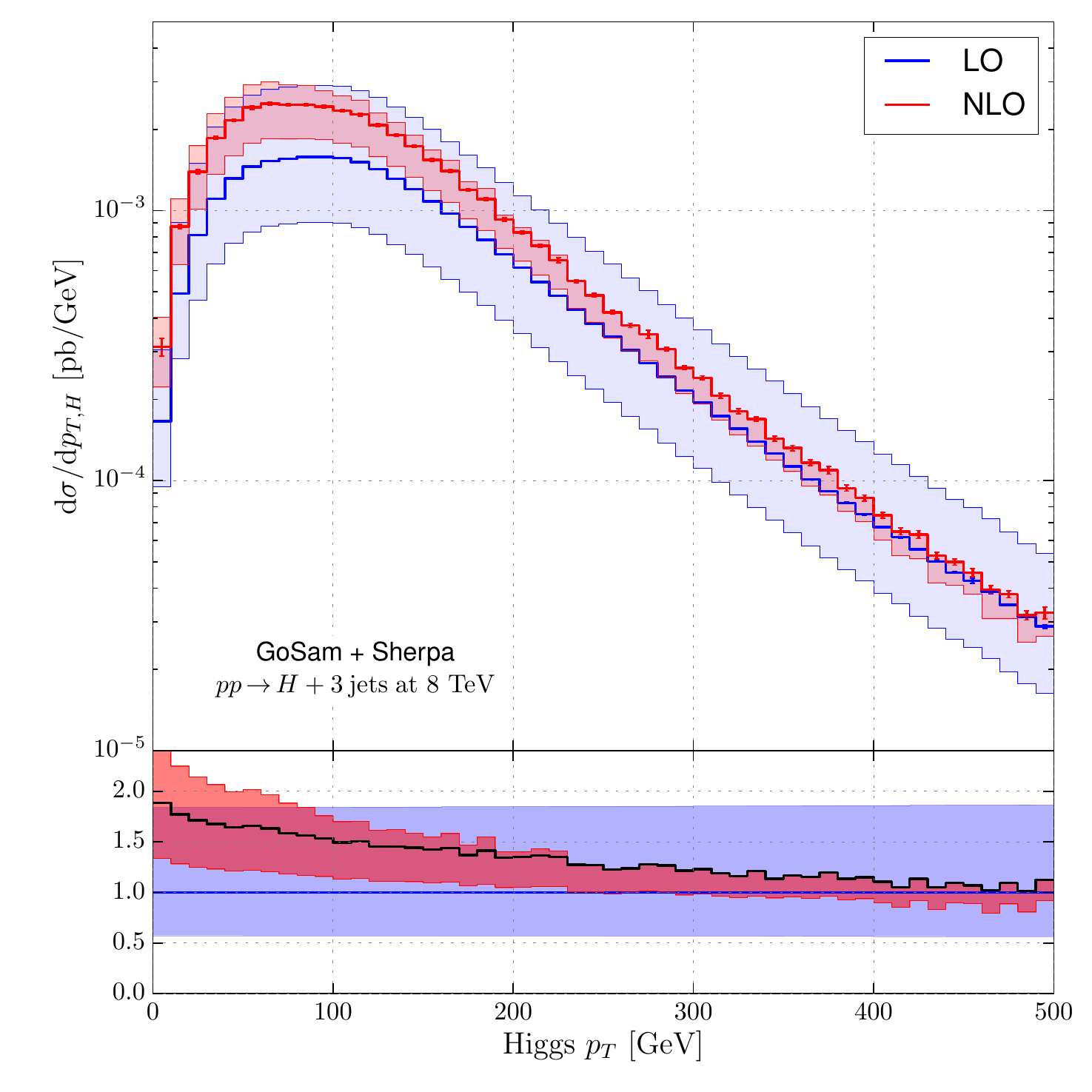}}
  \\
  \centering
  \subfloat[Scale choice B \eqref{scales:B}\label{fig:H3_higgs_scales:scaleB}]{\includegraphics[width=0.49\textwidth]{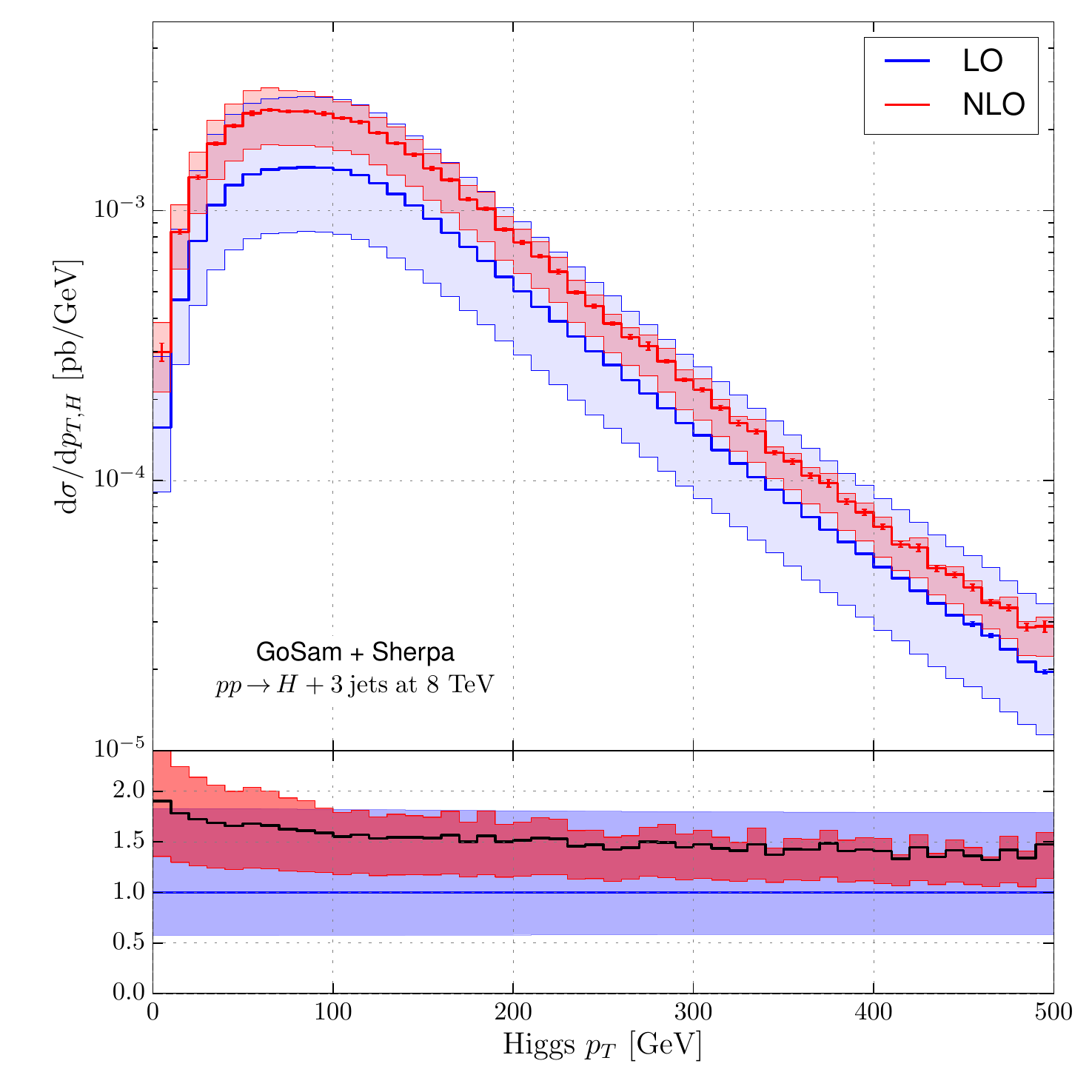}}
  \hfill
  \subfloat[Scale choice C \eqref{scales:C}\label{fig:H3_higgs_scales:scaleC}]{\includegraphics[width=0.49\textwidth]{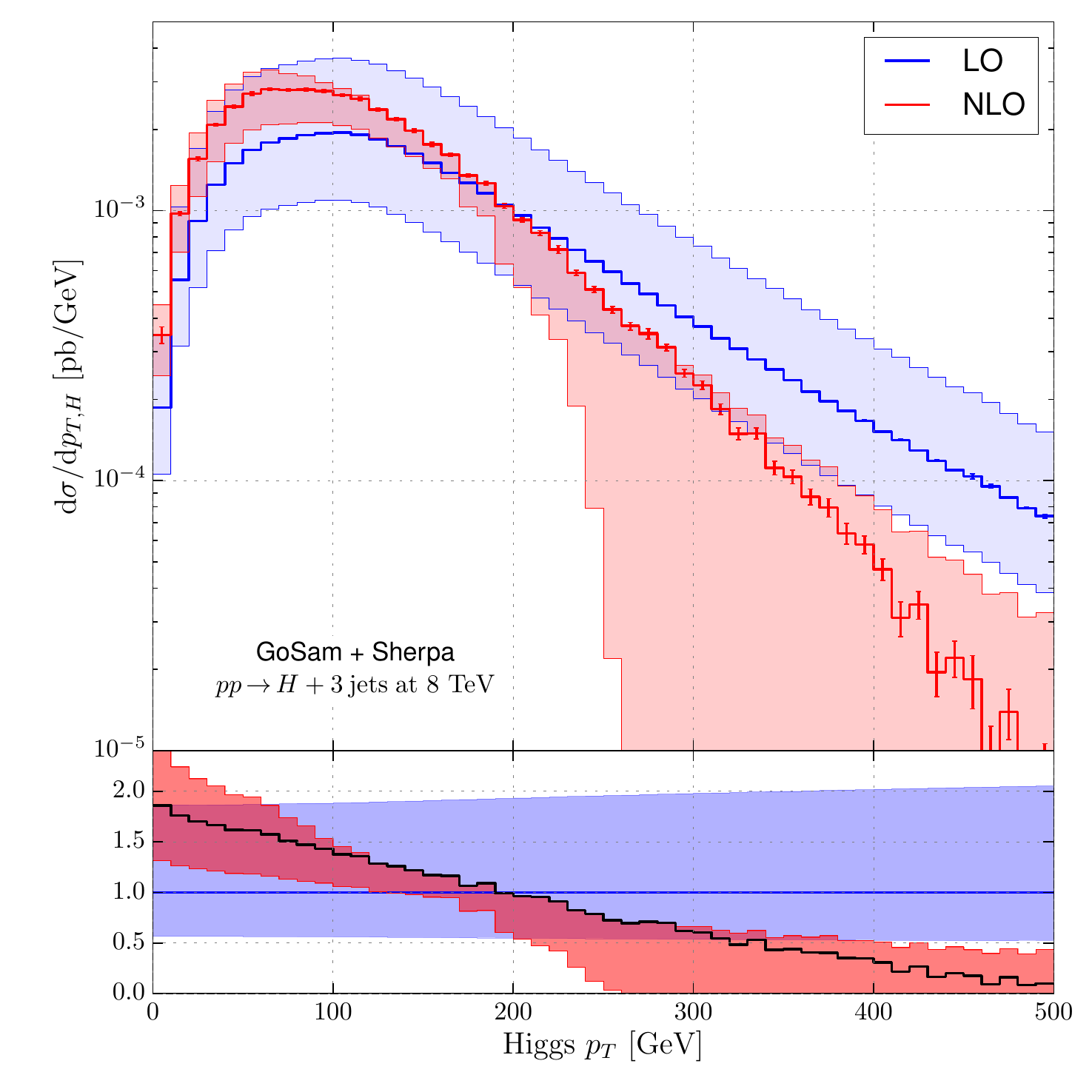}}
  \caption{The $p_T$-distribution of the Higgs in the \Hjjj process for 8 \TeV presented for
  the three scales A, B and C of Eq.~\eqref{scales}. The subplot \ref{fig:H3_higgs_scales:scaleRatio} shows the same distributions
  normalized to the NLO result for scale A.}
  \label{fig:H3_higgs_scales}
\end{figure}

We now turn to the discussion of one-particle or one-jet observables.
Figures~\ref{fig:H3_higgs_scales:scaleA}--\ref{fig:H3_higgs_scales:scaleC}
show the transverse momentum distribution of the Higgs in the \Hjjj process for 
the three different scale choices A, B and C of Eq.~\eqref{scales}, while Fig.~\ref{fig:H3_higgs_scales:scaleRatio}
shows the results for the different scales normalized to the NLO result for scale A.
The advantage of scale B is the flatness of the K-factor over the entire $p_T$ range.
This supports our choice to make
scale B the default scale. For the lower $p_T$ region up to $\sim 250$ \GeV also scale C seems to
be a sensible choice. However it completely breaks down for higher $p_T$, and the K-factor
can even become negative.\footnote{The same behavior has also been found for instance for \Wjjj
\cite{Berger:2009ep}.}
The different behaviors of the scales are more pronounced in Fig.~\ref{fig:H3_higgs_scales:scaleRatio}.
For scale C the leading order curve shows the opposite behavior compared to the NLO curve. For scale A
the situation is not as bad, but the K-factor still has a strong dependence on the
$p_T$ of the Higgs. This is much better for scale B, for this choice the LO and NLO curve are
almost parallel, which is a further confirmation that using $\hthatprime$ for all
factors of $\alphaS$ is a sensible choice.

\begin{figure}[t!]
  \centering
  \includegraphics[width=0.49\textwidth]{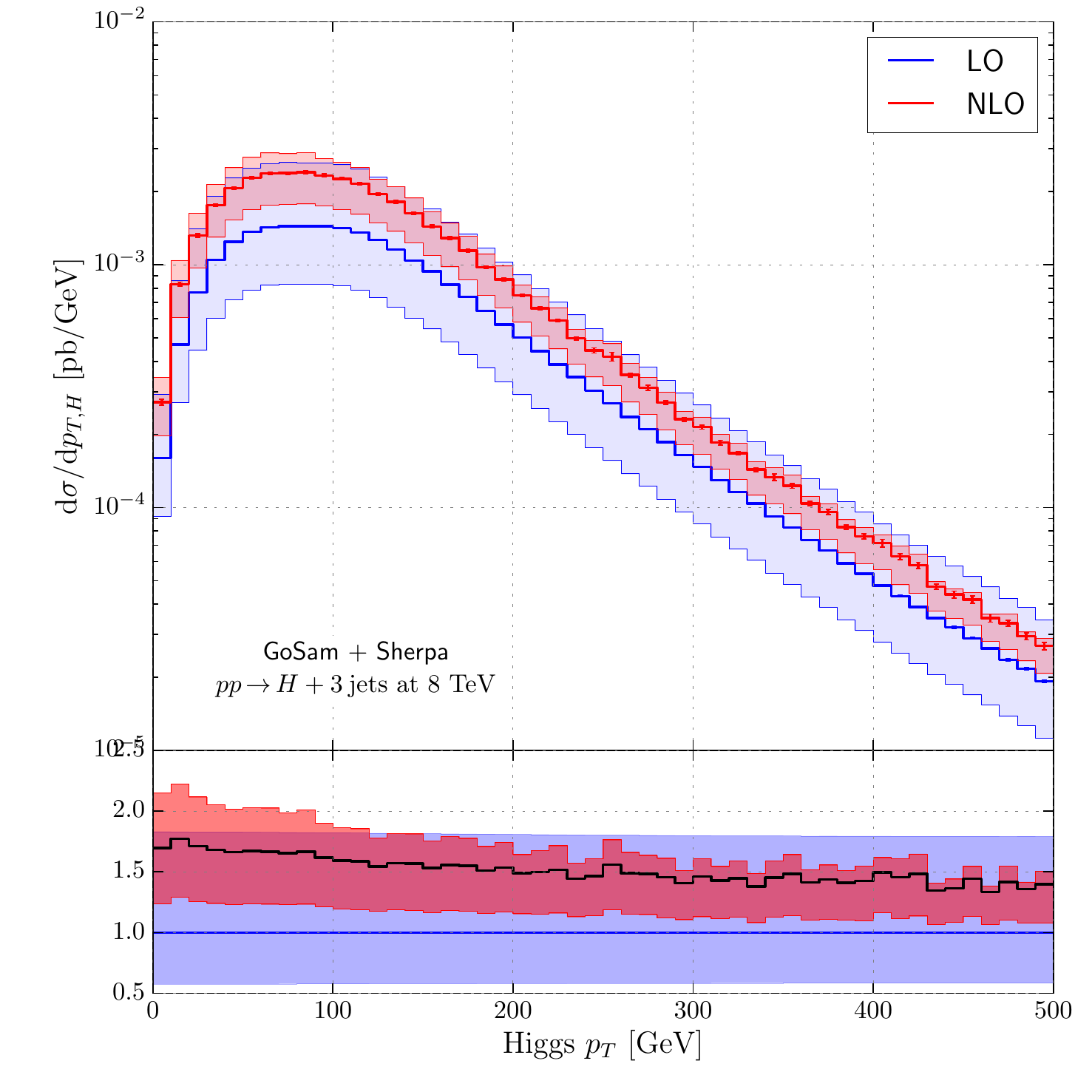}
  \hfill
  \includegraphics[width=0.49\textwidth]{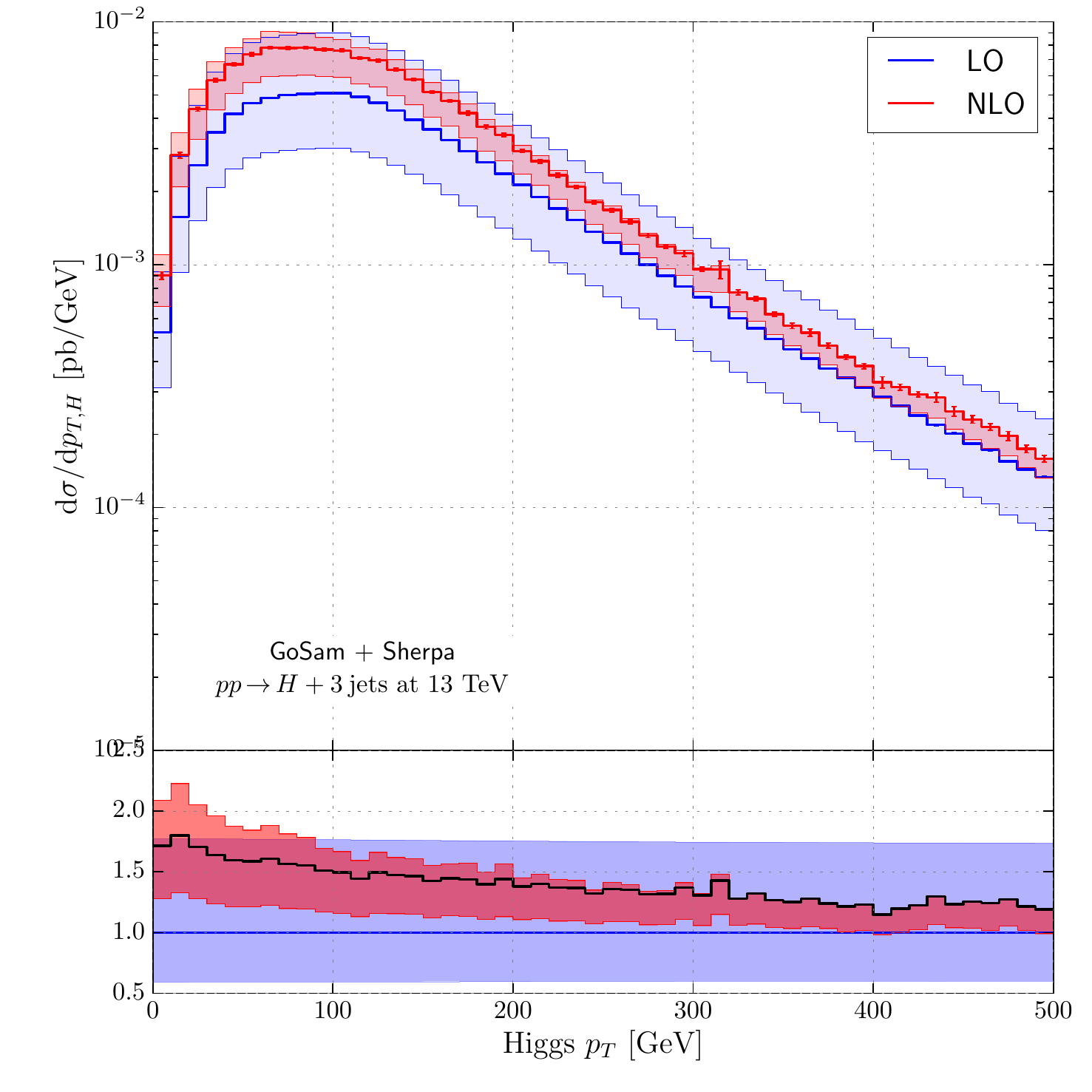}
  \\
  \includegraphics[width=0.49\textwidth]{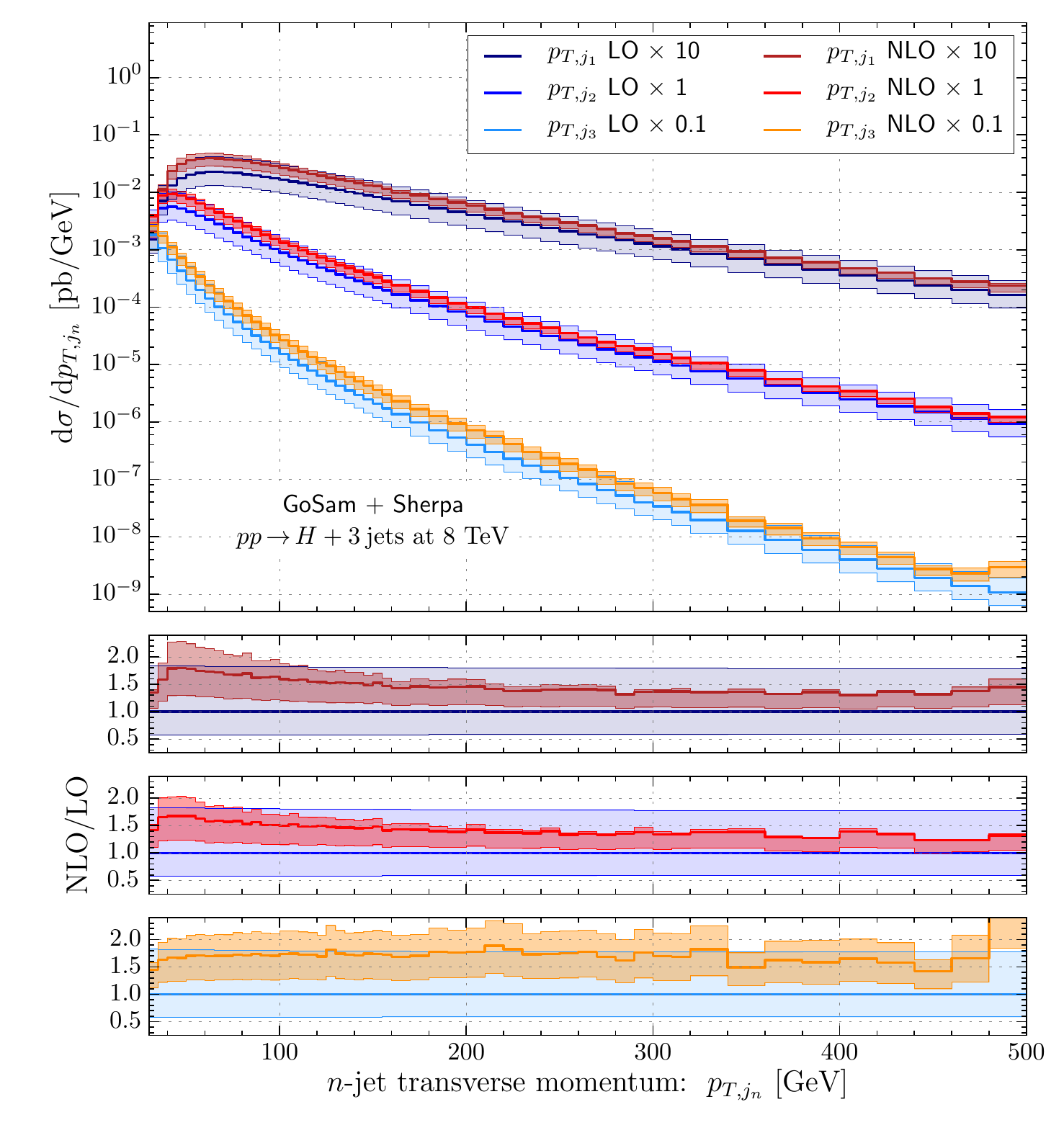}
  \hfill
  \includegraphics[width=0.49\textwidth]{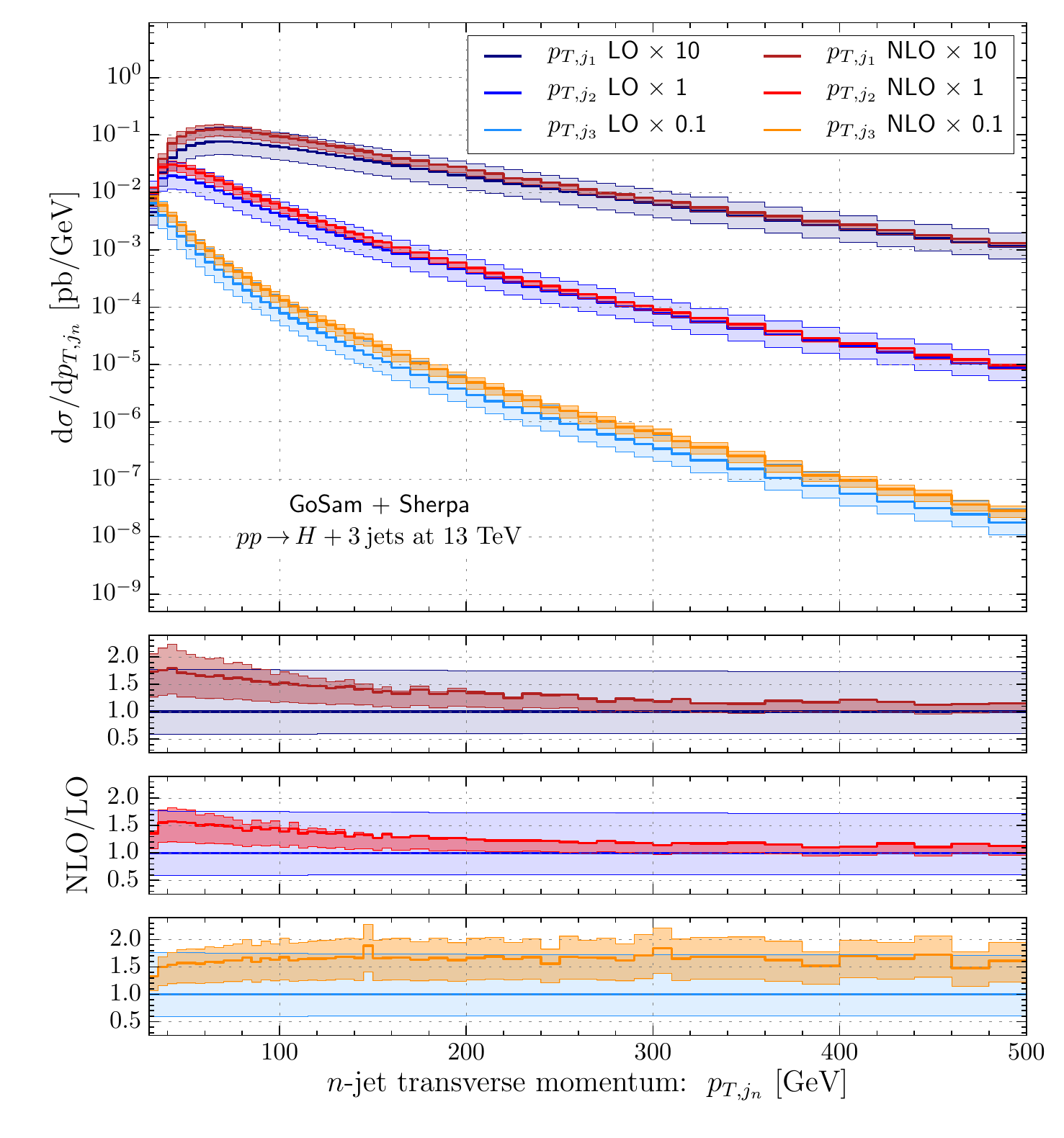}
  \caption{The $p_T$ distributions of the Higgs boson (upper row) and
    the three hardest jets (lower row) for 8 \TeV (l.h.s.)
    and 13 \TeV (r.h.s.). For the $p_T$ distribution of the jets, the
    curves for the first and the third jet
    have been rescaled by a factor of $10$ and $1/10$ respectively.
    }
  \label{fig:pt}
\end{figure}

In Fig.~\ref{fig:pt} we compare the $p_T$ distributions of the Higgs
and the associated jets at center of mass energy of 8 \TeV (left plots)
with the 13 \TeV (right plots) prediction.  The upper row shows the LO
and NLO result for the Higgs $p_T$. One observes that the NLO
corrections lead only to a mild change of the shape, with a small
decrease of the K-factor in the high-$p_T$ tail. Comparing the two
center of mass energies, we also see that the peak of the distribution
remains in the region slightly below 100 \GeV, for both LO and NLO.
The distribution is steeper for 8 \TeV, and the theoretical uncertainties 
at NLO are slightly smaller at 13 \TeV.

In the lower row we show the transverse momentum distribution for the three
leading-$p_T$ jets. For a better visibility of the different curves we
have rescaled the first jet by a factor of $10$, the third jet by a
factor of $0.1$, the second jet is shown unchanged. On a logarithmic
scale this leads to a vertical shift of the different curves,
preserving however their shapes and the size of the uncertainty
bands. It is therefore possible to better appreciate the different
behavior of the curves over the considered kinematical range. In the
ratio plots below we show the NLO/LO ratio for each pair of curves,
this means that each jet distribution is normalized to its own LO
distribution. Looking at the distributions we observe to a larger
extent the patterns seen for the Higgs-$p_T$: The peak of
the distribution is almost insensitive to a change of the center of
mass energy, but in the 8 \TeV result the tail of the curve decreases
faster. For the first and the second jet one also obtains a slight
decrease in the K-factor towards higher energies and a small reduction
of the theoretical uncertainties when going from 8 to 13 \TeV. The
third jet however behaves a bit differently. The K-factor is almost
flat and the size of the scale uncertainties remains constant
towards higher energies whereas for the first two jets one obtains a
slight reduction.

\begin{figure}[t!]
  \centering
  \includegraphics[width=0.49\textwidth]{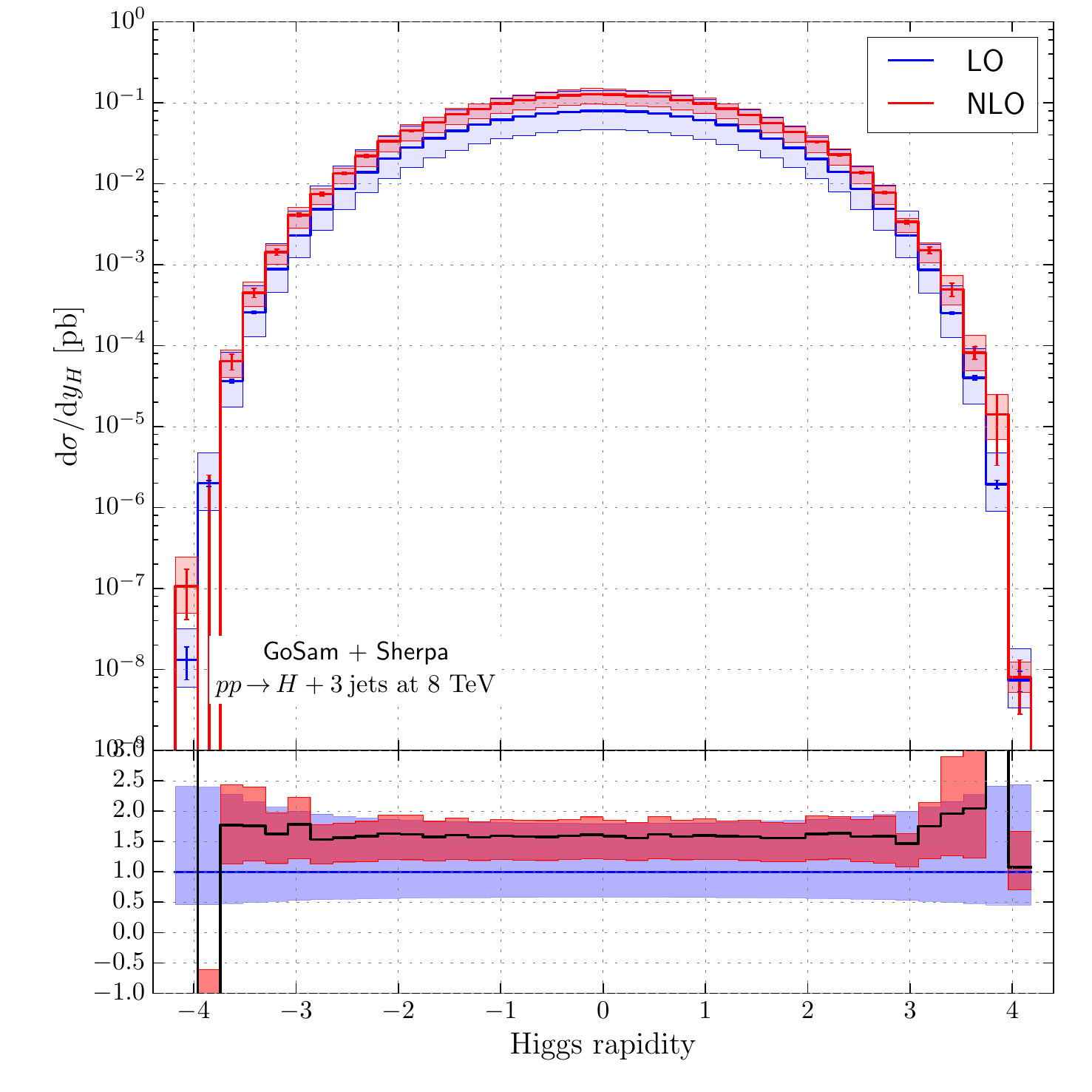}
  \hfill
  \includegraphics[width=0.49\textwidth]{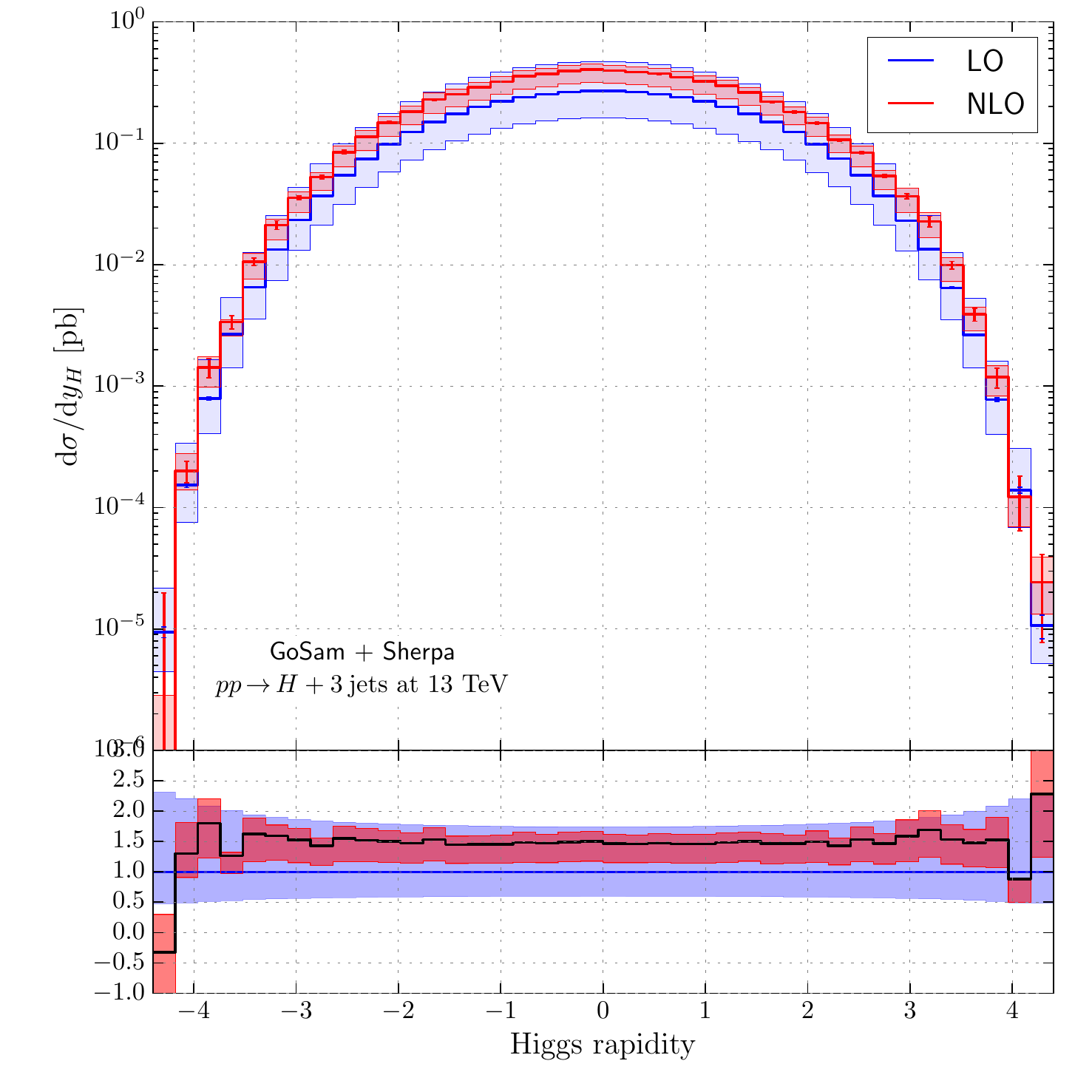}
  \\
  \includegraphics[width=0.49\textwidth]{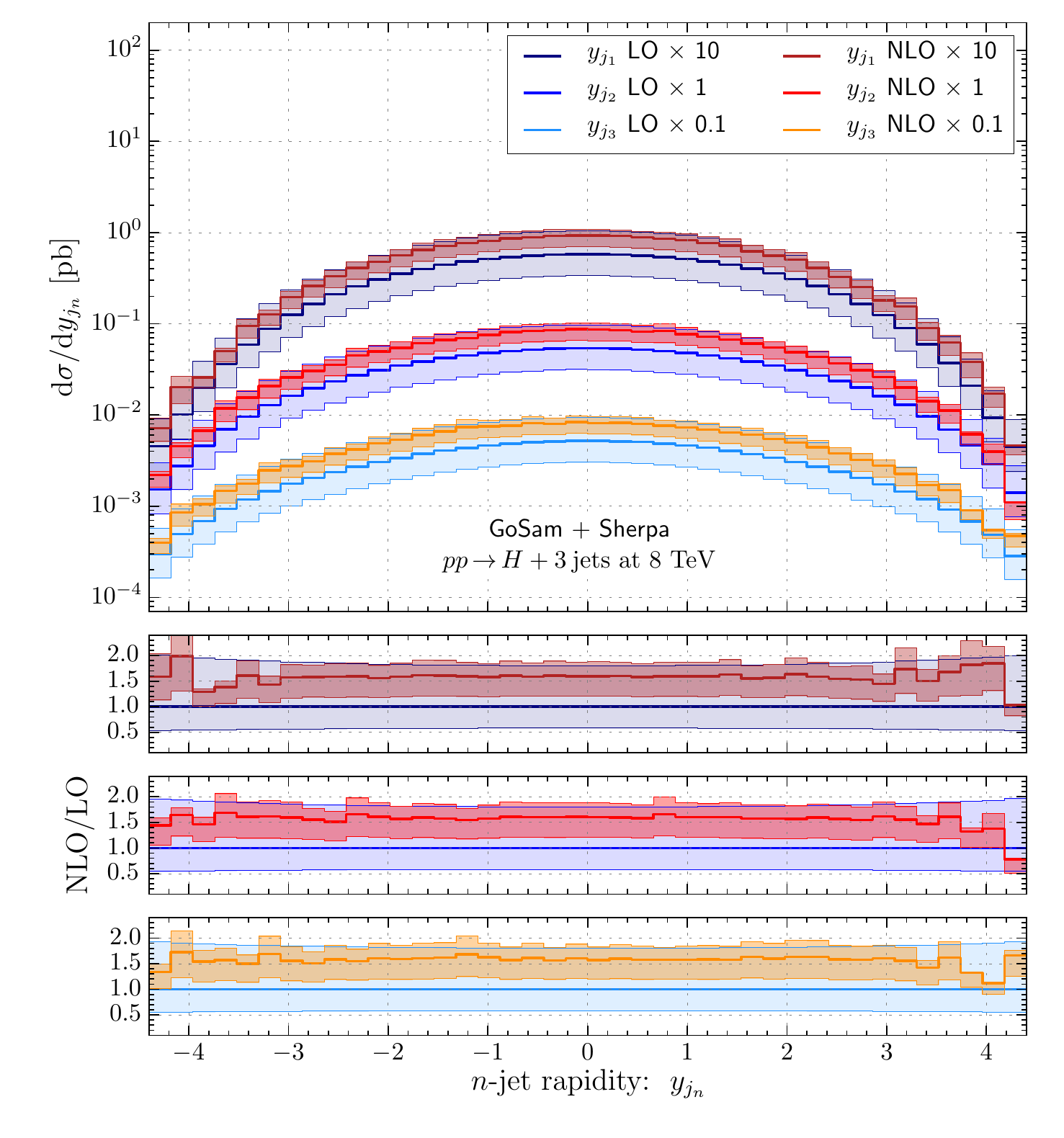}
  \hfill
  \includegraphics[width=0.49\textwidth]{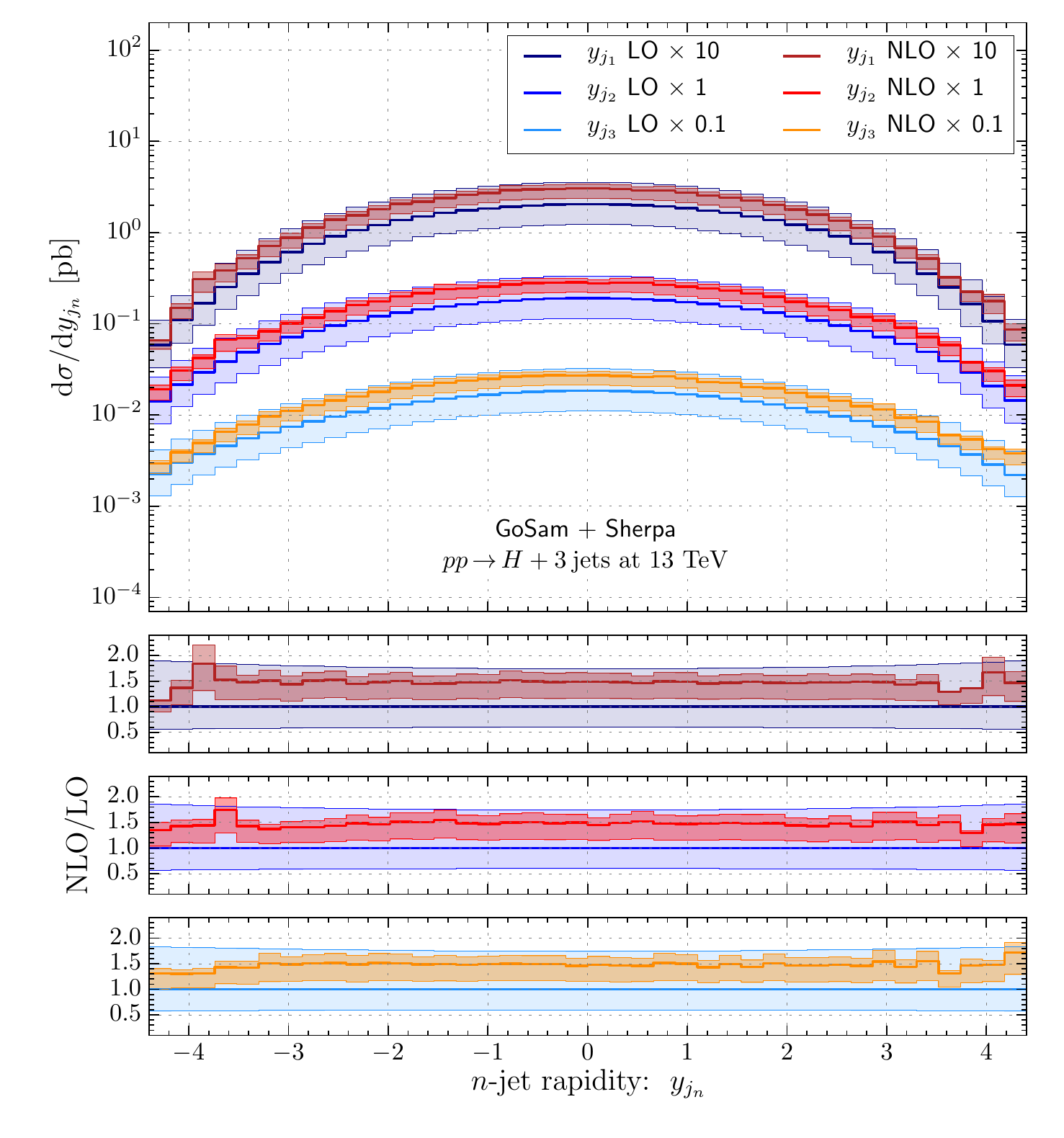}
  \caption{The rapidity distributions of the Higgs boson (upper row)
    and the three hardest jets (lower row) for 8 \TeV (l.h.s.)
    and 13 \TeV (r.h.s.). For the rapidity distribution of the jets, the curves for the first and the third jet
    have been rescaled by a factor of $10$ and $0.1$ respectively.
    }
  \label{fig:y}
\end{figure}

Another important observable is the rapidity distribution which is
shown in Fig.~\ref{fig:y} for the Higgs (upper row) and the three jets
(lower row) again for both 8 \TeV (left column) and 13 \TeV (right
column). Starting with the rapidity of the Higgs we first of all
observe a very flat K-factor across the whole range of the
distribution. Comparing the 8 \TeV with the 13 \TeV result this still
holds and is accompanied by a mild reduction of the scale uncertainty
in the 13 \TeV result, as already observed for the $p_T$
distributions. However the shape of the distribution changes for both
LO and NLO predictions when increasing the center of mass energy. For
8 \TeV the fraction of Higgs particles in the central region is higher
and we see a steeper decline of the cross section towards large values
of the rapidity. This is much less pronounced for the 13 \TeV case.
There we get a relative enhancement of the regions with a large
rapidity.

For the rapidity distribution of the jets in the lower row
of Fig.~\ref{fig:y} we have applied the same scaling procedure as for
the $p_T$ distribution in order to obtain a better readability of the
plots.  Also for the jets the K-factor is flat to a very good
approximation, with a small reduction of the scale uncertainties when
going from 8 to 13 \TeV. As for the Higgs, one can observe a relative
enhancement of the regions with large rapidities in the 13 \TeV
result. This is due to the fact that the increased center of mass
energy increases allows to fill more the phase space corners where the
particles are scattered towards the forward/backward regions, while
at the same time fulfilling the $p_T$ requirements on the jets.

\begin{figure}[t!]
  \centering
  \parbox[t]{\textwidth}{
  \includegraphics[width=0.49\textwidth]{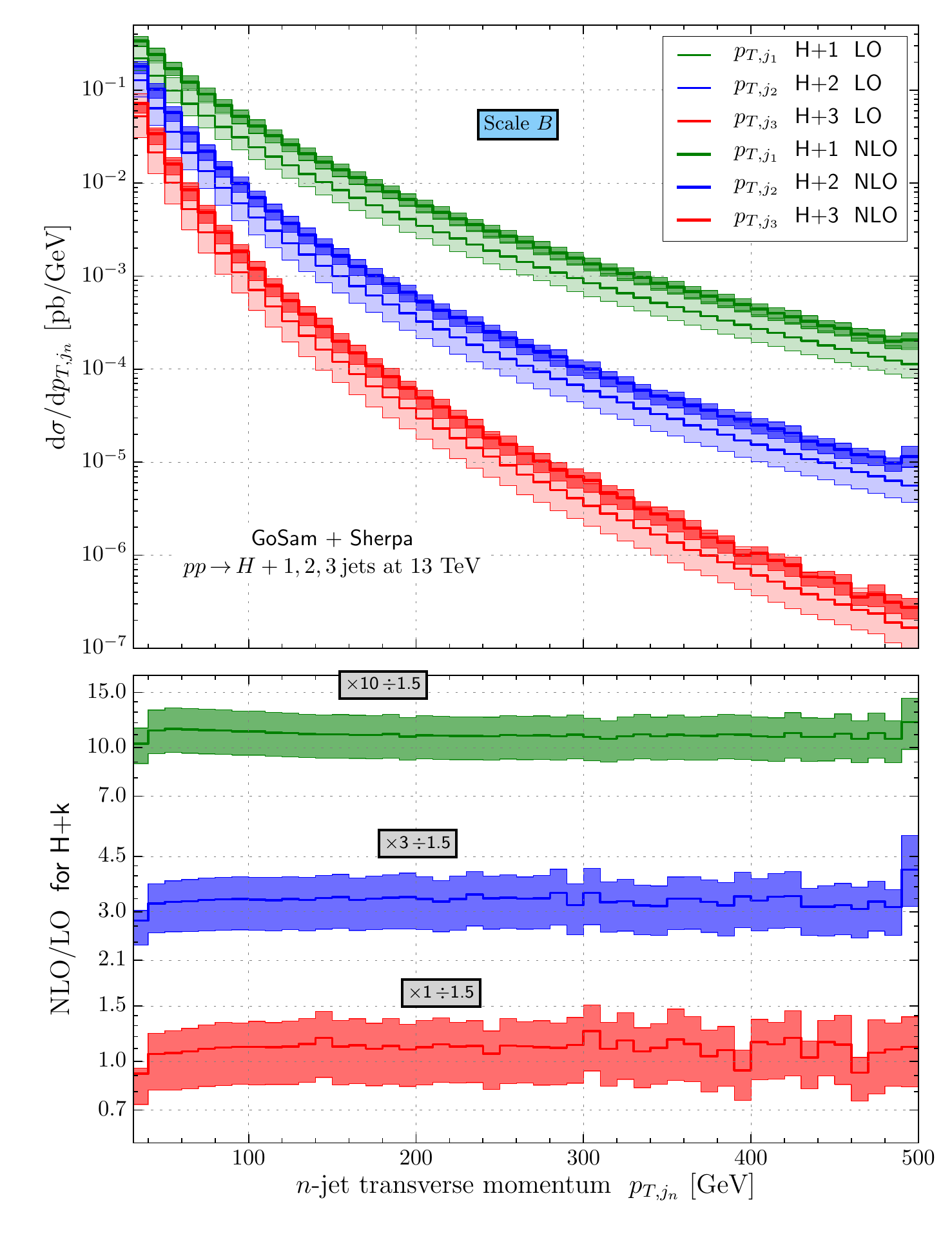}
  \hfill
  \includegraphics[width=0.49\textwidth]{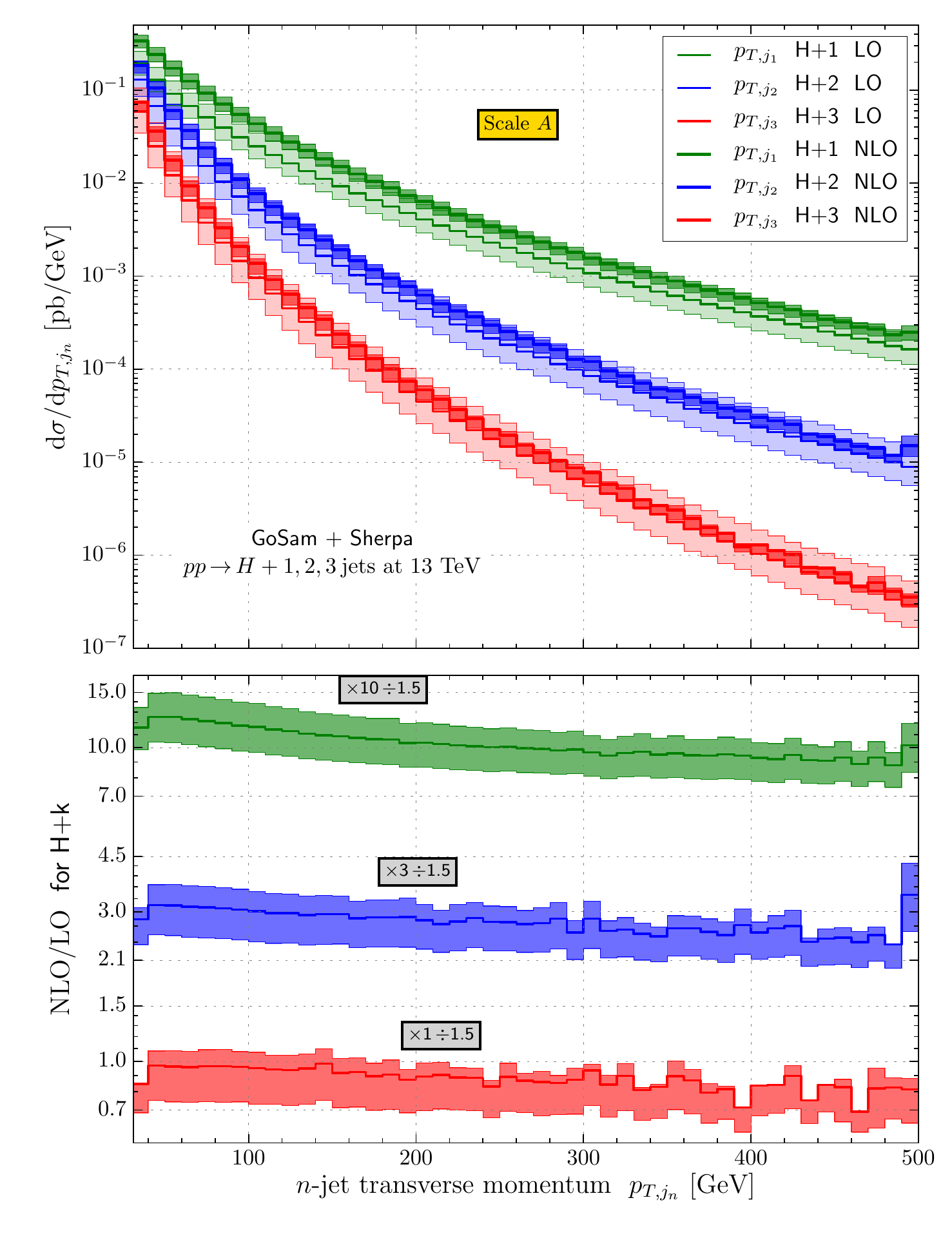}
  }
  \caption{\label{fig:subleadingjetpt}%
    Transverse momentum distribution of the first, second and third
    leading jet in \Hj, \Hjj and \Hjjj, respectively; on the left with
    the default scale choice B, on the right with the scale choice A.}
\end{figure}

We conclude this section by discussing the impact of higher-order corrections
on the wimpiest jet in \Hj, \Hjj and \Hjjj configurations. In NLO calculations
for \Wboson/\Zboson+jets performed with scale B, it was noted that the
transverse momentum spectrum
of this jet exhibits a flat K-factor~\cite{Joey}. We test for the effect in Higgs+jets
production for the first time, and we find a similar behavior, as exemplified in
Fig.~\ref{fig:subleadingjetpt} (left). The green curves show the first jet in \Hj,
the blue ones the second jet in \Hjj, and the red ones the third jet in \Hjjj.
The ratio plots show the transverse momentum dependent K-factors for
the three cases,
scaled by factors of 20/3 (\Hj), 2 (\Hjj), and 2/3 (\Hjjj). It is evident that the K-factors are not only
flat over the entire range, but they are also very similar for all three calculations.
Fig.~\ref{fig:subleadingjetpt} (right) shows the same analysis for the scale choice A.
In this case the K-factors have a larger transverse momentum dependence. Their average
also differs much more between the different jet multiplicities.

\subsection{Multi-particle observables and correlations}
\label{sec:gf:multiobs}

Multi-particle or multi-jet observables are at the core of any measurement
that involves many objects in the final state. They allow to test QCD
dynamics at the LHC to an unprecedented precision, and they often reveal
inappropriate modeling by LO calculations or by MC
event generators. 

\begin{figure}[t!]
  \centering
  \includegraphics[width=0.49\textwidth]{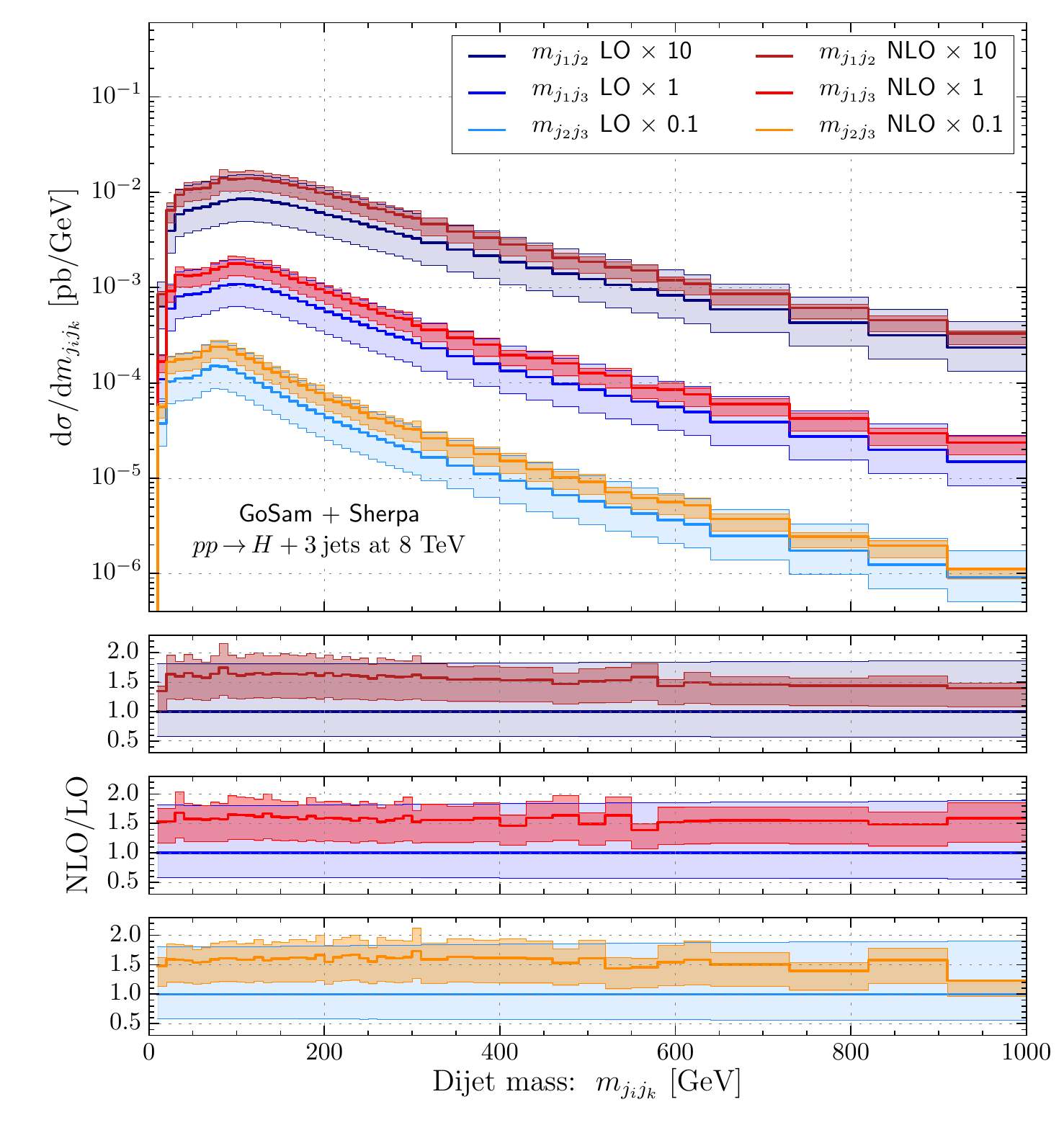}
  \hfill
  \includegraphics[width=0.49\textwidth]{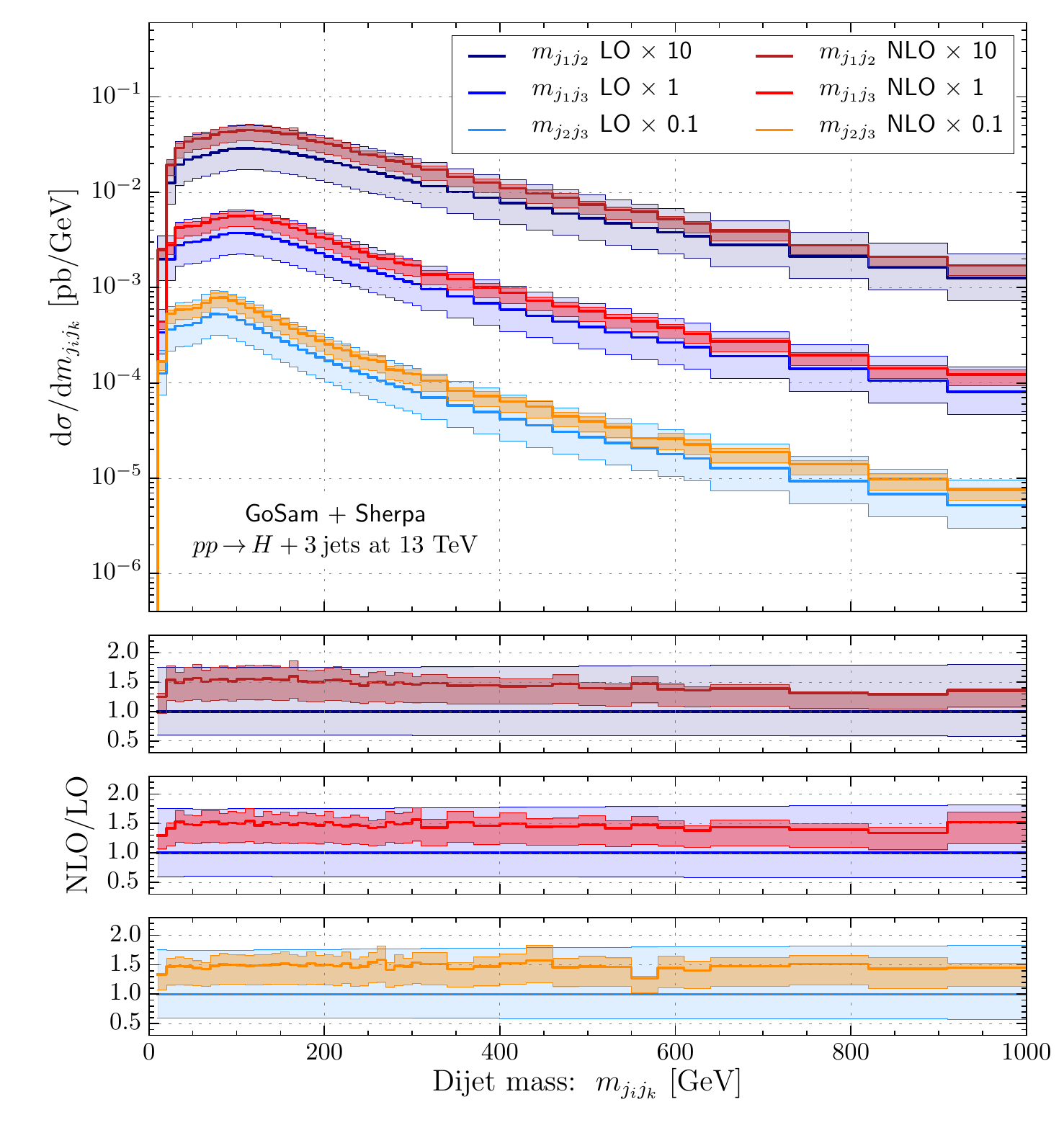}
  \caption{Invariant mass distribution for the dijet systems combined
    from the three hardest jets. Results are shown for 8 \TeV (l.h.s)
    and 13 \TeV (r.h.s). Jets are $p_T$-ordered.}
  \label{fig:mjj}
\end{figure}

Figure~\ref{fig:mjj} shows the dijet invariant mass 
distribution, $m_{j_ij_k}$, for each of the three
possible combination $(i=1,k=2),\,(i=1,k=3),\,(i=2,k=3)$, where the
jets are ordered in transverse momentum. The left panels show results 
for 8 \TeV, the right panels are for 13 \TeV. 
In order to avoid overlaps in the figure, the
curves for $m_{j_1j_2}$ are scaled by a factor of $10$, whereas
the curves for $m_{j_2j_3}$ are scaled by a factor $1/10$. We 
observe a steeper decrease of the distribution in the 8 \TeV case and
also for softer jets, as expected. Comparing the left and right panels, 
one observes that the maximum of the curves is to a good approximation 
independent of the collider energy.

In the lower part of the plots we show separately the K-factors for the
three distributions. Apart from the expected reduction of the
theoretical uncertainty we observe a K-factor that is approximately
constant for both energies and for all the three jet
combinations. Only the invariant mass of the two leading jets,
$m_{j_1j_2}$, shows a small decrease in the relative size of the NLO
corrections for higher values, in particular at 13 \TeV. This is of
course to a large extent due to the scale choice. 

\begin{figure}[t!]
  \centering
  \includegraphics[width=0.49\textwidth]{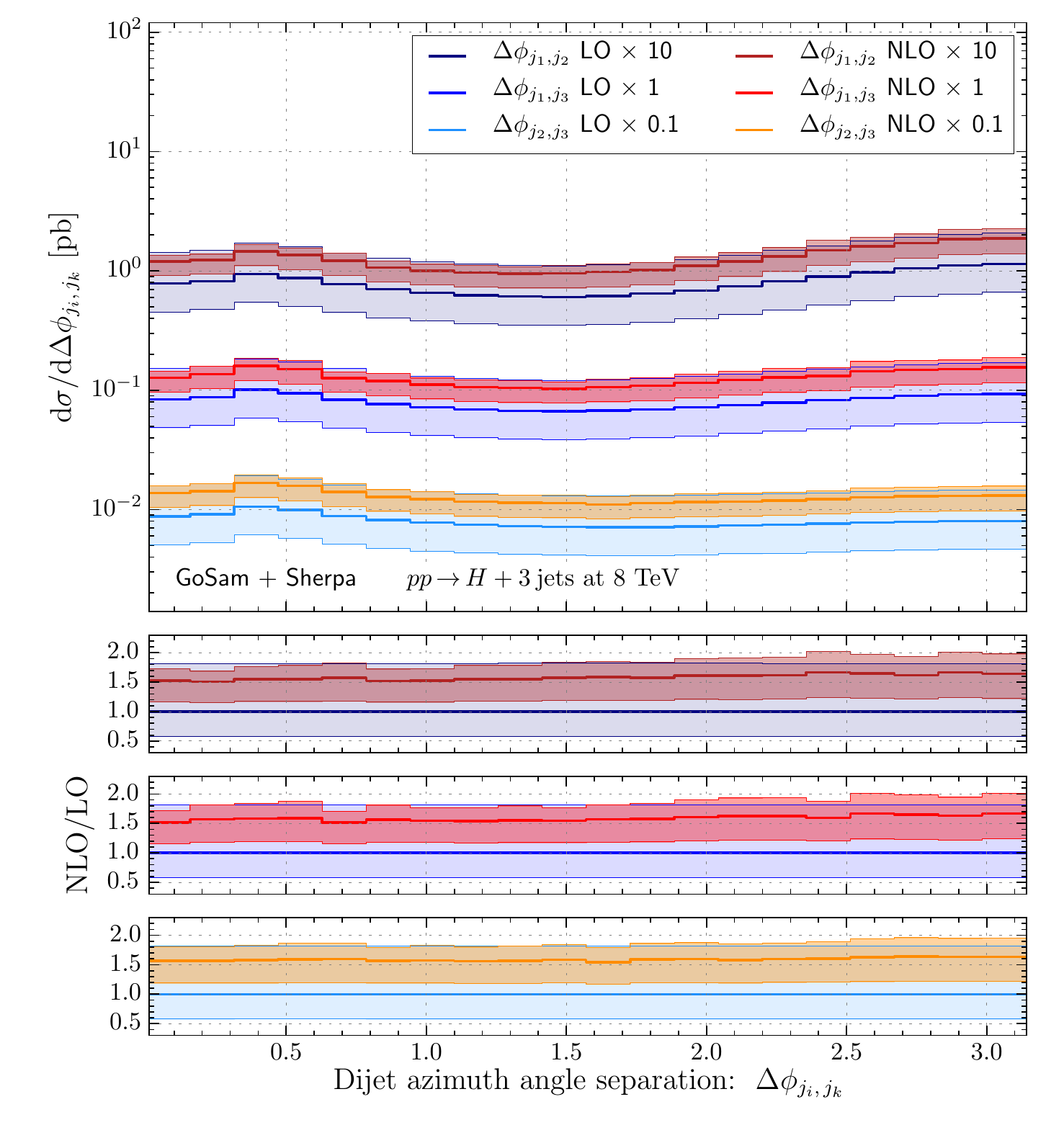}
  \hfill
  \includegraphics[width=0.49\textwidth]{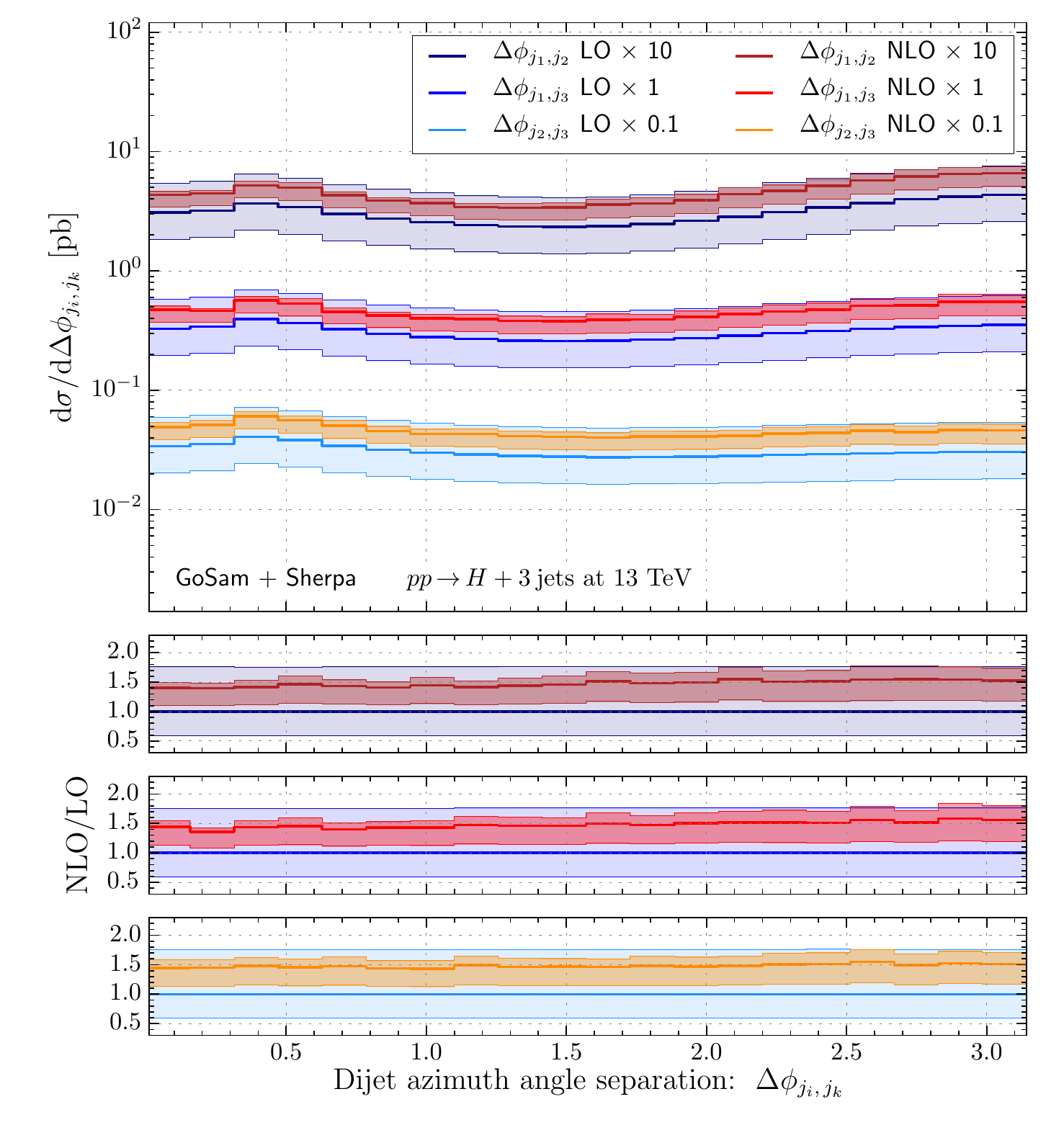}
  \caption{Distribution of the azimuthal angle $\Delta\phi$ between
    the first and the second jet (upper row),
    and between the second and the third jet (lower row). Results are shown for
    8~\TeV (l.h.s.) and 13~\TeV (r.h.s.).}
  \label{fig:dphi}
\end{figure}

A further observable that is particularly important in view of vector
boson fusion processes is the azimuthal angle~$\Delta\phi$ between
jets, as shown in Fig.~\ref{fig:dphi}.  Again we
give predictions for three different pairs of jets at both 8 \TeV, on
the left, and 13 \TeV, on the right. As for the invariant masses one
observes a flat K-factor for all combinations and for both energies. The
shape however changes when increasing the energy. This is particularly
visible in the peak regions, which are slightly more pronounced at 13
\TeV. Their position is related to the choice of the jet radius, and in
particular with configurations where the two jets have $\Delta y=0$
and the separation in $\phi$ exactly corresponds to the chosen value of $R$.

Further multi-particles observables, less related to multi-jet QCD
dynamics and more specific to Higgs production in association with
jets were already shown before in Section~\ref{sec:gf:xsecs} in
Figs.~\ref{fig:higgs12} and~\ref{fig:higgsjetdphi}.

\subsection{Multi-jet ratios at NLO}
\label{sec:inclusivevsexclusive}

In this section we ask the question how observables change in the
presence of additional QCD radiation, starting with a core process
of \Hj.  The VBF topology requires at least two jets, but our observations
hold in both cases, largely because the phase space available to QCD radiation
at the LHC is tremendous. This has been pointed out many times before, 
and a particularly nice example of the effects is given in
Ref.~\cite{Rubin:2010xp}.
In this section we use our NLO results for \Hjjj to make some of the
statements explicit.

\begin{figure}[t!]
  \centering~\hfill
  \includegraphics[width=0.439\textwidth]{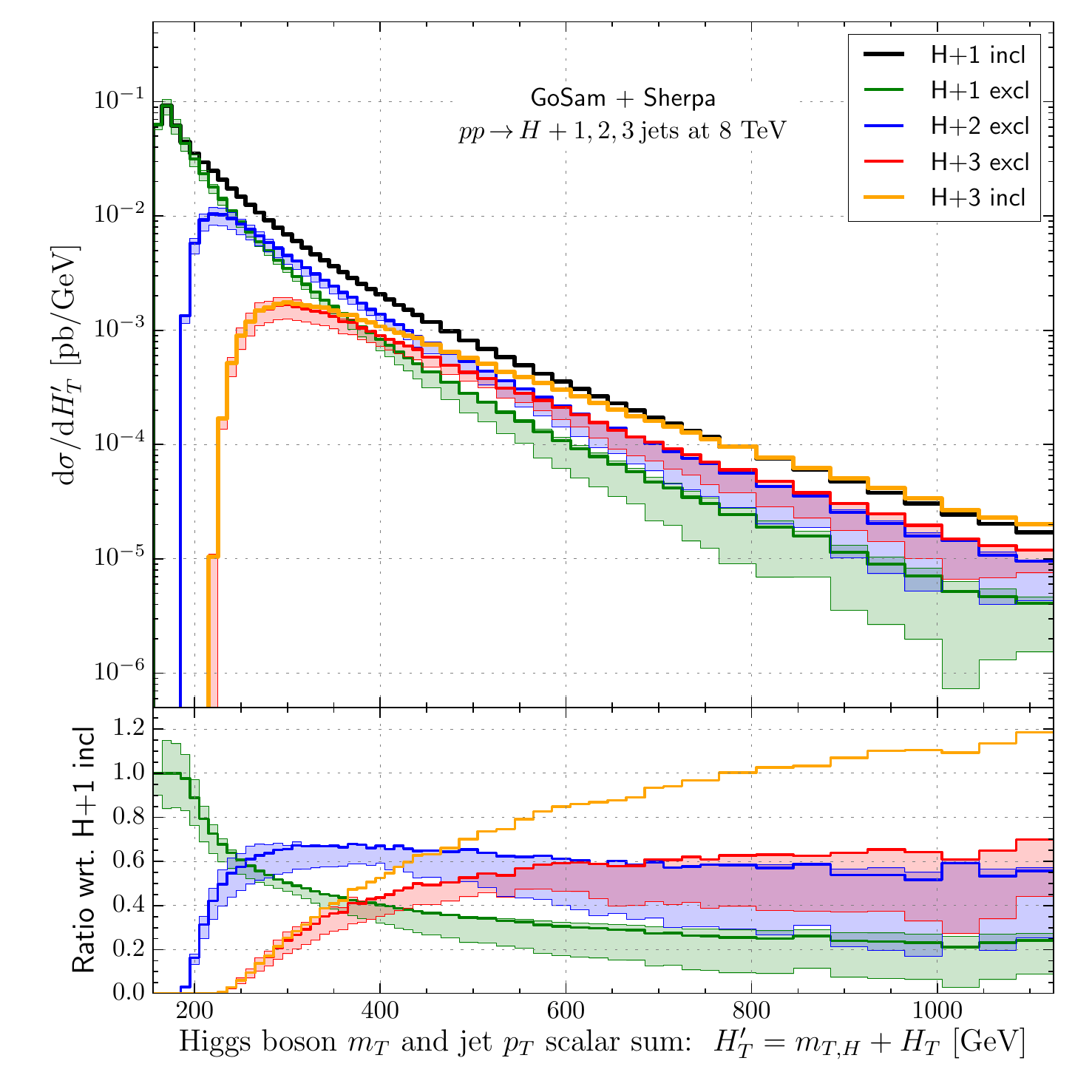}
  \hfill
  \includegraphics[width=0.439\textwidth]{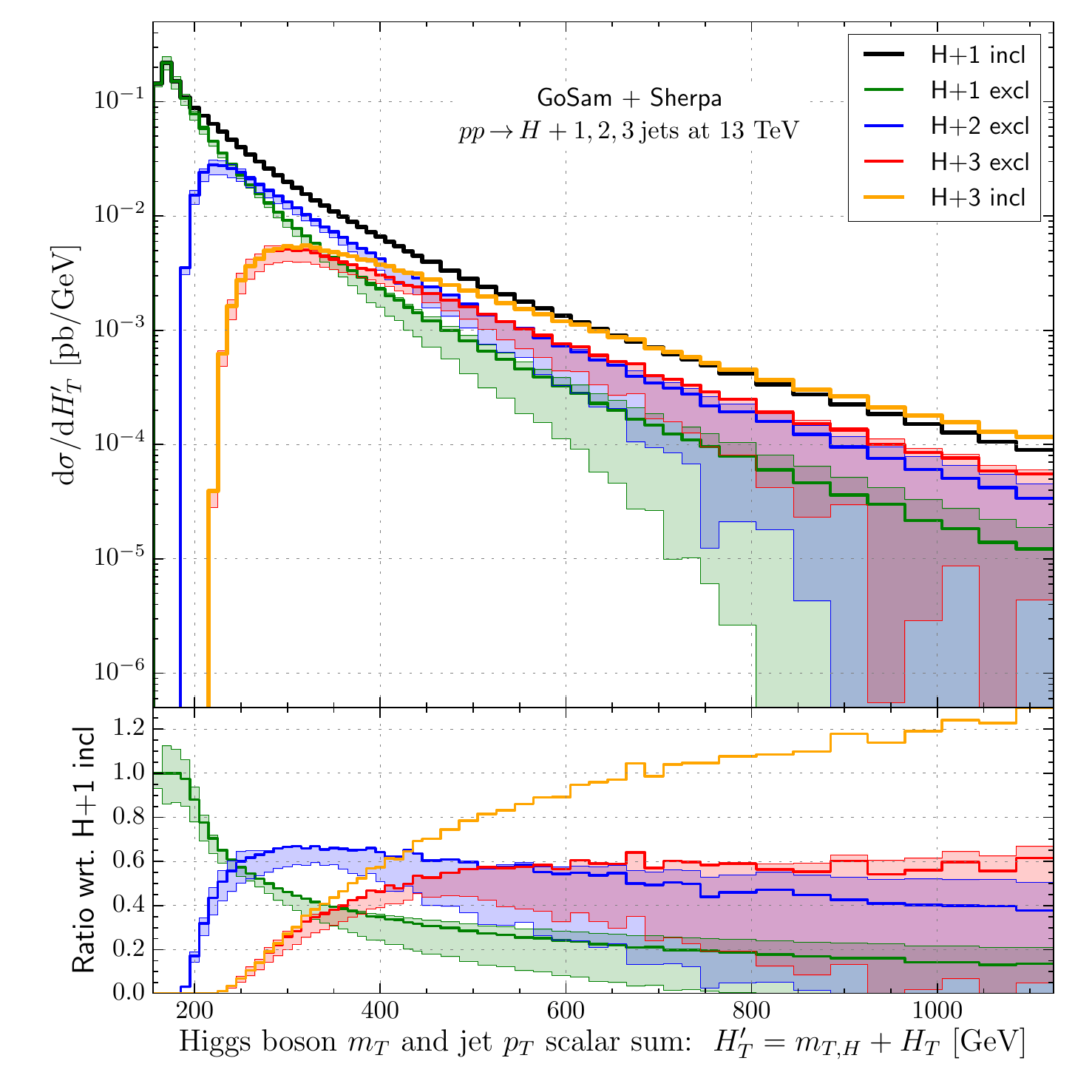}
  \hfill~\\~\hfill
  \includegraphics[width=0.439\textwidth]{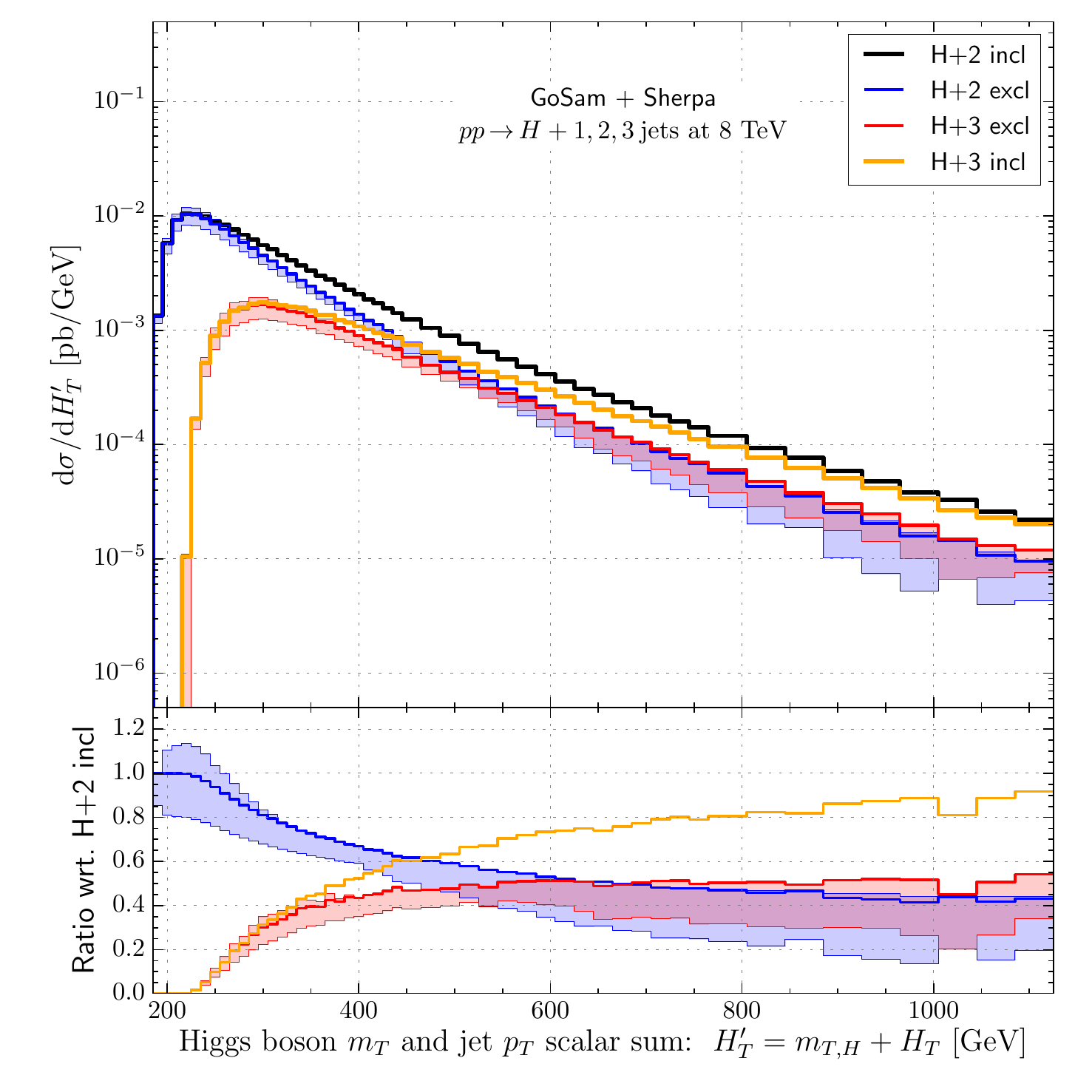}
  \hfill
  \includegraphics[width=0.439\textwidth]{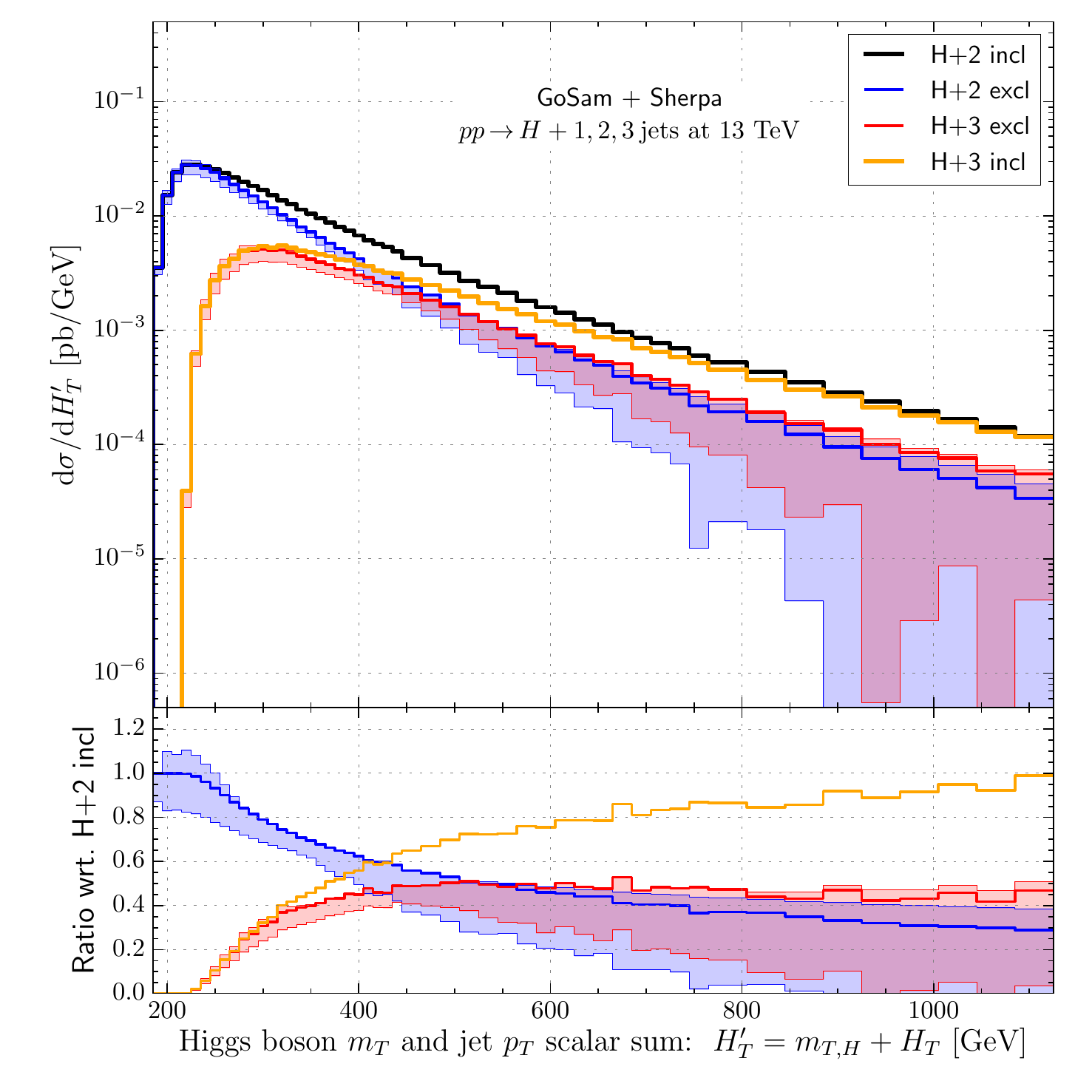}
  \hfill~\\~\hfill
  \includegraphics[width=0.439\textwidth]{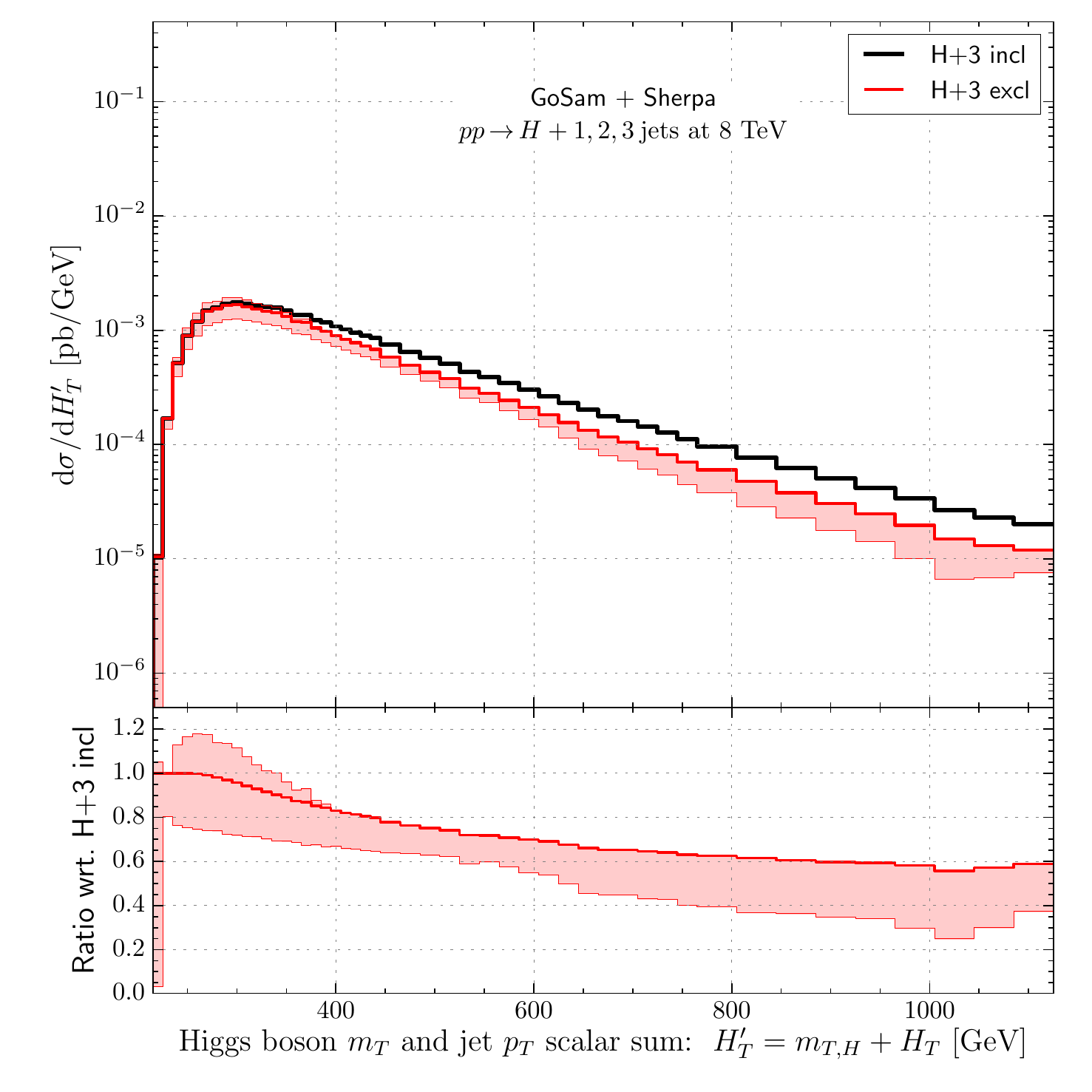}
  \hfill
  \includegraphics[width=0.439\textwidth]{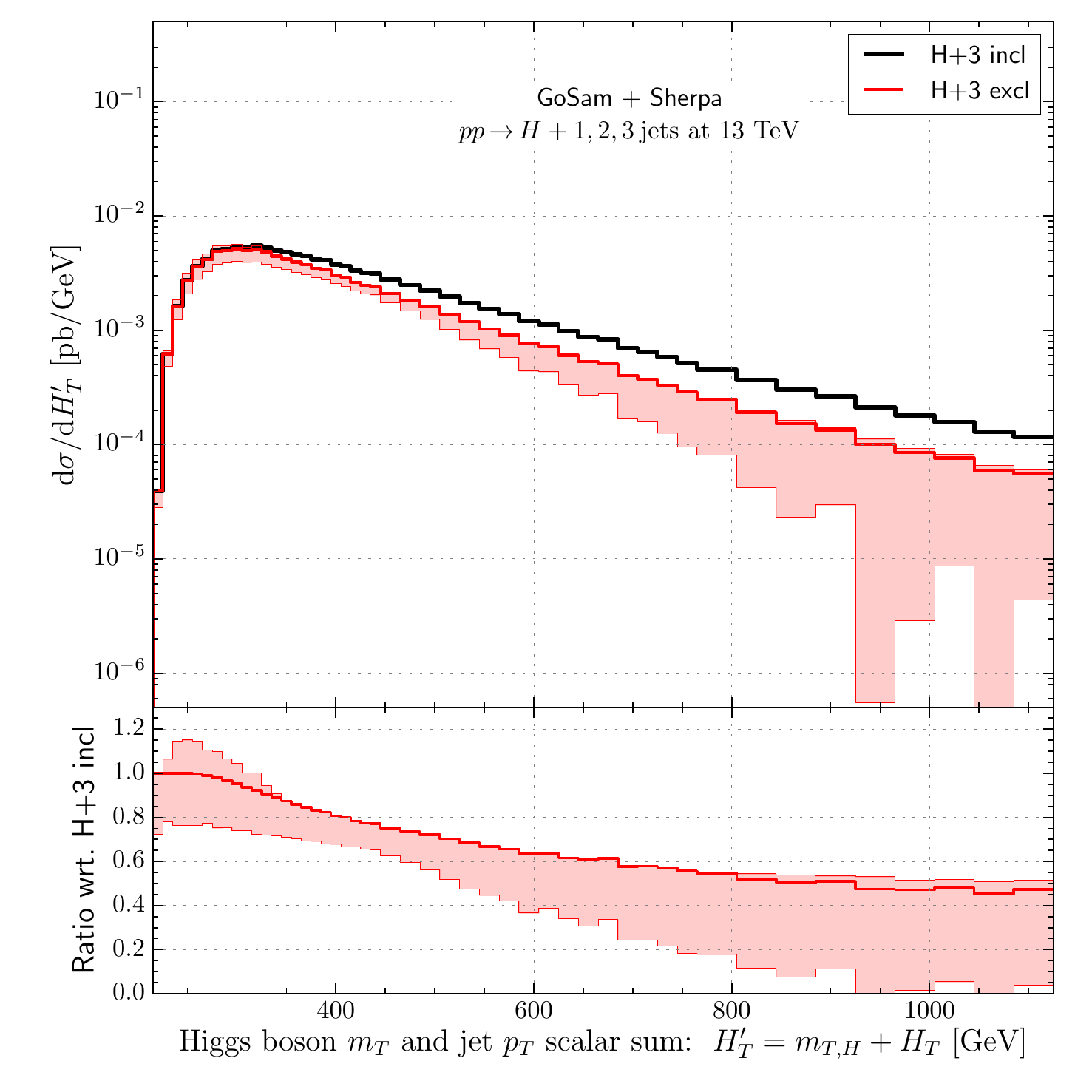}
  \hfill~
  \caption{\label{fig:xnlo-ratio-HTprime}%
    Inclusive and exclusive NLO rates for $H^\prime_T$ at 8 \TeV (left column) 
    and 13 \TeV (right column). The ratio plot details the magnitudes
    of the exclusive and higher multiplicity inclusive calculations relative to 
    the most inclusive calculation, \Hj. See text for more explanations.}
\end{figure}

\begin{figure}[t]
  \centering~\hfill
  \includegraphics[width=0.424\textwidth]{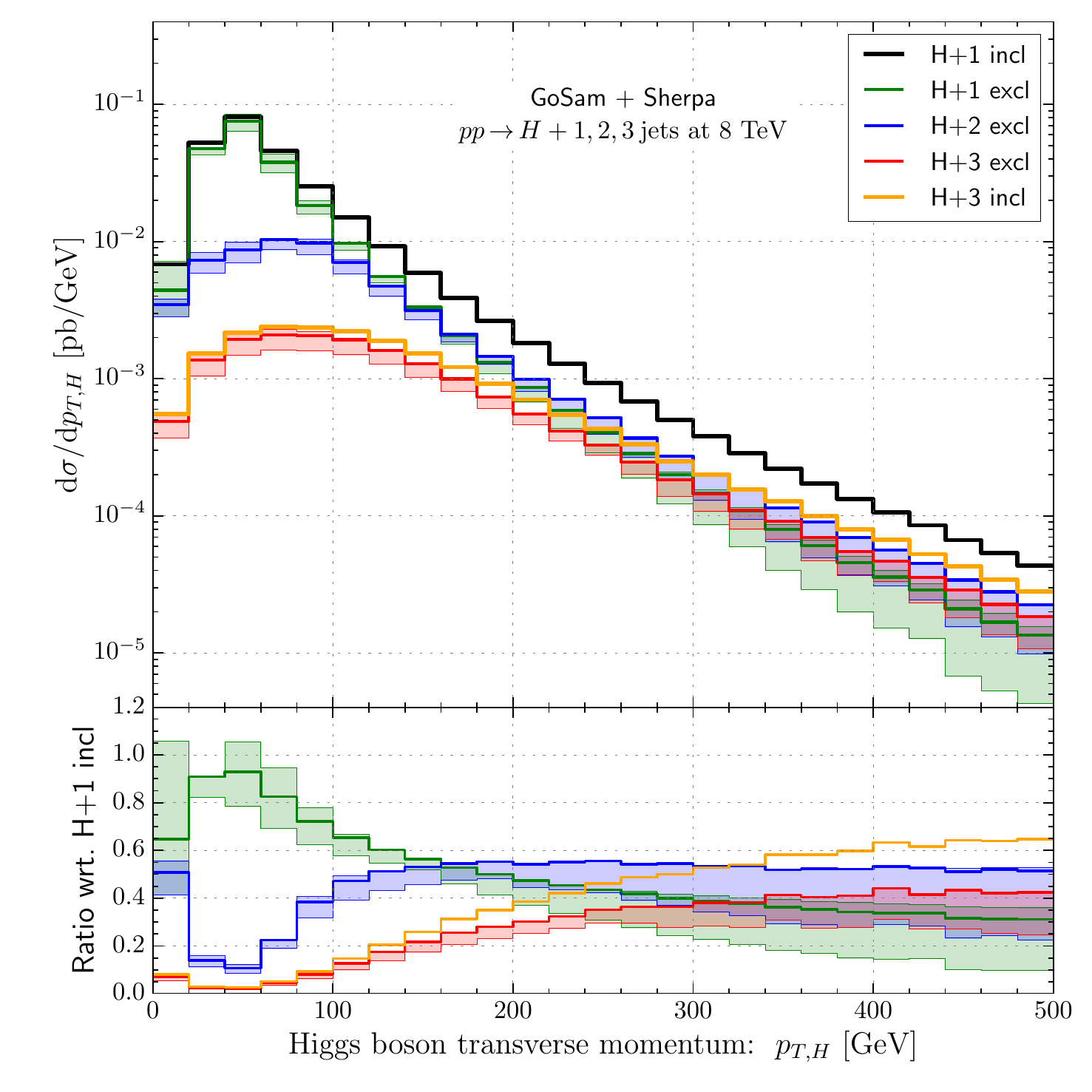}\hfill
  \includegraphics[width=0.424\textwidth]{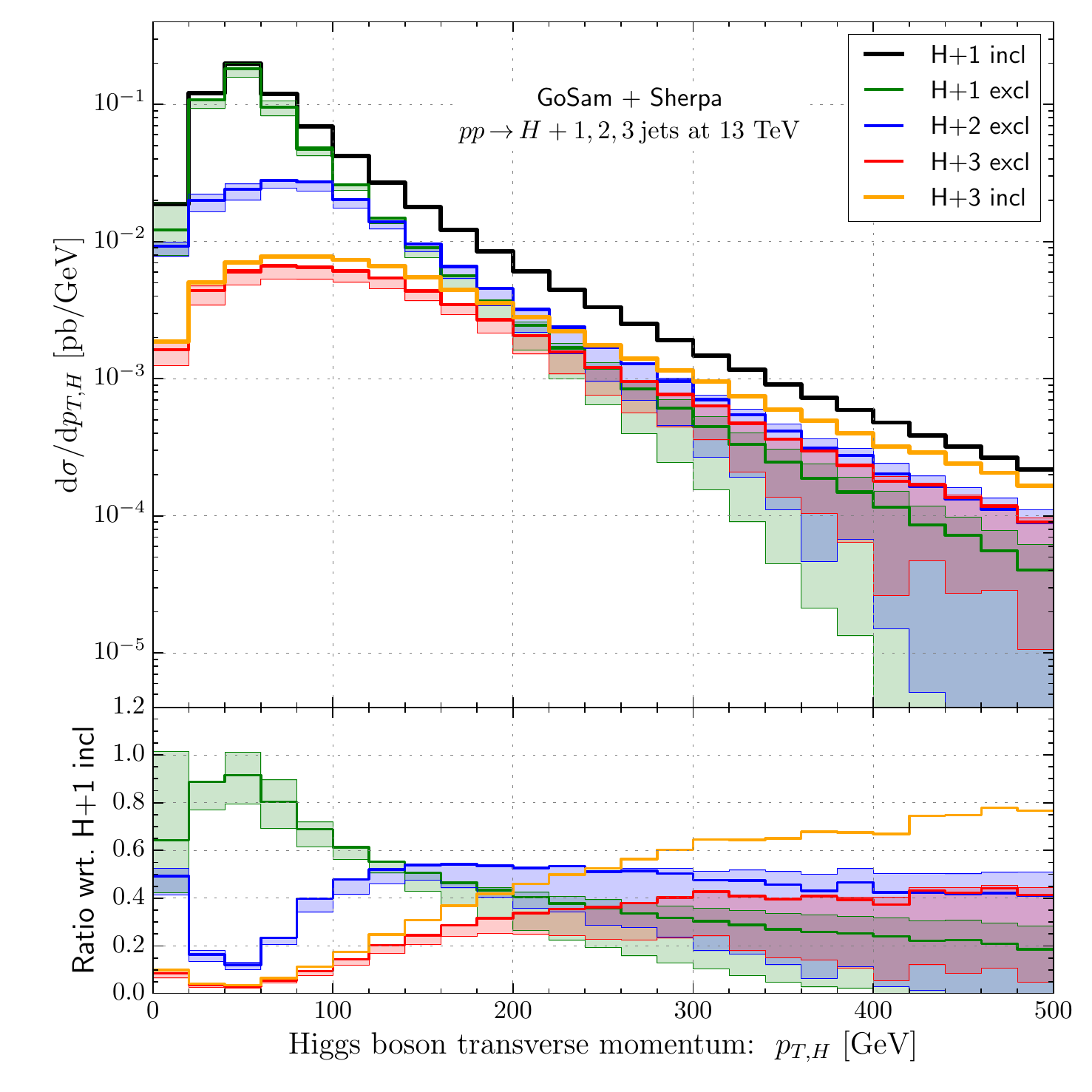}\hfill~
  \\~\hfill
  \includegraphics[width=0.424\textwidth]{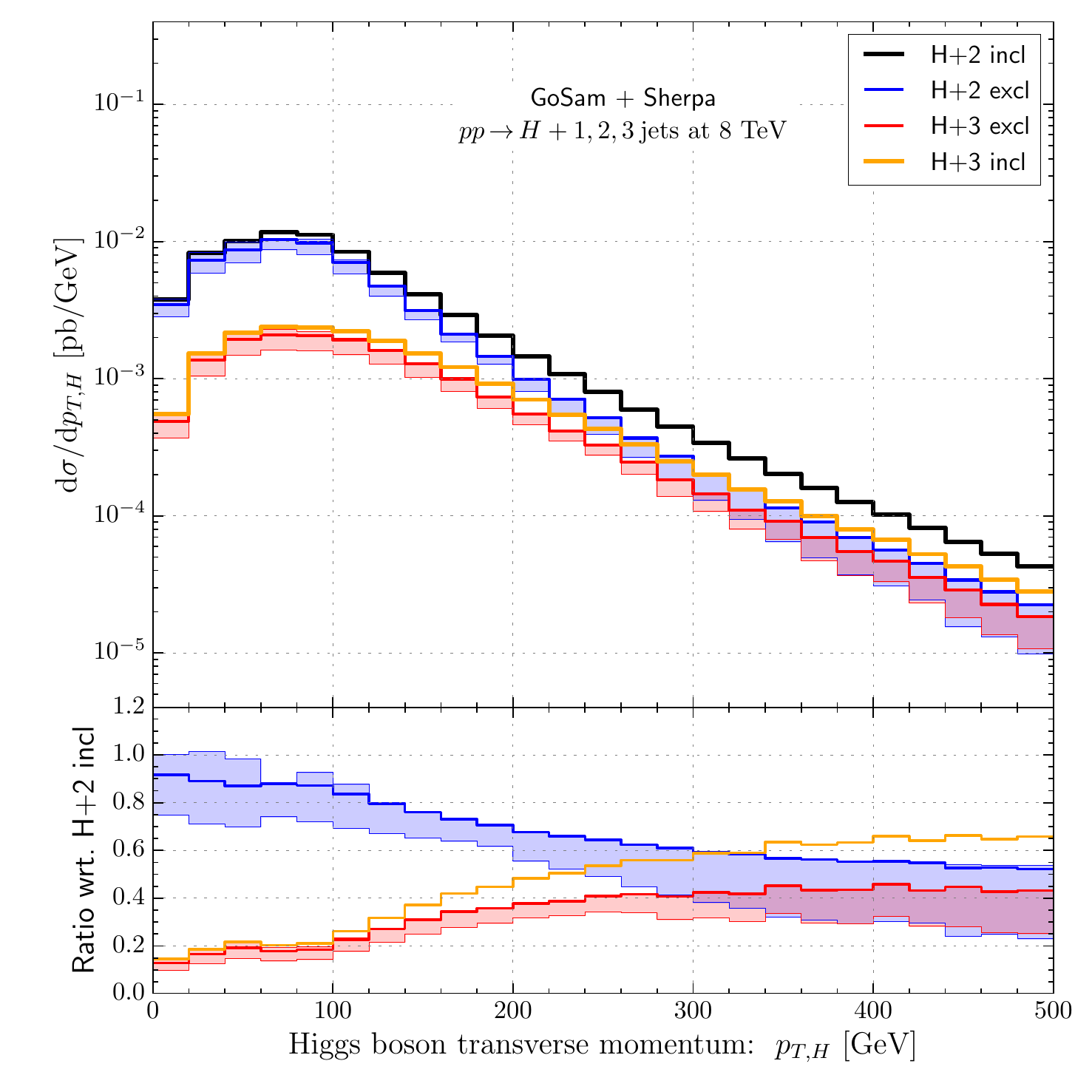}\hfill
  \includegraphics[width=0.424\textwidth]{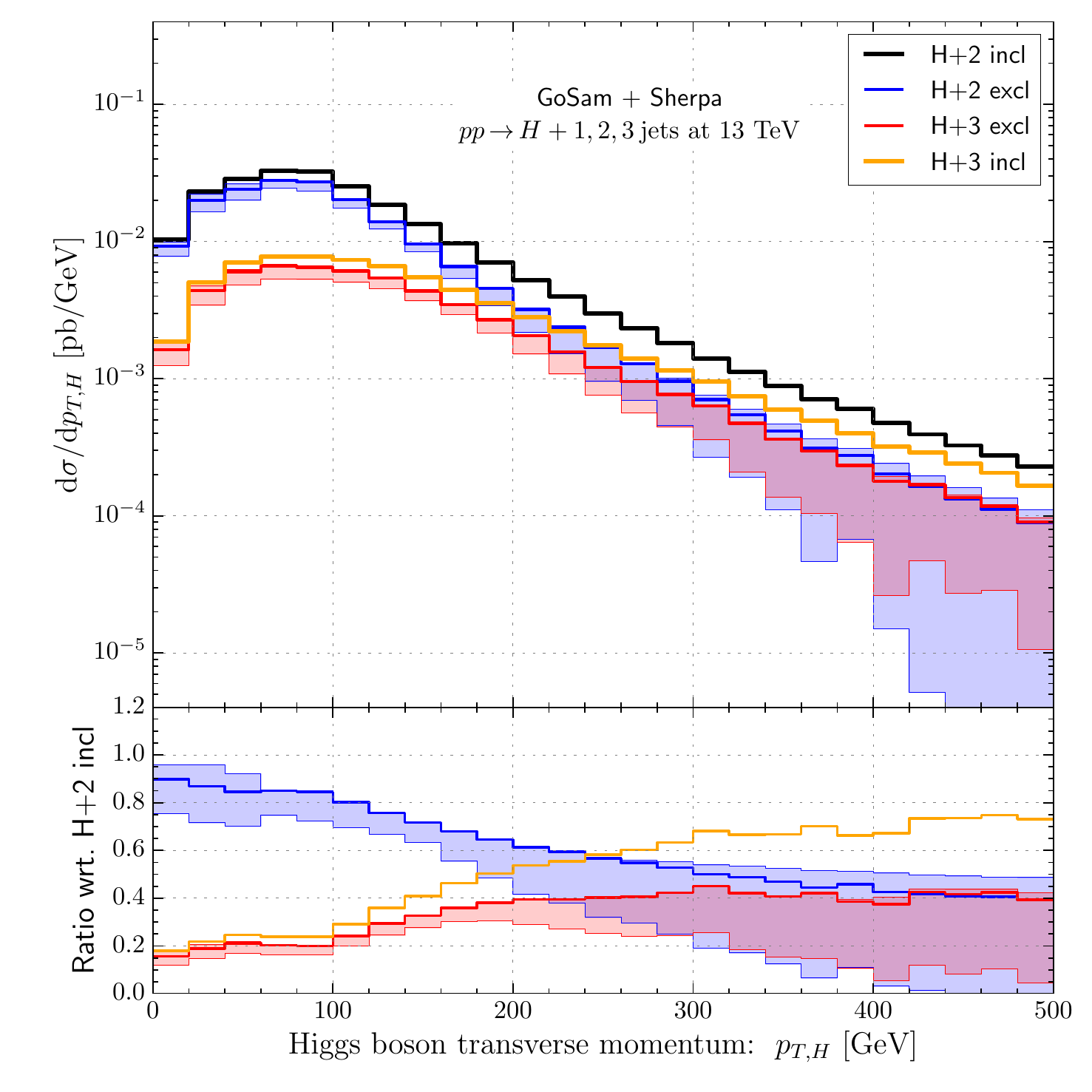}\hfill~
  \\~\hfill
  \includegraphics[width=0.424\textwidth]{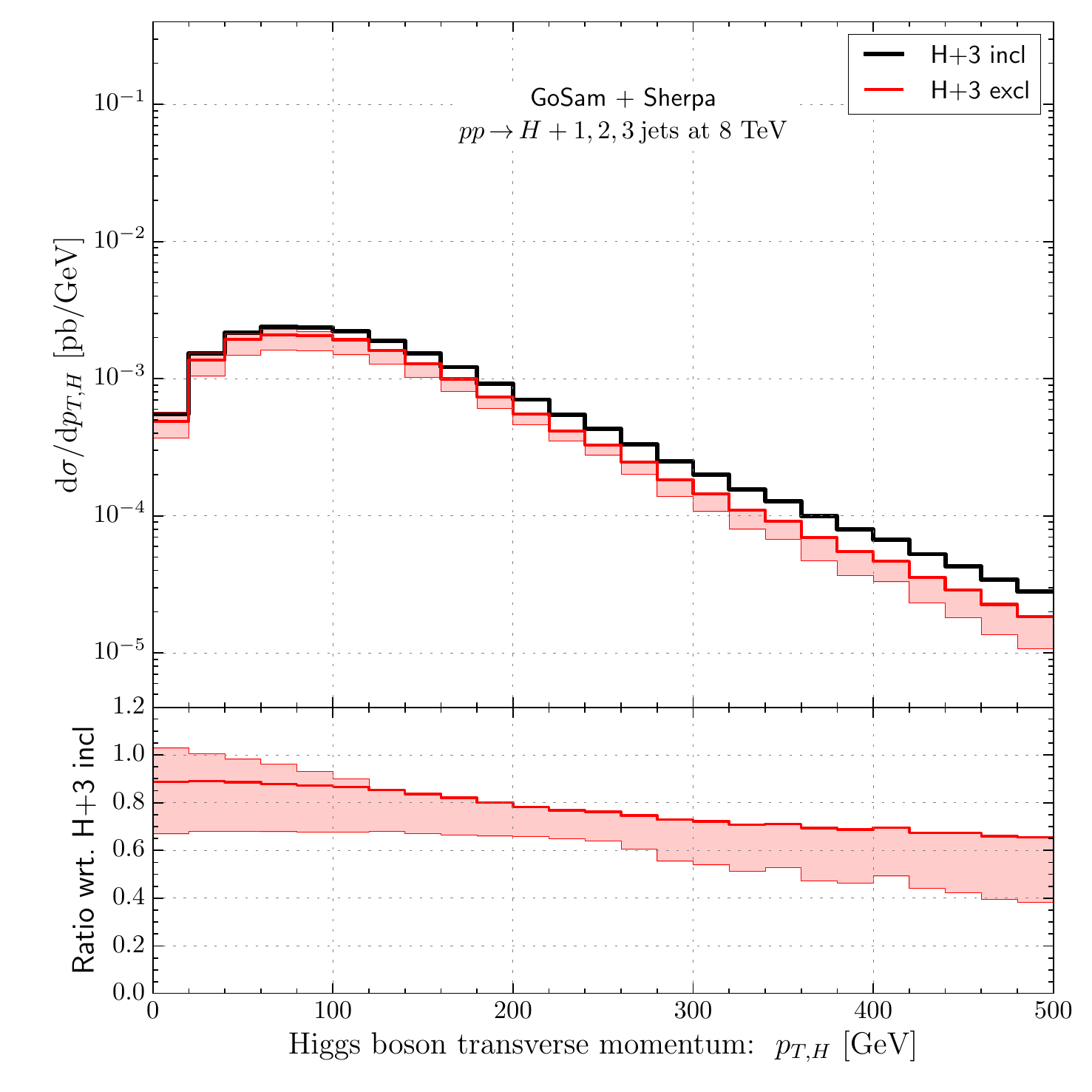}\hfill
  \includegraphics[width=0.424\textwidth]{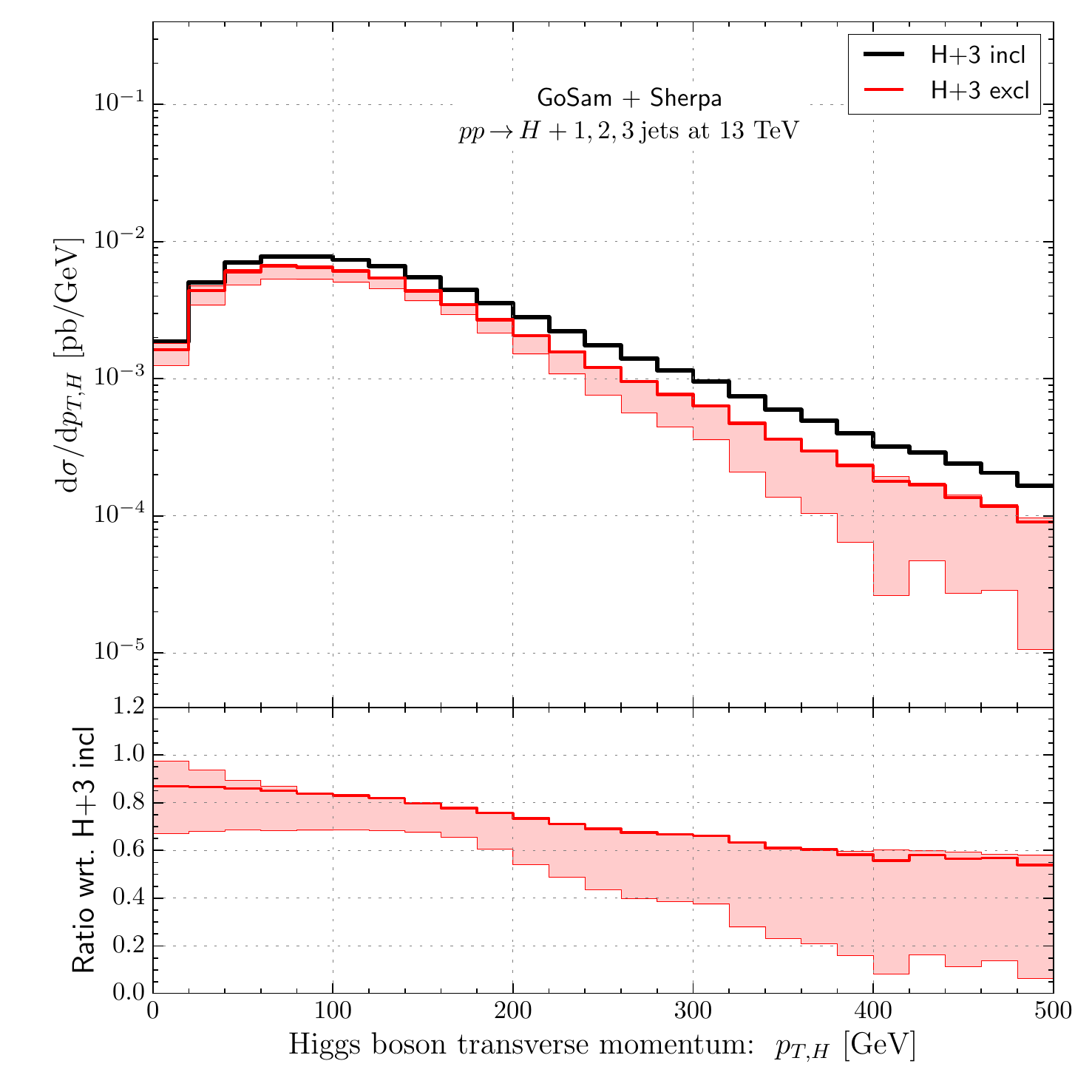}\hfill~
  \caption{\label{fig:xnlo-ratio-pTh}%
    Inclusive and exclusive NLO rates for for the transverse momentum of the
    Higgs boson at 8 \TeV (left column) 
    and 13 \TeV (right column). The ratio plot details the magnitudes
    of the exclusive and higher multiplicity inclusive calculations relative to 
    the most inclusive calculation, \Hj. See text for more explanations.}
\end{figure}

The visible energy, $H_T'$, is the classical example of a 1-jet inclusive observable
which is impacted by higher-order radiative effects, simply because it sums the Higgs
$p_T$ and all jet transverse momenta, irrespective of their correlations in azimuth.
The corresponding spectrum is shown in Fig.~\ref{fig:xnlo-ratio-HTprime} for 8 \TeV (left)
and 13 \TeV (right).  The upper plots show the distribution for one, two
and three jets. Unless stated otherwise, the jet multiplicity is
exclusive, i.e.~with a veto on an additional jet activity.  
The one jet and the three jet subprocesses are shown
twice, once for the exclusive case, and once for the inclusive case,
labelled by 'incl'. The lower panel shows the ratio of each contribution
to the inclusive \Hj prediction. In the following we will only
discuss the results for 13 \TeV. The exclusive \Hj contribution dominates
below 200 \GeV, but it falls off steeply towards higher values of $H_T'$.
The exclusive \Hjj contribution is negligible in the low $H_T'$ region, 
but just above 250 \GeV it takes over from the \Hj contribution and dominates
the $H_T'$ spectrum. The \Hjjj exclusive contribution shows a similar pattern.
Being completely negligible up to around 250 \GeV, its relative importance rises
quickly and it crosses the exclusive \Hj curve at around 350 \GeV.  
In the range of 500--600 \GeV it then becomes equally important 
as the exclusive \Hjj contribution. Comparing the exclusive \Hjjj 
to the inclusive \Hjjj result, we observe that even the fourth jet plays 
a very important role. Above 300 \GeV the inclusive \Hjjj result is the
second largest contribution to the $H_T'$ spectrum, and it rises
to 80 per cent of the inclusive \Hj at around 500 \GeV. In conclusion,
processes of higher multiplicity give rise to important contributions 
to the $H_T'$ spectrum. This does not only happen in the high $H_T'$ region
but already at around 250 \GeV. Including higher multiplicities is
therefore important for a reliable prediction of the distribution.

\begin{figure}[t!]
  \centering
  \includegraphics[width=0.49\textwidth]{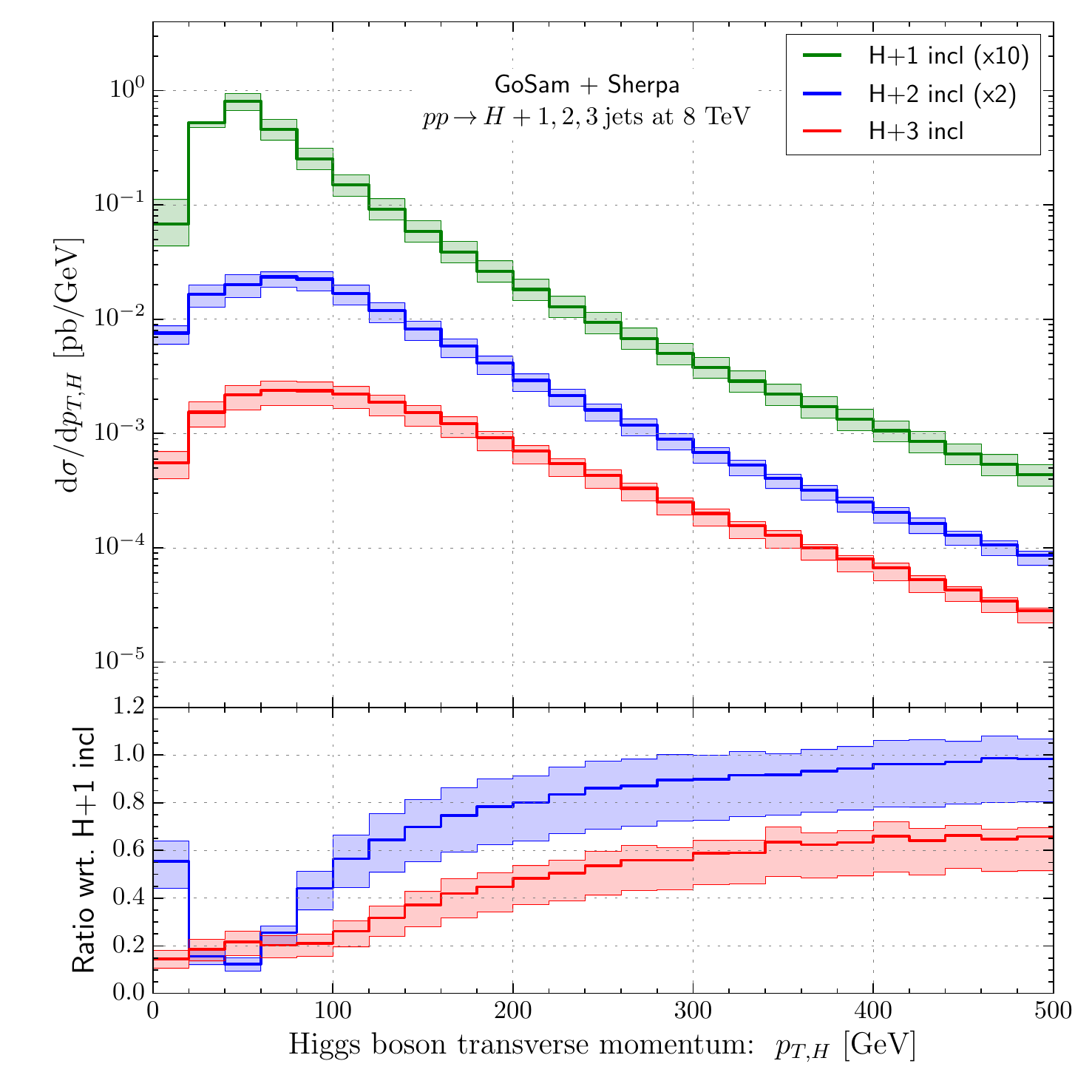}
  \hfill
  \includegraphics[width=0.49\textwidth]{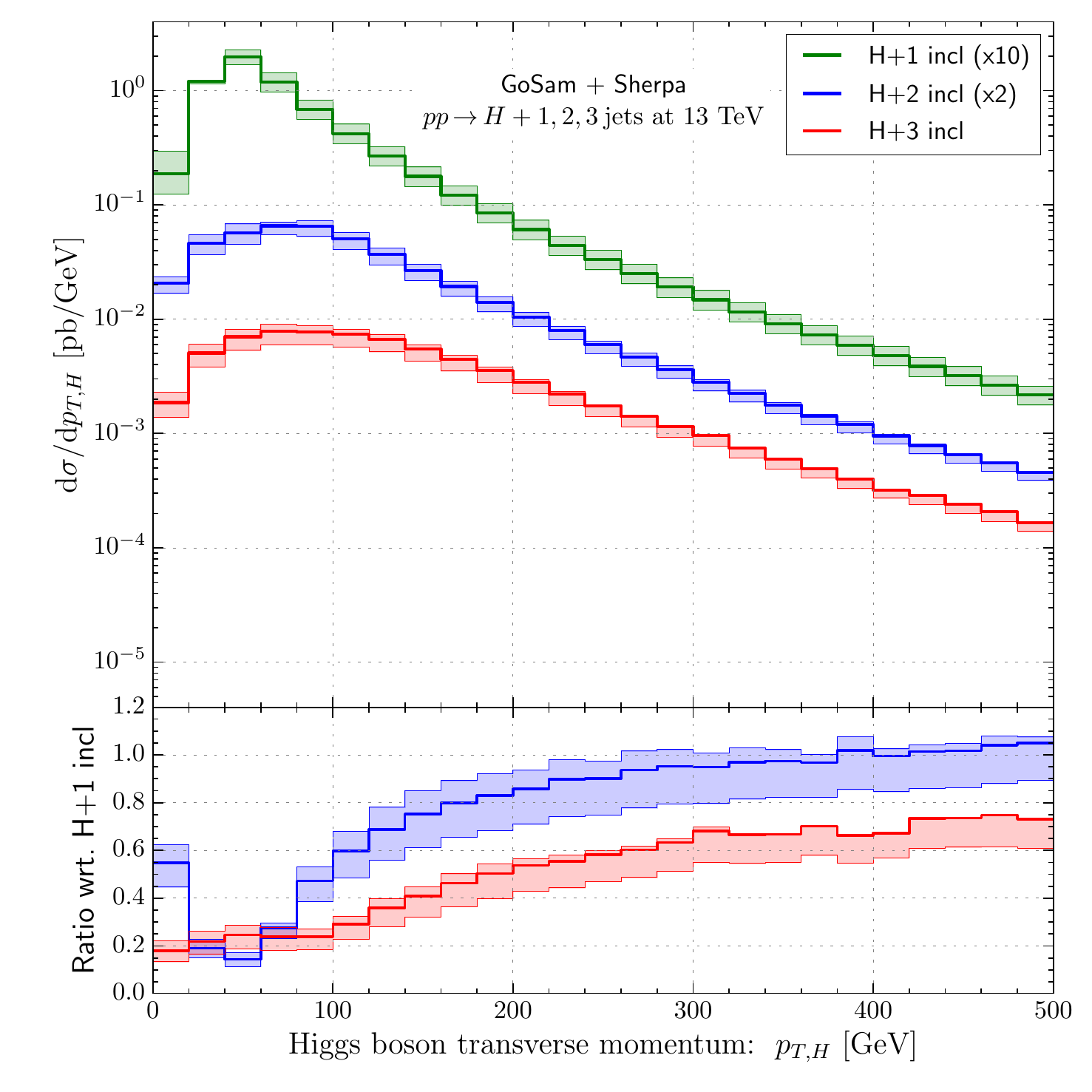}
  \\
  \includegraphics[width=0.49\textwidth]{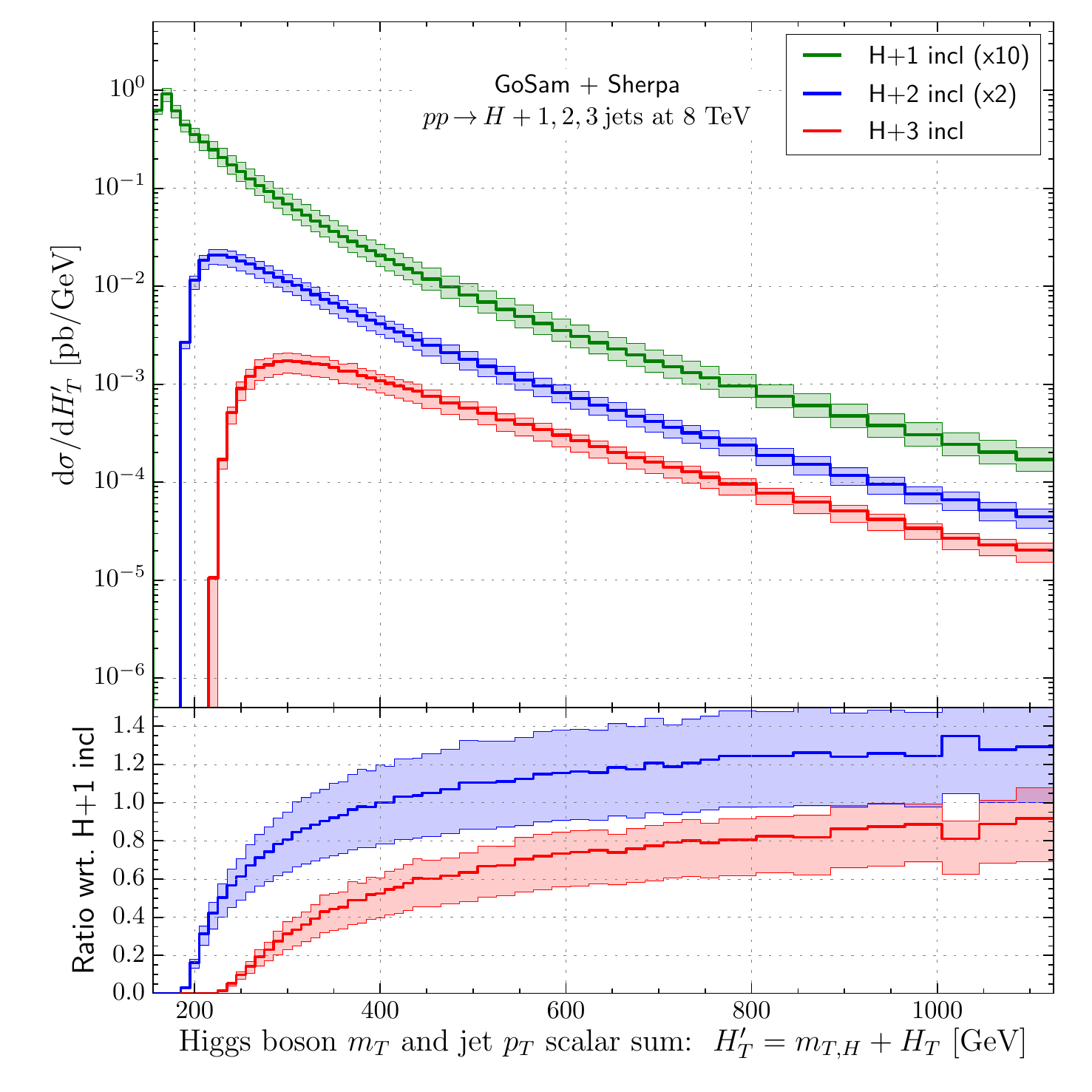}
  \hfill
  \includegraphics[width=0.49\textwidth]{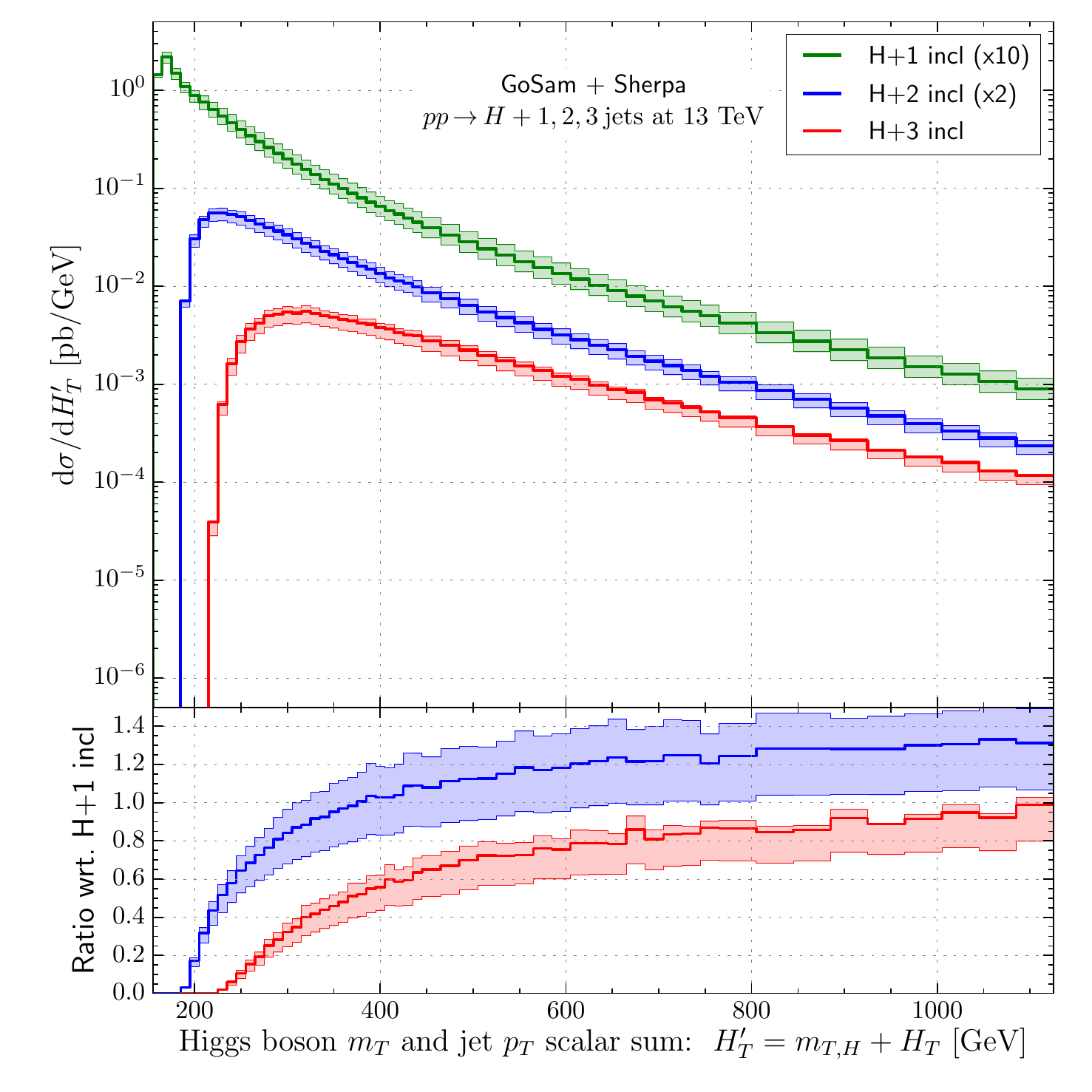}
  \caption{\label{fig:inlo-ratios}%
    Successive ratios of inclusive \Hnj differential cross
    sections at 8 and 13 \TeV for the transverse momentum of the Higgs boson 
    and the $H_T'$ distribution.}
\end{figure}

In the plots of the middle row of Fig.~\ref{fig:xnlo-ratio-HTprime} the
same constellation is shown without the \Hj process. Instead, all the
contributions are now normalized to the inclusive \Hjj result. Similar to the 
\Hj case, the \Hjj process the exclusive contribution is dominant in the 
low $H_T'$ range but its contribution drops fast with increasing $H_T'$. In
the range around 500 \GeV the exclusive \Hjjj is already of the same
size. This has to be kept in mind, particularly for VBF searches.

Finally the plots of the lowest row show the inclusive and exclusive 3
jet predictions and in the ratio plot the results were normalized to
the inclusive cross section. The pattern
is very similar to the previous cases, namely that the exclusive
contribution dominates the low $H_T'$ region and becomes increasingly
unimportant in the high $H_T'$ range.

\begin{figure}[t!]
  \centering
  \includegraphics[width=0.49\textwidth]{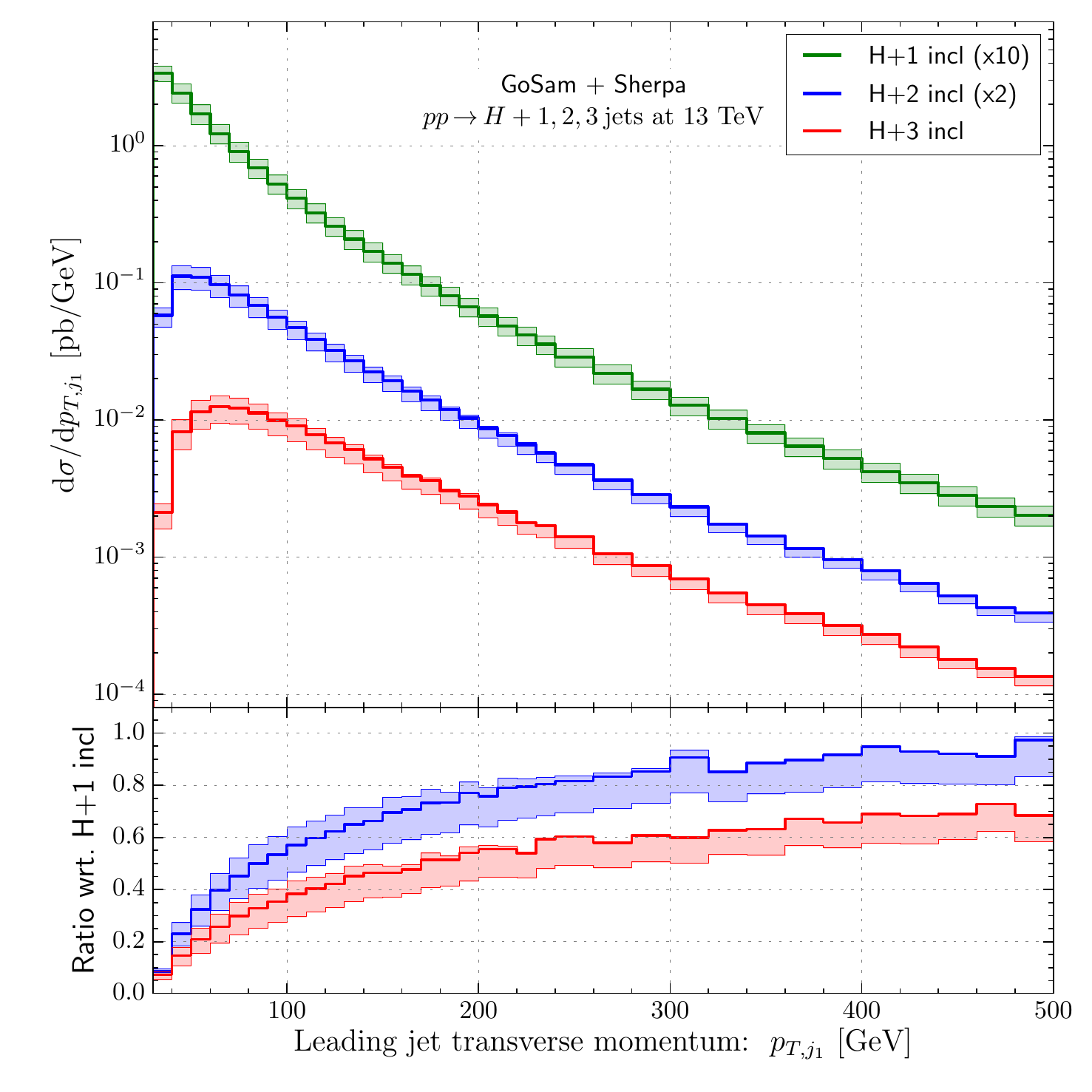}
  \hfill
  \includegraphics[width=0.49\textwidth]{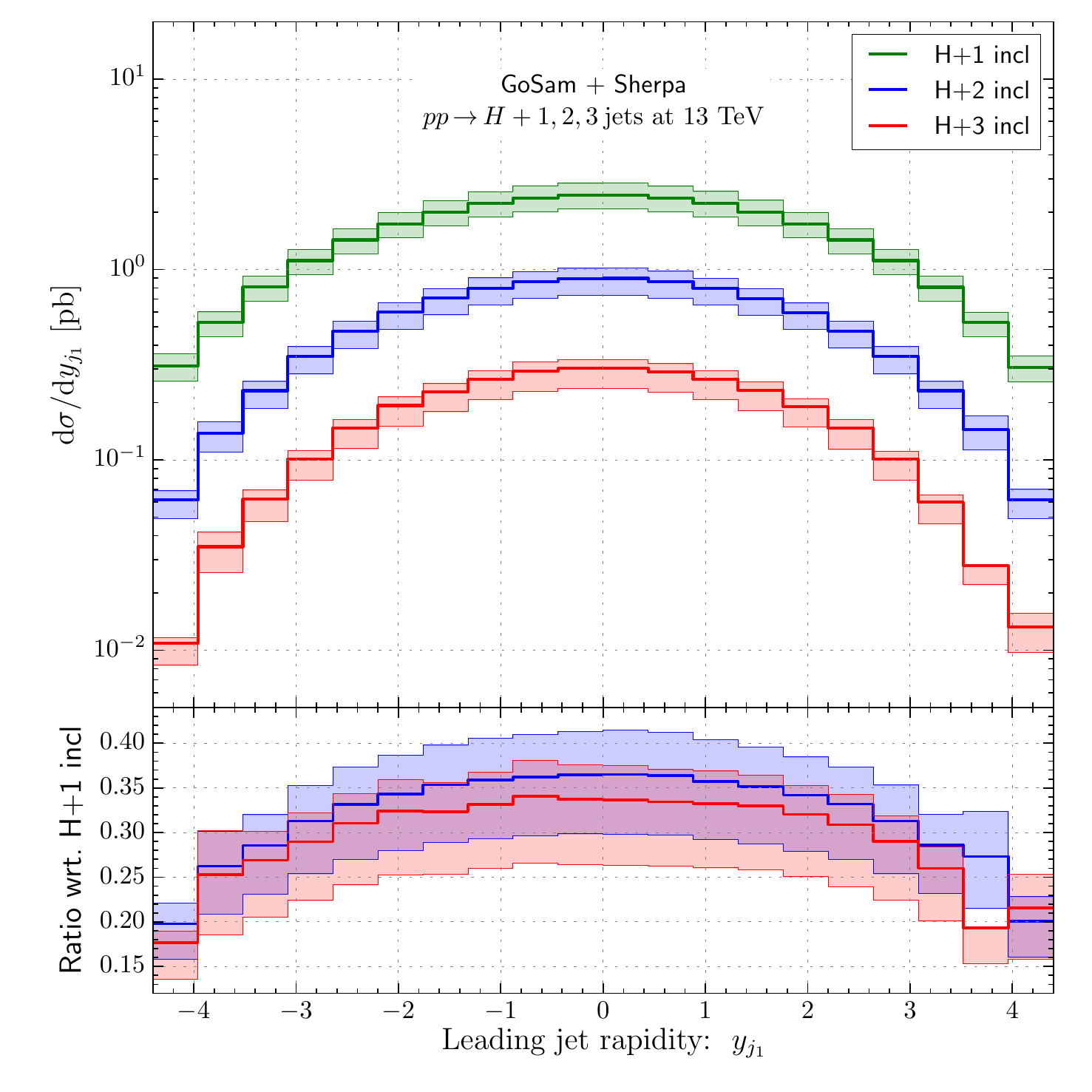}
  \caption{\label{fig:inlo-ratios-extras}%
    Successive ratios of inclusive \Hnj differential cross
    sections at 13 \TeV for the transverse momentum and rapidity
    distributions of the leading jet.}
\end{figure}

It is to be expected from the definition of the observable, that $H_T'$
is largely influenced by additional radiation, as shown above. 
What is more striking though, is that also more inclusive observables,
like the Higgs boson transverse momentum, are susceptible to the same
effect. This is exemplified in Fig.~\ref{fig:xnlo-ratio-pTh}.
We observe that higher-multiplicity processes play an equally important
role for the Higgs $p_T$ spectrum as they do for the visible energy.

A comparison of the inclusive differential cross sections for
both the Higgs $p_T$ and for the $H^{\prime}_T$ distribution is shown
in Fig.~\ref{fig:inlo-ratios}. The left panels show results for 8 \TeV,
the right panels for 13 \TeV. For better visibility, the \Hj contribution 
has been multiplied by a factor of $10$, and the
\Hjj result has been multiplied by a factor of $2$. The ratio plots
are not scaled. They show the ratio of the $(n+1)$-jet cross section 
normalized to the $n$-jet cross section.  This allows to judge
the relative importance of a contribution compared to the one is which
the jet multiplicity is one unit lower. Starting with the Higgs $p_T$
distribution we see the same qualitative behavior as in
Figs.~\ref{fig:xnlo-ratio-HTprime} and~\ref{fig:xnlo-ratio-pTh}. 
The sole difference is that the effects of
higher jet multiplicities are more pronounced in the case of
inclusive cross sections. In the high tail the \Hjj contribution
becomes as big as the \Hj contribution and the \Hjjj contribution
reaches about 60 per cent of the \Hjj contribution. Comparing 8 and 13
\TeV one also observes an enhancement of this effect for higher
energies.

The lower row shows the same for the $H^{\prime}_T$
observable. Again the pattern is qualitatively the same as for the
exclusive case shown in Fig.~\ref{fig:xnlo-ratio-HTprime} but also
more pronounced in the inclusive case. The \Hjj contribution supersedes
the \Hj already at moderate values of $H^{\prime}_T$ for both 8 and
13 \TeV and the \Hjjj contribution easily makes up 60-80 per cent of the
\Hjj result.

The same behavior as discussed in previous plots is found in the transverse
momentum distribution of the leading jet, shown on the l.h.s.\ of
Fig.~\ref{fig:inlo-ratios-extras}.
It can be compared to the transverse momentum of the Higgs, with an
increasing importance of the higher jet multiplicities when increasing
the transverse momentum. However for the rapidity of the
leading jet, shown on the r.h.s.\ of Fig.~\ref{fig:inlo-ratios-extras}
this is not the case.

\subsection{Comparing tagging jet selections and testing high-energy effects}
\label{sec:gf:hej}

Typically the definition of tagging jets is based on the jet
transverse momentum. An
alternative that is more suitable for the VBF Higgs analysis to be
investigated as well is to order the jets according to their rapidity
and choosing the most forward to the most backward jets.  We will
denote the first option as $p_T$-tagging and the latter as
$y$-tagging.  $y$-tagging is theoretically motivated not only because
of Higgs coupling measurements in the VBF channel, but also because it
allows to confirm the universal properties of QCD in the high-energy
limit. In this limit, $t$-channel gluon exchange dominates the cross
section. Jet production can then be described by Lipatov effective
vertices that are resummed in the BFKL
equation~\cite{Lipatov:1976zz,Kuraev:1977fs,Balitsky:1978ic}.  Event
generators based on a Monte-Carlo solution to this equation~\cite{
  Schmidt:1996fg,Orr:1997im,Andersen:2003an,Andersen:2003wy,Andersen:2006sp,Andersen:2006kp}
were constructed for the LHC in order to describe the relevant event
topologies at high
precision~\cite{Andersen:2008ue,Andersen:2009nu,Andersen:2009he,Andersen:2011hs,Andersen:2012gk}.
It is interesting to test how much phase space a calculation performed
in collinear factorization can cover before high-energy resummation
becomes relevant. Our calculation allows to study this question in
Higgs-boson production through gluon fusion for the very first time.

\begin{figure}[t!]
  \centering
  \includegraphics[width=0.49\textwidth]{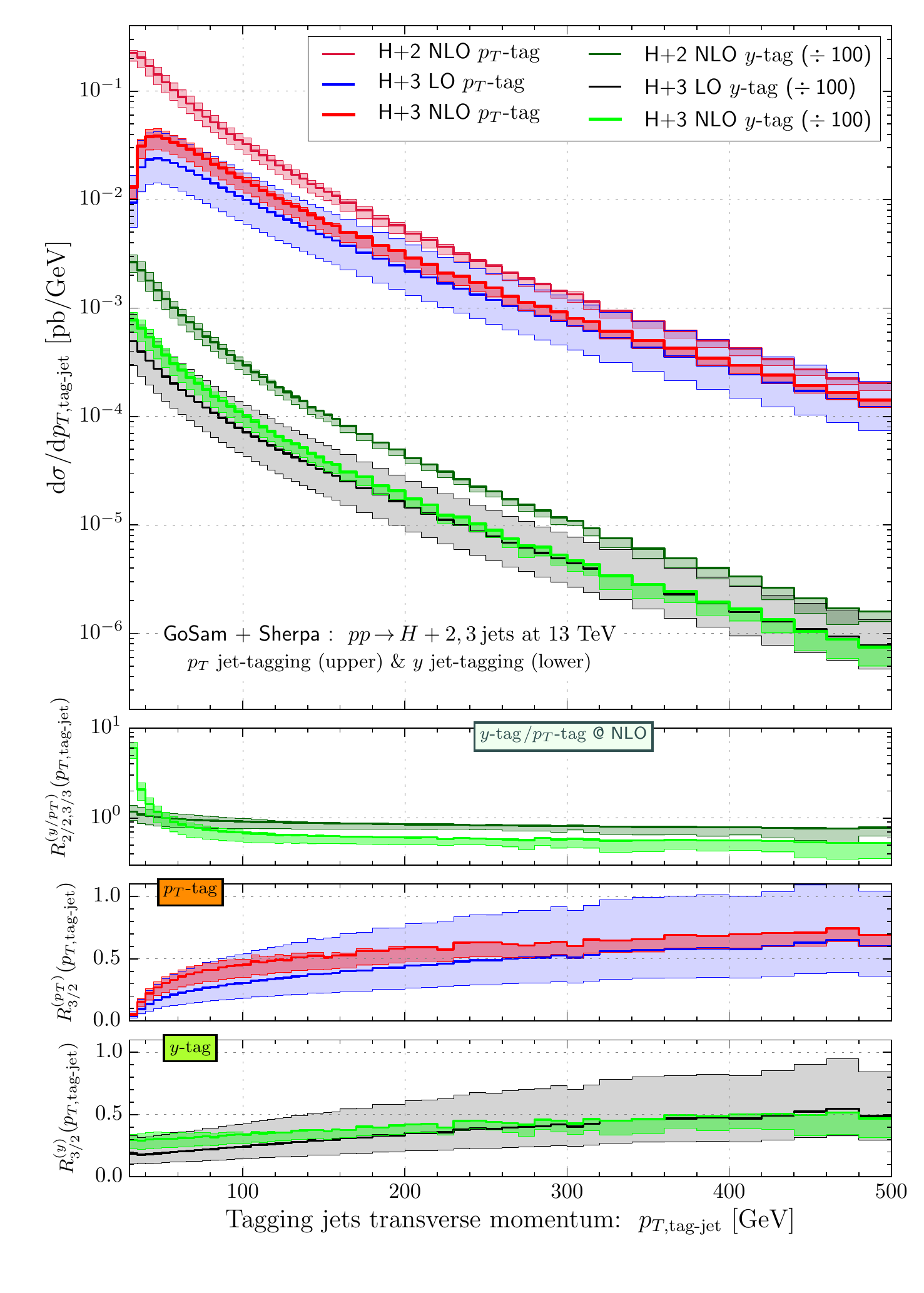}
  \hfill
  \includegraphics[width=0.49\textwidth]{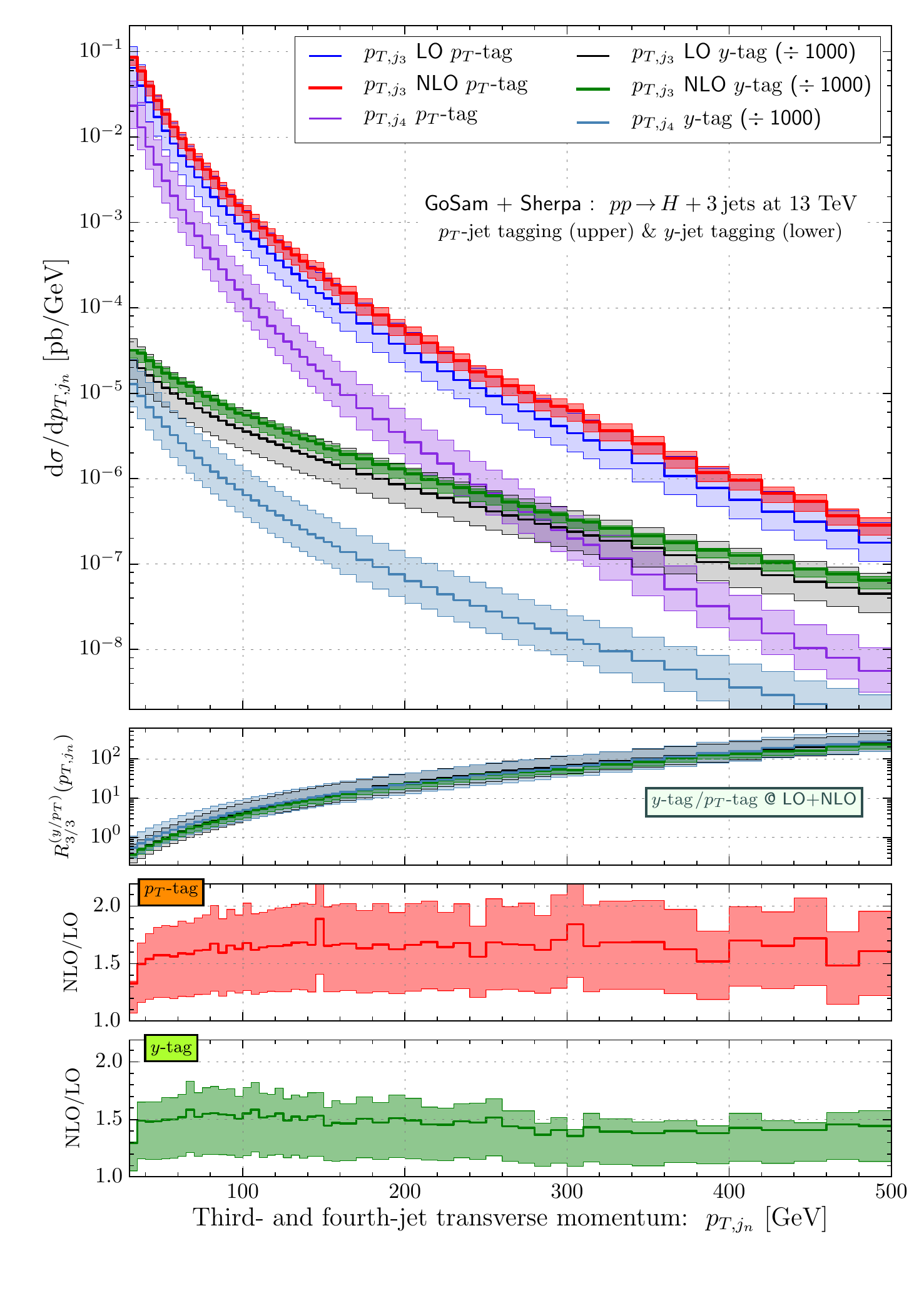}
  \caption{\label{fig:tag-sgljetpts}%
    Transverse momentum distribution of the tagging jets (left) and
    the subleading jets (right) at 13 \TeV. Distributions are shown
    for the two tagging jet definitions, $p_T$~jet-tagging and
    $y$~jet-tagging. See the text for more details, and definition of
    the depicted ratios.}
\end{figure}

\begin{figure}[t!]
  \centering
  \includegraphics[width=0.49\textwidth]{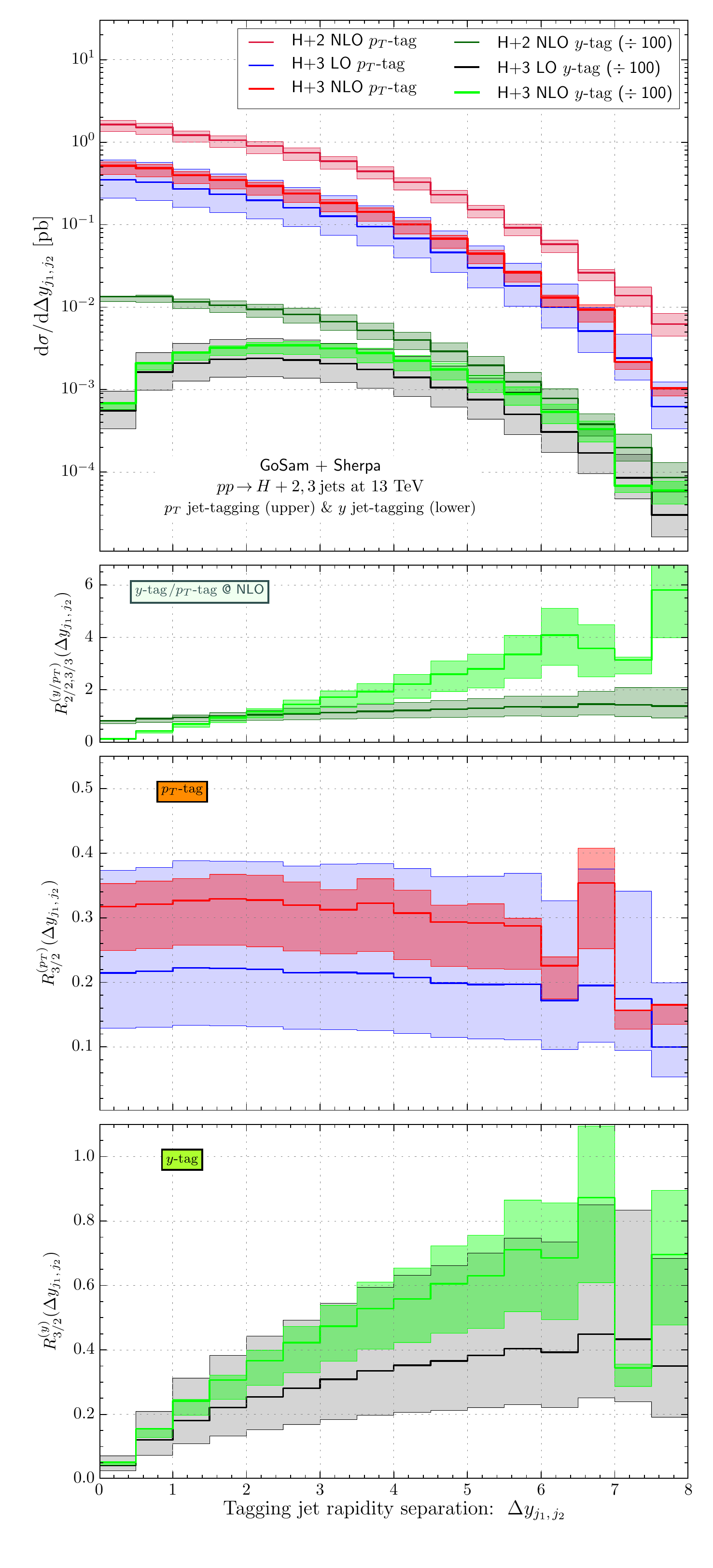}
  \hfill
  \includegraphics[width=0.49\textwidth]{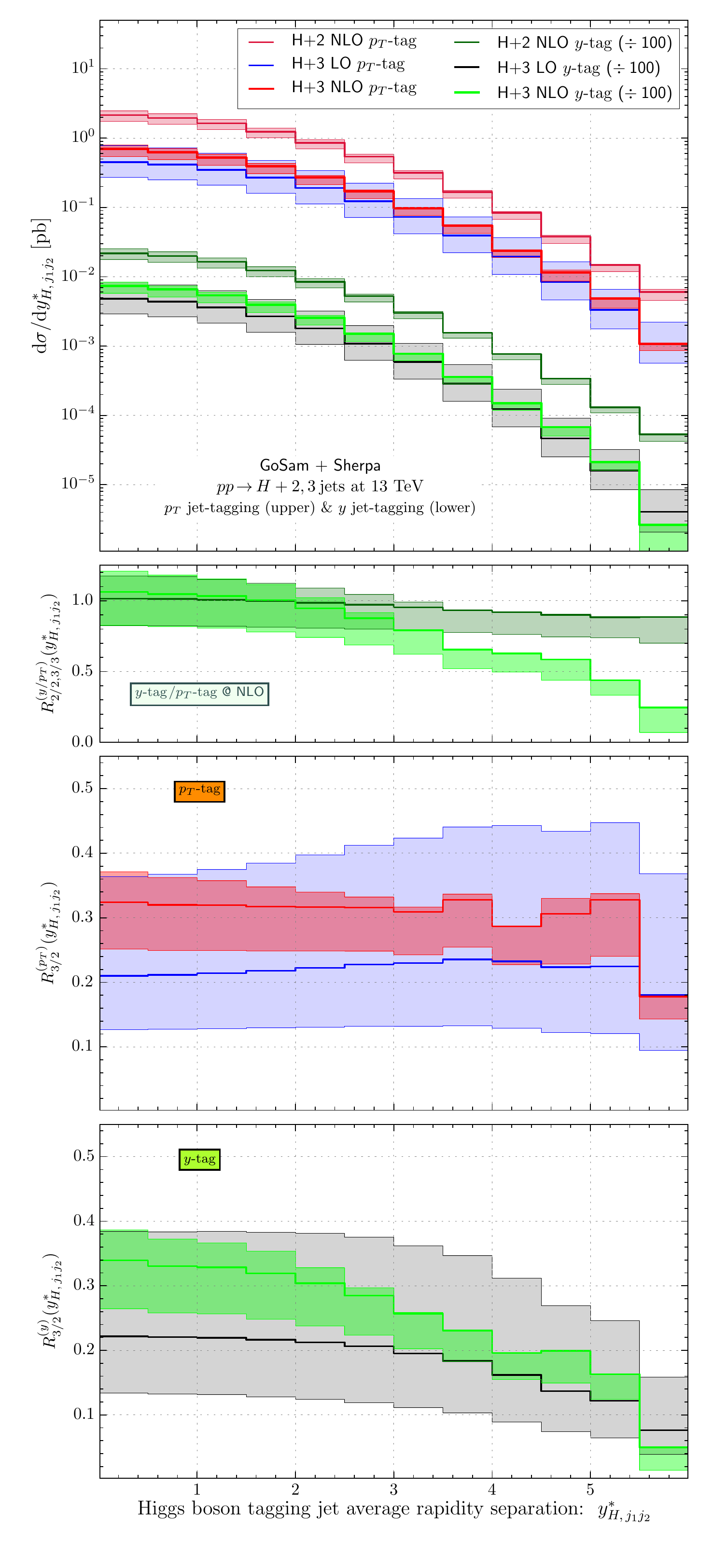}
  \caption{\label{fig:tag-sgljetraps}%
    NLO predictions for rapidity difference observables in \Hjj and
    \Hjjj production at the 13~\TeV LHC. The rapidity separation
    between the tagging jets is presented on the left while the $y^*$
    variable measuring the distance between the Higgs boson and the
    two tagging jets is depicted in the right panel. Distributions are
    shown for the two tagging jet definitions, $p_T$~jet tagging and
    $y$~jet-tagging. See text for more details, and definition of
    the depicted ratios.}
\end{figure}

Naturally, the observables most affected by the two tagging options
above are the rapidity distance of the two tag jets and their
inclusive transverse momentum. The spectrum for the latter is shown in
Fig.~\ref{fig:tag-sgljetpts} together with the transverse momentum
distribution of the subleading jets. The rapidity distance is instead
shown on the left in Fig.~\ref{fig:tag-sgljetraps} whereas on the right we
display the (averaged) rapidity difference between the tag jets and the
Higgs boson, $y_{\Higgs,j_1j_2}^{\ast}$. This latter obsevable is defined as
\begin{equation}\label{eq:ystarhjj}
y_{\Higgs,j_1j_2}^{\ast}=\left|y_\Higgs-(y_{j_1}+y_{j_2})/2\right|\,.
\end{equation}
The two tag jet selections are labelled by $p_T$-tag and $y$-tag.  The
upper panels show the NLO result for \Hjj as well as the LO and NLO
result for \Hjjj.  For better readability of the plots, all results
using $y$-tagging have been scaled by 1/100. The lower panels show
three different comparisons (top to bottom): The ratio of the NLO
result in $y$-tagging to the NLO result in $p_T$-tagging, both for
\Hjj (dark green) and \Hjjj (light green). The ratio of \Hjjj LO and
\Hjj NLO result (blue shaded region), and the ratio of \Hjjj NLO and
\Hjj NLO (red shaded area), both using $p_T$-tagging. And, finally,
the same two ratios, but this time using $y$-tagging. We will call
the latter two ratios $R_{3/2}^{(p_T)}$ and $R_{3/2}^{(y)}$,
respectively. Since the subleading jets are present only in \Hjjj
beyond LO, the right plot in Fig.~\ref{fig:tag-sgljetpts} does not
follow the same conventions. There the \Hjjj LO and NLO curves are
shown together with their ratio for the two tagging schemes.

Figure~\ref{fig:tag-sgljetraps} (left) shows a significant shift
towards larger rapidity difference in $y$-tagging, as compared to
$p_T$-tagging. Jets are still predominantly produced centrally,
because we apply democratic transverse momentum cuts, and the phase
space for centrally produced jets is
larger. Figure~\ref{fig:tag-sgljetpts} (left) exemplifies the change
in $p_T$-spectra, with the $y$-selection leading to softer tag jets
than the $p_T$-selection. This is most easily seen in the upper ratio
plot. For both $p_T$- and $y$-tagging the \Hjjj NLO results are very
similar to the LO results, albeit with reduced scale
uncertainty. Compared to the \Hjj NLO calculation, the shape of tag
jet distributions changes mostly in the low-$p_T$ region. This
indicates that the emission of a third jet, especially at large $p_T$,
is very likely. The large radiative corrections are described more
precisely by a \Hjjj than by a \Hjj calculation, leading to largely
reduced scale uncertainties. This confirms the findings of
Sec.~\ref{sec:inclusivevsexclusive}.

Another important feature appears in the lower ratio plots on the left
of Figs.~\ref{fig:tag-sgljetpts}-\ref{fig:tag-sgljetraps}. On the one
hand, $R_{3/2}^{(p_T)}(p_T)$ shows a roughly logarithmic rise, owing
to the large radiative corrections in the \Hjj process. At the same
time, $R_{3/2}^{(p_T)}(\Delta y)$ is approximately flat. On the other hand,
$R_{3/2}^{(y)}(\Delta y)$ is roughly proportional to $\Delta y$, with slightly
larger slope at NLO, indicating an increasing NLO over LO ratio.  We
can compare this, on a qualitative level, to the results presented in
Ref.~\cite{Campbell:2013qaa}, where $R_{3/2}^{(y)}(\Delta y)$ has also been
computed. This calculation was performed in an approach based on the
high-energy resummation (using \textsc{Hej}
\cite{Andersen:2009nu,Andersen:2011hs}) and compared to results from
collinear factorization (using MCFM \cite{Campbell:2006xx}). The
authors observed a considerable discrepancy between the two
calculations, particularly for large rapidity differences. In this
context it is important to stress that MCFM describes the \Hjjj
topology at LO accuracy only.  Comparing the findings with
Fig.~\ref{fig:tag-sgljetraps} (left) we note that the discrepancy
observed in \cite{Campbell:2013qaa} is largely reduced by the NLO
correction to the \Hjjj process.

It is interesting to consider the subleading jets shown in
Figure~\ref{fig:tag-sgljetpts} (right), which are defined for both
selections as the jets with highest transverse momentum, excluding tag
jets. We observe that the subleading jets have very different
$p_T$-spectra in the two selections, which is simply due to the fact
that the hardest jet is preferably produced at central rapidity,
making it the leading tag jet in the $p_T$-selection, but the first
subleading jet in the $y$-selection.  The differential K-factors shown
in the middle and lower ratio plots display only modest variation over
the entire kinematic range, with the theoretical uncertainty being
smaller for $y$-tagging.

Another important observable is $y_{\Higgs,j_1j_2}^{\ast}$ as shown in
Fig.~\ref{fig:tag-sgljetraps} (right) for \Hjjj at 13~\TeV.  The
overall behavior is comparable to the rapidity distance between the
tag jets shown on the left, however the inclusive $R_{3/2}$ ratios are
flatter for $y$-tagging and show a more pronounced decrease for $p_T$
tagging.


\section{Vector boson fusion phenomenology}
\label{sec:vbf}
The production of a Higgs boson in the VBF channel is phenomenologically 
highly relevant, as it allows to measure the couplings between electroweak 
gauge bosons and the Higgs boson. It also provides sensitivity to the CP 
structure of the Higgs couplings \cite{Plehn:2001nj}, as well as access to 
possible anomalous couplings in both the Higgs sector and the electroweak 
sector of the Standard Model.

As gluon fusion is an irreducible background to the VBF channel, the 
challenging task for theory is to provide a precise prediction of its rate 
compared to the signal. NLO precision for the signal (VBF with up to three 
jets) has already been achieved \cite{Rainwater:1998kj,Rainwater:1999sd,
  Figy:2007kv,Campanario:2013fsa}. In this section we therefore focus on 
the background.

The main obstacle is the extraction of the exclusive \Hjj cross section 
in the fiducial region of typical VBF analyses. We have already seen in 
Sec.~\ref{sec:inclusivevsexclusive} and Sec.~\ref{sec:gf:hej} that 
higher-multiplicity final states contribute sizeably to the inclusive 
cross section. If we extract the effect of a central or global jet veto 
on the \Hjj final state from the NLO \Hjj calculation, the prediction is 
of leading order accuracy and the associated theoretical uncertainty is 
therefore large. A more reliable fixed-order prediction is derived from 
a simultaneous calculation of \Hjj and \Hjjj. In this case, one obtains 
the exclusive \Hjj rate as the difference between the inclusive \Hjj 
result and the \Hjjj result in the vetoed region of the phase space, 
thus improving on the theoretical accuracy of logarithmically enhanced 
contributions related to the veto on additional jet activity \cite{Gangal:2013nxa}.
The kinematic distribution of \Hjjj events may 
also help to devise cuts for improving the purity of an LHC event 
sample. In this section we therefore provide results for the gluon 
fusion process when applying the typical VBF selection criteria as 
described in Eq.~\eqref{cuts:vbf},
\begin{equation}
  m_{j_1 j_2} > 400 \;\GeV,\quad \left|\Delta y_{j_1,j_2}\right| > 2.8\;.
  \label{cuts:vbf_restated}\;,
\end{equation}
and we focus in particular on observables where we expect different 
shapes between signal and background.

\subsection{Cross sections and scale dependence}
\label{sec:vbf:xs}

\begin{figure}[t!]
  \centering
  \includegraphics[width=0.49\textwidth]{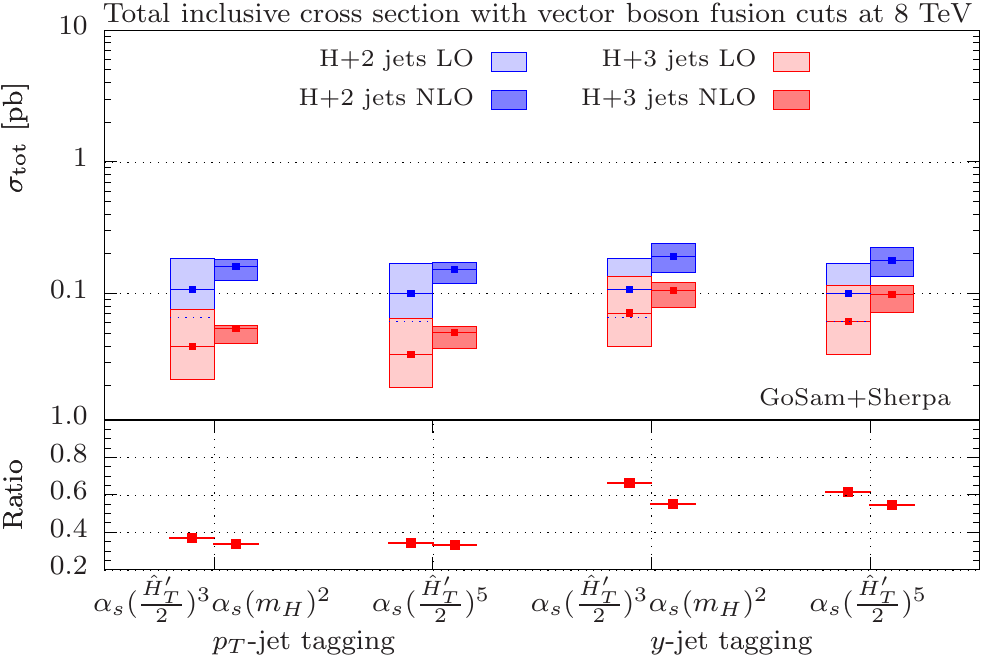}
  \hfill
  \includegraphics[width=0.49\textwidth]{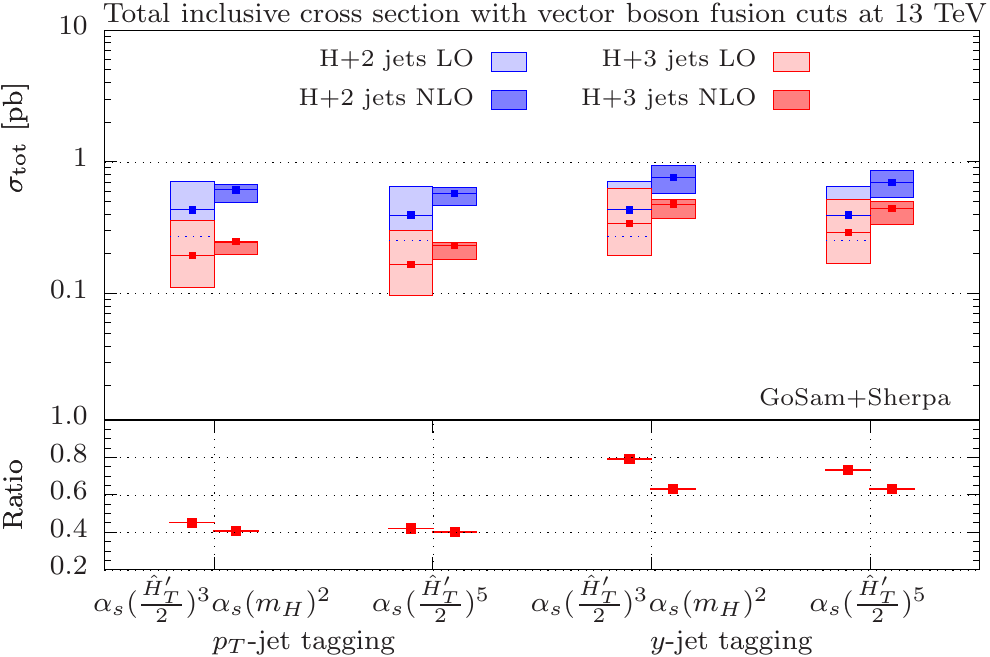}
  \caption{
    Total cross sections for \Hjj and \Hjjj using VBF cuts and two tagging jet
    definitions. Left plot is for 8 \TeV, right plot for 13 \TeV.
  }
  \label{fig:VBF_xsec}
\end{figure}

We start our discussion with the total cross sections as displayed in 
Fig.~\ref{fig:VBF_xsec}. In contrast to Sec.\ \ref{sec:gf:xsecs}, which 
implemented more generic multijet cuts, 
we refrain from including the \Hj result, as the VBF signal 
requires at least two jets. Having excluded the fixed scale as a 
sensible choice in the sections above we only show the two scale 
choices A and B for comparison. Simultaneously, we include two 
different definition of the tagging jets: defining them as the two 
jets with the largest transverse momenta, referred to in the following 
as \lq\lq $p_T$-tag\rq\rq, and defining them as the pair which spans 
the largest rapidity interval between them, referred to as \lq\lq 
$y$-tag\rq\rq. The third (and fourth) jet are then those among the 
remaining jets with the largest (second largest) transverse momentum. 

We observe that the choice of the tagging jet scheme has a considerable 
impact on the total cross section for the \Hjjj process whereas the 
\Hjj process is almost unaffected by it. This is easily understood as 
the latter is independent of the tagging scheme at leading order. At NLO 
a mild dependence is then introduced. This effect gets enhanced for the 
\Hjjj process, introducing a difference of the total cross section of 
almost a factor of two. This shows that the $y$-tag definition is much 
more sensitive to additional radiation beyond the two tagging jets.
However, independent both of the choice of the tagging scheme and the 
collider energy we see a good agreement between the LO ratios and the 
NLO ratios. Similarly, the effect of the scale choice is almost 
negligible.

\begin{figure}[t!]
  \centering
  \includegraphics[width=0.49\textwidth]{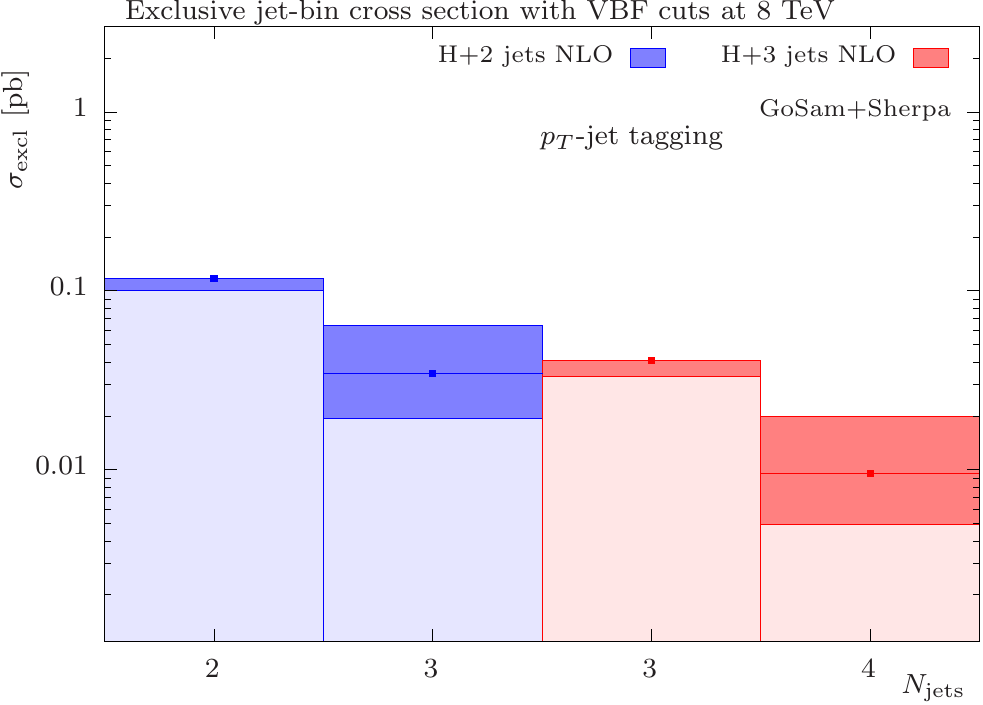}
  \hfill
  \includegraphics[width=0.49\textwidth]{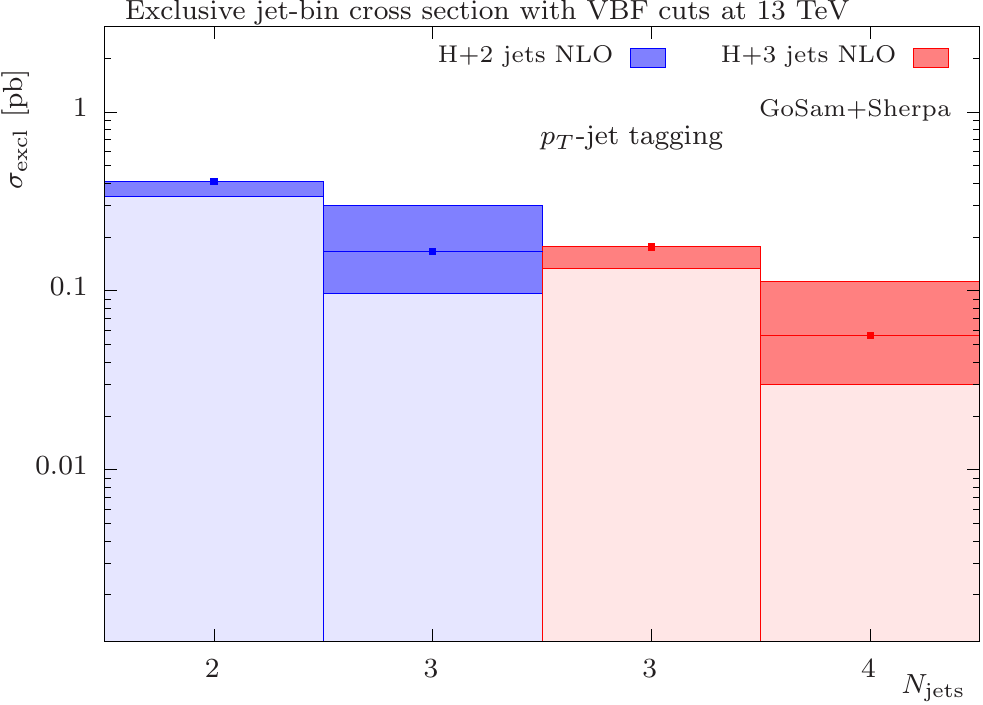}
  \\
  \includegraphics[width=0.49\textwidth]{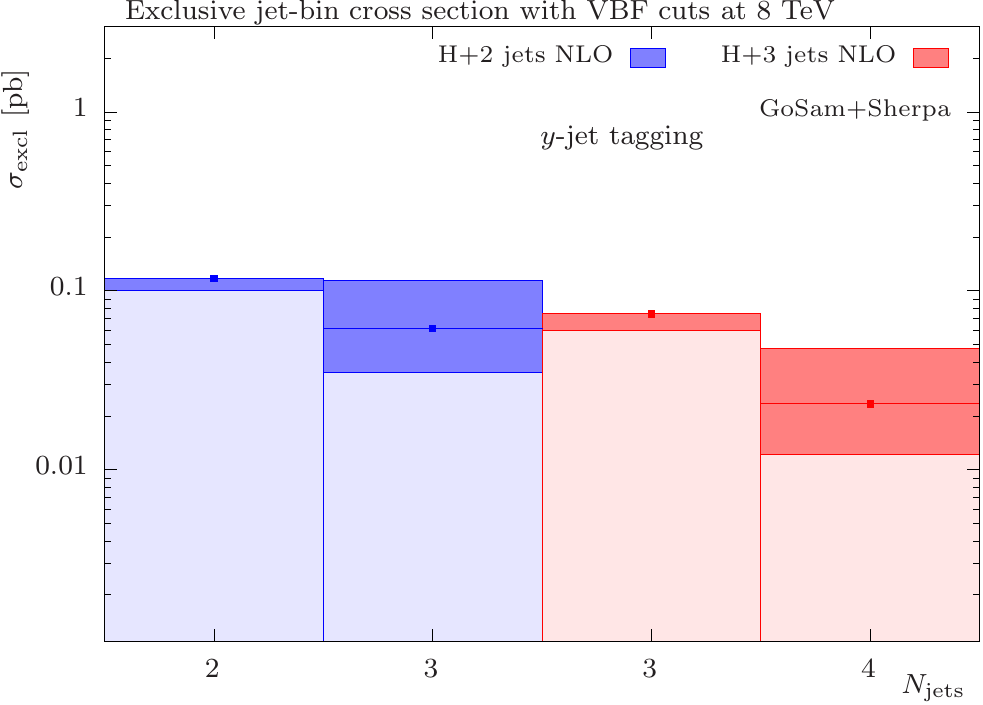}
  \hfill
  \includegraphics[width=0.49\textwidth]{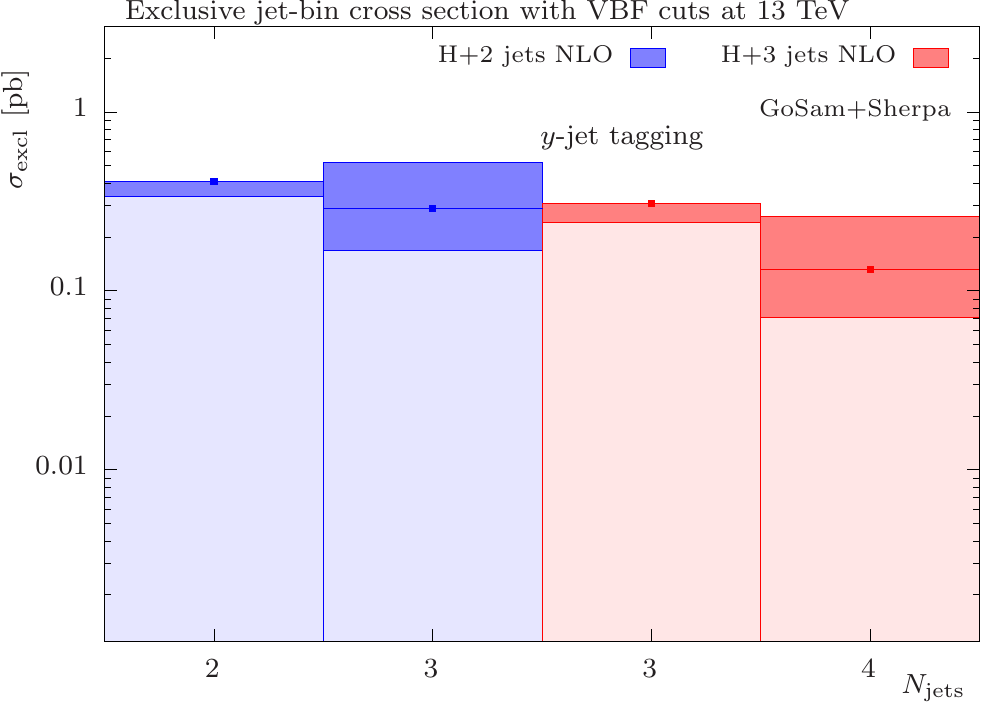}
  \caption{
    Exclusive jet cross sections for $p_T$ jet tagging (upper row) and 
    $y$-jet tagging (lower row) for 8 and 13 \TeV using VBF cuts.
  }
  \label{fig:VBF_nj}
\end{figure}

Fig.~\ref{fig:VBF_nj} shows the exclusive jet cross sections for the 
\Hjj and the \Hjjj process for the two tagging jet definitions at both 
center of mass energies considered in this paper. We show the 
contribution of the subprocess to the different jet multiplicities, 
i.e.\ the \Hjj NLO calculation contributes to the two-jet and the 
three-jet bins while the \Hjjj NLO calculation contributes to the 
three-jet and four-jet bins, where we have to keep in mind that the 
$(n+1)$ contribution is only present at leading order accuracy. 
Comparing these to the results presented in Fig.~\ref{fig:nj} where 
no topological cuts were applied it can be observed that the 
$(n+1)$-jet contribution is enhanced in the VBF fiducial region. In 
the $y$-tag scheme this effect is a somewhat stronger than for the 
$p_T$-tag scheme. The same is true comparing the 13 \TeV results to 
the 8 \TeV ones. The relative enhancement of the $(n+1)$-jet 
contribution implies that the contribution, which is only described 
with leading order accuracy, becomes more important. Turning this 
argument around, this means that the theoretical uncertainty is 
increased in the VBF fiducial region as the leading order pieces of 
the calculation have a larger contribution to the total result 
then in the simple dijet region of the previous section. 
This stresses the importance of the \Hjjj calculation also for the 
VBF region as it allows to determine the radiation of a third jet 
with NLO accuracy.
A detailed 
summary of the total cross sections of the various contributions, 
cross section ratios and jet fractions for 8 and 13 \TeV is listed in 
Tab.\ \ref{hj3nlo_tab_xsecs_vbf}. It shows values for both tagging 
schemes and gives the precise numbers for the respective $K$-factors.







%

\begin{table}[p]
  \centering\scriptsize
  \begin{tabular}{clrp{0.3mm}lrp{0.3mm}lr}
    \toprule\\[-8pt]
    Sample     & \multicolumn{6}{c}{Cross sections in pb for Higgs boson plus}\\[4pt]
    \phantom{$K$-factor} & $\ge2$ jets & $f_3$ && $\ge3$ jets & $f_4$ && $\ge4$ jets & $r_{n+1/n}$\\[4pt]
    \midrule\\[-7pt]
    \small{8 \TeV} \\[0pt]
    \midrule\\[-10pt]
    \multicolumn{4}{l}{LO \scriptsize(NLO PDFs)}\\[0pt]
    \midrule\\[-10pt]
    \Hjj ($p_T$-jet tagging)  & $0.100~^{+69\%}_{-38\%}$ &        &&                          &         &&                        & $0.344$ \\[8pt]
    \Hjjj ($p_T$-jet tagging) &                          & $1.00$ && $0.034~^{+87\%}_{-43\%}$ &         &&                        &         \\[5pt]
    \Hjj ($y$-jet tagging)    & $0.100~^{+69\%}_{-38\%}$ &        &&                          &         &&                        & $0.615$ \\[8pt]
    \Hjjj ($y$-jet tagging)   &                          & $1.00$ && $0.061~^{+86\%}_{-43\%}$ &         &&                        &         \\[5pt]
    \midrule\\[-10pt]
    \multicolumn{4}{l}{NLO}\\[0pt]
    \midrule\\[-10pt]
    \Hjj ($p_T$-jet tagging)  & $0.152~^{+14\%}_{-21\%}$ & $0.226$ && $0.034~^{+87\%}_{-44\%}$ &         &&                        & $0.333$ \\[8pt]
    \Hjjj ($p_T$-jet tagging) &                          &         && $0.051~^{+11\%}_{-24\%}$ & $0.190$ && $0.010~^{+105\%}_{-48\%}$ & $(0.190)$ \\[20pt]
    \Hjj ($y$-jet tagging)    & $0.179~^{+25\%}_{-25\%}$ & $0.343$ && $0.062~^{+87\%}_{-43\%}$ &         &&                        & $0.546$ \\[8pt]
    \Hjjj ($y$-jet tagging)   &                          &         && $0.098~^{+18\%}_{-25\%}$ & $0.239$ && $0.023~^{+105\%}_{-48\%}$ & $(0.239)$ \\[20pt]
    $K_2$, $K_3$ ($p_T$-jet tagging) & $1.52$                   &         && $1.47$                   & \\[5pt]
    $K_2$, $K_3$ ($y$-jet tagging)   & $1.79$                   &         && $1.59$                   & \\[5pt]
    \midrule\\[-14pt]
    \midrule\\[-7pt]
    \small{13 \TeV} \\[0pt]
    \midrule\\[-10pt]
    \multicolumn{4}{l}{LO \scriptsize(NLO PDFs)}\\[0pt]
    \midrule\\[-10pt]
    \Hjj ($p_T$-jet tagging)  & $0.395~^{+64\%}_{-36\%}$ &        &&                          &         &&                        & $0.421$ \\[8pt]
    \Hjjj ($p_T$-jet tagging) &                          & $1.00$ && $0.166~^{+81\%}_{-41\%}$ &         &&                        &         \\[5pt]
    \Hjj ($y$-jet tagging)    & $0.395~^{+64\%}_{-36\%}$ &        &&                          &         &&                        & $0.732$ \\[8pt]
    \Hjjj ($y$-jet tagging)   &                          & $1.00$ && $0.289~^{+81\%}_{-41\%}$ &         &&                        &         \\[5pt]
    \midrule\\[-10pt]
    \multicolumn{4}{l}{NLO}\\[0pt]
    \midrule\\[-10pt]
    \Hjj ($p_T$-jet tagging)  & $0.577~^{+11\%}_{-19\%}$ & $0.288$ && $0.166~^{+81\%}_{-42\%}$ &         &&                        & $0.403$ \\[8pt]
    \Hjjj ($p_T$-jet tagging) &                          &         && $0.233~^{+6\%}_{-22\%}$  & $0.243$ && $0.057~^{+99\%}_{-47\%}$ & $(0.243)$ \\[20pt]
    \Hjj ($y$-jet tagging)    & $0.700~^{+23\%}_{-23\%}$ & $0.412$ && $0.289~^{+81\%}_{-42\%}$ &         &&                        & $0.630$ \\[8pt]
    \Hjjj ($y$-jet tagging)   &                          &         && $0.441~^{+14\%}_{-24\%}$ & $0.299$ && $0.132~^{+98\%}_{-46\%}$ & $(0.299)$ \\[20pt]
    $K_2$, $K_3$ ($p_T$-jet tagging) & $1.46$                   &         && $1.40$                 & \\[5pt]
    $K_2$, $K_3$ ($y$-jet tagging)   & $1.77$                   &         && $1.53$                 & \\[5pt]
    \bottomrule
  \end{tabular}
  \caption{\label{hj3nlo_tab_xsecs_vbf} \small{ Cross sections in pb for the
    various parton-level Higgs boson plus jet samples used in this
    study with VBF type cuts and scale choice B (i.e. all scales are
    evaluated at $\hthatprime/2$). The upper and lower parts of
    the table show the LO and NLO results, respectively, together with
    their uncertainties (in percent) from varying scales by factors of
    two, up (subscript position) and down (superscript position). We
    report the results for both jet tagging definitions described in
    the text. NLO-to-LO $K$-factors, $K_n$, for the inclusive 2-jets
    ($n=2$) and 3-jets ($n=3$) bin, the cross section ratios
    $r_{3/2}$, $r_{4/3}$ and $m$-jet fractions, $f_m$, are given in
    addition. Since the predictions for $H$+4 jets are only LO
    accurate, $f_4$ and $r_{4/3}$ coincide.}}
\end{table}


\clearpage

\subsection{Differential observables}
\label{sec:vbf:diffobs}

In order to separate events tagged by the presence of a dijet
configuration which is compatible with a VBF process, experimental
analyses~\cite{Aad:2014eva,Aad:2014eha,Khachatryan:2014ira} rely on
multivariate discriminants which are based on boosted decision trees
(BDT). The typical discriminating variables used in these BDT are the
invariant mass of the tagging jet system $m_{j_1j_2}$, the
rapidity separation between the two tagging jets
$\Delta y_{j_1,\,j_2}$ and their transverse momenta, $p_{T,j_1}$ and
$p_{T,j_2}$. The rapidity of the leading tagging jet $y_{j_1}$ is also
taken into account as well as the azimuthal separation and the
rapidity separation between the Higgs boson and the tagging jet system,
$\Delta\phi_{\Higgs,\,j_1j_2}$ and $y_{\Higgs,\,j_1j_2}^{\ast}$, respectively.
Furthermore, in the measurements of the Higgs boson decaying to two
photons, one also uses the transverse momentum of the diphoton system
with respect to its thrust axis in the transverse plane,
$p_{T,\gamma\gamma,\mathrm{t}}$, and some observables directly related to one of
the two photons. The latter are not considered in the following since
the Higgs boson is not decayed in our analysis. Instead of
$p_{T,\gamma\gamma,\mathrm{t}}$ we will directly consider the transverse
momentum $p_{T,\Higgs}$ of the Higgs boson itself.

Because of the very peculiar signature of the VBF events, the tagging
jet invariant mass distribution $m_{j_1j_2}$ plays a key role in
determining whether an event could stem from a VBF process or not. For
this reason, it is interesting to consider a third jet tagging scheme
besides the $p_T$-tagging and the $y$-tagging introduced in
Sec.\ \ref{sec:gf:hej}: one can define the two tagging jets based on
the pair of jets that generates the largest invariant mass. In the
presence of three or more jets, the treatment of the subleading jets
is the same as in the other two schemes where they are ordered
according to their transverse momenta. Although closely related to the
$y$-tagging scheme, the new scheme referred to as $m_{jj}$-tagging
will serve as another benchmark scenario in the following discussion.

\begin{figure}[t!]
  \centering
  \includegraphics[width=0.49\textwidth]{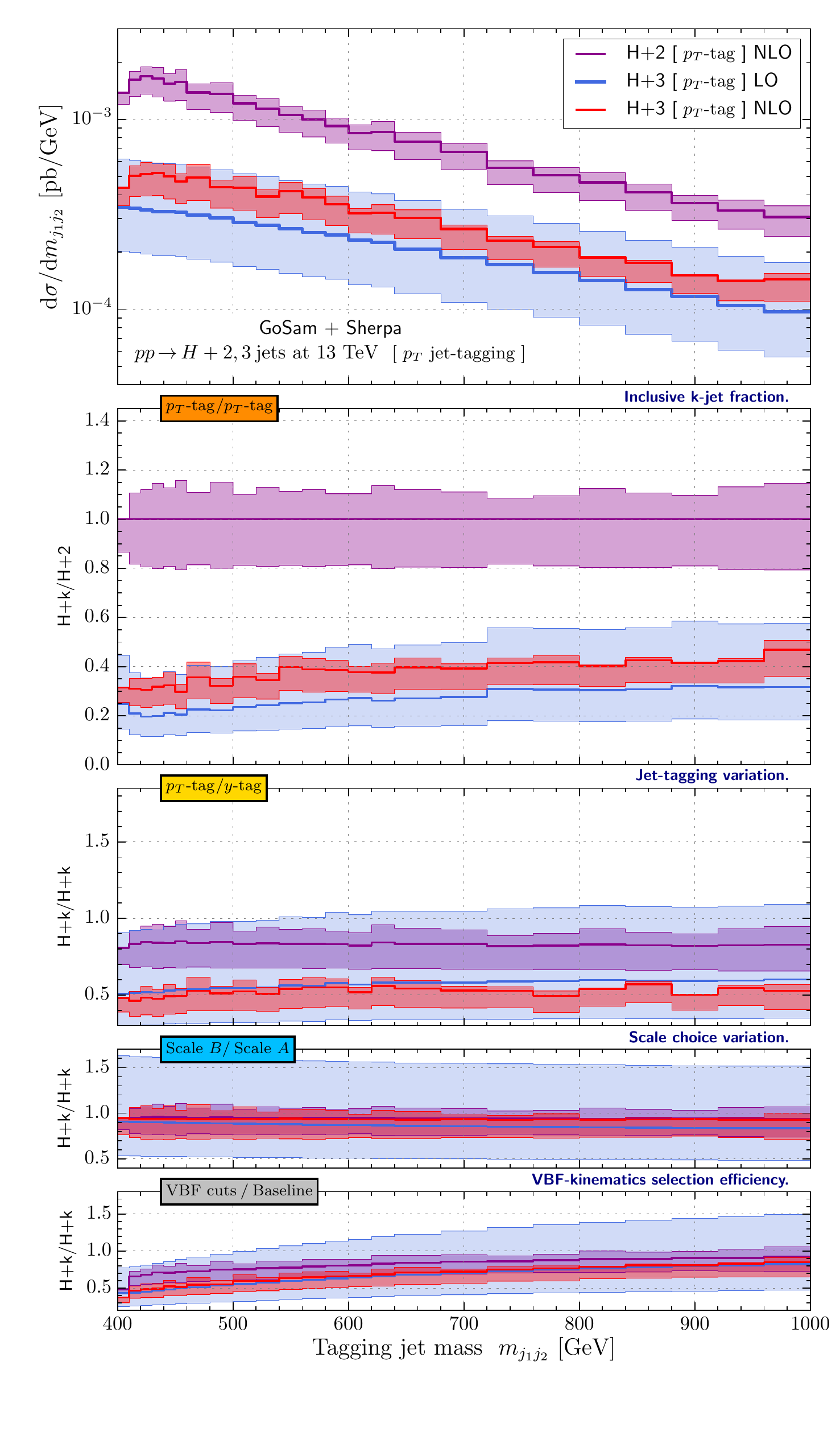}
  \hfill
  \includegraphics[width=0.49\textwidth]{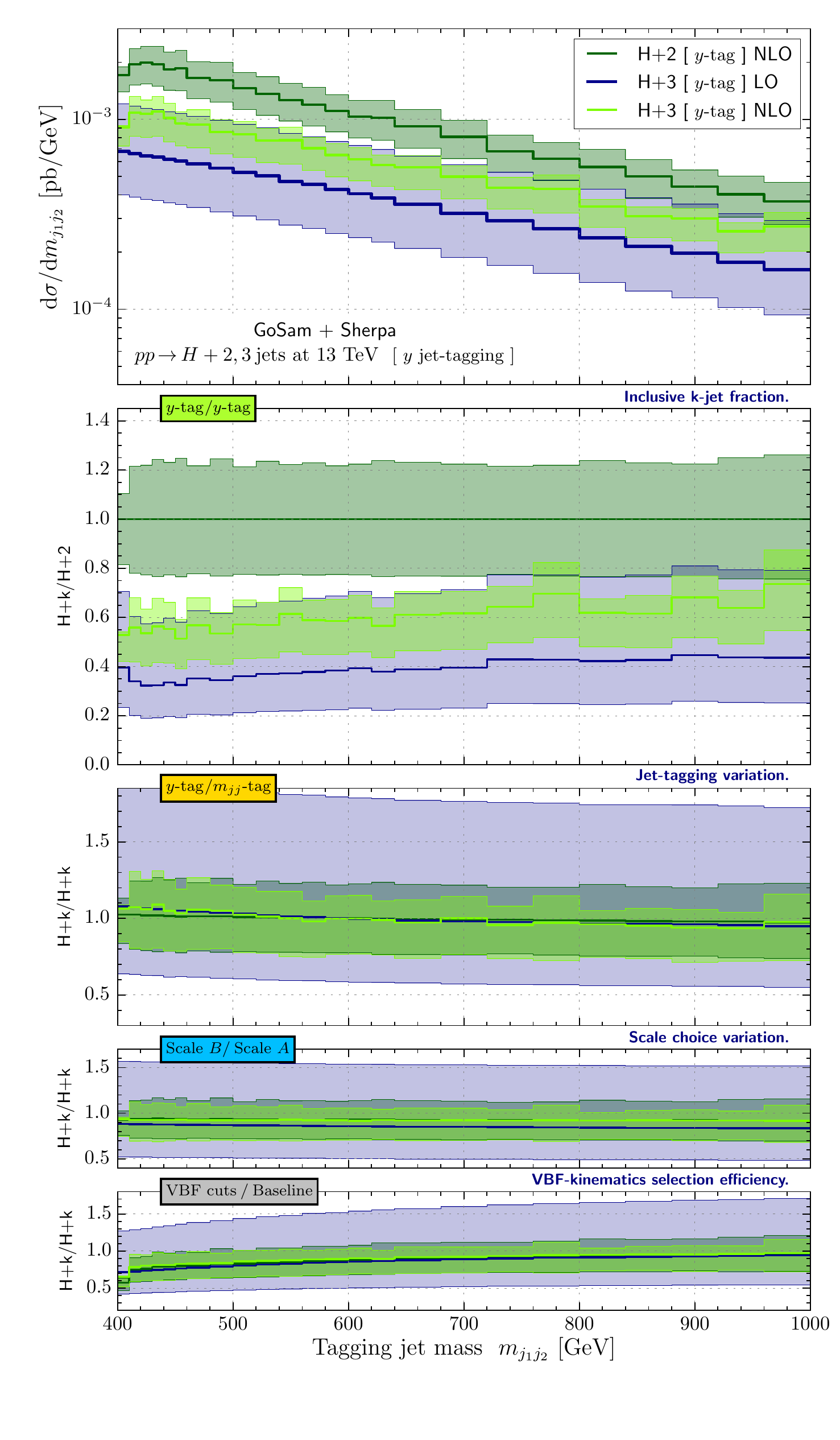}
  \caption{
    Impact of an \Hjjj description at NLO using the 
    scale choice B, cf.\ Eq.\ \eqref{scales:B}, on 
    the tagging jet invariant mass $m_{j_1j_2}$ 
    for the $p_T$-tagging (left) and $y$-tagging (right) jet selection 
    imposing VBF kinematic requirements at the LHC of 13 \TeV CM energy. 
    Displayed are the \Hjj NLO (purple/green), \Hjjj LO (blue/violet) and 
    \Hjjj NLO (red/yellow) predictions. The four ration plots now detail, 
    from top to bottom, the three-jet fraction, the difference between 
    tagging schemes, the difference between different functional forms of 
    the central scale choice, and the impact of the VBF cuts with respect 
    to the baseline dijet selection. Scale uncertainties with respect to 
    the central \Hjj NLO prediction are indicated by the shaded bands.
  }
  \label{fig:vbf-m12}
\end{figure}

All figures presented in this section will have the same structure:
they will show our results for \Hjj at NLO and \Hjjj at both LO and
NLO after the application of the VBF selection criteria. The main
plots on the left and on the right always contain the differential
distributions, which we obtained by utilizing the $p_T$-tagging and
$y$-tagging scheme, respectively. The differential cross sections of
each main plot are accompanied by four ratio plots. Starting from the
top we display ($i$) the three-jet fraction, ($ii$) the ratio to an
alternative tagging scheme definition ($p_T$-tagging/$y$-tagging on
the left and $y$-tagging/$m_{jj}$-tagging on the right), ($iii$) the
ratio to a different scale choice where instead of the default scale
B, we chose scale A and, finally, ($iv$) the reduction of the
respective baseline cross sections due to the VBF requirements given in
Eq.~\eqref{cuts:vbf}. Note that the basic gluon fusion cuts as stated
in Eq.~\eqref{cuts:basic} are used to define the baseline of the
respective \Hnj analysis. In the topmost subplot, all ratios are taken
with respect to the central \Hjj prediction at NLO accuracy using scale
choice B, cf.\ Eq.\ \eqref{scales:B}. The other three subplots show
the ratios between the respective \Hjj and \Hjjj samples that were
generated based on different ($ii$) jet tagging, ($iii$) scale setting
and ($iv$) selection cut level. In all cases, the shaded bands
indicate the respective standard scale uncertainties.

We start by considering the tagging jet invariant mass distribution
$m_{j_1j_2}$ reported in Fig.\ \ref{fig:vbf-m12}. After applying the VBF
cuts the three-jet fraction varies between 0.3 and 0.4 in the 
$p_T$-tagging scheme, increasing to 0.5-0.7 in the $y$-tagging scheme. The
contribution from \Hjjj is therefore non-negligible also for values
of $m_{j_1j_2}$ close to the cut. As already observed for the
inclusive cross section, the ratios between the results of different
tagging strategies shows a 25\% increase in the cross section for \Hjj
at NLO over the whole kinematic range and a 100\% increase for \Hjjj
both at LO and NLO when moving from $p_T$-tagging to $y$-tagging. The
results are instead almost identical for $y$-tagging and
$m_{jj}$-tagging. Also, varying the scale from choice B to choice A
does not have a big impact, in particular at NLO. Finally we observe
that the reduction in the cross section, due to remaining $\Delta y_{j_1,j_2}$ 
cut, is of about 50\% at around 400~\GeV and is, unsurprisingly, almost 
absent at 1~\TeV. There, almost all dijet configurations also fulfill the 
rapidity separation criterion. As we will see, this decrease is much 
more dramatic for other observables which are dominated by the bulk of 
the events just beyond the cut. 

\begin{figure}[t!]
  \centering
  \includegraphics[width=0.49\textwidth]{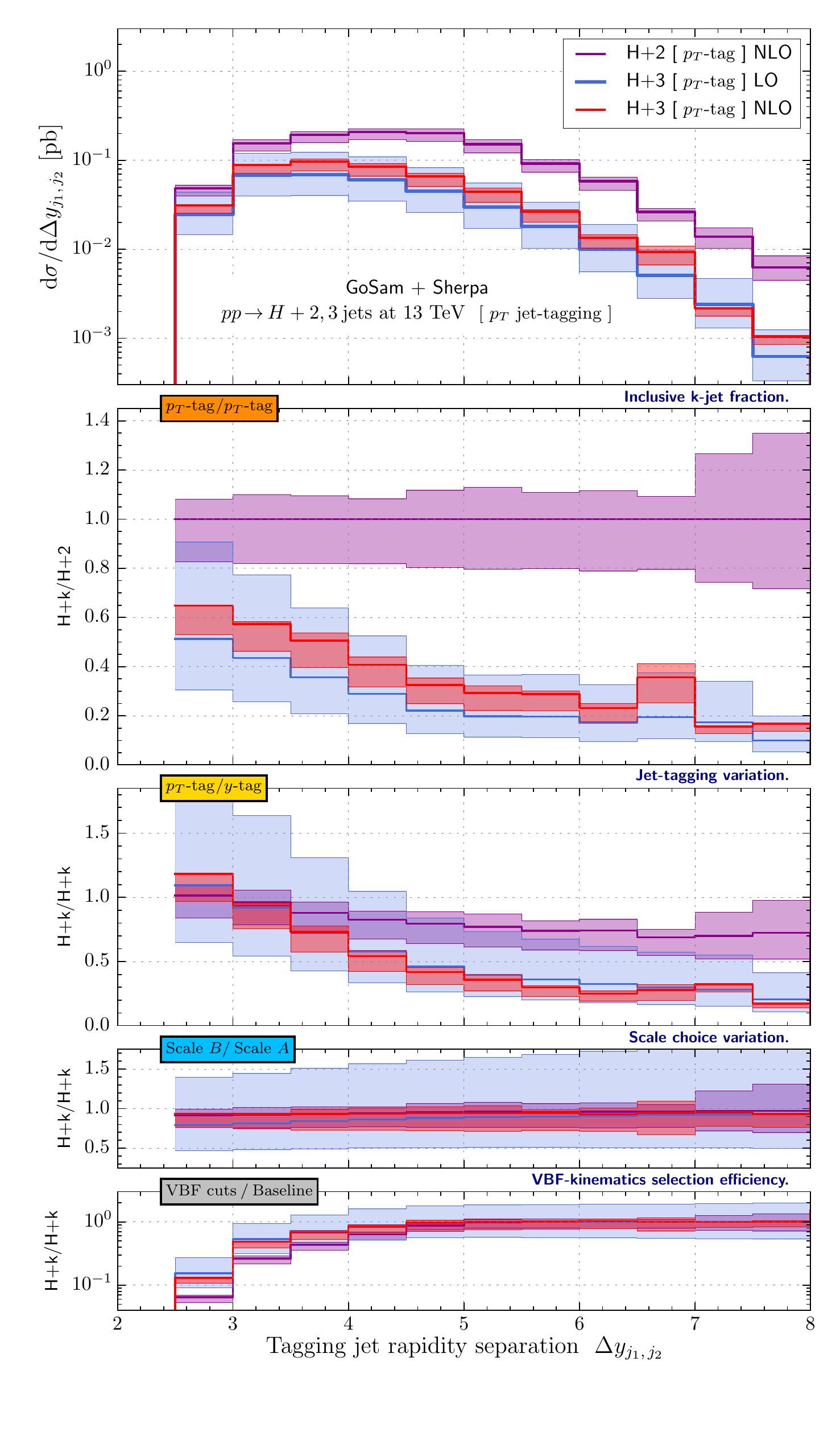}
  \hfill
  \includegraphics[width=0.49\textwidth]{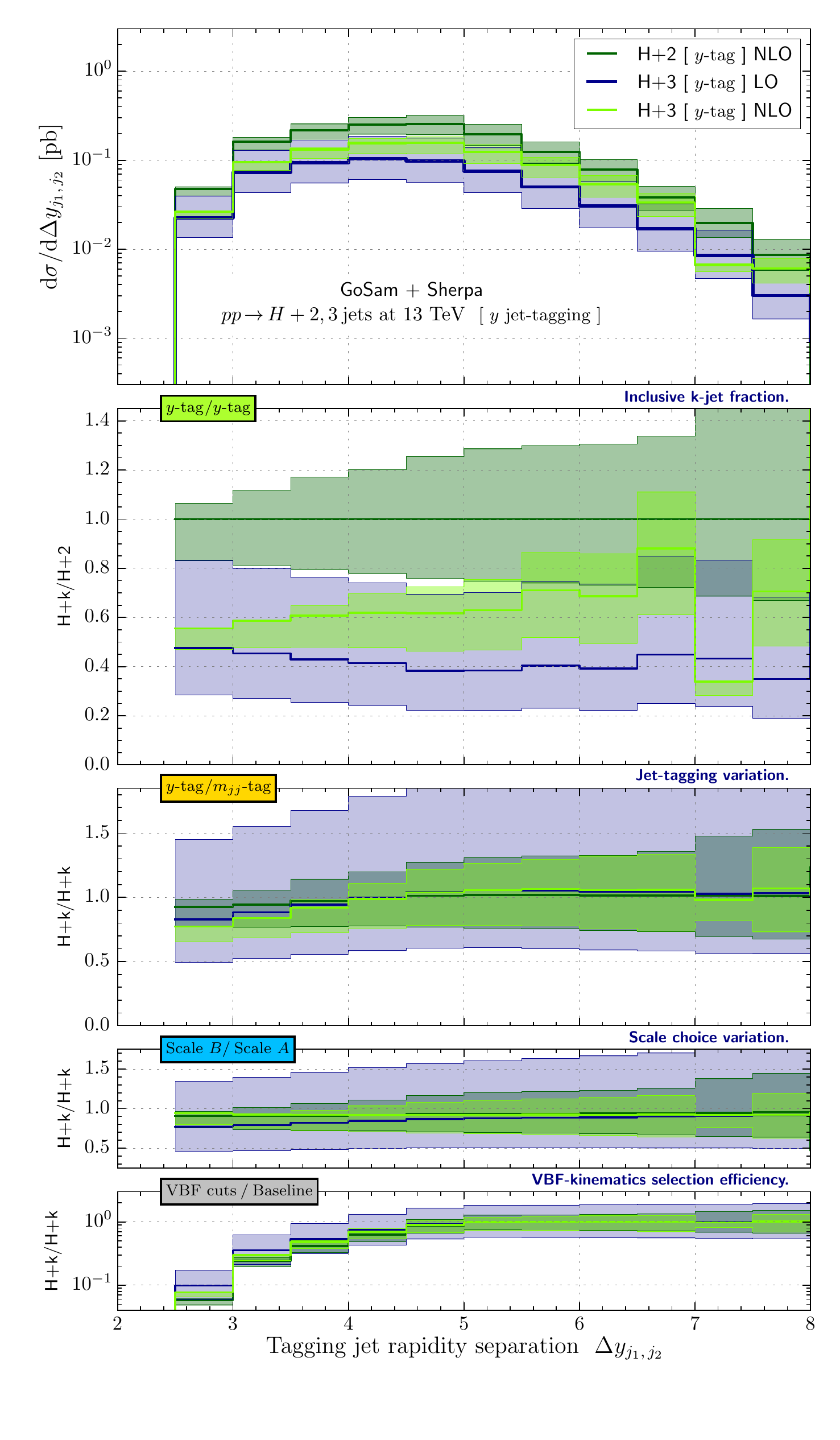}
  \caption{
    Impact of an \Hjjj description at NLO using the 
    scale choice B, cf.\ Eq.\ \eqref{scales:B}, on 
    the tagging rapidity separation $\Delta y_{j_1,j_2}$ 
    for the $p_T$-tagging (left) and $y$-tagging (right) jet selection 
    imposing VBF kinematic requirements at the LHC of 13 \TeV CM energy. 
    For details see Fig.\ \ref{fig:vbf-m12}.}
  \label{fig:vbf-dy12}
\end{figure}

The distribution of the tagging jet rapidity separation is in turn 
shown in Fig.\ \ref{fig:vbf-dy12}. On the left plot, in the 
$p_T$-tagging scheme, we observe an important change in the shape of 
the distribution going from \Hjj to \Hjjj. For small rapidity separations 
the presence of a further jet gives an additional contribution as 
large as 60\%. This decreases to less than 20\% for $\Delta y_{j_1,j_2}>7$. 
It instead remains approximately constant in the case of a $y$- or 
$m_{jj}$-tagging. Again, as discussed in Sec.\ \ref{sec:gf:hej} this 
reproduces the high energy behavior reported in \cite{Campbell:2013qaa}. 
Also in this case varying the scale choice has almost no impact. The 
lowest ratio leads to the conclusion that events with a rapidity interval 
of at least five units automatically fulfill the $m_{j_1j_2}$ cuts 
independent of the tagging scheme.

\begin{figure}[t!]
  \centering
  \includegraphics[width=0.49\textwidth]{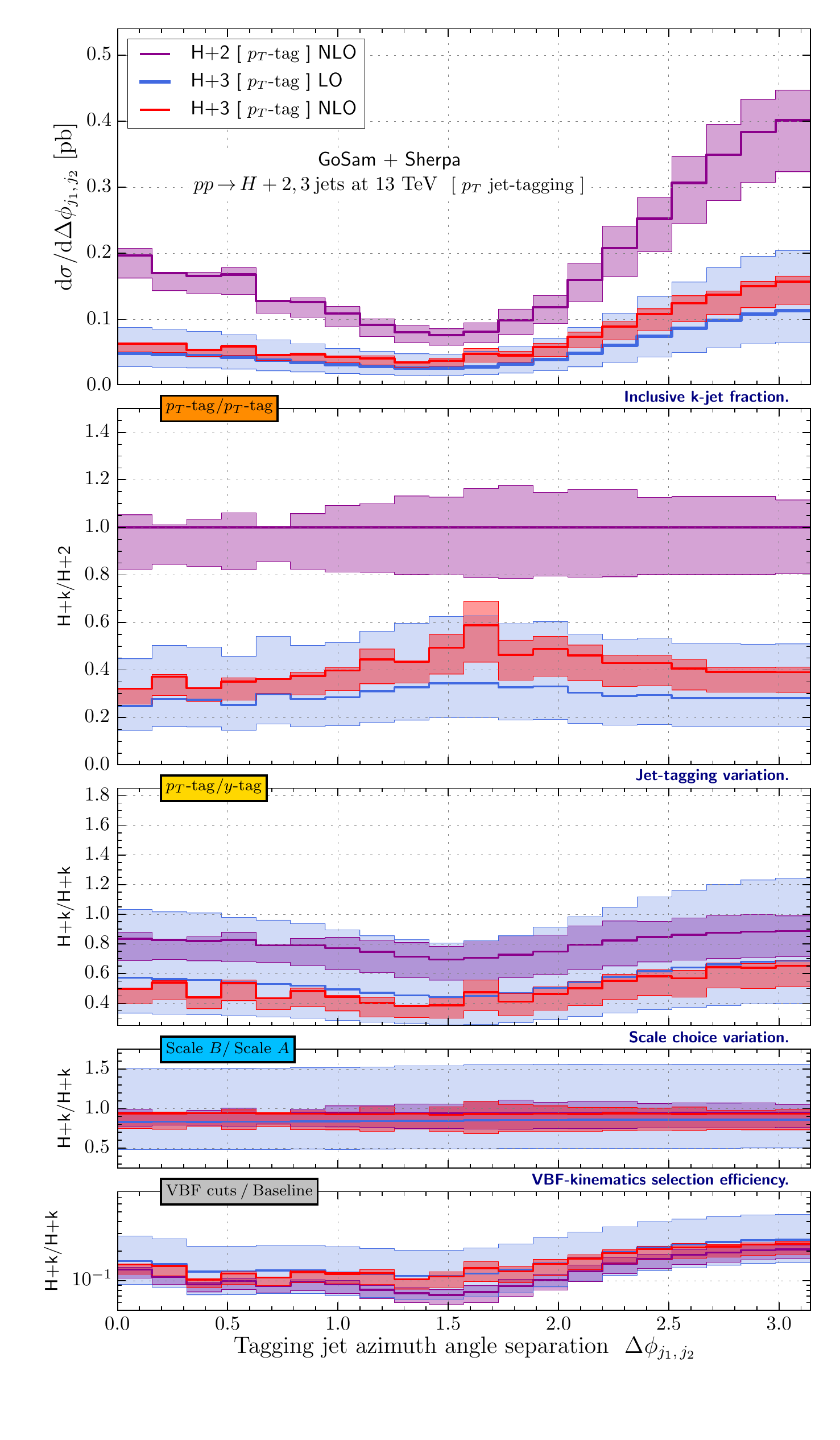}
  \hfill
  \includegraphics[width=0.49\textwidth]{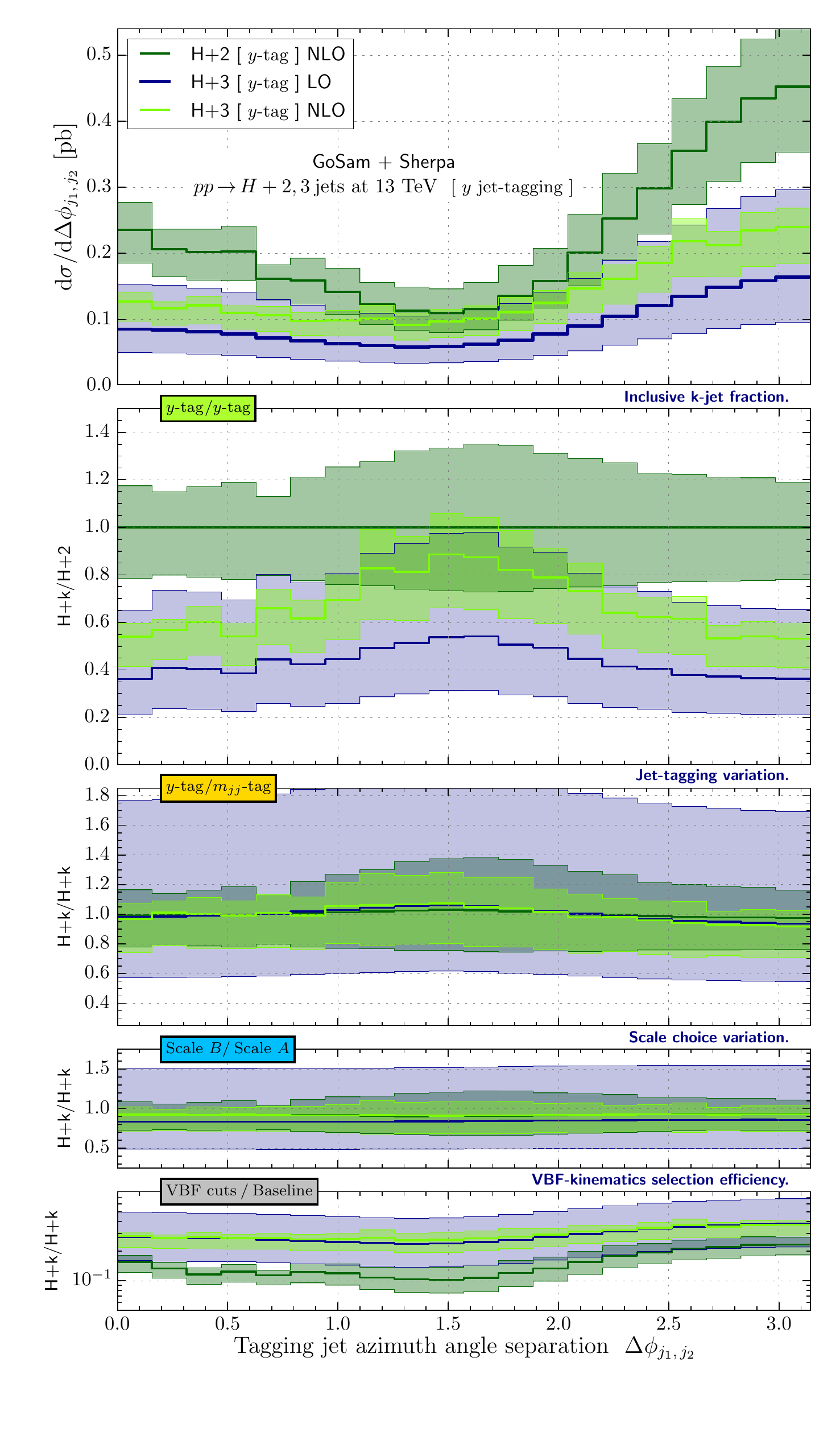}
  \caption{
    Impact of an \Hjjj description at NLO using the
    scale choice B, cf.\ Eq.\ \eqref{scales:B}, on 
    the tagging jet azimuthal separation $\Delta\phi_{j_1,j_2}$ 
    for the $p_T$-tagging (left) and $y$-tagging (right) jet selection 
    imposing VBF kinematic requirements at the LHC of 13 \TeV CM energy. 
    For details see Fig.\ \ref{fig:vbf-m12}.}
  \label{fig:vbf-dphi12}
\end{figure}

One of the most important distributions in the VBF process is the
difference in the azimuthal angle $\phi$ between the two tagging
jets. It allows a distinction between the
different possible CP-structures of the Higgs and is an interesting
channel to detect anomalous couplings.  We present the contribution
from the gluon fusion channel after VBF cuts in
Fig.\ \ref{fig:vbf-dphi12}. Comparing these plots with the ones
from the basic cuts, cf.\ Fig.\ \ref{fig:dphi}, a clear change in 
shape, in particular for high values of $\Delta\phi_{j_1,j_2}$, is 
evident. This can easily be understood by recalling that the
requirements of the VBF cuts, namely high invariant mass and a
considerable difference in rapidity forces the two jets into a
back-to-back configuration which is given at $\Delta\phi_{j_1,j_2}\lesssim\pi$. 
Interestingly, the largest scale dependence, originating in the large 
three-jet fraction (which approximately doubles here from 0.3 to 0.6) 
occurs at $\Delta\phi_{j_1,j_2}\approx\tfrac{\pi}{2}$.
This effect is even more pronounced in the $y$-tagging scheme, which 
largely coincides with the $m_{jj}$-tagging scheme, where the impact 
of the \Hjjj contribution is close to 90\% near the 
perpendicular azimuth.

\begin{figure}[t!]
  \centering \includegraphics[width=0.49\textwidth]{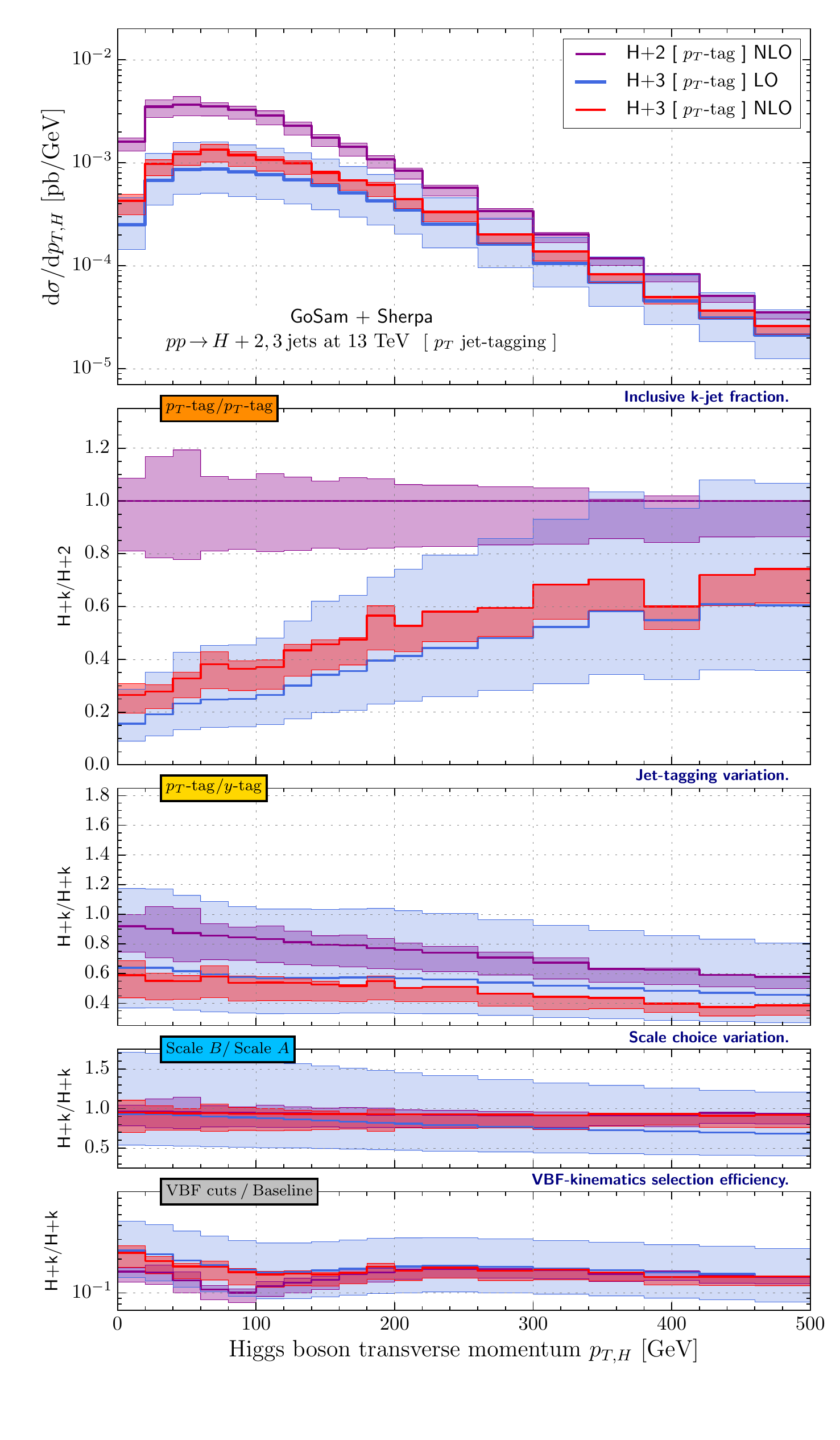}
  \hfill \includegraphics[width=0.49\textwidth]{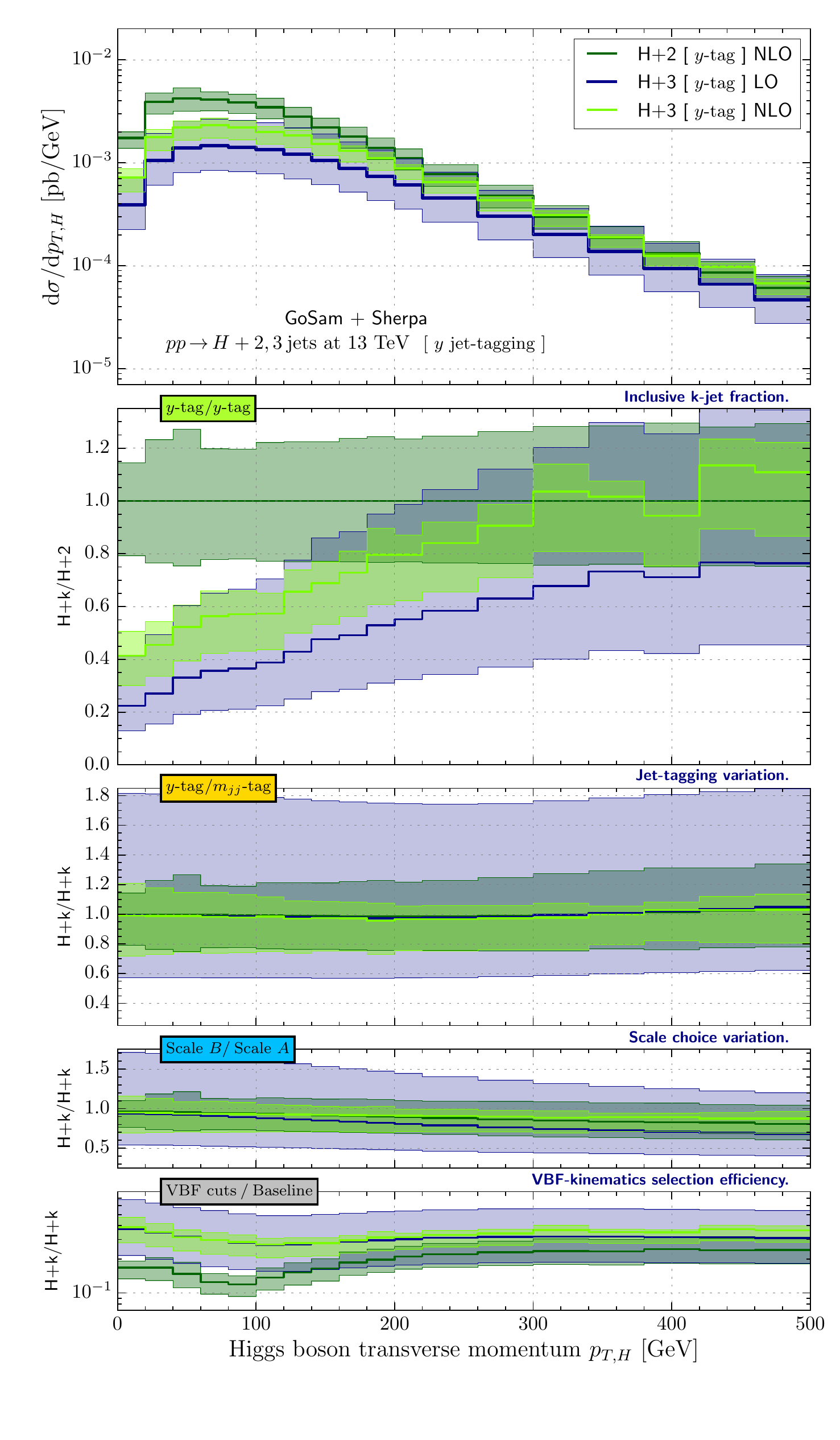}
  \caption{
    Impact of an \Hjjj description at NLO using the 
    scale choice B, cf.\ Eq.\ \eqref{scales:B}, on 
    the Higgs boson transverse momentum $p_{T,\Higgs}$ 
    for the $p_T$-tagging (left) and $y$-tagging (right) jet selection 
    imposing VBF kinematic requirements at the LHC of 13 \TeV CM energy. 
    For details see Fig.\ \ref{fig:vbf-m12}.
  }
  \label{fig:vbf-pTh}
\end{figure}

Turning now to the transverse momentum distribution of the Higgs boson, 
displayed in Fig.\ \ref{fig:vbf-pTh}, we observe that the shape remains 
largely unaffected by the more stringent VBF cuts wrt.\ the more liberal 
dijet selection (Fig.\ \ref{fig:xnlo-ratio-pTh}). This is no surprise 
as the additional VBF cuts do 
not act directly on the Higgs boson itself. The cross section, however, 
decreases by almost an order of magnitude over the whole kinematic range 
in the $p_T$-tagging scheme while the reduction is again only a factor of 
3 in the $y$-tagging and $m_{jj}$-tagging schemes. The choice of central 
renormalization scale introduces a slight tilt in the distributions 
irrespective of the tagging scheme. Consequently many observations 
done for Fig.\ \ref{fig:pt} still apply in this case. The contribution of 
\Hjjj at NLO becomes as large as 50\% of the \Hjj contribution already 
around 160~\GeV. This increases to 70\% when a $y$- or $m_{jj}$-tagging 
strategy is used, stressing the effective LO nature of the \Hjj NLO 
calculation in this region.

\begin{figure}[t!]
  \centering
  \includegraphics[width=0.49\textwidth]{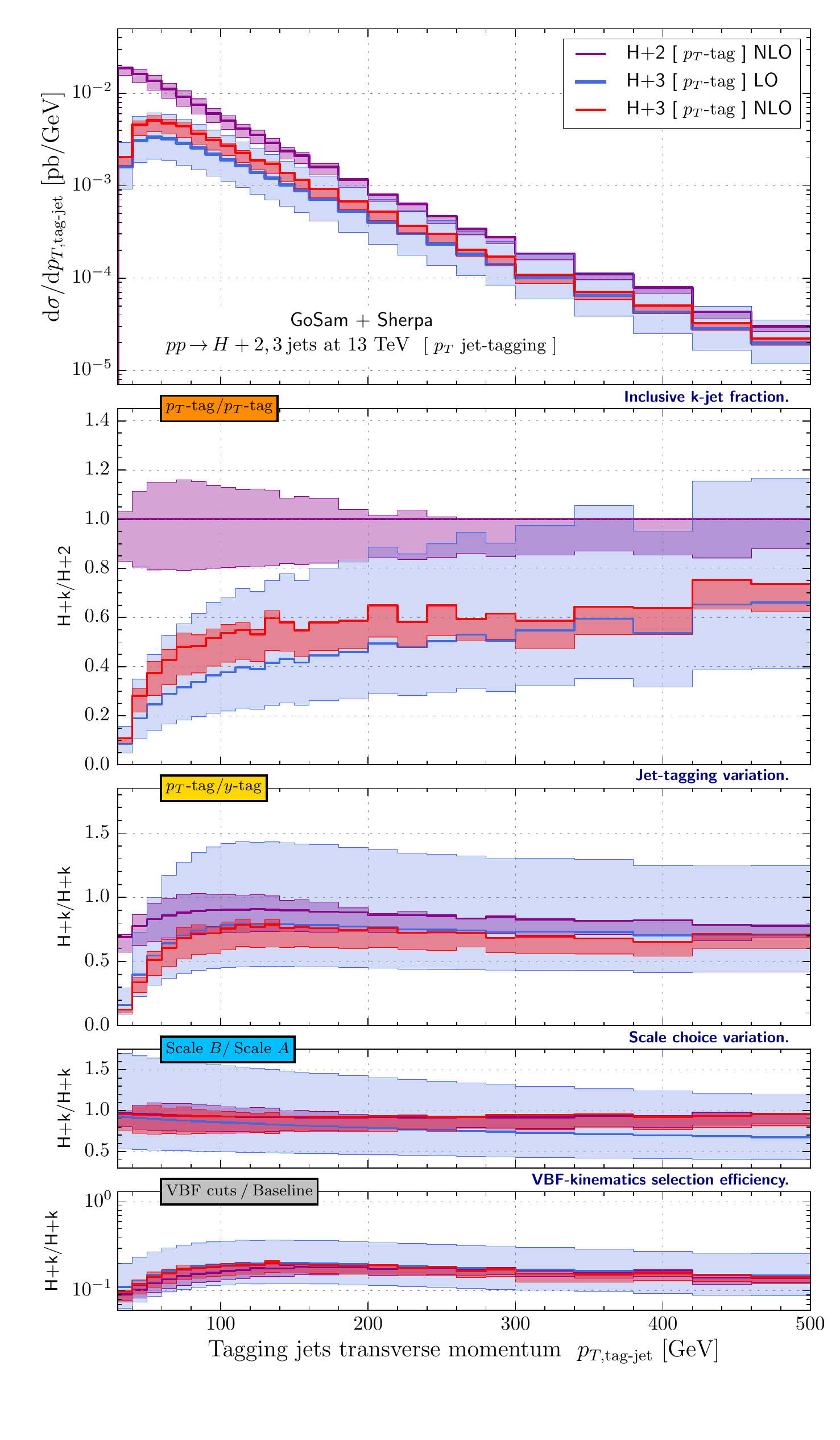}
  \hfill
  \includegraphics[width=0.49\textwidth]{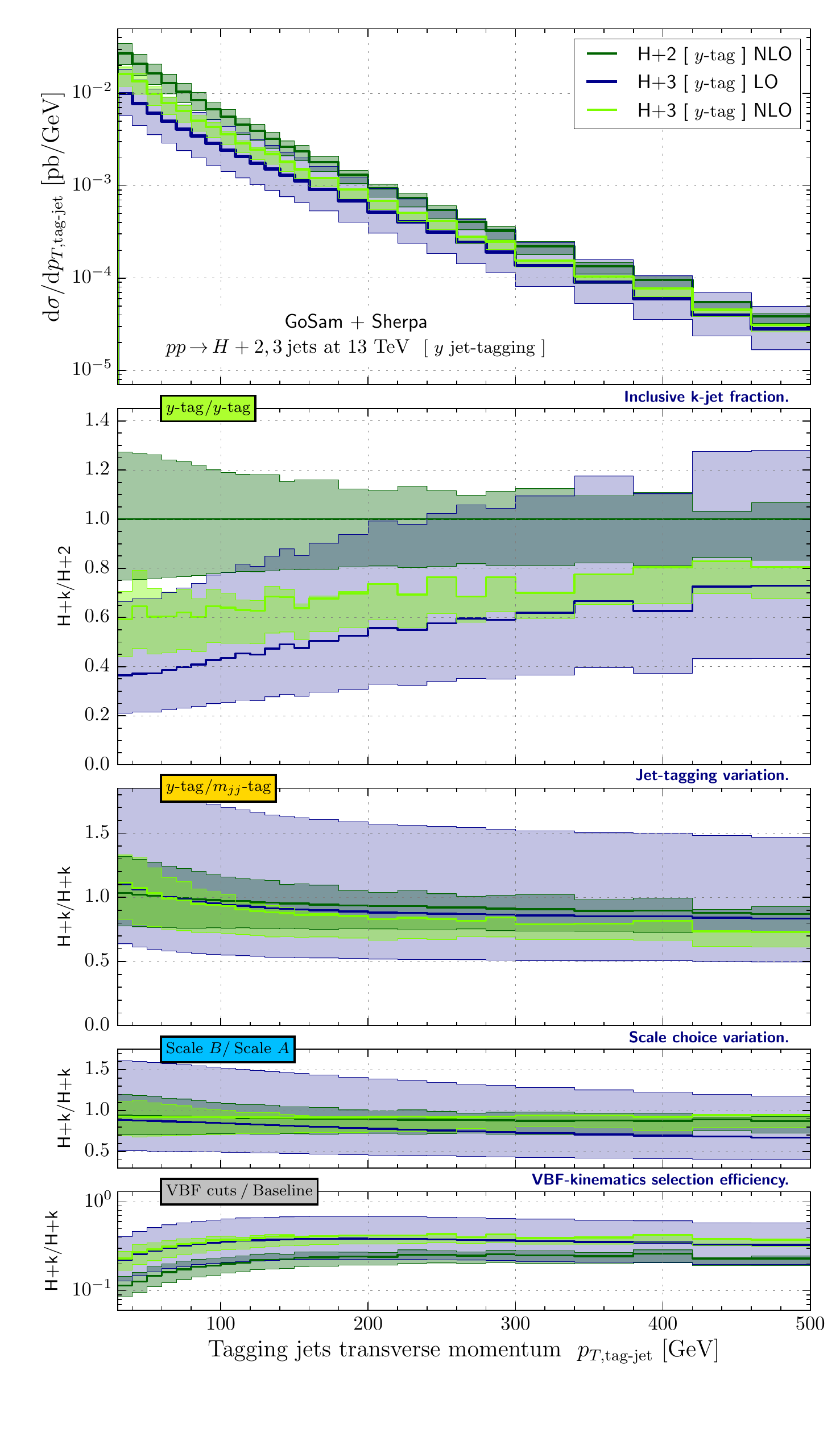}
  \caption{
    Impact of an \Hjjj description at NLO using the 
    scale choice B, cf.\ Eq.\ \eqref{scales:B}, on 
    the inclusive tagging jet transverse momentum $p_{T,\text{tag-jet}}$ 
    for the $p_T$-tagging (left) and $y$-tagging (right) jet selection 
    imposing VBF kinematic requirements at the LHC of 13 \TeV CM energy. 
    For details see Fig.\ \ref{fig:vbf-m12}.
  }
  \label{fig:vbf-pTj1}
\end{figure}

\begin{figure}[t!]
  \centering
  \includegraphics[width=0.49\textwidth]{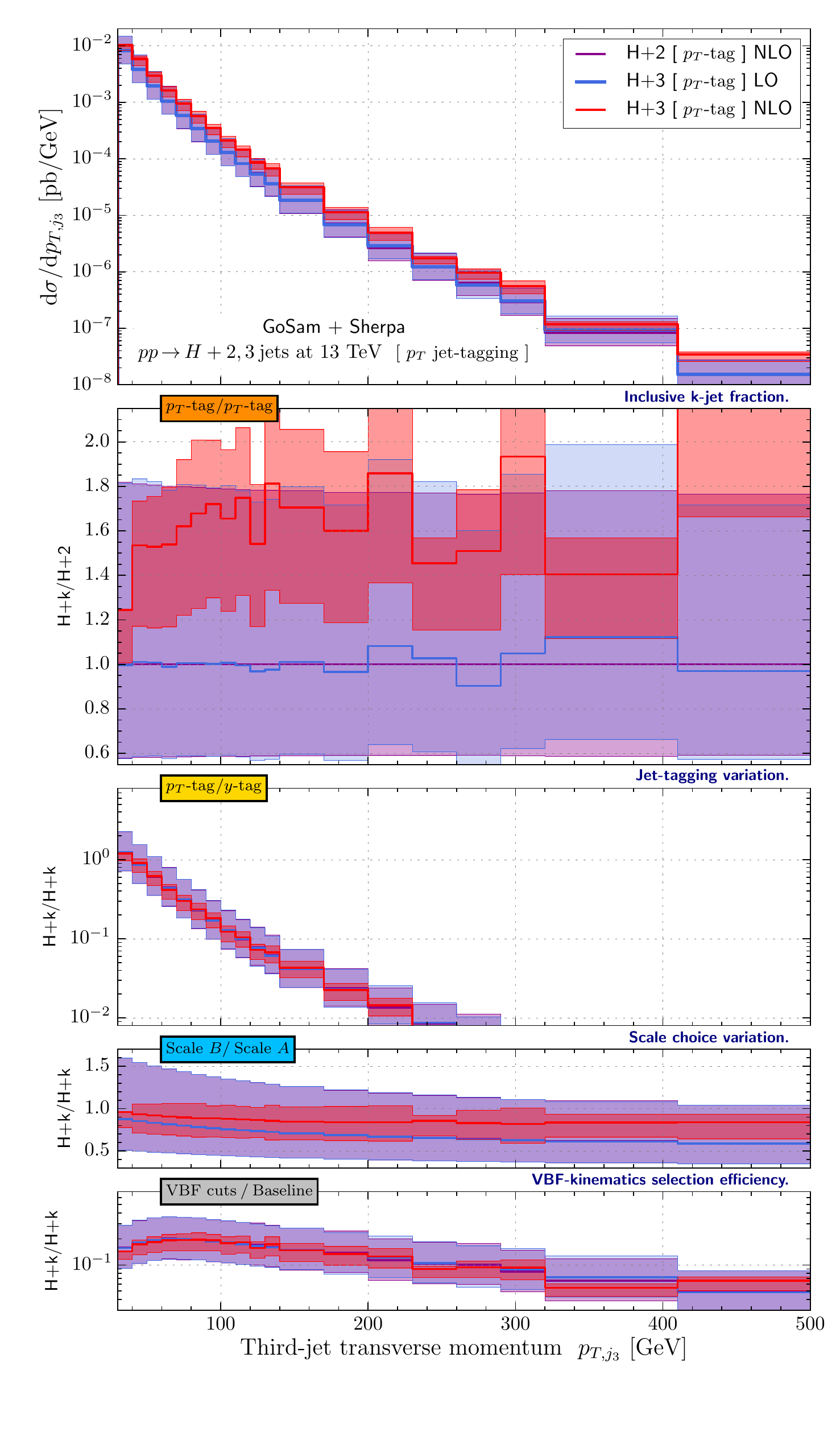}
  \hfill
  \includegraphics[width=0.49\textwidth]{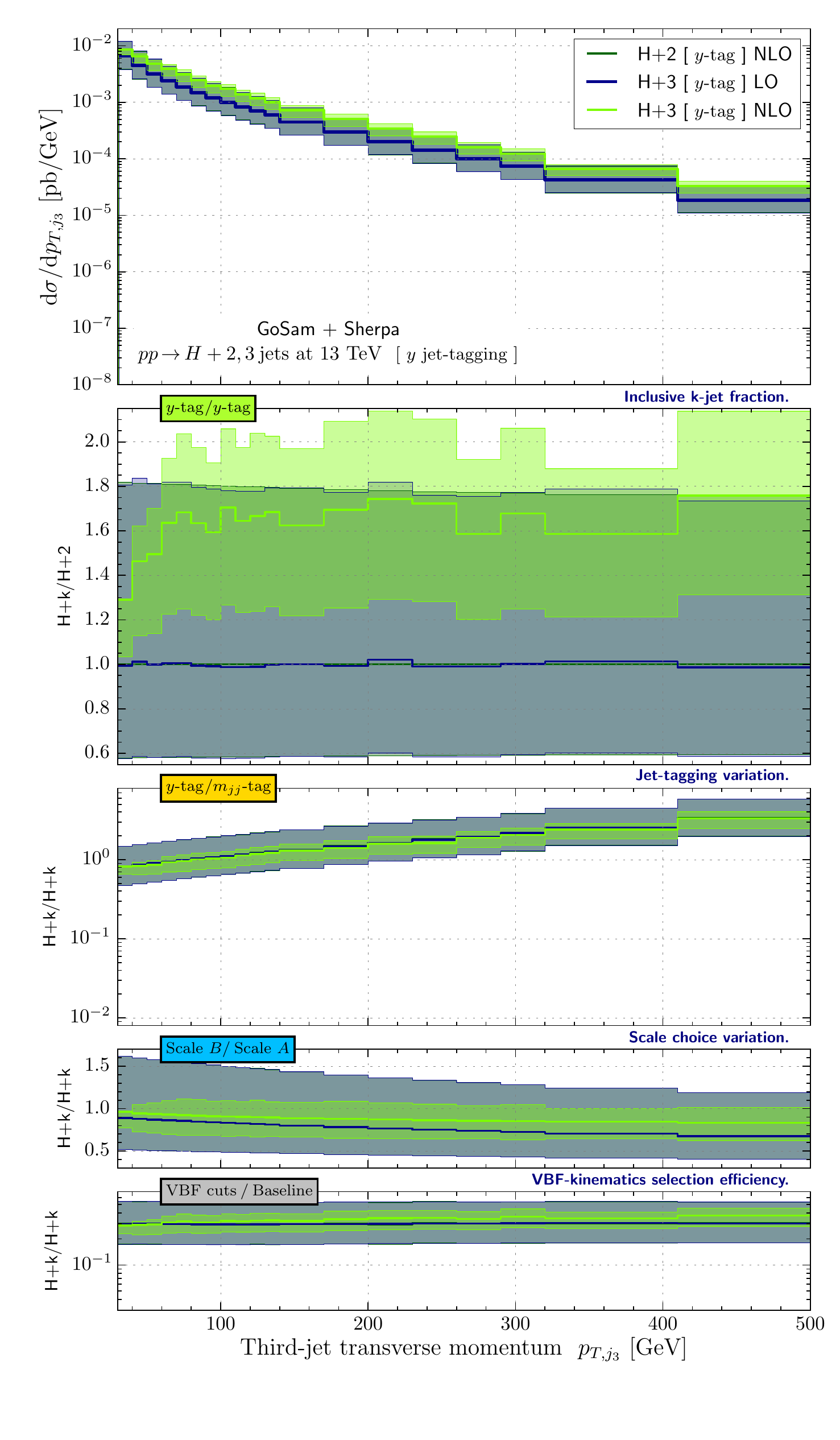}
  \caption{
    Impact of an \Hjjj description at NLO using the 
    scale choice B, cf.\ Eq.\ \eqref{scales:B}, on 
    the transverse momentum of the hardest non-tagging jet $p_{T,j_3}$ 
    for the $p_T$-tagging (left) and $y$-tagging (right) jet selection 
    imposing VBF kinematic requirements at the LHC of 13 \TeV CM energy. 
    For details see Fig.\ \ref{fig:vbf-m12}.
  }
  \label{fig:vbf-pTjsub}
\end{figure}

In Figs.\ \ref{fig:vbf-pTj1} and \ref{fig:vbf-pTjsub} the inclusive 
tagging jet transverse momentum and the transverse momentum of the 
leading non-tagging jet is shown. These plots can directly be
compared with Fig.\ \ref{fig:tag-sgljetpts}. Apart from the general 
decrease in the cross section, the curves are qualitatively very 
similar. Since \Hjj describes additional jet activity beyond the 
tagging jets only through the resolved real radiation contribution, 
the predictions from \Hjj at NLO and \Hjjj at LO are identical (up to 
statistical fluctuations) for $p_{T,j_3}$. The contribution from NLO 
corrections to \Hjjj therefore has a large impact leading to a large 
K-factor varying between 1.3 and 1.8. Consequently, as already seen in 
Sec.\ \ref{sec:gf:hej}, this observable also experiences a large impact 
on its properties by the choice of tagging scheme, leading to much 
steeper fall-off in the $p_T$-tagging scheme than in either the 
$y$-tagging scheme. The $m_{jj}$-tagging scheme here exhibits larger 
variations from the $y$-tagging scheme, distorting its behavior at 
large transverse momenta to smaller rates, while giving an overall 
similar behavior. The effect of the VBF cuts on the shape of the 
leading non-tagging jet transverse momentum is only moderate in the 
$y$-tagging scheme, while in the $p_T$-tagging scheme they reject more 
events with larger $p_{T,j_3}$ than with lower. This effect is more 
pronounced in the transverse momentum of the tagging jets themselves.
Again, the choice of functional form of the renormalization scale 
introduces a noticeable tilt into all observables.

\begin{figure}[t!]
  \centering
  \includegraphics[width=0.49\textwidth]{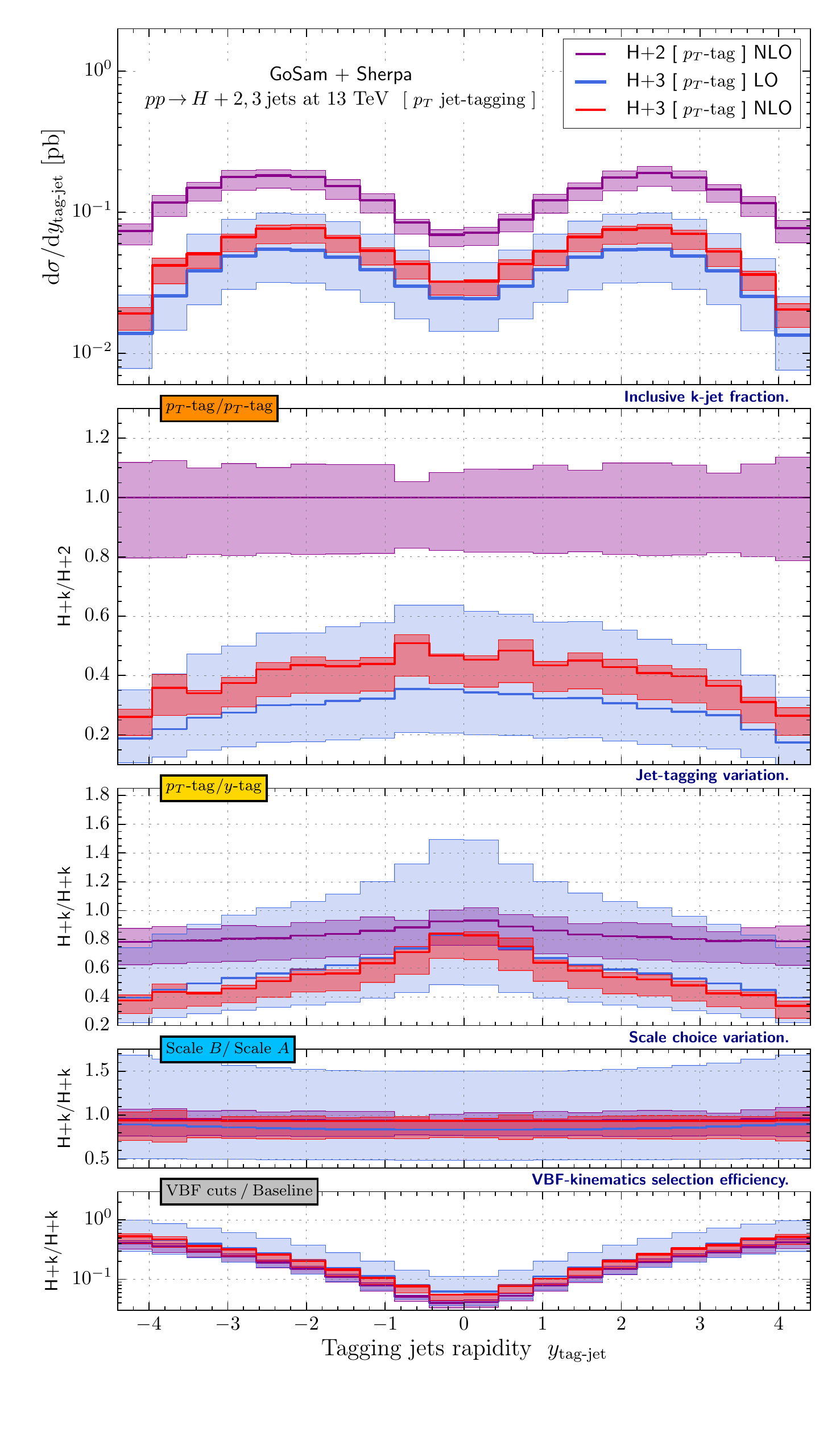}
  \hfill
  \includegraphics[width=0.49\textwidth]{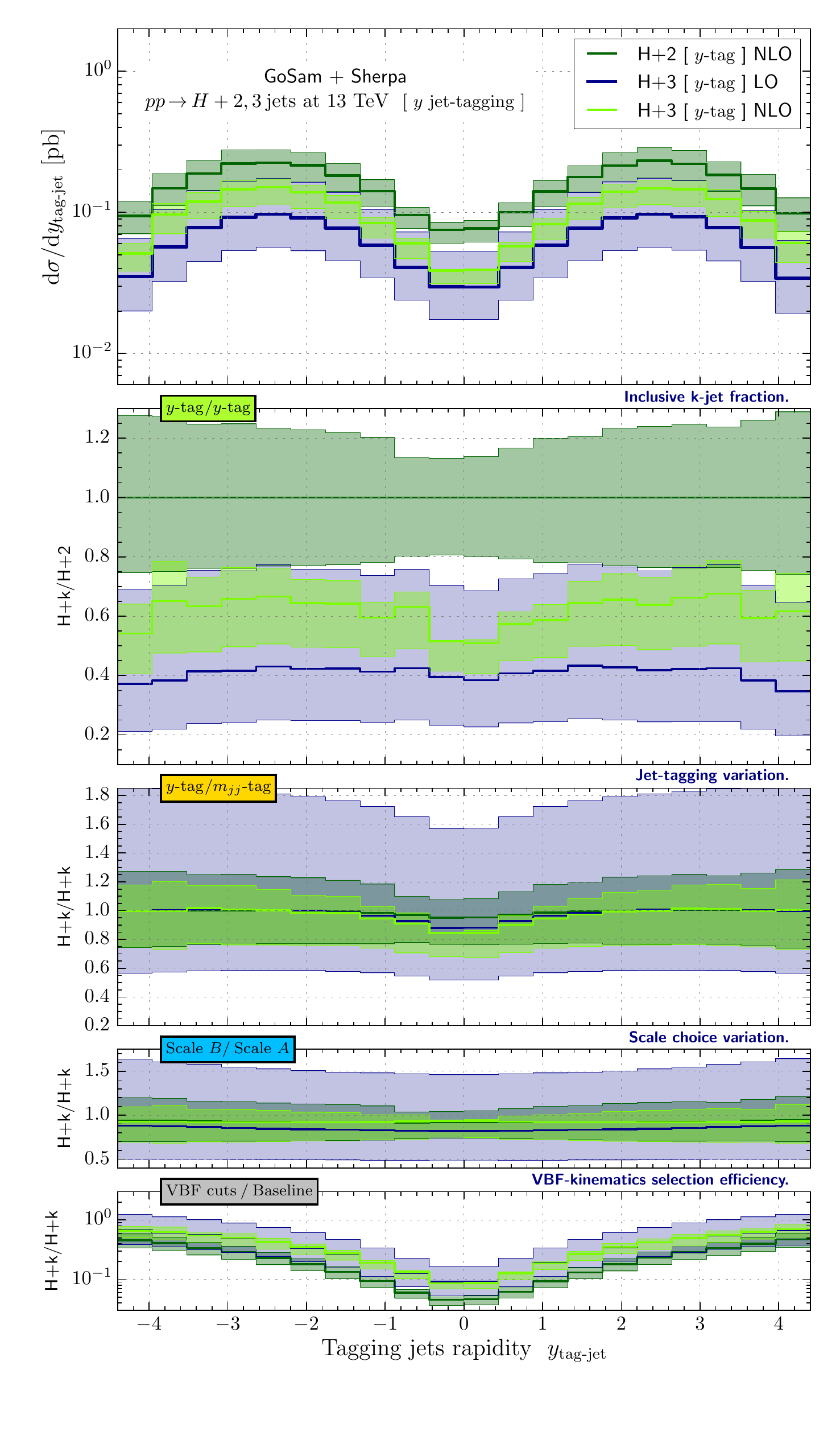}
  \caption{
    Impact of an \Hjjj description at NLO using the 
    scale choice B, cf.\ Eq.\ \eqref{scales:B}, on 
    the inclusive tagging jet rapidity $y_\text{tag-jet}$ 
    for the $p_T$-tagging (left) and $y$-tagging (right) jet selection 
    imposing VBF kinematic requirements at the LHC of 13 \TeV CM energy. 
    For details see Fig.\ \ref{fig:vbf-m12}.
  }
  \label{fig:vbf-rapj1}
\end{figure}

The situation changes for observables that are directly affected by the VBF
cuts.  In Fig.\ \ref{fig:vbf-rapj1} we show the inclusive rapidity
distributions of the tagging jets, which show the characteristic distinct dip 
in the central region which is not present in the case of basic cuts. The dip
is of course caused by the $\Delta y$ cut between the two tagging jets, that
forces the jets towards higher rapidities, leaving a gap in the
central region. The precise shape of this gap strongly depends on the 
choice of tagging scheme: using $y$-tagging it is somewhat wider than 
using $p_T$-tagging, while, again, $y$-tagging and $m_{jj}$-tagging 
give very similar results. Unsurprisingly, in three jet events in 
$p_T$-tagging, comprising on average about 40\% of all events, the dip 
is less pronounced. While in $y$ tagging the presence of such a third jet 
leads to a wider separation of the tagging jets, as always the most 
forward and backward ones are chosen. Here, the three jet fraction again is 
larger overall and ranges up to 60\%. The precise choice of scale hardly 
matters.

\begin{figure}[t!]
  \centering
  \includegraphics[width=0.49\textwidth]{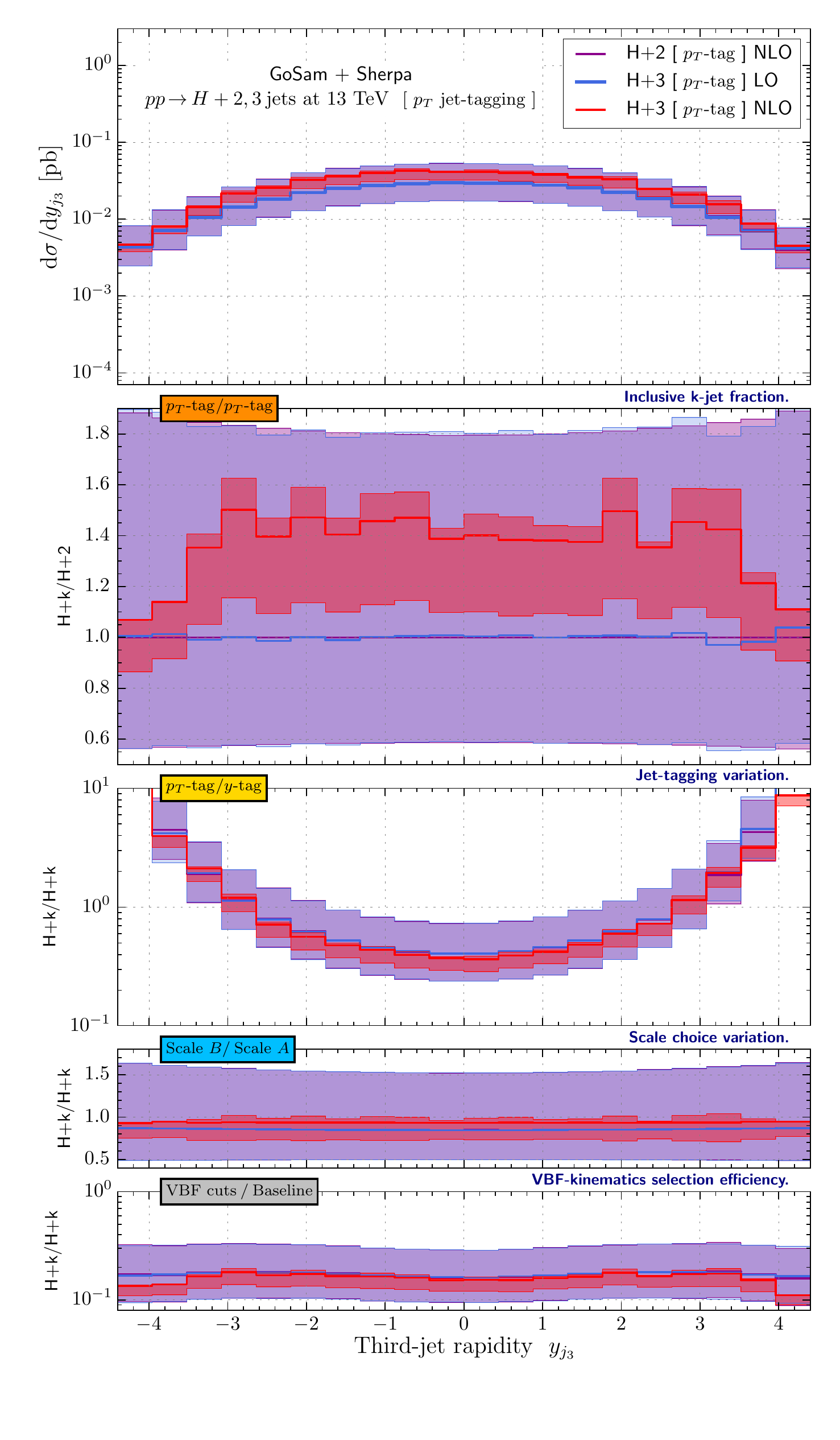}
  \hfill
  \includegraphics[width=0.49\textwidth]{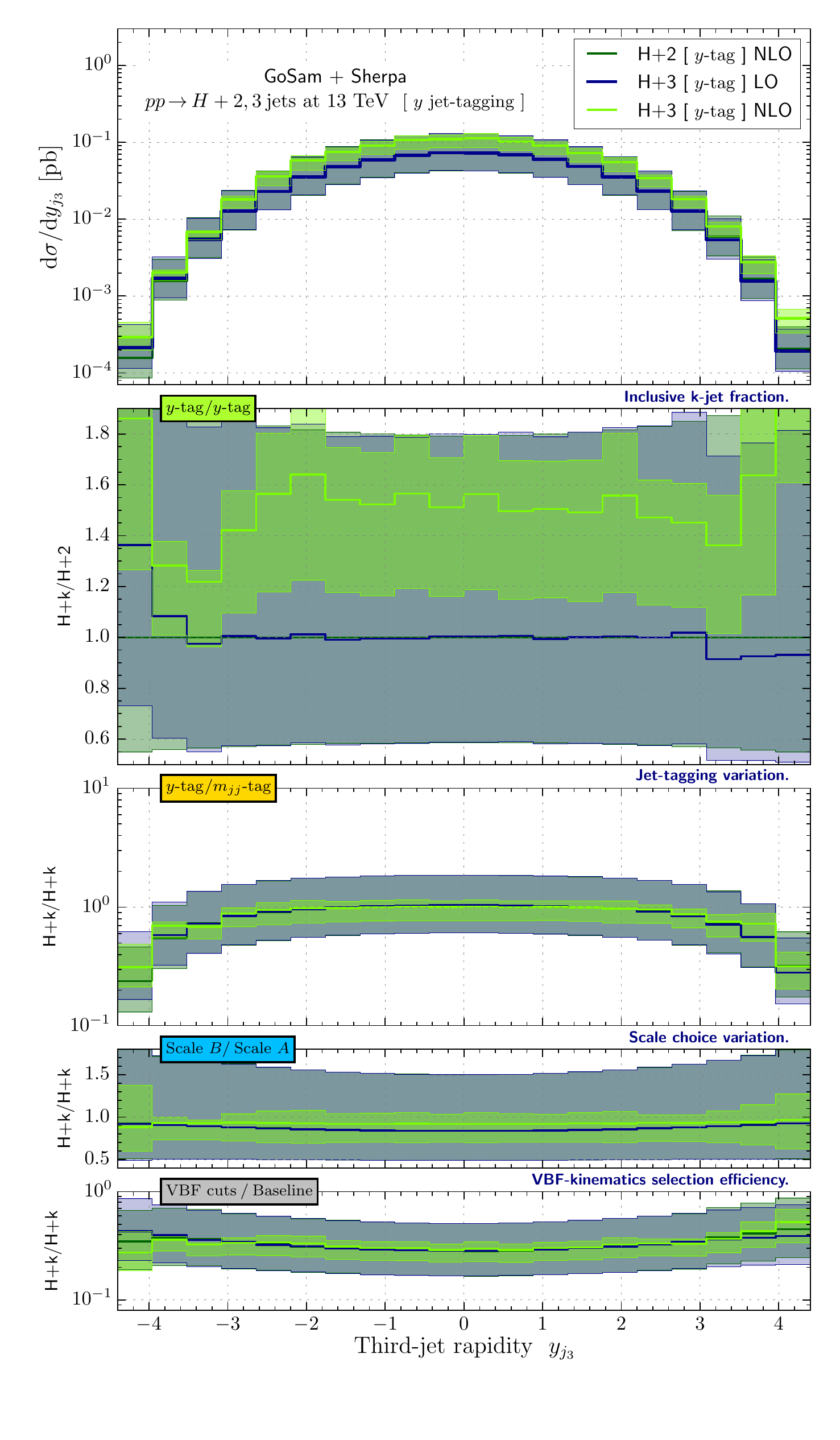}
  \caption{
    Impact of an \Hjjj description at NLO using the 
    scale choice B, cf.\ Eq.\ \eqref{scales:B}, on 
    the rapidity of the hardest non-tagging jet $y_{j_3}$ 
    for the $p_T$-tagging (left) and $y$-tagging (right) jet selection 
    imposing VBF kinematic requirements at the LHC of 13 \TeV CM energy. 
    For details see Fig.\ \ref{fig:vbf-m12}.
  }
  \label{fig:vbf-rapjsub}
\end{figure}

As the VBF cuts only act on the two tagging jets and Higgs boson production 
through gluon fusion, unlike production through weak boson fusion, comprises 
topologies with color connections between all colored partons, the above 
dip is not present neither in the rapidity distribution of any non-tagging 
jet nor for the Higgs boson, cf.\ Fig.\ \ref{fig:vbf-rapjsub}. Again, this 
observable is described at LO only by the \Hjj NLO calculation, coinciding 
with the \Hjjj LO calculation and necessitating the \Hjjj calculation at NLO 
accuracy. Characteristic differences are present between both tagging schemes. 
Besides the difference in the differential K-factor (1.4 for $p_T$-tagging 
and 1.5 for $y$-tagging), the shape is a direct consequence of which of the 
three jets are selected as tagging jets. Thus, in $y$-tagging the third jet 
is much more central than in $p_T$-tagging and only somewhat more central 
than in $m_{jj}$-tagging. As in Fig.\ \ref{fig:vbf-rapj1}, the choice of 
scale is almost inconsequential.

\begin{figure}[t!]
  \centering
  \includegraphics[width=0.49\textwidth]{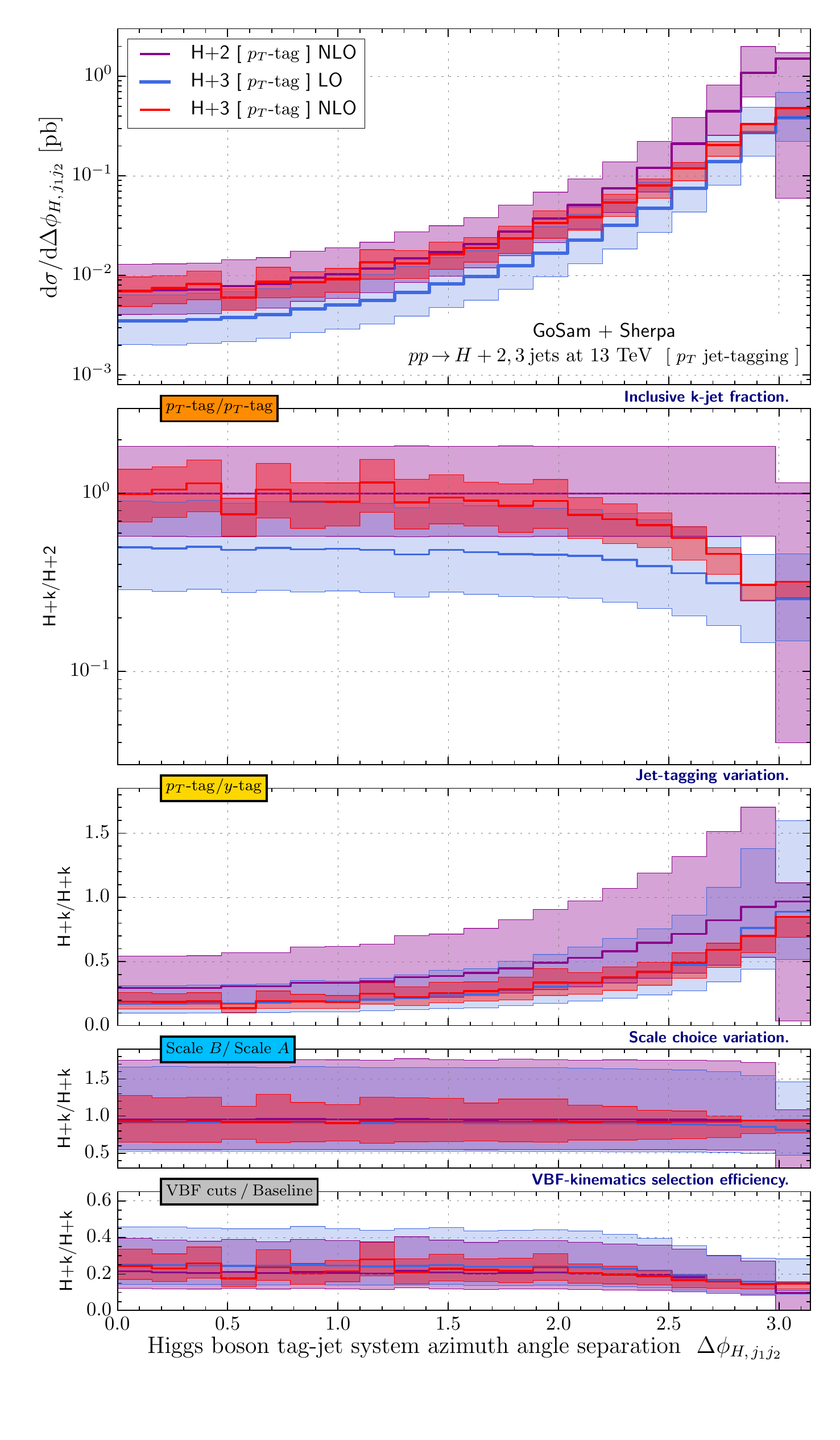}
  \hfill
  \includegraphics[width=0.49\textwidth]{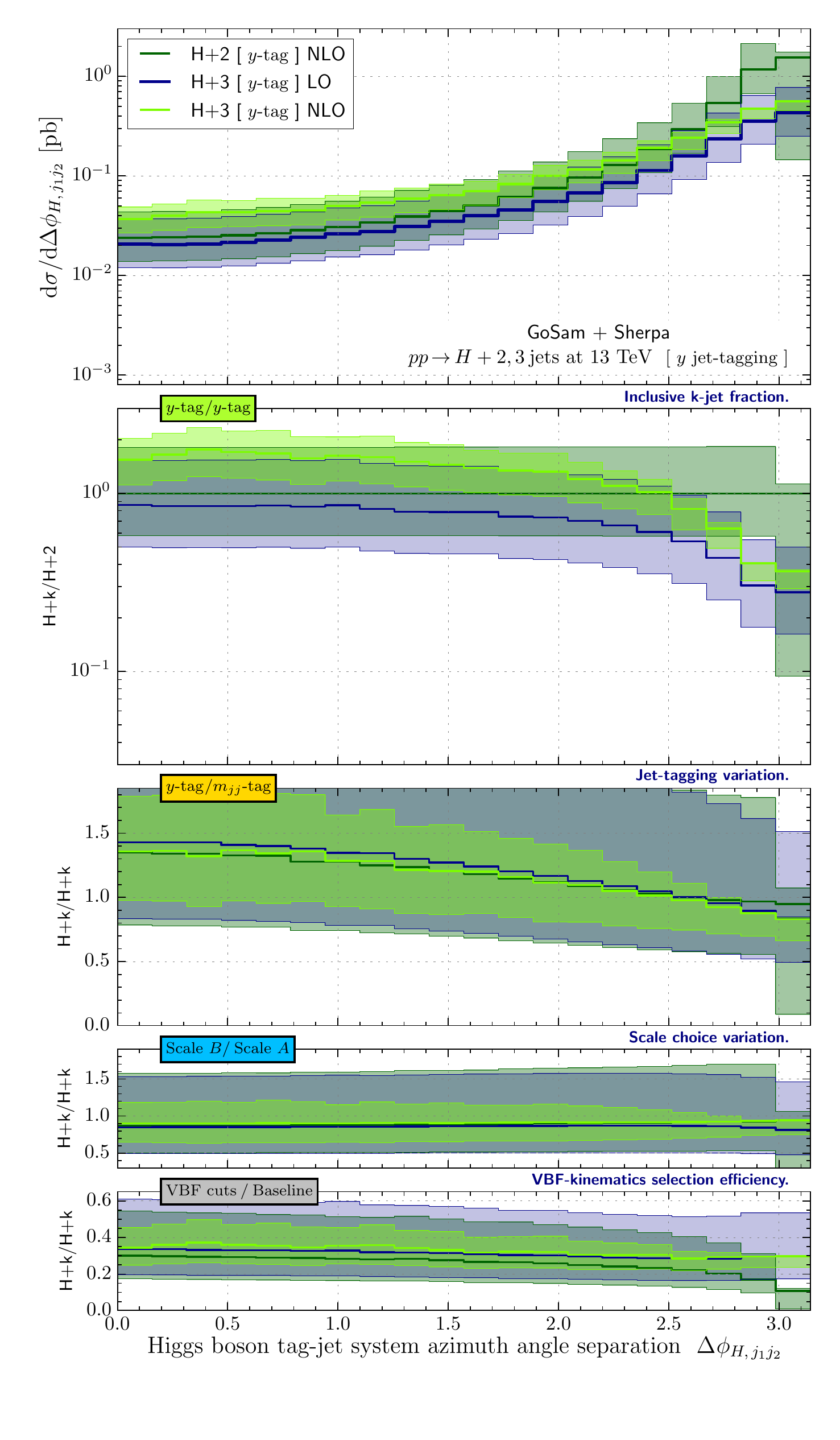}
  \caption{
    Impact of an \Hjjj description at NLO using the 
    scale choice B, cf.\ Eq.\ \eqref{scales:B}, on 
    the azimuthal separation of the Higgs boson and the tagging jet system 
    $\Delta\phi_{\Higgs,j_1j_2}$ 
    for the $p_T$-tagging (left) and $y$-tagging (right) jet selection 
    imposing VBF kinematic requirements at the LHC of 13 \TeV CM energy. 
    For details see Fig.\ \ref{fig:vbf-m12}.
  }
  \label{fig:vbf-dphi-h12}
\end{figure}

We finish discussing two observables which relate the Higgs boson to
the tagging jets. In Fig.\ \ref{fig:vbf-dphi-h12} we plot the azimuthal
separation between the Higgs boson and the dijet system defined by the
two tagging jets. This predictions can be compared with the ones in
Fig.\ \ref{fig:higgs12} (right), where they same observables is shown
with baseline cuts only. As shown in the bottom ratios, it is clear that the
shape of the predictions is very similar apart from a slightly milder
increase of the curve towards the back-to-back configuration. The
large uncertainty in the \Hjj curve reminds that this in only a LO
description. Therefore the contributions coming from \Hjjj NLO
corrections are particularly large and need to be taken into account
for a precise theoretical prediction. The previous considerations hold
both for $p_T$-tagging and for $y$-tagging. When directly comparing
the tagging schemes we observe that the predictions for the $y$-tagging 
scheme are flatter than both $p_T$- and $m_{jj}$-tagging scheme. While the 
$y$-tagging scheme favors configurations with the Higgs boson recoiling 
from both tagging jets as much as the $p_T$-tagging scheme and only 
somewhat less likely than the $m_{jj}$-tagging scheme, it allows for a 
factor of 3 (1.3) more events where the Higgs boson and the tagging jets 
recoil against the rest of the event. Again, this is easily understood as 
a consequence of the tagging jet selection process, leaving more or less 
energetic jets to recoil against and, thus, more or less opportunities for 
the Higgs and the tagging jet system to be boosted into the same direction.
Herein, scale choice A and B do not differ significantly.

\begin{figure}[t!]
  \centering
  \includegraphics[width=0.49\textwidth]{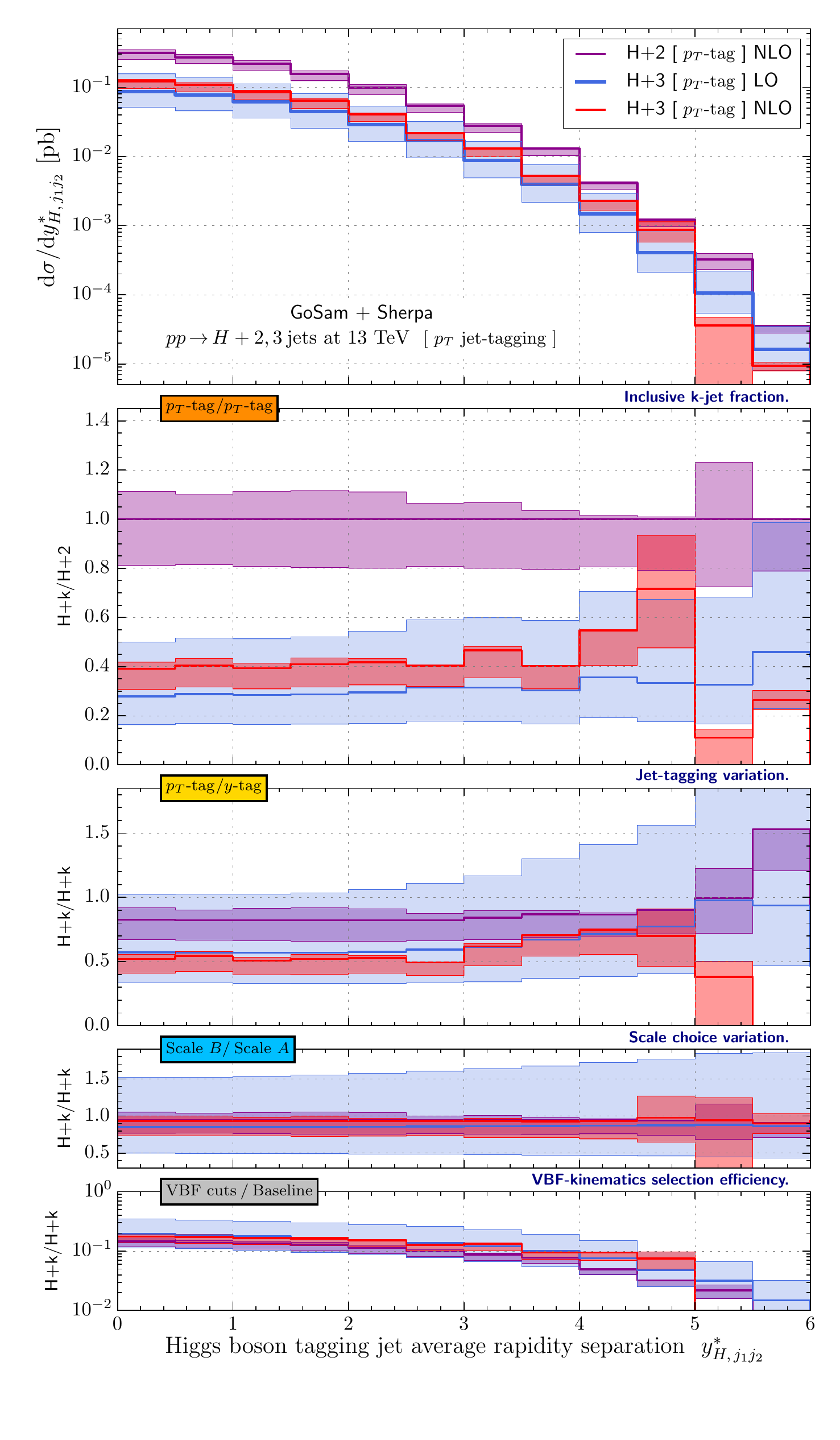}
  \hfill
  \includegraphics[width=0.49\textwidth]{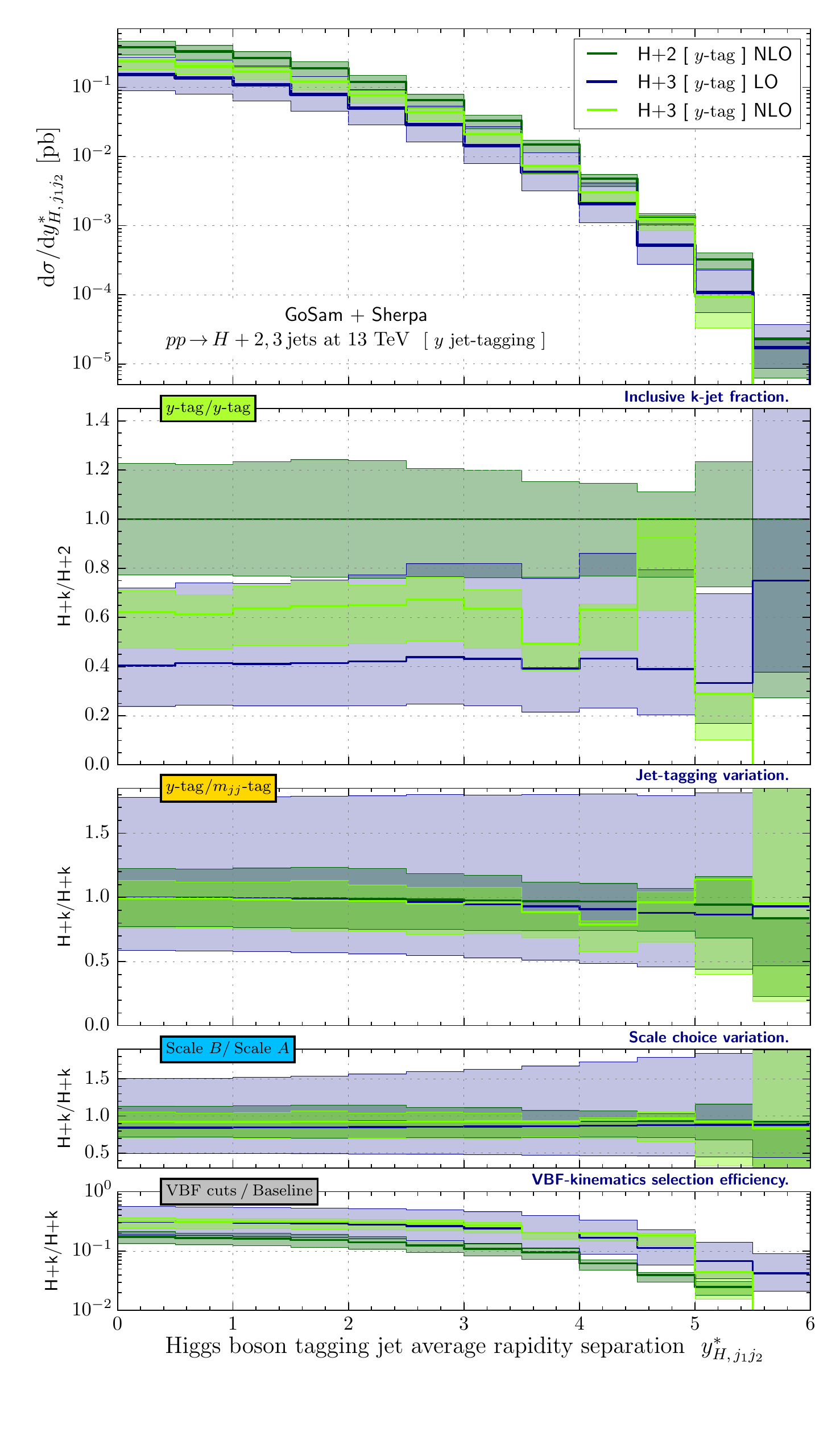}
  \caption{
    Impact of an \Hjjj description at NLO using the 
    scale choice B, cf.\ Eq.\ \eqref{scales:B}, on 
    the rapidity separation of the Higgs boson and tagging jets 
    $y_{\Higgs,j_1j_2}^*$ 
    for the $p_T$-tagging (left) and $y$-tagging (right) jet selection 
    imposing VBF kinematic requirements at the LHC of 13 \TeV CM energy. 
    For details see Fig.\ \ref{fig:vbf-m12}.
  }
  \label{fig:vbf-dy-h1}
\end{figure}

Finally, Fig.\ \ref{fig:vbf-dy-h1} shows again $\ystar{\Higgs}{j_1j_2}$,  
defined in Eq.\ \eqref{eq:ystarhjj}. Compared to the results with 
baseline cuts of Fig.~\ref{fig:tag-sgljetraps} (right), the distributions 
fall off a bit faster for very large rapidity separations, independently 
of the tagging method. Another difference which is worth mentioning is that
the differential three-jet fraction increases from about 40\% (60\%) at 
$\ystar{\Higgs}{j_1j_2}\lesssim 4$ to 80\% (90\%) at 
$\ystar{\Higgs}{j_1j_2}\approx 5$ in the $p_T$-tagging ($y$-tagging) 
scheme, making the contribution of \Hjjj NLO corrections even more 
important. Furthermore, in both cases the ratio compared to baseline cuts 
is more constant using the \Hjjj NLO calculation compared to the LO one.
Qualitatively, all tagging schemes predict the same shape for 
$\ystar{\Higgs}{j_1j_2}\lesssim 4$ with both the $y$- and $m_{jj}$-tagging 
scheme predicting a few more events at even larger $\ystar{\Higgs}{j_1j_2}$.
Again, scale choice A and B do not differ significantly.

\section{Conclusions}\label{sec:conclusions}

Gluon fusion is the dominant production mechanism for a Standard Model Higgs boson at the LHC.
The production of a Higgs boson in gluon fusion in association with jets also constitutes an irreducible
background to the vector boson fusion mechanism. Reliable predictions
for the Higgs boson plus jets processes are therefore indispensable
for a precise determination of the Higgs boson couplings and its
properties in the VBF signal.

In this paper we have presented a detailed phenomenological analysis of the gluon fusion contribution to Higgs boson
plus jets were we focused on two and three additional jets in the
final state. The calculations have been performed in the 
limit of an infinitely heavy top quark, at next-to-leading order in
QCD. Results for LHC collision energies of 8 \TeV and 13 \TeV 
have been obtained by the combination of the fully automated tools
\textsc{GoSam} and \textsc{Sherpa}.
The numerical results have been generated in two steps. First we have
produced sets of Ntuples
for the two energies and the three different jet multiplicities with a
minimal set of kinematic requirements, which in a second step, have
been analyzed for the particular scenarios.
The entire set of Ntuples will be made publicly available.

We have investigated two 
major scenarios, one defined by applying only basic selection cuts, and the second by applying the 
considerably more constraining VBF cuts where we also investigated alternative tagging jet selections.
We found that independent of the final state jet multiplicity the NLO QCD corrections remain sizeable and are therefore an
important prerequisite for a reliable prediction. In particular in the VBF scenario, for both the 
two jet as well as the three jet bin, the additional jet production
accounts for a considerable fraction of the 
total cross section which means that the results, to a large extent, are
only given with leading order accuracy. However, if one considers a
veto on the third jet in a two jet calculation, this again would
introduce large theoretical uncertainties. Therefore the calculation
of the three jet process with NLO accuracy provides important
information also for the exclusive two jet result.

For inclusive observables, i.e.\ observables that are not a priori
dependent on a specific number of jets, such as the transverse
momentum of the Higgs boson, we find that the higher jet
multiplicities are important 
for the correct description of the shape of the observables. In particular in the tails of the distributions,
which might be sensitive to new physics, they can make up the dominant contribution. Also here, the inclusion
of the NLO corrections of \Hjjj leads to an improvement of the
theoretical prediction.

We discussed a large variety of differential distributions which are
suitable to distinguish the gluon fusion process from that of the
vector boson fusion. Some of these observables are also used as input
variables for the boosted decision trees in the experiment. We
particularly described the effects of a third jet as well as the
impact of the NLO corrections.

Further improvements could certainly be achieved by providing a merged
NLO result of the different jet multiplicities, but also through the
inclusion of top-quark mass effects as well as the matching of the
\Hjjj NLO result with a parton shower.
Due to the complexity of these improvements they are however beyond
the scope of this paper and we leave them for future work.

\section*{Acknowledgments}
We thank Joey Huston for several fruitful discussions and Pierpaolo
Mastrolia for his encouragement during several stages of this project.
This work was supported by the US Department of Energy under
contract DE--AC02--76SF00515 and the Swiss National Foundation (SNF) 
under contract PP00P2--128552. M.S.\ would like to thank the University
Duisburg--Essen for the kind hospitality during the final stages of
the project. The work of G.L.\ and V.Y.\ was supported by the Alexander
von Humboldt Foundation, in the framework of the Sofja Kovaleskaja
Award 2010, endowed by the German Federal Ministry of Education and
Research.

\bibliographystyle{JHEP}

\begin{thebibliography}{100}

\bibitem{Aad:2012tfa}
{\bf ATLAS Collaboration} Collaboration, G.~Aad et~al., {\it {Observation of a
  new particle in the search for the Standard Model Higgs boson with the ATLAS
  detector at the LHC}},  {\em Phys.Lett.} {\bf B716} (2012) 1--29,
  [\href{http://xxx.lanl.gov/abs/1207.7214}{{\tt arXiv:1207.7214}}].

\bibitem{Chatrchyan:2012ufa}
{\bf CMS Collaboration} Collaboration, S.~Chatrchyan et~al., {\it {Observation
  of a new boson at a mass of 125 GeV with the CMS experiment at the LHC}},
  {\em Phys.Lett.} {\bf B716} (2012) 30--61,
  [\href{http://xxx.lanl.gov/abs/1207.7235}{{\tt arXiv:1207.7235}}].

\bibitem{Aad:2013xqa}
{\bf ATLAS} Collaboration, G.~Aad et~al., {\it {Evidence for the spin-0 nature
  of the Higgs boson using ATLAS data}},  {\em Phys.Lett.} {\bf B726} (2013)
  120--144, [\href{http://xxx.lanl.gov/abs/1307.1432}{{\tt arXiv:1307.1432}}].

\bibitem{Chatrchyan:2013iaa}
{\bf CMS} Collaboration, S.~Chatrchyan et~al., {\it {Measurement of Higgs boson
  production and properties in the WW decay channel with leptonic final
  states}},  {\em JHEP} {\bf 1401} (2014) 096,
  [\href{http://xxx.lanl.gov/abs/1312.1129}{{\tt arXiv:1312.1129}}].

\bibitem{Chatrchyan:2013mxa}
{\bf CMS} Collaboration, S.~Chatrchyan et~al., {\it {Measurement of the
  properties of a Higgs boson in the four-lepton final state}},  {\em
  Phys.Rev.} {\bf D89} (2014), no.~9 092007,
  [\href{http://xxx.lanl.gov/abs/1312.5353}{{\tt arXiv:1312.5353}}].

\bibitem{Aad:2013wqa}
{\bf ATLAS} Collaboration, G.~Aad et~al., {\it {Measurements of Higgs boson
  production and couplings in diboson final states with the ATLAS detector at
  the LHC}},  {\em Phys.Lett.} {\bf B726} (2013) 88--119,
  [\href{http://xxx.lanl.gov/abs/1307.1427}{{\tt arXiv:1307.1427}}].

\bibitem{Aad:2014eha}
{\bf ATLAS} Collaboration, G.~Aad et~al., {\it {Measurement of Higgs boson
  production in the diphoton decay channel in pp collisions at center-of-mass
  energies of 7 and 8 TeV with the ATLAS detector}},  {\em Phys.Rev.} {\bf D90}
  (2014), no.~11 112015, [\href{http://xxx.lanl.gov/abs/1408.7084}{{\tt
  arXiv:1408.7084}}].

\bibitem{Aad:2014lwa}
{\bf ATLAS} Collaboration, G.~Aad et~al., {\it {Measurements of fiducial and
  differential cross sections for Higgs boson production in the diphoton decay
  channel at $\sqrt{s}=8$ TeV with ATLAS}},  {\em JHEP} {\bf 1409} (2014) 112,
  [\href{http://xxx.lanl.gov/abs/1407.4222}{{\tt arXiv:1407.4222}}].

\bibitem{Aad:2014aba}
{\bf ATLAS} Collaboration, G.~Aad et~al., {\it {Measurement of the Higgs boson
  mass from the $H\rightarrow \gamma\gamma$ and $H \rightarrow ZZ^{*}
  \rightarrow 4\ell$ channels with the ATLAS detector using 25 fb$^{-1}$ of
  $pp$ collision data}},  {\em Phys.Rev.} {\bf D90} (2014), no.~5 052004,
  [\href{http://xxx.lanl.gov/abs/1406.3827}{{\tt arXiv:1406.3827}}].

\bibitem{Aad:2014fia}
{\bf ATLAS} Collaboration, G.~Aad et~al., {\it {Search for Higgs boson decays
  to a photon and a Z boson in pp collisions at $\sqrt{s}$=7 and 8 TeV with the
  ATLAS detector}},  {\em Phys.Lett.} {\bf B732} (2014) 8--27,
  [\href{http://xxx.lanl.gov/abs/1402.3051}{{\tt arXiv:1402.3051}}].

\bibitem{Aad:2015vsa}
{\bf ATLAS} Collaboration, G.~Aad et~al., {\it {Evidence for the Higgs-boson
  Yukawa coupling to tau leptons with the ATLAS detector}},
  \href{http://xxx.lanl.gov/abs/1501.0494}{{\tt arXiv:1501.0494}}.

\bibitem{ATLAS:2014aga}
{\bf ATLAS} Collaboration, G.~Aad et~al., {\it {Observation and measurement of
  Higgs boson decays to $WW^{\ast}$ with the ATLAS detector}},
  \href{http://xxx.lanl.gov/abs/1412.2641}{{\tt arXiv:1412.2641}}.

\bibitem{Chatrchyan:2014nva}
{\bf CMS} Collaboration, S.~Chatrchyan et~al., {\it {Evidence for the 125 GeV
  Higgs boson decaying to a pair of $\tau$ leptons}},  {\em JHEP} {\bf 1405}
  (2014) 104, [\href{http://xxx.lanl.gov/abs/1401.5041}{{\tt
  arXiv:1401.5041}}].

\bibitem{Chatrchyan:2014vua}
{\bf CMS} Collaboration, S.~Chatrchyan et~al., {\it {Evidence for the direct
  decay of the 125 GeV Higgs boson to fermions}},  {\em Nature Phys.} {\bf 10}
  (2014) 557--560, [\href{http://xxx.lanl.gov/abs/1401.6527}{{\tt
  arXiv:1401.6527}}].

\bibitem{Khachatryan:2014ira}
{\bf CMS} Collaboration, V.~Khachatryan et~al., {\it {Observation of the
  diphoton decay of the Higgs boson and measurement of its properties}},  {\em
  Eur.Phys.J.} {\bf C74} (2014), no.~10 3076,
  [\href{http://xxx.lanl.gov/abs/1407.0558}{{\tt arXiv:1407.0558}}].

\bibitem{Khachatryan:2014jba}
{\bf CMS} Collaboration, V.~Khachatryan et~al., {\it {Precise determination of
  the mass of the Higgs boson and tests of compatibility of its couplings with
  the standard model predictions using proton collisions at 7 and 8 TeV}},
  \href{http://xxx.lanl.gov/abs/1412.8662}{{\tt arXiv:1412.8662}}.

\bibitem{Englert:1964et}
F.~Englert and R.~Brout, {\it {Broken Symmetry and the Mass of Gauge Vector
  Mesons}},  {\em Phys.Rev.Lett.} {\bf 13} (1964) 321--323.

\bibitem{Higgs:1964pj}
P.~W. Higgs, {\it {Broken Symmetries and the Masses of Gauge Bosons}},  {\em
  Phys.Rev.Lett.} {\bf 13} (1964) 508--509.

\bibitem{Guralnik:1964eu}
G.~Guralnik, C.~Hagen, and T.~Kibble, {\it {Global Conservation Laws and
  Massless Particles}},  {\em Phys.Rev.Lett.} {\bf 13} (1964) 585--587.

\bibitem{Kibble:1967sv}
T.~Kibble, {\it {Symmetry breaking in nonAbelian gauge theories}},  {\em
  Phys.Rev.} {\bf 155} (1967) 1554--1561.

\bibitem{Cahn:1983ip}
R.~Cahn and S.~Dawson, {\it {Production of Very Massive Higgs Bosons}},  {\em
  Phys.Lett.} {\bf B136} (1984) 196.

\bibitem{Kane:1984bb}
G.~L. Kane, W.~Repko, and W.~Rolnick, {\it {The Effective W+-, Z0 Approximation
  for High-Energy Collisions}},  {\em Phys.Lett.} {\bf B148} (1984) 367--372.

\bibitem{Rainwater:1997dg}
D.~L. Rainwater and D.~Zeppenfeld, {\it {Searching for $H\to\gamma\gamma$ in
  weak boson fusion at the LHC}},  {\em JHEP} {\bf 9712} (1997) 005,
  [\href{http://xxx.lanl.gov/abs/hep-ph/9712271}{{\tt hep-ph/9712271}}].

\bibitem{Rainwater:1998kj}
D.~L. Rainwater, D.~Zeppenfeld, and K.~Hagiwara, {\it {Searching for
  $H\to\tau^+\tau^-$ in weak boson fusion at the CERN LHC}},  {\em Phys.Rev.}
  {\bf D59} (1998) 014037, [\href{http://xxx.lanl.gov/abs/hep-ph/9808468}{{\tt
  hep-ph/9808468}}].

\bibitem{Rainwater:1999sd}
D.~L. Rainwater and D.~Zeppenfeld, {\it {Observing $H\to W^*W^* \to e^\pm
  \mu\mp \not{p}_T$ in weak boson fusion with dual forward jet tagging at the
  CERN LHC}},  {\em Phys.Rev.} {\bf D60} (1999) 113004,
  [\href{http://xxx.lanl.gov/abs/hep-ph/9906218}{{\tt hep-ph/9906218}}].

\bibitem{Plehn:1999nw}
T.~Plehn, D.~L. Rainwater, and D.~Zeppenfeld, {\it {Probing the MSSM Higgs
  sector via weak boson fusion at the LHC}},  {\em Phys.Lett.} {\bf B454}
  (1999) 297--303, [\href{http://xxx.lanl.gov/abs/hep-ph/9902434}{{\tt
  hep-ph/9902434}}].

\bibitem{Eboli:2000ze}
O.~J. Eboli and D.~Zeppenfeld, {\it {Observing an invisible Higgs boson}},
  {\em Phys.Lett.} {\bf B495} (2000) 147--154,
  [\href{http://xxx.lanl.gov/abs/hep-ph/0009158}{{\tt hep-ph/0009158}}].

\bibitem{Han:1992hr}
T.~Han, G.~Valencia, and S.~Willenbrock, {\it {Structure function approach to
  vector boson scattering in p p collisions}},  {\em Phys.Rev.Lett.} {\bf 69}
  (1992) 3274--3277, [\href{http://xxx.lanl.gov/abs/hep-ph/9206246}{{\tt
  hep-ph/9206246}}].

\bibitem{Figy:2003nv}
T.~Figy, C.~Oleari, and D.~Zeppenfeld, {\it {Next-to-leading order jet
  distributions for Higgs boson production via weak boson fusion}},  {\em
  Phys.Rev.} {\bf D68} (2003) 073005,
  [\href{http://xxx.lanl.gov/abs/hep-ph/0306109}{{\tt hep-ph/0306109}}].

\bibitem{Bolzoni:2010xr}
P.~Bolzoni, F.~Maltoni, S.-O. Moch, and M.~Zaro, {\it {Higgs production via
  vector-boson fusion at NNLO in QCD}},  {\em Phys.Rev.Lett.} {\bf 105} (2010)
  011801, [\href{http://xxx.lanl.gov/abs/1003.4451}{{\tt arXiv:1003.4451}}].

\bibitem{Ciccolini:2007jr}
M.~Ciccolini, A.~Denner, and S.~Dittmaier, {\it {Strong and electroweak
  corrections to the production of Higgs + 2jets via weak interactions at the
  LHC}},  {\em Phys.Rev.Lett.} {\bf 99} (2007) 161803,
  [\href{http://xxx.lanl.gov/abs/0707.0381}{{\tt arXiv:0707.0381}}].

\bibitem{Ciccolini:2007ec}
M.~Ciccolini, A.~Denner, and S.~Dittmaier, {\it {Electroweak and QCD
  corrections to Higgs production via vector-boson fusion at the LHC}},  {\em
  Phys.Rev.} {\bf D77} (2008) 013002,
  [\href{http://xxx.lanl.gov/abs/0710.4749}{{\tt arXiv:0710.4749}}].

\bibitem{Dawson:1990zj}
S.~Dawson, {\it {Radiative corrections to Higgs boson production}},  {\em
  Nucl.Phys.} {\bf B359} (1991) 283--300.

\bibitem{Djouadi:1991tka}
A.~Djouadi, M.~Spira, and P.~Zerwas, {\it {Production of Higgs bosons in proton
  colliders: QCD corrections}},  {\em Phys.Lett.} {\bf B264} (1991) 440--446.

\bibitem{deFlorian:1999zd}
D.~de~Florian, M.~Grazzini, and Z.~Kunszt, {\it {Higgs production with large
  transverse momentum in hadronic collisions at next-to-leading order}},  {\em
  Phys.Rev.Lett.} {\bf 82} (1999) 5209--5212,
  [\href{http://xxx.lanl.gov/abs/hep-ph/9902483}{{\tt hep-ph/9902483}}].

\bibitem{Campbell:2010cz}
J.~M. Campbell, R.~K. Ellis, and C.~Williams, {\it {Hadronic production of a
  Higgs boson and two jets at next-to-leading order}},  {\em Phys.Rev.} {\bf
  D81} (2010) 074023, [\href{http://xxx.lanl.gov/abs/1001.4495}{{\tt
  arXiv:1001.4495}}].

\bibitem{Campbell:2006xx}
J.~M. Campbell, R.~K. Ellis, and G.~Zanderighi, {\it {Next-to-Leading order
  Higgs + 2 jet production via gluon fusion}},  {\em JHEP} {\bf 0610} (2006)
  028, [\href{http://xxx.lanl.gov/abs/hep-ph/0608194}{{\tt hep-ph/0608194}}].

\bibitem{Ravindran:2002dc}
V.~Ravindran, J.~Smith, and W.~Van~Neerven, {\it {Next-to-leading order QCD
  corrections to differential distributions of Higgs boson production in hadron
  hadron collisions}},  {\em Nucl.Phys.} {\bf B634} (2002) 247--290,
  [\href{http://xxx.lanl.gov/abs/hep-ph/0201114}{{\tt hep-ph/0201114}}].

\bibitem{Dixon:2009uk}
L.~J. Dixon and Y.~Sofianatos, {\it {Analytic one-loop amplitudes for a Higgs
  boson plus four partons}},  {\em JHEP} {\bf 0908} (2009) 058,
  [\href{http://xxx.lanl.gov/abs/0906.0008}{{\tt arXiv:0906.0008}}].

\bibitem{DelDuca:2004wt}
V.~Del~Duca, A.~Frizzo, and F.~Maltoni, {\it {Higgs boson production in
  association with three jets}},  {\em JHEP} {\bf 0405} (2004) 064,
  [\href{http://xxx.lanl.gov/abs/hep-ph/0404013}{{\tt hep-ph/0404013}}].

\bibitem{Dixon:2004za}
L.~J. Dixon, E.~N. Glover, and V.~V. Khoze, {\it {MHV rules for Higgs plus
  multi-gluon amplitudes}},  {\em JHEP} {\bf 0412} (2004) 015,
  [\href{http://xxx.lanl.gov/abs/hep-th/0411092}{{\tt hep-th/0411092}}].

\bibitem{Badger:2004ty}
S.~Badger, E.~N. Glover, and V.~V. Khoze, {\it {MHV rules for Higgs plus
  multi-parton amplitudes}},  {\em JHEP} {\bf 0503} (2005) 023,
  [\href{http://xxx.lanl.gov/abs/hep-th/0412275}{{\tt hep-th/0412275}}].

\bibitem{Ellis:2005qe}
R.~K. Ellis, W.~Giele, and G.~Zanderighi, {\it {Virtual QCD corrections to
  Higgs boson plus four parton processes}},  {\em Phys.Rev.} {\bf D72} (2005)
  054018, [\href{http://xxx.lanl.gov/abs/hep-ph/0506196}{{\tt
  hep-ph/0506196}}].

\bibitem{Ellis:2005zh}
R.~K. Ellis, W.~Giele, and G.~Zanderighi, {\it {Semi-numerical evaluation of
  one-loop corrections}},  {\em Phys.Rev.} {\bf D73} (2006) 014027,
  [\href{http://xxx.lanl.gov/abs/hep-ph/0508308}{{\tt hep-ph/0508308}}].

\bibitem{Berger:2006sh}
C.~F. Berger, V.~Del~Duca, and L.~J. Dixon, {\it {Recursive Construction of
  Higgs-Plus-Multiparton Loop Amplitudes: The Last of the Phi-nite Loop
  Amplitudes}},  {\em Phys.Rev.} {\bf D74} (2006) 094021,
  [\href{http://xxx.lanl.gov/abs/hep-ph/0608180}{{\tt hep-ph/0608180}}].

\bibitem{Badger:2006us}
S.~Badger and E.~N. Glover, {\it {One-loop helicity amplitudes for H $\to$
  gluons: The All-minus configuration}},  {\em Nucl.Phys.Proc.Suppl.} {\bf 160}
  (2006) 71--75, [\href{http://xxx.lanl.gov/abs/hep-ph/0607139}{{\tt
  hep-ph/0607139}}].

\bibitem{Badger:2007si}
S.~Badger, E.~N. Glover, and K.~Risager, {\it {One-loop phi-MHV amplitudes
  using the unitarity bootstrap}},  {\em JHEP} {\bf 0707} (2007) 066,
  [\href{http://xxx.lanl.gov/abs/0704.3914}{{\tt arXiv:0704.3914}}].

\bibitem{Glover:2008ffa}
E.~N. Glover, P.~Mastrolia, and C.~Williams, {\it {One-loop phi-MHV amplitudes
  using the unitarity bootstrap: The General helicity case}},  {\em JHEP} {\bf
  0808} (2008) 017, [\href{http://xxx.lanl.gov/abs/0804.4149}{{\tt
  arXiv:0804.4149}}].

\bibitem{Badger:2009hw}
S.~Badger, E.~Nigel~Glover, P.~Mastrolia, and C.~Williams, {\it {One-loop Higgs
  plus four gluon amplitudes: Full analytic results}},  {\em JHEP} {\bf 1001}
  (2010) 036, [\href{http://xxx.lanl.gov/abs/0909.4475}{{\tt
  arXiv:0909.4475}}].

\bibitem{Badger:2009vh}
S.~Badger, J.~M. Campbell, R.~K. Ellis, and C.~Williams, {\it {Analytic results
  for the one-loop NMHV Hqqgg amplitude}},  {\em JHEP} {\bf 0912} (2009) 035,
  [\href{http://xxx.lanl.gov/abs/0910.4481}{{\tt arXiv:0910.4481}}].

\bibitem{vanDeurzen:2013rv}
H.~van Deurzen, N.~Greiner, G.~Luisoni, P.~Mastrolia, E.~Mirabella, et~al.,
  {\it {NLO QCD corrections to the production of Higgs plus two jets at the
  LHC}},  {\em Phys.Lett.} {\bf B721} (2013) 74--81,
  [\href{http://xxx.lanl.gov/abs/1301.0493}{{\tt arXiv:1301.0493}}].

\bibitem{Campbell:2012am}
J.~M. Campbell, R.~K. Ellis, R.~Frederix, P.~Nason, C.~Oleari, et~al., {\it
  {NLO Higgs Boson Production Plus One and Two Jets Using the POWHEG BOX,
  MadGraph4 and MCFM}},  {\em JHEP} {\bf 1207} (2012) 092,
  [\href{http://xxx.lanl.gov/abs/1202.5475}{{\tt arXiv:1202.5475}}].

\bibitem{Hoeche:2014lxa}
S.~Hoeche, F.~Krauss, and M.~Schonherr, {\it {Uncertainties in MEPS@NLO
  calculations of h+jets}},  {\em Phys.Rev.} {\bf D90} (2014), no.~1 014012,
  [\href{http://xxx.lanl.gov/abs/1401.7971}{{\tt arXiv:1401.7971}}].

\bibitem{Cullen:2013saa}
G.~Cullen, H.~van Deurzen, N.~Greiner, G.~Luisoni, P.~Mastrolia, et~al., {\it
  {Next-to-Leading-Order QCD Corrections to Higgs Boson Production Plus Three
  Jets in Gluon Fusion}},  {\em Phys.Rev.Lett.} {\bf 111} (2013), no.~13
  131801, [\href{http://xxx.lanl.gov/abs/1307.4737}{{\tt arXiv:1307.4737}}].

\bibitem{Mastrolia:2012bu}
P.~Mastrolia, E.~Mirabella, and T.~Peraro, {\it {Integrand reduction of
  one-loop scattering amplitudes through Laurent series expansion}},  {\em
  JHEP} {\bf 1206} (2012) 095, [\href{http://xxx.lanl.gov/abs/1203.0291}{{\tt
  arXiv:1203.0291}}].

\bibitem{vanDeurzen:2013saa}
H.~van Deurzen, G.~Luisoni, P.~Mastrolia, E.~Mirabella, G.~Ossola, et~al., {\it
  {Multi-leg One-loop Massive Amplitudes from Integrand Reduction via Laurent
  Expansion}},  {\em JHEP} {\bf 1403} (2014) 115,
  [\href{http://xxx.lanl.gov/abs/1312.6678}{{\tt arXiv:1312.6678}}].

\bibitem{Peraro:2014cba}
T.~Peraro, {\it {Ninja: Automated Integrand Reduction via Laurent Expansion for
  One-Loop Amplitudes}},  \href{http://xxx.lanl.gov/abs/1403.1229}{{\tt
  arXiv:1403.1229}}.

\bibitem{Butterworth:2014efa}
J.~Butterworth, G.~Dissertori, S.~Dittmaier, D.~de~Florian, N.~Glover, et~al.,
  {\it {Les Houches 2013: Physics at TeV Colliders: Standard Model Working
  Group Report}},  \href{http://xxx.lanl.gov/abs/1405.1067}{{\tt
  arXiv:1405.1067}}.

\bibitem{Boughezal:2013uia}
R.~Boughezal, F.~Caola, K.~Melnikov, F.~Petriello, and M.~Schulze, {\it {Higgs
  boson production in association with a jet at next-to-next-to-leading order
  in perturbative QCD}},  {\em JHEP} {\bf 1306} (2013) 072,
  [\href{http://xxx.lanl.gov/abs/1302.6216}{{\tt arXiv:1302.6216}}].

\bibitem{Chen:2014gva}
X.~Chen, T.~Gehrmann, E.~Glover, and M.~Jaquier, {\it {Precise QCD predictions
  for the production of Higgs + jet final states}},  {\em Phys.Lett.} {\bf
  B740} (2014) 147--150, [\href{http://xxx.lanl.gov/abs/1408.5325}{{\tt
  arXiv:1408.5325}}].

\bibitem{Boughezal:2015dra}
R.~Boughezal, F.~Caola, K.~Melnikov, F.~Petriello, and M.~Schulze, {\it {Higgs
  Boson Production in Association with a Jet at Next-to-Next-to-Leading
  Order}},  \href{http://xxx.lanl.gov/abs/1504.0792}{{\tt arXiv:1504.0792}}.

\bibitem{Boughezal:2015aha}
R.~Boughezal, C.~Focke, W.~Giele, X.~Liu, and F.~Petriello, {\it {Higgs boson
  production in association with a jet using jettiness subtraction}},
  \href{http://xxx.lanl.gov/abs/1505.0389}{{\tt arXiv:1505.0389}}.

\bibitem{Harlander:2002wh}
R.~V. Harlander and W.~B. Kilgore, {\it {Next-to-next-to-leading order Higgs
  production at hadron colliders}},  {\em Phys.Rev.Lett.} {\bf 88} (2002)
  201801, [\href{http://xxx.lanl.gov/abs/hep-ph/0201206}{{\tt
  hep-ph/0201206}}].

\bibitem{Anastasiou:2005qj}
C.~Anastasiou, K.~Melnikov, and F.~Petriello, {\it {Fully differential Higgs
  boson production and the di-photon signal through next-to-next-to-leading
  order}},  {\em Nucl.Phys.} {\bf B724} (2005) 197--246,
  [\href{http://xxx.lanl.gov/abs/hep-ph/0501130}{{\tt hep-ph/0501130}}].

\bibitem{Grazzini:2008tf}
M.~Grazzini, {\it {NNLO predictions for the Higgs boson signal in the H $\to$
  WW $\to$ lnu lnu and H $\to$ ZZ $\to$ 4l decay channels}},  {\em JHEP} {\bf
  0802} (2008) 043, [\href{http://xxx.lanl.gov/abs/0801.3232}{{\tt
  arXiv:0801.3232}}].

\bibitem{Anastasiou:2013mca}
C.~Anastasiou, C.~Duhr, F.~Dulat, F.~Herzog, and B.~Mistlberger, {\it
  {Real-virtual contributions to the inclusive Higgs cross-section at
  $N^3LO$}},  {\em JHEP} {\bf 12} (2013) 088,
  [\href{http://xxx.lanl.gov/abs/1311.1425}{{\tt arXiv:1311.1425}}].

\bibitem{Kilgore:2013gba}
W.~B. Kilgore, {\it {One-Loop Single-Real-Emission Contributions to $pp\to H +
  X$ at Next-to-Next-to-Next-to-Leading Order}},  {\em Phys.Rev.} {\bf D89}
  (2014) 073008, [\href{http://xxx.lanl.gov/abs/1312.1296}{{\tt
  arXiv:1312.1296}}].

\bibitem{Ball:2013bra}
R.~D. Ball, M.~Bonvini, S.~Forte, S.~Marzani, and G.~Ridolfi, {\it {Higgs
  production in gluon fusion beyond NNLO}},  {\em Nucl.Phys.} {\bf B874} (2013)
  746--772, [\href{http://xxx.lanl.gov/abs/1303.3590}{{\tt arXiv:1303.3590}}].

\bibitem{Li:2013lsa}
Y.~Li and H.~X. Zhu, {\it {Single soft gluon emission at two loops}},  {\em
  JHEP} {\bf 11} (2013) 080, [\href{http://xxx.lanl.gov/abs/1309.4391}{{\tt
  arXiv:1309.4391}}].

\bibitem{Duhr:2013msa}
C.~Duhr and T.~Gehrmann, {\it {The two-loop soft current in dimensional
  regularization}},  {\em Phys.Lett.} {\bf B727} (2013) 452--455,
  [\href{http://xxx.lanl.gov/abs/1309.4393}{{\tt arXiv:1309.4393}}].

\bibitem{Anastasiou:2014vaa}
C.~Anastasiou, C.~Duhr, F.~Dulat, E.~Furlan, T.~Gehrmann, et~al., {\it {Higgs
  boson gluon-fusion production at threshold in $N^3LO$ $QCD$}},  {\em
  Phys.Lett.} {\bf B737} (2014) 325--328,
  [\href{http://xxx.lanl.gov/abs/1403.4616}{{\tt arXiv:1403.4616}}].

\bibitem{Anastasiou:2014lda}
C.~Anastasiou, C.~Duhr, F.~Dulat, E.~Furlan, T.~Gehrmann, et~al., {\it {Higgs
  boson gluon-fusion production beyond threshold in N3LO QCD}},
  \href{http://xxx.lanl.gov/abs/1411.3584}{{\tt arXiv:1411.3584}}.

\bibitem{Li:2014bfa}
Y.~Li, A.~von Manteuffel, R.~M. Schabinger, and H.~X. Zhu, {\it {N$^3$LO Higgs
  and Drell-Yan production at threshold: the one-loop two-emission
  contribution}},  {\em Phys.Rev.} {\bf D90} (2014), no.~5 053006,
  [\href{http://xxx.lanl.gov/abs/1404.5839}{{\tt arXiv:1404.5839}}].

\bibitem{Ahmed:2014uya}
T.~Ahmed, M.~Mandal, N.~Rana, and V.~Ravindran, {\it {Rapidity distributions in
  Drell-Yan and Higgs productions at threshold in N$^3$LO QCD}},  {\em
  Phys.Rev.Lett.} {\bf 113} (2014) 212003,
  [\href{http://xxx.lanl.gov/abs/1404.6504}{{\tt arXiv:1404.6504}}].

\bibitem{Anastasiou:2015ema}
C.~Anastasiou, C.~Duhr, F.~Dulat, F.~Herzog, and B.~Mistlberger, {\it {Higgs
  boson gluon-fusion production in N3LO QCD}},
  \href{http://xxx.lanl.gov/abs/1503.0605}{{\tt arXiv:1503.0605}}.

\bibitem{Andersen:2009he}
J.~R. Andersen and J.~M. Smillie, {\it {The Factorisation of the t-channel Pole
  in Quark-Gluon Scattering}},  {\em Phys.Rev.} {\bf D81} (2010) 114021,
  [\href{http://xxx.lanl.gov/abs/0910.5113}{{\tt arXiv:0910.5113}}].

\bibitem{Andersen:2011hs}
J.~R. Andersen and J.~M. Smillie, {\it {Multiple Jets at the LHC with High
  Energy Jets}},  {\em JHEP} {\bf 1106} (2011) 010,
  [\href{http://xxx.lanl.gov/abs/1101.5394}{{\tt arXiv:1101.5394}}].

\bibitem{Bern:2013gka}
Z.~Bern, L.~Dixon, F.~Febres~Cordero, S.~H{\"o}che, H.~Ita, D.~Kosower,
  D.~Ma{\^i}tre, and K.~Ozeren, {\it {Next-to-Leading Order $W + 5$-Jet
  Production at the LHC}},  {\em Phys.Rev.} {\bf D88} (2013), no.~1 014025,
  [\href{http://xxx.lanl.gov/abs/1304.1253}{{\tt arXiv:1304.1253}}].

\bibitem{Bern:2014fea}
Z.~Bern, L.~Dixon, F.~F. Cordero, S.~H{\"o}che, H.~Ita, D.~Kosower, and
  D.~Ma{\^i}tre, {\it {Extrapolating W-Associated Jet-Production Ratios at the
  LHC}},  \href{http://xxx.lanl.gov/abs/1412.4775}{{\tt arXiv:1412.4775}}.

\bibitem{Gerwick:2012hq}
E.~Gerwick, T.~Plehn, S.~Schumann, and P.~Schichtel, {\it {Scaling Patterns for
  QCD Jets}},  {\em JHEP} {\bf 1210} (2012) 162,
  [\href{http://xxx.lanl.gov/abs/1208.3676}{{\tt arXiv:1208.3676}}].

\bibitem{Gerwick:2012fw}
E.~Gerwick, S.~Schumann, B.~Gripaios, and B.~Webber, {\it {QCD Jet Rates with
  the Inclusive Generalized kt Algorithms}},  {\em JHEP} {\bf 1304} (2013) 089,
  [\href{http://xxx.lanl.gov/abs/1212.5235}{{\tt arXiv:1212.5235}}].

\bibitem{DelDuca:2001eu}
V.~Del~Duca, W.~Kilgore, C.~Oleari, C.~Schmidt, and D.~Zeppenfeld, {\it {Higgs
  + 2 jets via gluon fusion}},  {\em Phys.Rev.Lett.} {\bf 87} (2001) 122001,
  [\href{http://xxx.lanl.gov/abs/hep-ph/0105129}{{\tt hep-ph/0105129}}].

\bibitem{DelDuca:2001fn}
V.~Del~Duca, W.~Kilgore, C.~Oleari, C.~Schmidt, and D.~Zeppenfeld, {\it {Gluon
  fusion contributions to H + 2 jet production}},  {\em Nucl.Phys.} {\bf B616}
  (2001) 367--399, [\href{http://xxx.lanl.gov/abs/hep-ph/0108030}{{\tt
  hep-ph/0108030}}].

\bibitem{Campanario:2013mga}
F.~Campanario and M.~Kubocz, {\it {Higgs boson production in association with
  three jets via gluon fusion at the LHC: Gluonic contributions}},  {\em
  Phys.Rev.} {\bf D88} (2013), no.~5 054021,
  [\href{http://xxx.lanl.gov/abs/1306.1830}{{\tt arXiv:1306.1830}}].

\bibitem{Cullen:2011ac}
G.~Cullen, N.~Greiner, G.~Heinrich, G.~Luisoni, P.~Mastrolia, et~al., {\it
  {Automated One-Loop Calculations with GoSam}},  {\em Eur.Phys.J.} {\bf C72}
  (2012) 1889, [\href{http://xxx.lanl.gov/abs/1111.2034}{{\tt
  arXiv:1111.2034}}].

\bibitem{Cullen:2014yla}
G.~Cullen, H.~van Deurzen, N.~Greiner, G.~Heinrich, G.~Luisoni, et~al., {\it
  {G$\scriptsize{O}$S$\scriptsize{AM}$-2.0: a tool for automated one-loop
  calculations within the Standard Model and beyond}},  {\em Eur.Phys.J.} {\bf
  C74} (2014), no.~8 3001, [\href{http://xxx.lanl.gov/abs/1404.7096}{{\tt
  arXiv:1404.7096}}].

\bibitem{Gleisberg:2008ta}
T.~Gleisberg, S.~Hoeche, F.~Krauss, M.~Schonherr, S.~Schumann, et~al., {\it
  {Event generation with SHERPA 1.1}},  {\em JHEP} {\bf 0902} (2009) 007,
  [\href{http://xxx.lanl.gov/abs/0811.4622}{{\tt arXiv:0811.4622}}].

\bibitem{Binoth:2010xt}
T.~Binoth, F.~Boudjema, G.~Dissertori, A.~Lazopoulos, A.~Denner, et~al., {\it
  {A Proposal for a standard interface between Monte Carlo tools and one-loop
  programs}},  {\em Comput.Phys.Commun.} {\bf 181} (2010) 1612--1622,
  [\href{http://xxx.lanl.gov/abs/1001.1307}{{\tt arXiv:1001.1307}}].

\bibitem{Alioli:2013nda}
S.~Alioli, S.~Badger, J.~Bellm, B.~Biedermann, F.~Boudjema, et~al., {\it
  {Update of the Binoth Les Houches Accord for a standard interface between
  Monte Carlo tools and one-loop programs}},  {\em Comput.Phys.Commun.} {\bf
  185} (2014) 560--571, [\href{http://xxx.lanl.gov/abs/1308.3462}{{\tt
  arXiv:1308.3462}}].

\bibitem{Nogueira:1991ex}
P.~Nogueira, {\it {Automatic Feynman graph generation}},  {\em J.Comput.Phys.}
  {\bf 105} (1993) 279--289.

\bibitem{Vermaseren:2000nd}
J.~Vermaseren, {\it {New features of FORM}},
  \href{http://xxx.lanl.gov/abs/math-ph/0010025}{{\tt math-ph/0010025}}.

\bibitem{Kuipers:2012rf}
J.~Kuipers, T.~Ueda, J.~Vermaseren, and J.~Vollinga, {\it {FORM version 4.0}},
  {\em Comput.Phys.Commun.} {\bf 184} (2013) 1453--1467,
  [\href{http://xxx.lanl.gov/abs/1203.6543}{{\tt arXiv:1203.6543}}].

\bibitem{Cullen:2010jv}
G.~Cullen, M.~Koch-Janusz, and T.~Reiter, {\it {Spinney: A Form Library for
  Helicity Spinors}},  {\em Comput.Phys.Commun.} {\bf 182} (2011) 2368--2387,
  [\href{http://xxx.lanl.gov/abs/1008.0803}{{\tt arXiv:1008.0803}}].

\bibitem{Reiter:2009ts}
T.~Reiter, {\it {Optimising Code Generation with haggies}},  {\em
  Comput.Phys.Commun.} {\bf 181} (2010) 1301--1331,
  [\href{http://xxx.lanl.gov/abs/0907.3714}{{\tt arXiv:0907.3714}}].

\bibitem{Ossola:2006us}
G.~Ossola, C.~G. Papadopoulos, and R.~Pittau, {\it {Reducing full one-loop
  amplitudes to scalar integrals at the integrand level}},  {\em Nucl.Phys.}
  {\bf B763} (2007) 147--169,
  [\href{http://xxx.lanl.gov/abs/hep-ph/0609007}{{\tt hep-ph/0609007}}].

\bibitem{Mastrolia:2008jb}
P.~Mastrolia, G.~Ossola, C.~Papadopoulos, and R.~Pittau, {\it {Optimizing the
  Reduction of One-Loop Amplitudes}},  {\em JHEP} {\bf 0806} (2008) 030,
  [\href{http://xxx.lanl.gov/abs/0803.3964}{{\tt arXiv:0803.3964}}].

\bibitem{Ossola:2008xq}
G.~Ossola, C.~G. Papadopoulos, and R.~Pittau, {\it {On the Rational Terms of
  the one-loop amplitudes}},  {\em JHEP} {\bf 0805} (2008) 004,
  [\href{http://xxx.lanl.gov/abs/0802.1876}{{\tt arXiv:0802.1876}}].

\bibitem{Mastrolia:2010nb}
P.~Mastrolia, G.~Ossola, T.~Reiter, and F.~Tramontano, {\it {Scattering
  AMplitudes from Unitarity-based Reduction Algorithm at the Integrand-level}},
   {\em JHEP} {\bf 1008} (2010) 080,
  [\href{http://xxx.lanl.gov/abs/1006.0710}{{\tt arXiv:1006.0710}}].

\bibitem{Heinrich:2010ax}
G.~Heinrich, G.~Ossola, T.~Reiter, and F.~Tramontano, {\it {Tensorial
  Reconstruction at the Integrand Level}},  {\em JHEP} {\bf 1010} (2010) 105,
  [\href{http://xxx.lanl.gov/abs/1008.2441}{{\tt arXiv:1008.2441}}].

\bibitem{Binoth:2008uq}
T.~Binoth, J.-P. Guillet, G.~Heinrich, E.~Pilon, and T.~Reiter, {\it {Golem95:
  A Numerical program to calculate one-loop tensor integrals with up to six
  external legs}},  {\em Comput.Phys.Commun.} {\bf 180} (2009) 2317--2330,
  [\href{http://xxx.lanl.gov/abs/0810.0992}{{\tt arXiv:0810.0992}}].

\bibitem{Cullen:2011kv}
G.~Cullen, J.~P. Guillet, G.~Heinrich, T.~Kleinschmidt, E.~Pilon, et~al., {\it
  {Golem95C: A library for one-loop integrals with complex masses}},  {\em
  Comput.Phys.Commun.} {\bf 182} (2011) 2276--2284,
  [\href{http://xxx.lanl.gov/abs/1101.5595}{{\tt arXiv:1101.5595}}].

\bibitem{vanHameren:2010cp}
A.~van Hameren, {\it {OneLOop: For the evaluation of one-loop scalar
  functions}},  {\em Comput.Phys.Commun.} {\bf 182} (2011) 2427--2438,
  [\href{http://xxx.lanl.gov/abs/1007.4716}{{\tt arXiv:1007.4716}}].

\bibitem{Catani:1996vz}
S.~Catani and M.~Seymour, {\it {A General algorithm for calculating jet
  cross-sections in NLO QCD}},  {\em Nucl.Phys.} {\bf B485} (1997) 291--419,
  [\href{http://xxx.lanl.gov/abs/hep-ph/9605323}{{\tt hep-ph/9605323}}].

\bibitem{Gleisberg:2008fv}
T.~Gleisberg and S.~Hoeche, {\it {Comix, a new matrix element generator}},
  {\em JHEP} {\bf 0812} (2008) 039,
  [\href{http://xxx.lanl.gov/abs/0808.3674}{{\tt arXiv:0808.3674}}].

\bibitem{Hoeche:2014xx}
S.~H{\"o}che, {\it {Efficient dipole subtraction with Comix}}, .

\bibitem{Stelzer:1994ta}
T.~Stelzer and W.~Long, {\it {Automatic generation of tree level helicity
  amplitudes}},  {\em Comput.Phys.Commun.} {\bf 81} (1994) 357--371,
  [\href{http://xxx.lanl.gov/abs/hep-ph/9401258}{{\tt hep-ph/9401258}}].

\bibitem{Alwall:2007st}
J.~Alwall, P.~Demin, S.~de~Visscher, R.~Frederix, M.~Herquet, et~al., {\it
  {MadGraph/MadEvent v4: The New Web Generation}},  {\em JHEP} {\bf 0709}
  (2007) 028, [\href{http://xxx.lanl.gov/abs/0706.2334}{{\tt
  arXiv:0706.2334}}].

\bibitem{Frederix:2008hu}
R.~Frederix, T.~Gehrmann, and N.~Greiner, {\it {Automation of the Dipole
  Subtraction Method in MadGraph/MadEvent}},  {\em JHEP} {\bf 0809} (2008) 122,
  [\href{http://xxx.lanl.gov/abs/0808.2128}{{\tt arXiv:0808.2128}}].

\bibitem{Frederix:2010cj}
R.~Frederix, T.~Gehrmann, and N.~Greiner, {\it {Integrated dipoles with
  MadDipole in the MadGraph framework}},  {\em JHEP} {\bf 1006} (2010) 086,
  [\href{http://xxx.lanl.gov/abs/1004.2905}{{\tt arXiv:1004.2905}}].

\bibitem{Maltoni:2002qb}
F.~Maltoni and T.~Stelzer, {\it {MadEvent: Automatic event generation with
  MadGraph}},  {\em JHEP} {\bf 0302} (2003) 027,
  [\href{http://xxx.lanl.gov/abs/hep-ph/0208156}{{\tt hep-ph/0208156}}].

\bibitem{Wilczek:1977zn}
F.~Wilczek, {\it {Decays of Heavy Vector Mesons Into Higgs Particles}},  {\em
  Phys.Rev.Lett.} {\bf 39} (1977) 1304.

\bibitem{Hamilton:2012np}
K.~Hamilton, P.~Nason, and G.~Zanderighi, {\it {MINLO: Multi-Scale Improved
  NLO}},  {\em JHEP} {\bf 1210} (2012) 155,
  [\href{http://xxx.lanl.gov/abs/1206.3572}{{\tt arXiv:1206.3572}}].

\bibitem{Bern:2013zja}
Z.~Bern, L.~Dixon, F.~Febres~Cordero, S.~Hoeche, H.~Ita, et~al., {\it {Ntuples
  for NLO Events at Hadron Colliders}},  {\em Comput.Phys.Commun.} {\bf 185}
  (2014) 1443--1460, [\href{http://xxx.lanl.gov/abs/1310.7439}{{\tt
  arXiv:1310.7439}}].

\bibitem{Cacciari:2005hq}
M.~Cacciari and G.~P. Salam, {\it {Dispelling the $N^{3}$ myth for the $k_t$
  jet-finder}},  {\em Phys.Lett.} {\bf B641} (2006) 57--61,
  [\href{http://xxx.lanl.gov/abs/hep-ph/0512210}{{\tt hep-ph/0512210}}].

\bibitem{Cacciari:2008gp}
M.~Cacciari, G.~P. Salam, and G.~Soyez, {\it {The Anti-k(t) jet clustering
  algorithm}},  {\em JHEP} {\bf 0804} (2008) 063,
  [\href{http://xxx.lanl.gov/abs/0802.1189}{{\tt arXiv:0802.1189}}].

\bibitem{Cacciari:2011ma}
M.~Cacciari, G.~P. Salam, and G.~Soyez, {\it {FastJet User Manual}},  {\em
  Eur.Phys.J.} {\bf C72} (2012) 1896,
  [\href{http://xxx.lanl.gov/abs/1111.6097}{{\tt arXiv:1111.6097}}].

\bibitem{Lai:2010vv}
H.-L. Lai, M.~Guzzi, J.~Huston, Z.~Li, P.~M. Nadolsky, et~al., {\it {New parton
  distributions for collider physics}},  {\em Phys.Rev.} {\bf D82} (2010)
  074024, [\href{http://xxx.lanl.gov/abs/1007.2241}{{\tt arXiv:1007.2241}}].

\bibitem{Berger:2009zg}
C.~Berger, Z.~Bern, L.~J. Dixon, F.~Febres~Cordero, D.~Forde, et~al., {\it
  {Precise Predictions for $W$ + 3 Jet Production at Hadron Colliders}},  {\em
  Phys.Rev.Lett.} {\bf 102} (2009) 222001,
  [\href{http://xxx.lanl.gov/abs/0902.2760}{{\tt arXiv:0902.2760}}].

\bibitem{Martin:2009iq}
A.~Martin, W.~Stirling, R.~Thorne, and G.~Watt, {\it {Parton distributions for
  the LHC}},  {\em Eur.Phys.J.} {\bf C63} (2009) 189--285,
  [\href{http://xxx.lanl.gov/abs/0901.0002}{{\tt arXiv:0901.0002}}].

\bibitem{Ball:2012cx}
R.~D. Ball, V.~Bertone, S.~Carrazza, C.~S. Deans, L.~Del~Debbio, et~al., {\it
  {Parton distributions with LHC data}},  {\em Nucl.Phys.} {\bf B867} (2013)
  244--289, [\href{http://xxx.lanl.gov/abs/1207.1303}{{\tt arXiv:1207.1303}}].

\bibitem{Lykken:2011uv}
J.~D. Lykken, A.~Martin, and J.-C. Winter, {\it {Semileptonic Decays of the
  Higgs Boson at the Tevatron}},  {\em JHEP} {\bf 1208} (2012) 062,
  [\href{http://xxx.lanl.gov/abs/1111.2881}{{\tt arXiv:1111.2881}}].

\bibitem{Berger:2009ep}
C.~Berger, Z.~Bern, L.~J. Dixon, F.~Febres~Cordero, D.~Forde, et~al., {\it
  {Next-to-Leading Order QCD Predictions for W+3-Jet Distributions at Hadron
  Colliders}},  {\em Phys.Rev.} {\bf D80} (2009) 074036,
  [\href{http://xxx.lanl.gov/abs/0907.1984}{{\tt arXiv:0907.1984}}].

\bibitem{Joey}
J.~Huston. {Private communication}.

\bibitem{Rubin:2010xp}
M.~Rubin, G.~P. Salam, and S.~Sapeta, {\it {Giant QCD K-factors beyond NLO}},
  {\em JHEP} {\bf 1009} (2010) 084,
  [\href{http://xxx.lanl.gov/abs/1006.2144}{{\tt arXiv:1006.2144}}].

\bibitem{Lipatov:1976zz}
L.~Lipatov, {\it {Reggeization of the Vector Meson and the Vacuum Singularity
  in Nonabelian Gauge Theories}},  {\em Sov.J.Nucl.Phys.} {\bf 23} (1976)
  338--345.

\bibitem{Kuraev:1977fs}
E.~Kuraev, L.~Lipatov, and V.~S. Fadin, {\it {The Pomeranchuk Singularity in
  Nonabelian Gauge Theories}},  {\em Sov.Phys.JETP} {\bf 45} (1977) 199--204.

\bibitem{Balitsky:1978ic}
I.~Balitsky and L.~Lipatov, {\it {The Pomeranchuk Singularity in Quantum
  Chromodynamics}},  {\em Sov.J.Nucl.Phys.} {\bf 28} (1978) 822--829.

\bibitem{Schmidt:1996fg}
C.~R. Schmidt, {\it {A Monte Carlo solution to the BFKL equation}},  {\em
  Phys.Rev.Lett.} {\bf 78} (1997) 4531--4535,
  [\href{http://xxx.lanl.gov/abs/hep-ph/9612454}{{\tt hep-ph/9612454}}].

\bibitem{Orr:1997im}
L.~H. Orr and W.~J. Stirling, {\it {Dijet production at hadron hadron colliders
  in the BFKL approach}},  {\em Phys.Rev.} {\bf D56} (1997) 5875--5884,
  [\href{http://xxx.lanl.gov/abs/hep-ph/9706529}{{\tt hep-ph/9706529}}].

\bibitem{Andersen:2003an}
J.~R. Andersen and A.~Sabio~Vera, {\it {Solving the BFKL equation in the
  next-to-leading approximation}},  {\em Phys.Lett.} {\bf B567} (2003)
  116--124, [\href{http://xxx.lanl.gov/abs/hep-ph/0305236}{{\tt
  hep-ph/0305236}}].

\bibitem{Andersen:2003wy}
J.~R. Andersen and A.~Sabio~Vera, {\it {The Gluon Green's function in the BFKL
  approach at next-to-leading logarithmic accuracy}},  {\em Nucl.Phys.} {\bf
  B679} (2004) 345--362, [\href{http://xxx.lanl.gov/abs/hep-ph/0309331}{{\tt
  hep-ph/0309331}}].

\bibitem{Andersen:2006sp}
J.~R. Andersen, {\it {On the role of NLL corrections and energy conservation in
  the high energy evolution of QCD}},  {\em Phys.Lett.} {\bf B639} (2006)
  290--293, [\href{http://xxx.lanl.gov/abs/hep-ph/0602182}{{\tt
  hep-ph/0602182}}].

\bibitem{Andersen:2006kp}
J.~R. Andersen, {\it {The Quark-Antiquark Contribution to the Fully Exclusive
  BFKL Evolution at NLL Accuracy}},  {\em Phys.Rev.} {\bf D74} (2006) 114008,
  [\href{http://xxx.lanl.gov/abs/hep-ph/0611011}{{\tt hep-ph/0611011}}].

\bibitem{Andersen:2008ue}
J.~R. Andersen and C.~D. White, {\it {A New Framework for Multijet Predictions
  and its application to Higgs Boson production at the LHC}},  {\em Phys.Rev.}
  {\bf D78} (2008) 051501, [\href{http://xxx.lanl.gov/abs/0802.2858}{{\tt
  arXiv:0802.2858}}].

\bibitem{Andersen:2009nu}
J.~R. Andersen and J.~M. Smillie, {\it {Constructing All-Order Corrections to
  Multi-Jet Rates}},  {\em JHEP} {\bf 1001} (2010) 039,
  [\href{http://xxx.lanl.gov/abs/0908.2786}{{\tt arXiv:0908.2786}}].

\bibitem{Andersen:2012gk}
J.~R. Andersen, T.~Hapola, and J.~M. Smillie, {\it {W Plus Multiple Jets at the
  LHC with High Energy Jets}},  {\em JHEP} {\bf 1209} (2012) 047,
  [\href{http://xxx.lanl.gov/abs/1206.6763}{{\tt arXiv:1206.6763}}].

\bibitem{Campbell:2013qaa}
J.~Campbell, K.~Hatakeyama, J.~Huston, F.~Petriello, J.~R. Andersen, et~al.,
  {\it {Working Group Report: Quantum Chromodynamics}},
  \href{http://xxx.lanl.gov/abs/1310.5189}{{\tt arXiv:1310.5189}}.

\bibitem{Plehn:2001nj}
T.~Plehn, D.~L. Rainwater, and D.~Zeppenfeld, {\it {Determining the structure
  of Higgs couplings at the LHC}},  {\em Phys.Rev.Lett.} {\bf 88} (2002)
  051801, [\href{http://xxx.lanl.gov/abs/hep-ph/0105325}{{\tt
  hep-ph/0105325}}].

\bibitem{Figy:2007kv}
T.~Figy, V.~Hankele, and D.~Zeppenfeld, {\it {Next-to-leading order QCD
  corrections to Higgs plus three jet production in vector-boson fusion}},
  {\em JHEP} {\bf 0802} (2008) 076,
  [\href{http://xxx.lanl.gov/abs/0710.5621}{{\tt arXiv:0710.5621}}].

\bibitem{Campanario:2013fsa}
F.~Campanario, T.~M. Figy, S.~Plätzer, and M.~Sjödahl, {\it {Electroweak
  Higgs Boson Plus Three Jet Production at Next-to-Leading-Order QCD}},  {\em
  Phys.Rev.Lett.} {\bf 111} (2013), no.~21 211802,
  [\href{http://xxx.lanl.gov/abs/1308.2932}{{\tt arXiv:1308.2932}}].

\bibitem{Gangal:2013nxa}
S.~Gangal and F.~J. Tackmann, {\it {Next-to-leading-order uncertainties in
  Higgs+2 jets from gluon fusion}},  {\em Phys.Rev.} {\bf D87} (2013), no.~9
  093008, [\href{http://xxx.lanl.gov/abs/1302.5437}{{\tt arXiv:1302.5437}}].

\bibitem{Aad:2014eva}
{\bf ATLAS} Collaboration, G.~Aad et~al., {\it {Measurements of Higgs boson
  production and couplings in the four-lepton channel in pp collisions at
  center-of-mass energies of 7 and 8 TeV with the ATLAS detector}},  {\em
  Phys.Rev.} {\bf D91} (2015), no.~1 012006,
  [\href{http://xxx.lanl.gov/abs/1408.5191}{{\tt arXiv:1408.5191}}].

\end{thebibliography}
\providecommand{\href}[2]{#2}\begingroup\raggedright\endgroup

\end{document}